\pdfoutput=1
\newcommand*{\ATLASLATEXPATH}{latex/}
\documentclass[cernpreprint, UKenglish, texlive=2016, subfig, subcaption]{\ATLASLATEXPATH atlasdoc}
\usepackage{\ATLASLATEXPATH atlaspackage}
\usepackage{\ATLASLATEXPATH atlasbiblatex}

\usepackage{\ATLASLATEXPATH atlasphysics}
\graphicspath{{logos/}{figures/}}

\usepackage{IBLPaper-defs}

\AtlasTitle{Production and Integration of the ATLAS
Insertable B-Layer}

\AtlasAbstract{
During the shutdown of the CERN Large Hadron Collider 
in 2013-2014, an additional pixel layer was installed  
between the existing Pixel detector of the ATLAS experiment and a new, smaller radius beam pipe.  
The motivation for this new pixel layer, the Insertable B-Layer (IBL), was to maintain or improve the robustness and performance of the ATLAS tracking system, given the higher instantaneous and integrated luminosities realised following the shutdown. 
Because of the extreme radiation and collision rate environment, several new radiation-tolerant sensor and electronic technologies were utilised for this layer. This paper reports on the IBL construction and integration prior to its operation in the ATLAS detector.  
}

\author{The ATLAS IBL Collaboration}

\AtlasDate{\today}

\AtlasDOI{10.1088/1748-0221/13/05/T05008}

\renewcommand\Authands{, } % avoid ``. and'' for last author
\renewcommand\Affilfont{\itshape\small} % affiliation formatting
\newcommand*{\e}{$e^-$}                % e will be e^- in italic everywhere
\newcommand*{\lumi}{\centi\meter$^{-2}$\second$^{-1}$}
\DeclareSIUnit \nq {\text{n}_{\text{eq}} \per \centi \meter ^{2}}
\DeclareSIUnit \lumi {\centi\meter^{-2}\second^{-1}}
\DeclareSIUnit \xzero {\%\,X_0}
\DeclareSIUnit \e {e^-}

\newcommand*{\bit}{Bit}
\newcommand*{\barn}{b}
\newcommand*{\ppm}{ppm}
\newcommand*{\pixel}{pixel}
\newcommand*{\percent}{\%}

\newcommand*{\degree}{\,$^{\circ}$}

\hypersetup{pdftitle={ATLAS document},pdfauthor={The ATLAS IBL Collaboration}}

\begin{document}

\maketitle

% IBL Collaboration author list - created 2012-01-24

\begin{flushleft}
{\Large The ATLAS IBL Collaboration}

\bigskip

B.~Abbott$^\mathrm{33}$,
J.~Albert$^\mathrm{47}$,
F.~Alberti$^\mathrm{28a}$,
M.~Alex$^\mathrm{13}$,
G.~Alimonti$^\mathrm{28a}$,
S.~Alkire$^\mathrm{11}$,
P.~Allport$^\mathrm{23}$$^{, \mathrm{101}}$,
S.~Altenheiner$^\mathrm{13}$,
L.S.~Ancu$^\mathrm{7}$$^{, \mathrm{14}}$,
E.~Anderssen$^\mathrm{4}$
% ...

\bigskip

% ...
$^{4}$ Physics Division, Lawrence Berkeley National Laboratory and University of California, Berkeley CA, United States of America

% ...
$^{7}$ Albert Einstein Centre for Fundamental Physics and Laboratory for High Energy Physics, University of Bern, Bern, Switzerland

% ...
$^{11}$ Nevis Laboratory, Columbia University, Irvington NY, United States of America

%$^{12}$ Department of Physics, Brandeis University, Waltham MA, United States of America

% ...
$^{13}$ Fakult\"{a}t Physik, Technische Universit\"{a}t Dortmund, Dortmund, Germany

$^{14}$ Section de Physique, Universit\'e de Gen\`eve, Geneva, Switzerland

% ...
$^{23}$ Oliver Lodge Laboratory, University of Liverpool, Liverpool, United Kingdom

% ...
$^{28}$ $^{(a)}$INFN Sezione di Milano; $^{(b)}$Dipartimento di Fisica, Universit\`a di Milano, Milano, Italy

% ...
$^{33}$ Homer L. Dodge Department of Physics and Astronomy, University of Oklahoma, Norman OK, United States of America

% ...
$^{47}$ Department of Physics and Astronomy, University of Victoria, Victoria, BC, Canada

% ...
$^{101}$ now at School of Physics and Astronomy, University of Birmingham, Birmingham, United Kingdom

% ...

\end{flushleft}

%\newpage\tableofcontents\newpage
\newpage

%-------------------------------------------------------------------------------
%\section{Introduction}
%\label{sec:intro}
%-------------------------------------------------------------------------------

%Place your introduction here.~\cite{Aad:1129811}
\section{Introduction}
\label{section:introduction}

The ATLAS~\cite{ATLAS:2008} general purpose detector is used for the study of proton-proton ($pp$) and heavy-ion collisions at the CERN Large Hadron Collider (LHC)~\cite{Evans:2008zzb}. It successfully collected data at $pp$ collision energies of 7 and 8\,TeV in the period of 2010-2012, known as Run~1. Following an LHC shutdown in 2013-2014 (LS1), it has collected data since 2015 at a $pp$ collision energy of 13 TeV (the so-called Run~2). 

The ATLAS inner tracking detector (ID)~\cite{ATLAS:2008, Aad:2010bx} provides charged particle tracking with high efficiency 
in the pseudorapidity\footnote{%
ATLAS uses a right-handed coordinate system with its origin at the nominal interaction point (IP) in the centre of the detector and the $z$-axis along the beam pipe. The $x$-axis points from the IP to the centre of the LHC ring, and the $y$-axis points upward. Cylindrical coordinates $(r,\phi)$ are used in the transverse plane, $\phi$ being the azimuthal angle around the $z$-axis. The pseudorapidity is defined in terms of the polar angle $\theta$ as $\eta=-\ln\tan(\theta/2)$. An ATLAS convention refers to the (-z) side of the detector as the C-side, and the (+z) side of the detector as the A-side.}
range of $|\eta|\,<\,2.5$. With increasing radial distance from the interaction region, it  consists of silicon pixel and micro-strip detectors, followed by a transition radiation tracker (TRT) detector, all surrounded by a superconducting solenoid providing a 2\,T magnetic field. 

The original ATLAS pixel detector~\cite{Aad:2008zz, HEIM2014227},  referred to in this paper as the Pixel detector, was the innermost part of the ID during Run 1. It consists of three barrel layers (named the B-Layer, Layer 1 and Layer 2 with increasing radius) and three disks on each side of the interaction region, to guarantee  at least three space points over the full tracking $|\eta|$ range. 
It was designed to operate for the Phase-I period of the LHC, that is with a peak luminosity of \SI{1e34}{\lumi} and an integrated luminosity of approximately \SI{340}{\femto\barn^{-1}} corresponding to a TID of up to 50\,MRad\footnote
   {The Total Ionising Dose (TID) in silicon is a measure of the radiation dose for the front-end electronics. For silicon sensors, a more relevant measure of the radiation dose is the non-ionising energy loss (NIEL), normally expressed as the equivalent damage of a fluence of 1~MeV neutrons (\SI{}{\nq}).} and a fluence of up to \SI{1e15}{\nq} NIEL.
However, for luminosities exceeding \SI{2e34}{\lumi}, which are now expected during the Phase-I operation, the read-out efficiency of the Pixel layers will deteriorate. 

This paper describes the construction and surface integration of an additional pixel layer, the Insertable B-Layer (IBL)~\cite{Capeans:1291633}, installed during the LS1 shutdown between the B-Layer and a new smaller radius beam pipe. The main motivations of the IBL were to maintain the full ID tracking performance and robustness during Phase-I operation, despite read-out bandwidth limitations of the Pixel layers (in particular the B-Layer) at the expected Phase-I peak luminosity, and accumulated radiation damage to the silicon sensors and front-end electronics. The IBL is designed to operate until the end of Phase-I, when a full tracker upgrade is planned~\cite{Collaboration:2257755} for high luminosity LHC (HL-LHC) operation from approximately  2025.

The IBL is a small detector that was constructed on a short timescale using the results from sensor, electronic and mechanical R\&D programs, to operate over an extended period in a hostile environment. The emphasis was to construct the detector on time, while identifying and understanding the various production and quality assurance (QA) issues with the R\&D groups and industrial partners. Some choices during the IBL construction were consequently influenced by the schedule. 
The procurement, QA and assembly of the different IBL components into loaded staves were undertaken 
at the participating institutes. The staves were then transported to CERN, where the final IBL integration and testing was 
made before installation in the ATLAS experiment.

The motivations and performance of the IBL are briefly described in Section {\ref{section:detector}} together with a brief introduction to the detector layout and the electronic system design. 
Section {\ref{section:modules}} describes the production and QA of the individual pixel module components (the sensors, front-end electronics, and module hybrids). This is followed by a discussion of the module assembly and tests to ensure the required 
electrical and mechanical quality of the modules. 
The technical specification and fabrication of local support staves and their associated  electrical services are discussed in Section {\ref{section:stavecomponent}}. In Section {\ref{section:staveload}} the loading of accepted pixel modules on the staves is described, together with a discussion of the module and stave QA at successive steps in the loading process. Section {\ref{section:electronicinterfaces}} briefly describes the off-detector services, including the detector control, interlock and power supply systems, and the data acquisition. The integration of the staves and their services around the beam pipe is presented in Section {\ref{section:integration}}. 
Finally, Section {\ref{section:conclusion}} lists the most critical aspects of the IBL project, together with a short summary of the IBL status following its successful installation in ATLAS.

%-------------------------------------------------------------------------------
%\section{ATLAS detector}
%\label{sec:detector}
%-------------------------------------------------------------------------------

%The ATLAS detector~\cite{PERF-2007-01} ...
% \input{atlas-detector}

\section{Detector overview and physics motivations}
\label{section:detector}

% Chapter coordination:

\subsection{Layout overview}
\label{sec:layout_introduction}

The %Insertable Pixel Layer 
IBL is a new layer of pixel sensors designed to fit between the B-Layer of the existing Pixel detector and a new beam pipe
of reduced inner radius of \SI{23.5}{\milli\meter}. It consists of 14 carbon composite %AGC 160118 carbon-fibre 
staves, providing full azimuthal ($\phi$) hermeticity for high
transverse momentum ($\pt$~>~1~GeV) particles and longitudinal coverage up to $|\eta|$ of 3. 
Each stave supports 20 pixel sensor modules together with their electrical services and a cooling pipe.
Each module is constructed from a pixel sensor (Section {\ref{sec:mod_sensor}}) with each pixel of nominal size $\SI{250}{} \times \SI{50}{\micro\meter\squared}$ electrically bonded (Section {\ref{sec:mod_bumping}}) to a channel of a read-out chip (the FE-I4B chip described below and in Section {\ref{sec:overview_electronic}}).
The IBL volume contains the staves and the services in the space between an %external 
inner support tube (IST) fixed on the Pixel structure and an inner positioning tube (IPT) 
with an inner radius of 29 mm. A key feature is that independent radial volumes are installed, allowing for the removal of the beam pipe with respect to the IBL package, or the IBL and beam pipe with respect to the Pixel package.

\begin{figure}[!htb]
\begin{center}
\includegraphics[width=0.71\textwidth, angle = -90 ]{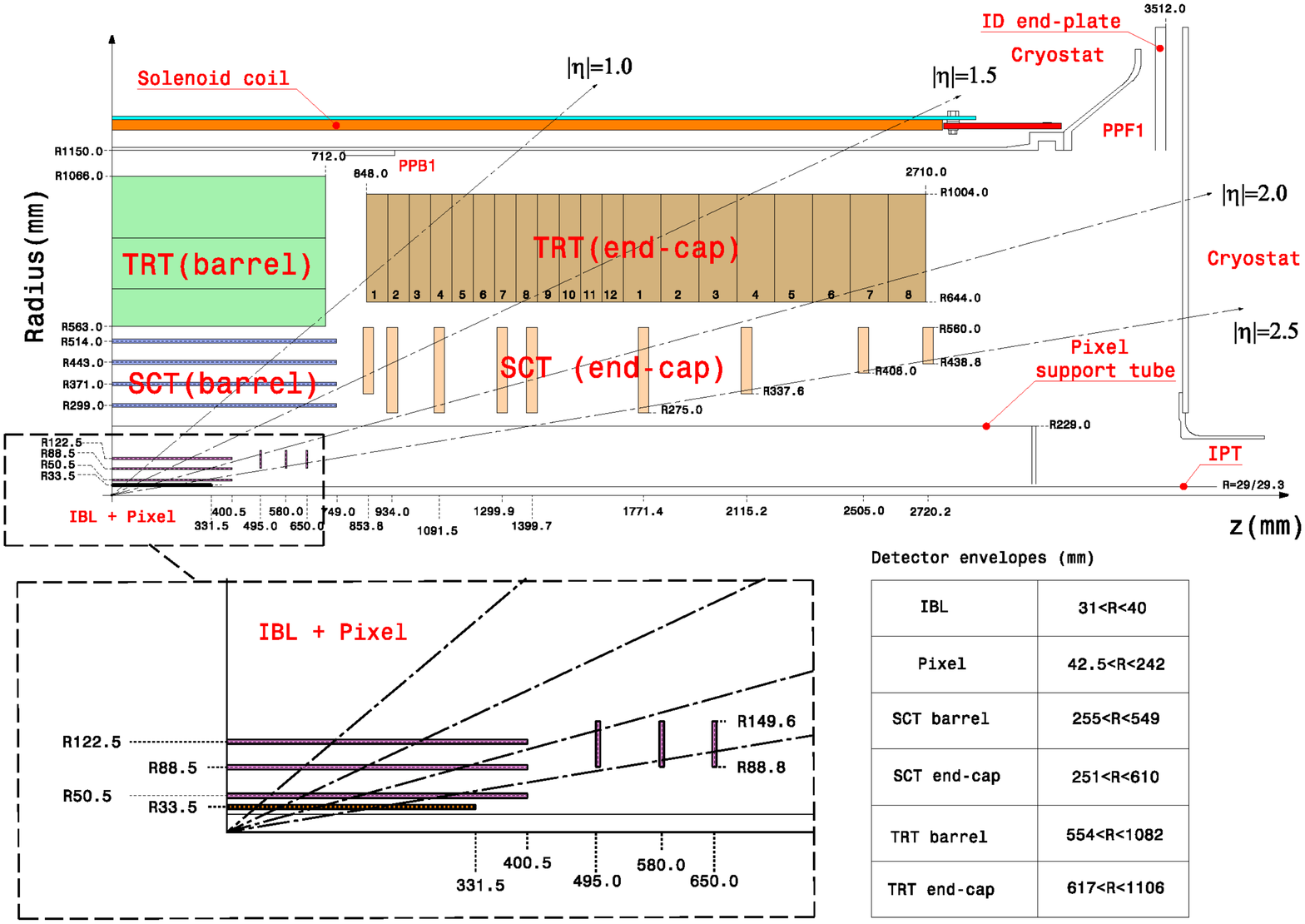}
\caption{The layout of the ATLAS inner tracking detector, including the additional IBL detector layer. The inner positioning tube (IPT) supports the IBL staves and separates them from the beam pipe.}
\label{fig:NewID}
\end{center}
\end{figure}

\begin{figure}[!htb]
\begin{center}
 \includegraphics[width=0.8\textwidth]{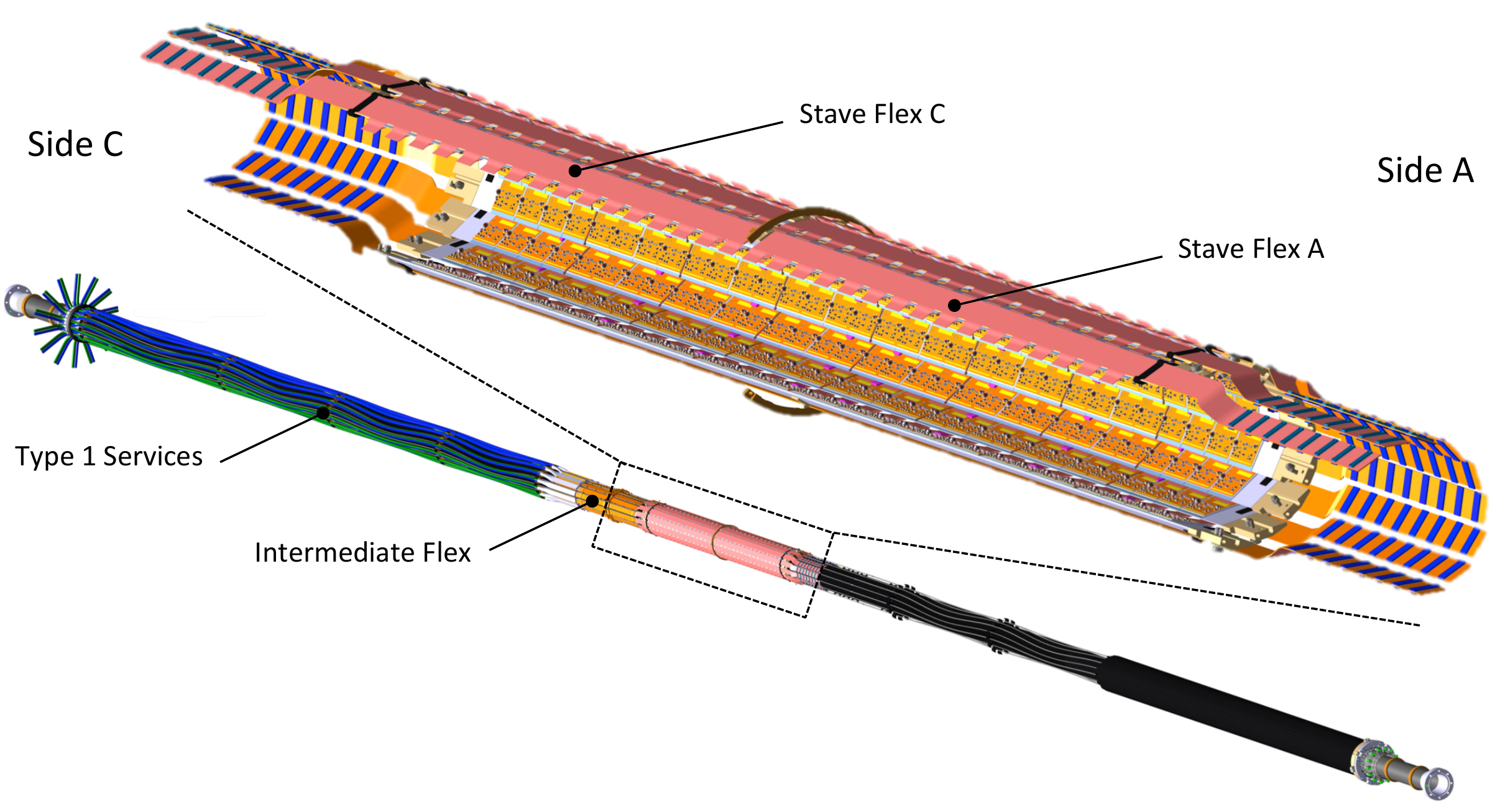}
\caption{Longitudinal view of the IBL detector and its services. 
The insert shows an enlarged 3-dimensional view of the detector with its modules arranged cylindrically around the  beam pipe.}
\label{fig:IBLView2}
\end{center}
\end{figure}

The ATLAS ID, including the IBL detector and its envelope, is shown in Figure~\ref{fig:NewID}.
The %detailed 
3-dimensional structure of the IBL detector with its services is shown in Figure~\ref{fig:IBLView2}.

The main IBL layout parameters are summarised in Table~\ref{tab:LayoutTable} and a comparison between the technical
characteristics of the IBL and the Pixel detector is shown in Table~\ref{tab:TableComparison}.
With a mean sensor radius of \SI{33.5}{\milli\meter} (compared with \SI{50.5}{\milli\meter} for the Pixel B-Layer), the
IBL sensors and front-end electronics must cope with a much higher hit rate and radiation doses of %AGC respectively
\SI{5e15}{\nq} NIEL and 250\,MRad TID during Phase-I operation.
To address these requirements, a new front-end read-out chip, the FE-I4B~\cite{FE_I4:2010}, was developed in \SI{130}{\nano\meter} CMOS technology 
satisfying the ATLAS requirements of radiation tolerance and read-out efficiency at high luminosity.
In addition, the FE-I4B chip has a substantially larger active area compared to
the FE-I3 front-end chip~\cite{Aad:2008zz} of the Pixel detector, and a cell size reduced to $\SI{250}{} \times \SI{50}{\micro\meter\squared}$ from $\SI{400}{} \times \SI{50}{\micro\meter\squared}$, the shorter side being in the transverse plane. The smaller layer radius and the reduced pixel cell length are crucial parameters in defining the performance improvement
of the ID, in particular the track-extrapolation resolution.

\begin{table}[!htb]
\centering
\begin{tabular}{lc}
\hline \hline
Item         & Value\\
\hline
Number of staves & 14 \\
Number of physical modules per stave & 20  (12 planar, 8 3D)\\
%Number of DCS modules per stave & 16 \\
Number of FEs per stave & 32 \\
Coverage in $\eta$, no vertex spread & $|\eta|\,<\,3.0$ \\
Coverage in $\eta$, 2$\sigma$ (\SI{122}{\milli\meter}) vertex spread & $|\eta|\,<\,2.58$ \\
Active $|z|$ stave length (\SI{}{\milli\meter}) & 330.15 \\
%Geometrical acceptance in $z$ min, max (\%) & 97.4, 98.8 \\
Stave tilt in $\phi$ (degree) & 14 \\
Overlap in $\phi$ (degree) & 1.82 \\
Center of the sensor radius (\SI{}{\milli\meter}) & 33.5 \\
%Radiation length at $z$ = 0  (\SI{}{\xzero}) & 1.88\\
\hline \hline
\end{tabular}  
\caption{Main layout parameters for the IBL detector.
\label{tab:LayoutTable}}
\end{table}

\begin{figure}[!htb]
\begin{center}
\includegraphics[width=0.85\textwidth]{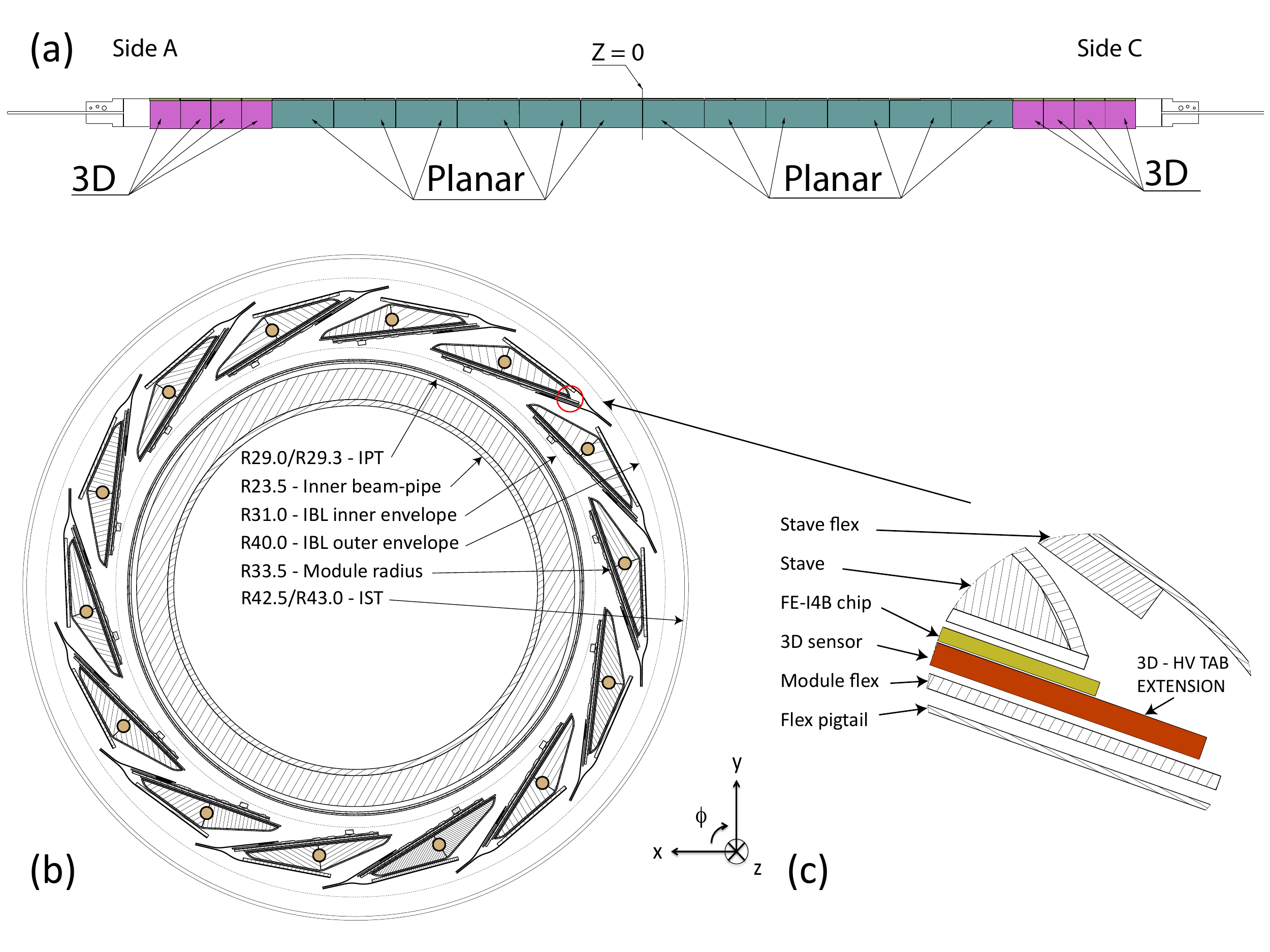}
\caption{ IBL detector layout: (a) Longitudinal layout of planar and 3D modules on a stave. (b) An $r-\phi$
              section showing the  beam pipe, the inner positioning tube (IPT), the staves of the IBL detector and the inner support tube (IST), as viewed from the C-side.
              (c) An expanded $r-\phi$ view of the corner of a 3D module fixed to the stave.}
\label{fig:IBLLayout}
\end{center}
\end{figure}

The IBL stave configuration is shown in Figure~\ref{fig:IBLLayout}.  Two module types~\cite{IBL_mod_proto} are installed on
each stave. A total of 12 double-chip %(DC) 
planar n-in-n sensors similar to those equipping the Pixel detector,
each bump-bonded to two FE-I4B read-out chips, populate the central stave region. Four single-chip %(SC) 
3D sensors, adopted for the first time in a collider tracking detector and each bump-bonded to one FE-I4B chip, populate each end of the stave.
The staves are mounted with the sensors facing the beam pipe and are inclined in azimuth by \SI{14}{\degree} 
%with respect to the radial direction in order 
to achieve an overlap of the active area. This tilt also compensates for the Lorentz angle of drifting charges in the case of planar sensors, and the effect of partial column inefficiency for normal incidence tracks in the case of 3D sensors.
Owing to space constraints, the sensors are not shingled along the stave (in $z$). %In order 
To minimise the dead region, modules are glued on the stave with a physical gap of \SI{200}{\micro\meter}.

\begin{table}[tbph]
        \centering
        \begin{tabular}{lcc}
        \hline \hline
         Technical characteristic & Pixel & IBL \\
        \hline
Active surface (\SI{}{\meter\squared}) &  1.73 &  0.15 \\
Number of channels (x 10$^{6}$) & 80.36 & 12.04 \\
        \hline
        Pixel size (\SI{}{\micro\meter\squared}) & 50$\times$400 & 50$\times$250 \\
        Pixel array (columns $\times$ rows) & 160$\times$18 & 336$\times$80 \\
        Front-end chip size (\SI{}{\milli\meter\squared}) & 7.6$\times$10.8 & 20.2$\times$19.0 \\
        Active surface fraction (\SI{}{\percent}) & 74 & 89 \\
        Analog current (\SI{}{\micro\ampere\per\pixel}) & 26 & 10 \\
        Digital current (\SI{}{\micro\ampere\per\pixel}) & 17 & 10 \\
        Analog voltage (\SI{}{\volt}) & 1.6 & 1.4 \\
        Digital voltage (\SI{}{\volt}) & 2.0 & 1.2 \\
        Data out transmission (\SI{}{\mega\bit\per\second}) & 40-160 & 160 \\
        \hline
Sensor type & planar & planar / 3D \\
Sensor thickness  (\SI{}{\micro\meter}) & 250 & 200 / 230 \\
Layer thickness  (\SI{}{\xzero})  & 2.8  & 1.88  \\
Cooling fluid & C$_3$ F$_8$ & CO$_2$  \\
\hline \hline
        \end{tabular}
        \caption{Comparison of the main characteristics of the Pixel and IBL detectors.}
        \label{tab:TableComparison}
\end{table}

\begin{figure}[!htb]
        \centering
        \begin{subfigure}[t]{0.47\textwidth}
                \includegraphics[width=\textwidth]{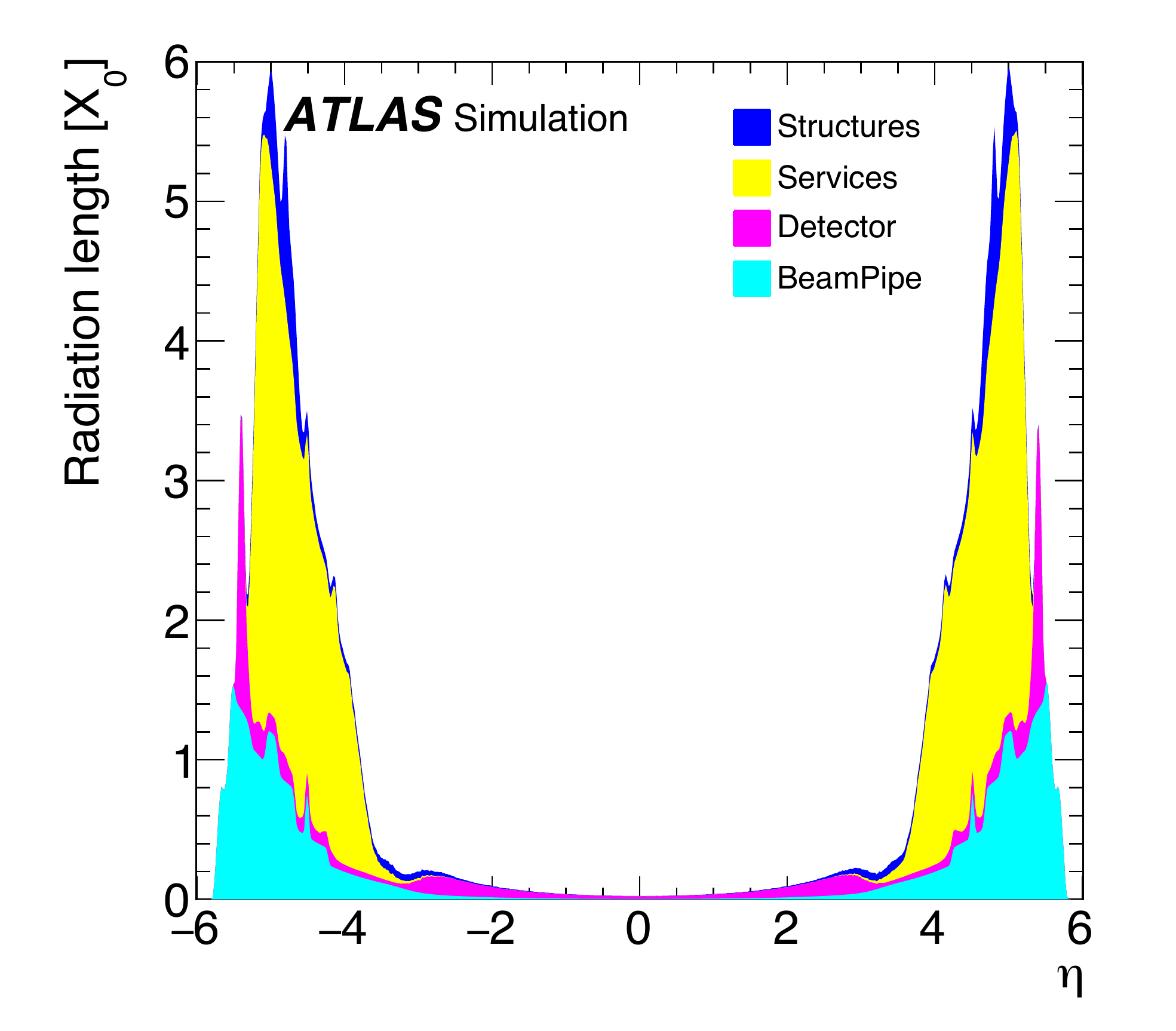}
                \caption{}
                \label{fig:xo_linear}
        \end{subfigure}
        \begin{subfigure}[t]{0.47\textwidth}
                \includegraphics[width=\textwidth]{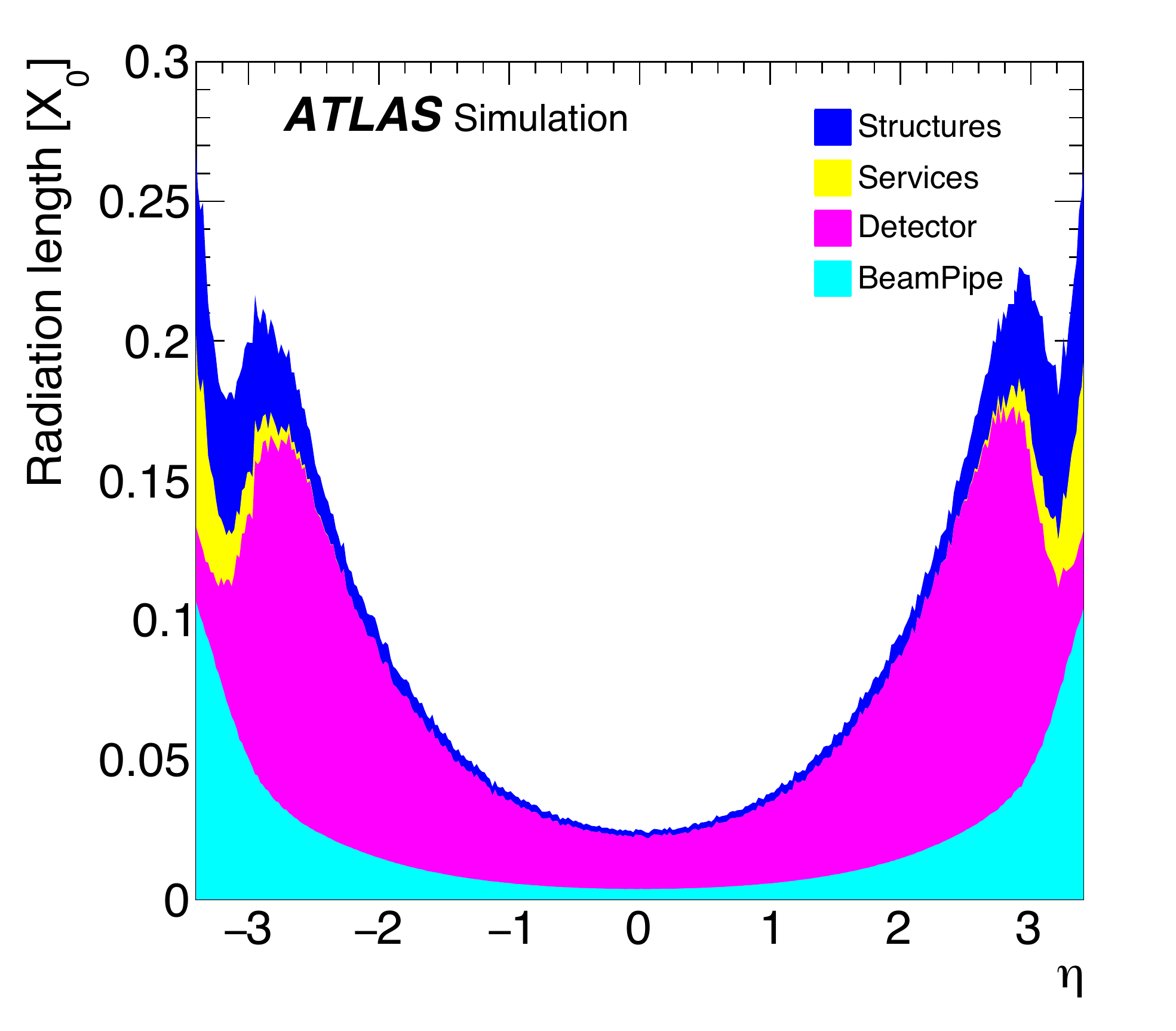}
                \caption{}
                \label{fig:xo_log}
        \end{subfigure}
\caption{Material budget of the IBL detector as a fraction of X$_0$, as implemented in the ATLAS geometry model using straight tracks originating from the nominal beam line at z = 0. Different components are shown:  beam pipe, detector (IBL staves, modules, inner positioning tube (IPT)), services (cooling and cables) and structures (stave rings, end-blocks, sealing ring area, inner support tube (IST)). \subref{fig:xo_linear}  Dependence on $\eta$, averaged over $\phi$. \subref{fig:xo_log} A zoomed view of the central  |$\eta$| region where precise tracking (|$\eta$|~<~2.5) is performed.}
\label{fig:IBLMaterialBudget}
\end{figure}

Minimising the material budget is very important for the optimisation of the tracking and vertex performance.
The IBL radiation length, averaged over azimuth and taking into account the stave tilt and the overlap between staves, is estimated to be 1.88\%\,X$_0$ for %straight 
tracks produced perpendicular to the beam axis at $z = 0$. This is $\sim$\SI{30}{\percent} less than that of the 
Pixel B-Layer\footnote{The as-built IBL radiation length was evaluated using the ATLAS geometry model, as discussed in the IBL TDR~\cite{Capeans:1291633}. The difference with respect to the value reported in the IBL TDR is mainly due to an  initial underestimation of the module material and the addition of the IPT. A recent description of the ATLAS ID material and its comparison with Run~2 collision data is now available~\cite{Aaboud:2017pjd}.}. 
The reduced thickness was achieved by using more advanced technologies as
 discussed in the following sections. These include: a new low-mass module design;
  local support structures (staves) made of low density, thermally conductive carbon foam; the use of CO$_2$ evaporative cooling, which is more efficient in terms of mass flow and pipe size; and electrical power services using aluminium conductors. Table~\ref{tab:MaterialBudget} reports the main contributions to the IBL material budget. Figure~\ref{fig:IBLMaterialBudget} shows the material traversed by a straight track  originating in $z = 0$ as a function of $\eta$, smeared over the azimuthal angle. 

\begin{table}[tbph]
        \centering
        \begin{tabular}{lc}
        \hline \hline
        Item & Value (\SI{}{\xzero})\\
        \hline
        Beam pipe & 0.32 \\
\hline
        IPT & 0.12 \\
        Module &  0.76 \\
        Stave   &  0.60 \\
        Services   & 0.19  \\
        IST   &  0.21 \\
        \hline
         IBL total   & 1.88 \\
        \hline \hline
        \end{tabular}
        \caption{IBL material budget as a fraction of X$_0$, averaged over the azimuthal angle $\phi$ for straight tracks produced perpendicular to the beam axis at $z = 0$, as implemented in the ATLAS geometry model. The beam pipe material is excluded from the IBL total.
        \label{tab:MaterialBudget}}
\end{table}
  
\subsection{System overview}
\label{sec:electronic_introduction}

The IBL electronic system includes the FE-I4B read-out chip, the off-detector read-out boards (the read-out driver (ROD)~\cite{Gabrielli:2015} and the back-of-crate board (BOC)~\cite{BOC:2014, Wensing:2012ff}), the detector control system (DCS) \cite{DCS:2011}, the electronics and sensor power supplies, and all of their associated electrical and optical services. The data acquisition (DAQ) \cite{Polini:2011wna} controls the transfer of data to and from the off-detector read-out boards, while the DCS controls the electrical and environmental monitoring of the detector as well as the power distribution  to the pixel sensors and FE-I4B chips. 

\begin{figure}[htb]
	\centering	 \includegraphics[width=0.95\textwidth]{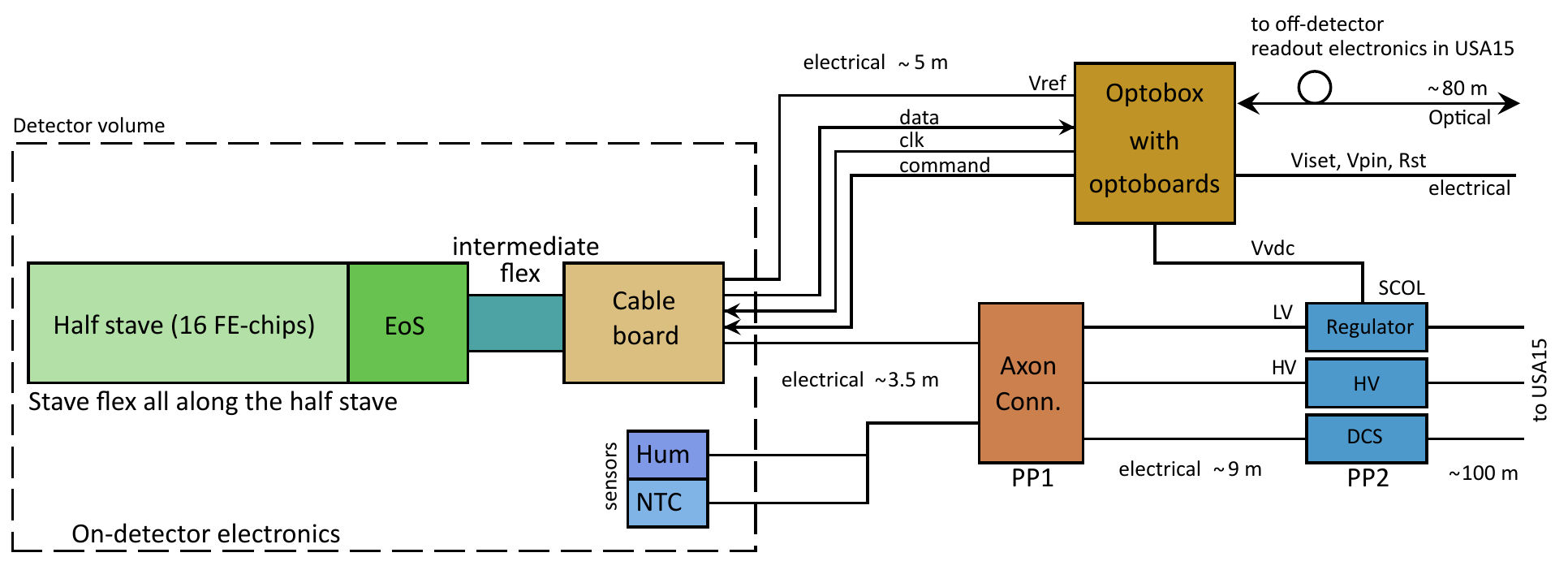}
	\caption{Block diagram of on-detector and off-detector electrical services for one half-stave of the IBL detector. The on-detector front-end read-out and services are described in Sections {\ref{section:modules}} and {\ref{section:stavecomponent}}. The off-detector services (Type~1 cables) reaching the PP1 patch panel and opto-box are described in Section {\ref{section:electronicinterfaces}}.}
	\label{fig:sec7_1_IBL-electrical-services}
\end{figure}

The electrical service design was driven by physical space constraints, especially in the inner region where all services and connectors must fit into the narrow IBL envelope over a length of approximately \SI{3}{\meter}, and by the conflicting requirements of material budget, radiation hardness and electrical performance. Optical transmission is excluded in the IBL envelope because of the high radiation level. 
%The physical dimensions of the connectors must be compact enough to allow the IBL insertion through the Pixel detector.
Figure~\ref{fig:sec7_1_IBL-electrical-services} shows a block diagram of electrical services for one half-stave (the services are symmetrical, at each end of a stave). 
Each module of a given half-stave is connected electrically by a stave flex to an end-of-stave (EoS) card. 
The stave flex transfers the data from the half-stave, as well as control signals from the DAQ and DCS, and the power distribution. 
In the EoS region the detector services are connected via intermediate flexes to a cable board. The cable board connects the flexes to $\sim$\SI{3}{\meter}-long extensions (Type 1 cables) that reach the ID end-plate where the first Patch Panel (PP1) is located to allow for electrical and optical connections to the external services after installation in ATLAS. A second break of the electrical services occurs at another Patch Panel (PP2) on the detector periphery that is accessible during a short shut-down. 

%The design, procurement and quality assurance (QA) of both %the design, production and QA
Details of the FE-I4B chip and the module flex hybrid that connect a module to the stave flex are described in Section {\ref{section:modules}}, while 
those of the stave flexes are described in Section {\ref{section:stavecomponent}}. The off-detector electronics and power-supplies, as well as the electrical and optical services, are described in Section {\ref{section:electronicinterfaces}}. 

\subsection{Tracking and flavour tagging performance}
\label{sec:performance}

The ATLAS ID provides charged particle tracking with high efficiency in the $|\eta| < 2.5$ range over the full azimuthal range.
The pixel layers are crucial for the  reconstruction of charged particles trajectories, for their extrapolation to the production point
and for the reconstruction of multiple collision and decay vertices which occur in each bunch crossing.
The pixels are therefore of crucial importance to the flavour tagging performance. 
Any inefficiency of the innermost B-Layer would result in a degradation of that performance.
 
A first assessment of the expected improvements in tracking and vertex reconstruction performance was performed for the IBL
Technical Design Report (TDR)~\cite{Capeans:1291633}. Since then, the ATLAS simulation, digitisation and cluster reconstruction
algorithms have been refined and improved.

The IBL improves the track extrapolation resolution with respect to the Pixel detector of Run~1 by providing
an additional high-precision hit closer to the interaction point.
This is particularly important for low \pt\ particles,
where it mitigates the effect of multiple scattering in the detector material on the track extrapolation,
thus improving the impact parameter resolution in both the transverse ($d_0$) and longitudinal ($z_0$)
projections. The smaller pixel pitch of the IBL in the longitudinal direction contributes %additionally
to improving the resolution in $z_0$ across the full \pt\ spectrum.

The track reconstruction performance has been evaluated using Monte Carlo simulations of \ttbar\ events,
comparing the Run~1 detector geometry to a geometry including the IBL, while keeping all other conditions
unchanged. An improvement in %the longitudinal impact parameter 
the $z_0$ resolution of %a factor of around  
approximately 2 (1.5) for tracks
with \pT\  %around 
of 1 (100) GeV is observed following the addition of the IBL. %For 
In the transverse direction,
the addition of the IBL improves the $d_0$ resolution by a factor of %around 
approximately 2 for tracks with \pT\ of 1~GeV,
with the resolutions for the two geometries converging beyond 10~GeV. These results are confirmed by
comparing the track impact parameter resolution measured in Run 1 (2012) data with that in Run 2 (2015) 
data~\cite{Potamianos:2016ptf}.

In addition to %improving the reconstruction of 
the charged particle track reconstruction, %themselves, 
these improvements enhance %also enhance among other things 
the primary vertex reconstruction and resolution, the secondary vertex finding,
and the flavour tagging performance, hence considerably extending the physics reach of ATLAS analyses.

%The IBL also helps to maintain the tracking performance and to ensure robustness at high luminosity,
%when the B-Layer starts to deteriorate from radiation damage, high pile-up occupancy or the irreparable
%failures of its chips or modules.
The IBL also helps to maintain the performance and robustness of the ID track reconstruction when the B-Layer read-out efficiency  
deteriorates at high peak luminosity, or after a large integrated 
luminosity (radiation damage to the sensors and front-end electronics as well as possible irreparable failures of its chips and modules).

 The flavour tagging performance expected with the addition of the IBL is evaluated using a more realistic simulation of the ATLAS ID based on the final IBL geometry, an updated digitisation model and improved reconstruction algorithms with respect to the IBL TDR. The latter include a refined neural network clustering algorithm~\cite{Aad:2014yva}, a new tracking configuration, which improves the treatment of shared clusters in the core of a dense jet environment~\cite{ATL-PHYS-PUB-2015-006} and new flavour tagging algorithms. These results supersede those presented in the IBL TDR. Results are based on fully simulated \ttbar\ production events at a collision energy of 13\,TeV. The average level of pile-up is approximately 20, reflecting the Run~1 luminosity profile. Jets used for flavour tagging are reconstructed using the anti-$k_t$ algorithm~\cite{AntiKt} with radius $R=0.4$.
ATLAS combines the discriminating variables obtained from impact parameter, inclusive secondary vertex and multi-vertex reconstruction
algorithms. A detailed description of these algorithms can be found in reference~\cite{btagRun1}.
 
\begin{figure}[!htb]
        \centering
        \begin{subfigure}[t]{0.47\textwidth}
                \includegraphics[width=\textwidth]{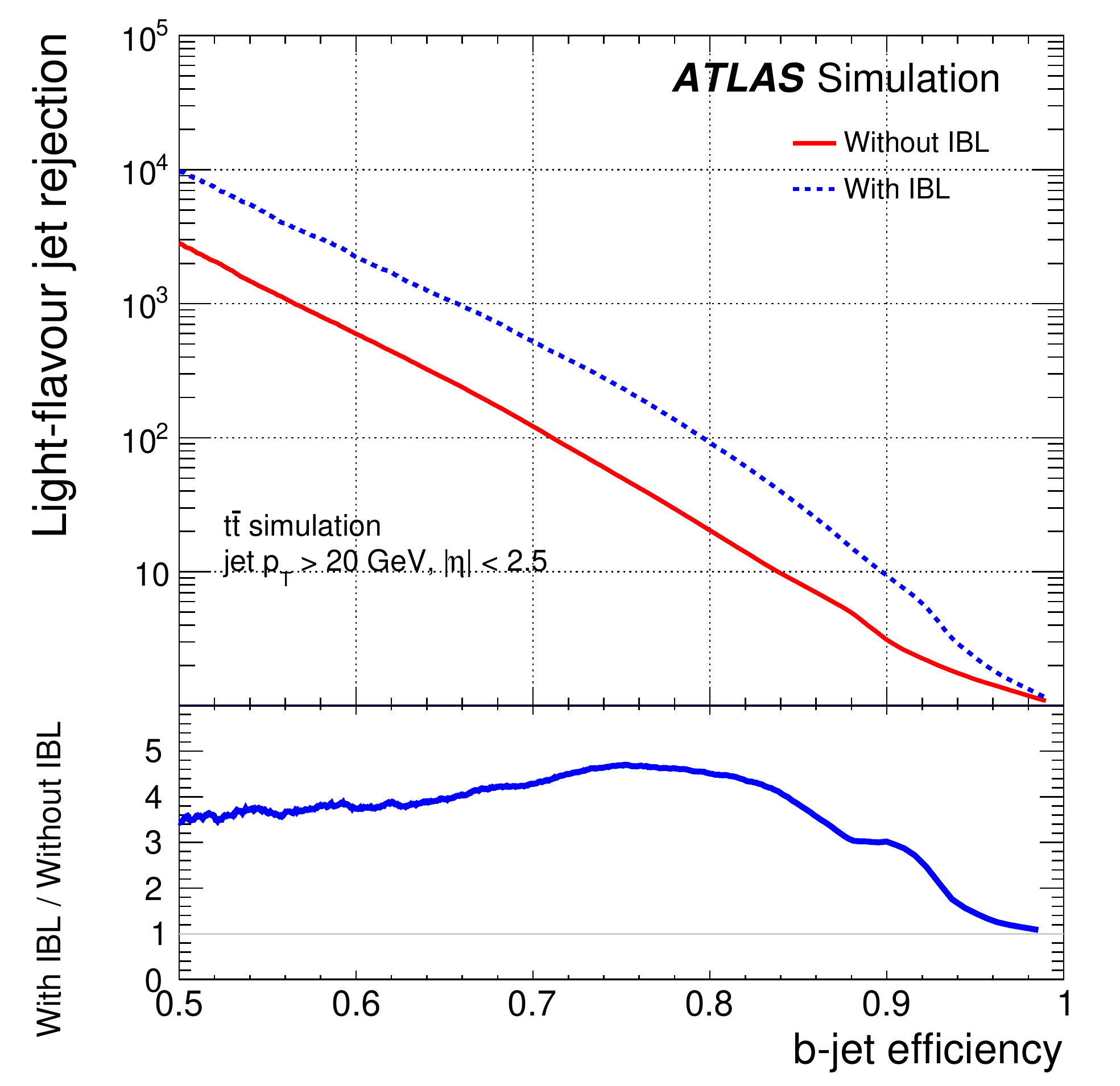}
%\vspace {-3.5cm}
                \caption{}
                \label{fig:d0f}
        \end{subfigure}
        \begin{subfigure}[t]{0.47\textwidth}
                \includegraphics[width=\textwidth]{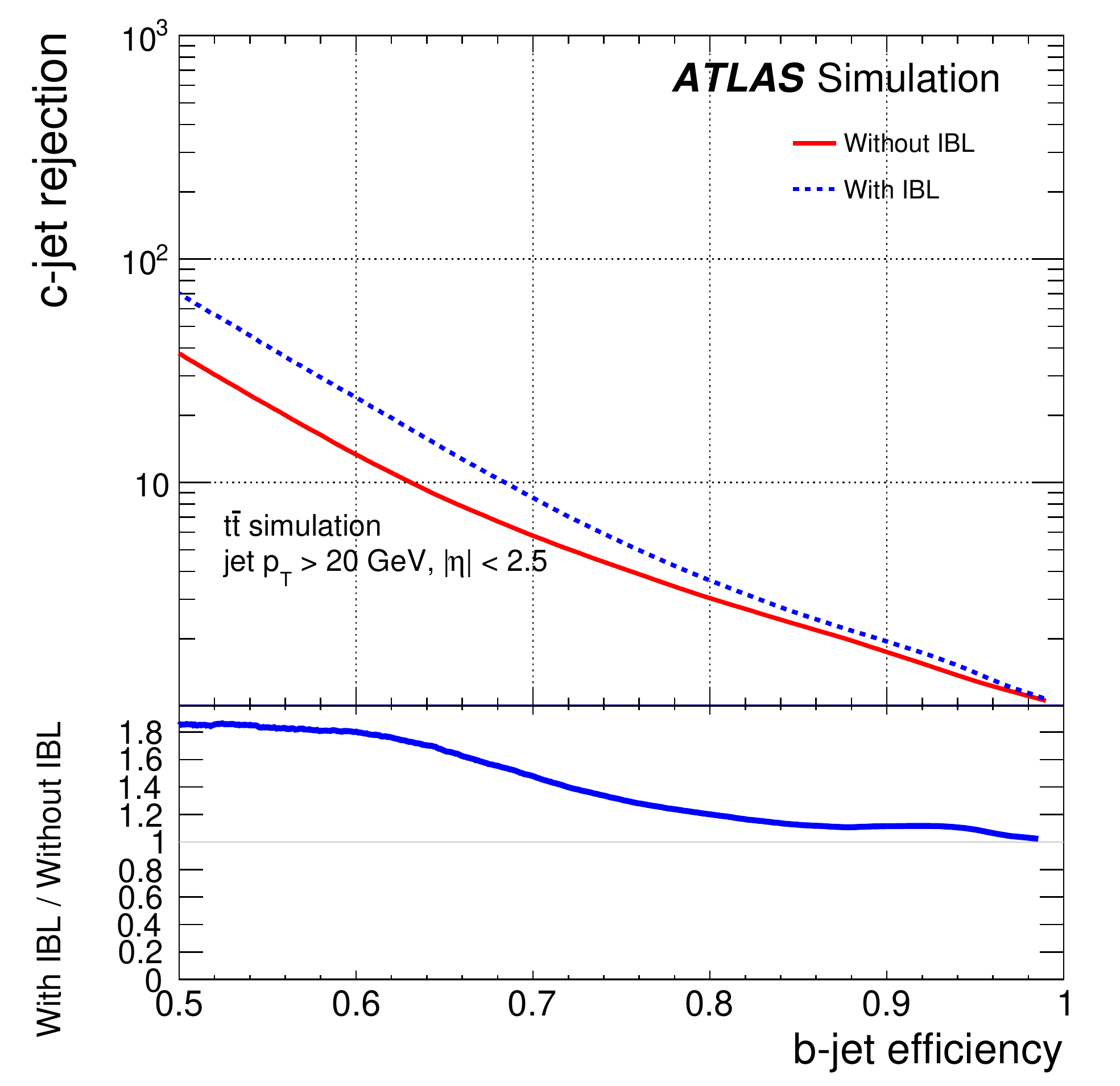}
%\vspace {-3.5cm}
                \caption{}
                \label{fig:z0f}
        \end{subfigure}
%\vspace {-2cm}
\caption{Comparison of \subref{fig:d0f} light-jet and \subref{fig:z0f} $c$-jet rejection as a function of $b$-jet tagging efficiency for the Run~1 (without IBL) and Run~2 (with IBL) detector layouts under the same conditions, obtained with the MV2c20 algorithm. The rejection is defined as the reciprocal of the tagging efficiency. Results are derived from jets %from 
produced in \ttbar\ events, with jets  passing the $\pt$~>~20~GeV and $|\eta|<2.5$ selection.}
\label{fig:btagging_inclusiveROCcurve}
\end{figure}
 
The combination of the input variables obtained from these algorithms is obtained using a boosted decision tree (MV2c20)~\cite{btagRun2}) that returns a continuum variable peaked around 1 for jets likely to contain a $b$-flavoured hadron and around $-1$ for those likely
to originate from light-flavoured quarks. This MV2c20 is an evolution of the neural network  algorithm used during Run~1~\cite{btagRun1}. In order to perform an useful comparison, the MV2c20 algorithm has been separately re-trained for the $t \bar t$ sample generated using the ATLAS Run~1 geometry, without the IBL, and the ATLAS Run~2 geometry, which includes the IBL.
 
Figure~\ref{fig:btagging_inclusiveROCcurve} shows the light-jet and $c$-jet rejection as a function of the $b$-jet purity obtained with the two configurations. The addition of the IBL improves the light-jet ($c$-jet) rejection by a factor up to 4 (1.8) for $b$-jet
tagging efficiencies up to 85\%.  Physics analyses will most often profit from the improved performance by re-tuning their $b$-tagging requirements in such a way to keep a similar background rejection with an increased signal efficiency. The improvement in performance at constant rejection is summarised in Table~\ref{tab:btagging_fixedRejection} for different working points.
 
\begin{table}[htb!]
\begin{center}
\begin{tabular}{lccc}
\hline \hline
light-jet rejection                                     & $b$-jet efficiency		 & $b$-jet efficiency \\ 
			                                      & without IBL (\%) 		 & with IBL (\%) \\ \hline
1000                                                       & 57     				& 65 \\
100                                                         & 71    				& 79 \\
10                                                           & 84   				& 90 \\ \hline
\hline
$c$-jet rejection                                     & $b$-jet efficiency 		& $b$-jet efficiency \\ 
			                                      & without IBL (\%) 		& with IBL (\%) \\ \hline
20                                                           & 56     				& 62 \\
10                                                           & 63  				& 68 \\
5                                                             & 72     				& 76 \\ \hline \hline
\end{tabular}
\caption{\label{tab:btagging_fixedRejection} Comparison of the $b$-jet tagging efficiency for fixed light- or $c$-jet rejection
for the Run~1 (without IBL) and Run~2 (with IBL) detector layouts under the same conditions. Results are obtained for jets in simulated \ttbar\ events satisfying $\pt$~>~20~GeV and $|\eta|<2.5$.}
\end{center}
\end{table}
 
\begin{figure}[!htb]
        \centering
        \begin{subfigure}[t]{0.47\textwidth}
                \includegraphics[width=\textwidth]{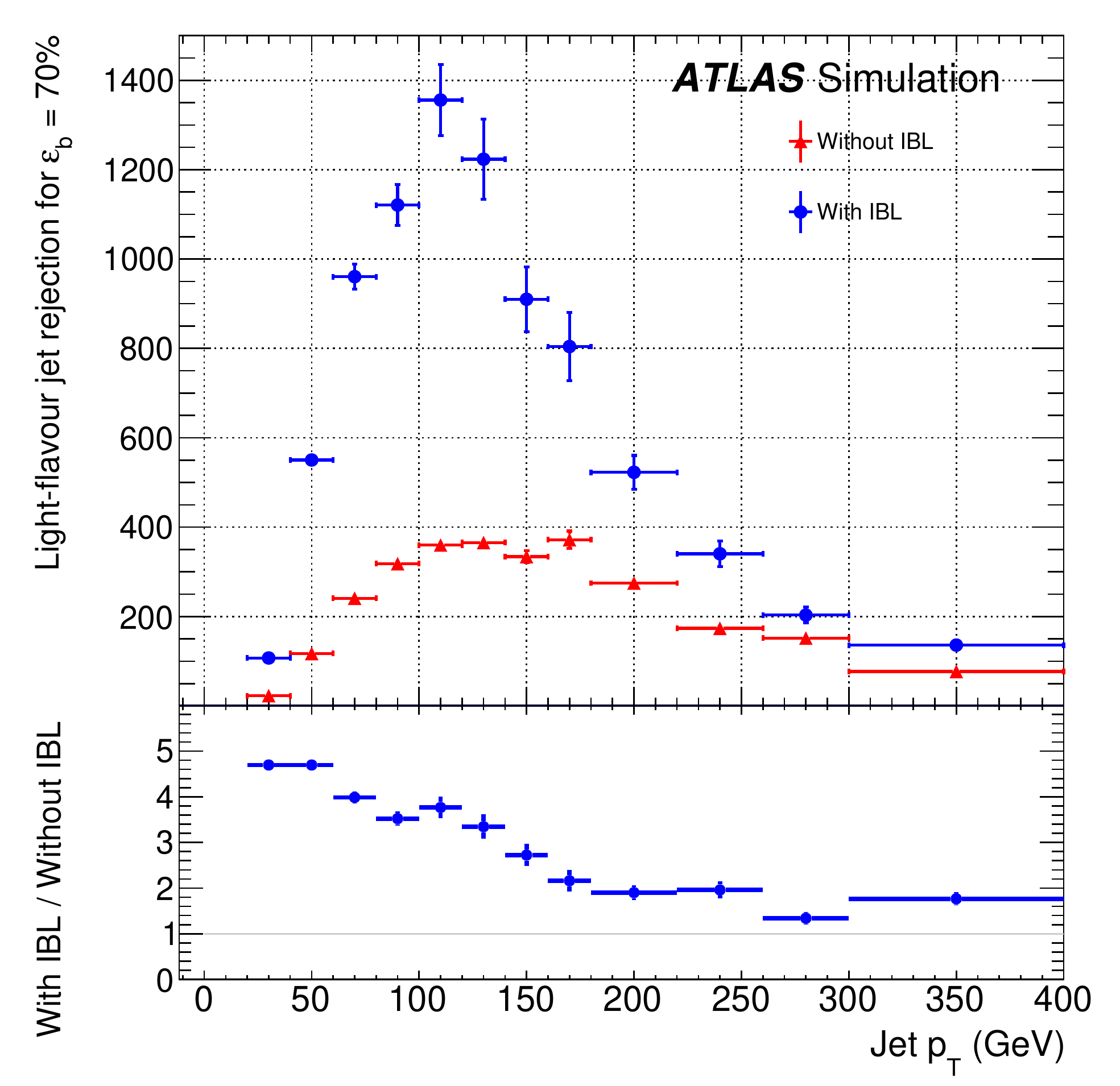}
%\vspace {-3.5cm}
                \caption{}
                \label{fig:d0b}
        \end{subfigure}
        \begin{subfigure}[t]{0.47\textwidth}
                \includegraphics[width=\textwidth]{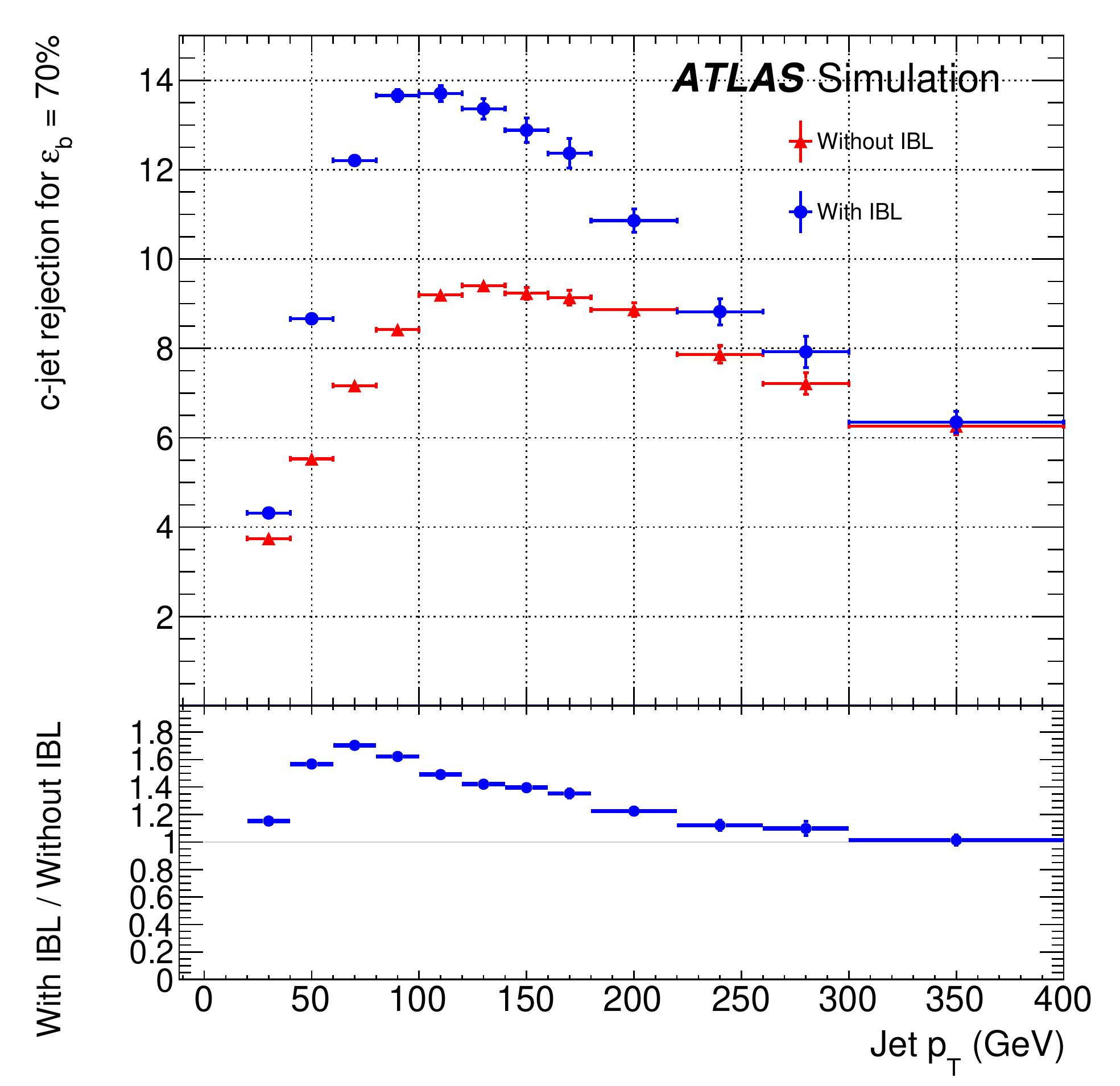}
%\vspace {-3.5cm}
                \caption{}
                \label{fig:z0b}
        \end{subfigure}
%\vspace {-2cm}
\caption{Comparison of \subref{fig:d0b} light-jet and  \subref{fig:z0b} $c$-jet rejection
as a function of jet transverse momentum, while keeping the $b$-tagging efficiency fixed at 70\% in each $p_{T}$ bin
for the Run~1 (without IBL) and Run~2 (with IBL) detector layouts under the same conditions, obtained with the MV2c20 algorithm. The rejection is defined as the reciprocal of  the tagging efficiency. Results are derived using jets produced in  \ttbar\ events and passing the $\pt$~>~20~GeV and $|\eta|<2.5$ selection.}
\label{fig:btagging_vsPT}
\end{figure}

The $b$-tagging  performance as a function of  jet \pt\ is shown in Figure~\ref{fig:btagging_vsPT}.
The largest improvements are seen at low values of the jet \pt\, where the proximity of the IBL to the interaction region 
%allows to reduce significantly 
significantly reduces 
the impact of multiple scattering in the track reconstruction. The improvement in light-jet ($c$-jet) rejection
ranges reaches a factor 4 (1.6) 
for jet $\pt$~$\lesssim$~100~GeV while at higher \pt\ the tracking
performance gain is limited by shared clusters from collimated tracks produced in the core of high \pt\ jets.
 
%-------------------------------------------------------------------------------
%\section{Analysis}
%\label{sec:analysis}
%-------------------------------------------------------------------------------

%You can find some text snippets that can be used in papers in \texttt{template/atlas-snippets.tex}.
%Some of the snippets need the \texttt{jetetmiss} option passed to \texttt{atlasphysics}.
%\input{atlas-snippets}

%-------------------------------------------------------------------------------
%\section{Results}
%\label{sec:result}
%-------------------------------------------------------------------------------

%Place your results here.

% All figures and tables should appear before the summary and conclusion.
% The package placeins provides the macro \FloatBarrier to achieve this.
% \FloatBarrier

\section{Modules}
\label{section:modules}

The basic building block of the IBL detector is the module. For each beam crossing an FE-I4B read-out chip records, digitises and locally stores the  data from a silicon sensor that is connected to it. Two sensor technologies are used: planar and 3D. 
A planar silicon wafer contains four sensor tiles, each of nominal dimension \SI{41340}{\micro\meter} $\times$ \SI{18600}{\micro\meter} (\SI{41315}{\micro\meter} $\times$ \SI{18585}{\micro\meter} in the production process after dicing).
A 3D silicon wafer contains eight sensor tiles, each of dimension \SI{20400}{\micro\meter} $\times$ \SI{18700}{\micro\meter}. 
There are consequently two module types, planar and 3D:

\begin{itemize}
\item[-] {A planar module consists of a planar sensor tile connected to two FE-I4B chips. Each chip consists 26880 pixel cells having analog and digital 
circuitry arranged in a matrix of 80 columns  of \SI{250}{\micro\meter} pitch and 336 rows of  \SI{50}{\micro\meter} pitch. Each FE-I4B cell is bonded using Sn/Ag bumps to a corresponding cell of the planar tile;}
\item[-] {A 3D module consists of a 3D sensor tile connected to a single FE-I4B chip with each cell of the FE-I4B chip bonded to a corresponding cell of the 3D tile;}
\item[-] {A double-sided, flexible printed circuit (the module flex hybrid) connects the module to external electrical services.}
\end{itemize}

The sensor design, production and yield is discussed in Section~\ref{sec:mod_sensor}. This is followed by a discussion of the FE-I4B production and yield in Section~\ref{sec:overview_electronic}. The module hybridisation, that is the bump-bonding of the FE-I4B chip(s) and a wafer to produce a bare module, is made industrially. A module flex hybrid is then attached at module production sites to the bare module, prior to detailed performance studies of the final (dressed) module. The module hybridisation, the module flex hybrid connectivity and the final performance are described in Sections~\ref{sec:module_assembly} and \ref{sec:module_qa}. Finally, the overall module production yield is summarised in Section~\ref{sec:prod_yield}. 

\subsection{Sensors}
\label{sec:mod_sensor}

Two sensor technologies are used for IBL modules. The planar sensor 
is a development of the Pixel detector sensor design, with several improvements. 
Most notably, since the limited IBL clearance precludes sensor shingling along the staves (as in the Pixel detector), the inactive sensor edges are substantially reduced to minimise efficiency losses. The 3D sensor design~\cite{original_3d_paper} is a new technology developed for increased radiation hardness, and relies on columnar electrodes penetrating the substrate, reducing the drift path with respect to the planar approach while keeping a similar thickness and thus signal size. As discussed in detail in reference~\cite{IBL_mod_proto}, both sensor types show satisfactory test-beam performance in terms of noise, hit efficiency and hit uniformity for a fluence of up to \SI{5e15}{\nq}. An effective inactive edge width of 
\SI{215}{\micro\meter} (\SI{175}{\micro\meter}) was measured for planar (3D) sensors.

The planar n$^+$-in-n sensors have proven their excellent performance during the Run~1 operation of the Pixel detector and are a 
well-developed technology. 
Nevertheless, the 3D sensors have a potentially important advantage in terms of power consumption after high radiation because of their lower operating voltage. 

Double-chip planar sensor modules cover the central region of the detector, \SI{75}{\percent} of the active area, while the high $\eta$ regions are populated by single-chip 3D sensor modules. This mitigates the reduced efficiency measured for normal incidence in the region of the 3D sensor electrodes.

\subsubsection{Planar design}
\label{sec:planar_design}

\begin{figure}[h]
\centering
(a)~\includegraphics[width=0.90\textwidth]{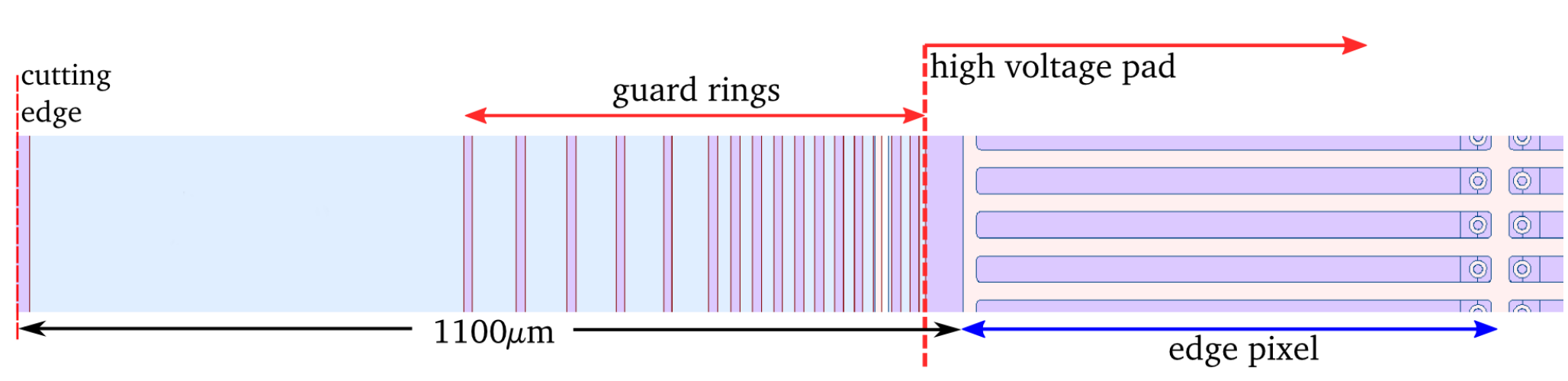}\\
(b)~\includegraphics[width=0.90\textwidth]{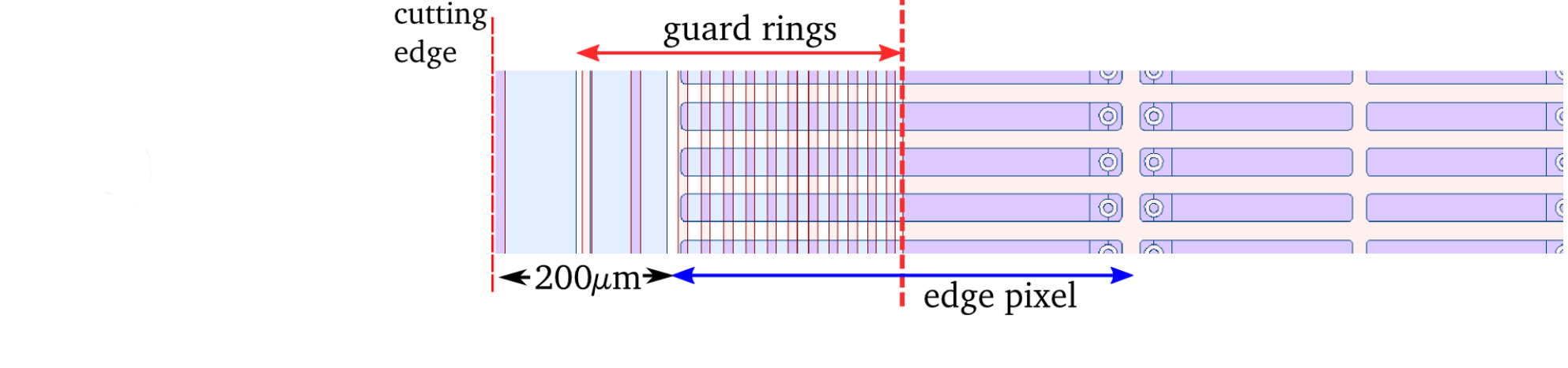}\\
  \caption{Comparison of the edge designs of (a) the ATLAS Pixel detector sensor and (b) the planar IBL pixel sensor. The inactive edge has been reduced from 1100 to \SI{200}{\micro\meter}. 
Blue shades represent the n-implantation on the front-side of the sensor. Purple shades represent the blue n-implantation on the front-side of the sensor superimposed with red shading for the p-implantation on the back-side. The HV backplane area is metalised and is indicated by a dashed red line and arrow.}
\label{planar-figure-edge}
\end{figure}

The design of the planar IBL sensor is an evolution of  the Pixel detector sensor~\cite{Aad:2008zz} with n$^+$-in-n pixels. The n-side segmentation matches in size the FE-I4B read-out electronics connected via bump-bonds; a guard-ring structure is placed on the p-side. The planar IBL double-chip sensors are produced at CiS\footnote{CiS Forschungsinstitut
f$\ddot{\mathrm u}$r
 Mikrosensorik und Photovoltaik GmbH, Erfurt (Germany).}, 
 using n-type wafers  of \SI{100}{\milli\meter} diameter and \SI{200}{\micro\meter} thickness, with resistivity in the range \SI{}{2-5} \SI{}{\kilo\ohm\centi\meter} and a $<\!\!111\!\!>$ crystal orientation. Each wafer contains four sensor tiles of 
  mean dimension \SI{41315}{\micro\meter} $\times$ \SI{18585}{\micro\meter} after dicing.
 Details of the sensor design can be found in reference~\cite{WittigPhd}. 
 Key features include slim edges achieved by stretching the edge pixel size opposite to the guard-rings to \SI{500}{\micro\meter}, possible in n$^+$-in-n sensors because of the double-sided process; this option was 
implemented after %suggested by AGC-200817 
extensive studies of the sensor efficiency in the peripheral area~\cite{planar-slim-edge-2012}. The number of guard-rings was optimised based on a complementary study, which evaluated the breakdown behaviour after partial guard-ring removal~\cite{planar-guard-ring-removal-2010}. Compared to the Pixel detector sensor, the number of guard-rings has been reduced from 16 to 13 and the cutting edge has been moved closer to them, 
as indicated in Figure \ref{planar-figure-edge} where the overall reduction of the inactive edge (from 1100 to \SI{200}{\micro\meter}) is shown. 

The nominal pixel size is \SI{250}{\micro\meter} by \SI{50}{\micro\meter} pitch, matched to that of the FE-I4B chip. The two central columns of these double-chip sensors are extended to \SI{450}{\micro\meter} rather than \SI{250}{\micro\meter} to cover the gap between the two adjacent FE-I4B chips. 

\subsubsection{3D design}
\label{sec:3d_design}

In 3D pixel sensors, the columnar %column-like 
electrodes penetrate the substrate instead of being implanted on the wafer surface. 
% AGC 171117 The depletion region therefore grows parallel to the wafer surface. 
The depletion electric field is therefore parallel to the wafer surface.
The position and doping of the $\sim$\SI{10}{\micro\meter} wide %-thick 
 columns define the pixel configuration; the distance between electrodes can be 
 % AGC 171117 five to ten times smaller than the detector thickness (typically a few hundred microns), 
 typically fives times smaller than the $\sim$\SI{230}{\micro\meter} sensor thickness,
 thereby dramatically reducing the charge-collection distance and depletion voltage. 
Although the fabrication process of 3D sensors is more complex, significant advantages can potentially be realised by independently controlling the drift distance and the sensor thickness. Because of the low depletion voltage, the power dissipation per unit leakage current is reduced. The cooling requirements are therefore less demanding. The signal size is determined by the sensor thickness, independently of the small drift distance. Furthermore, the drift perpendicular to the track direction results in fast signals, which are robust against charge trapping caused by heavy radiation damage~\cite{original_3d_paper}.

\begin{figure}[htbp]
\centering
\includegraphics[width=\textwidth]{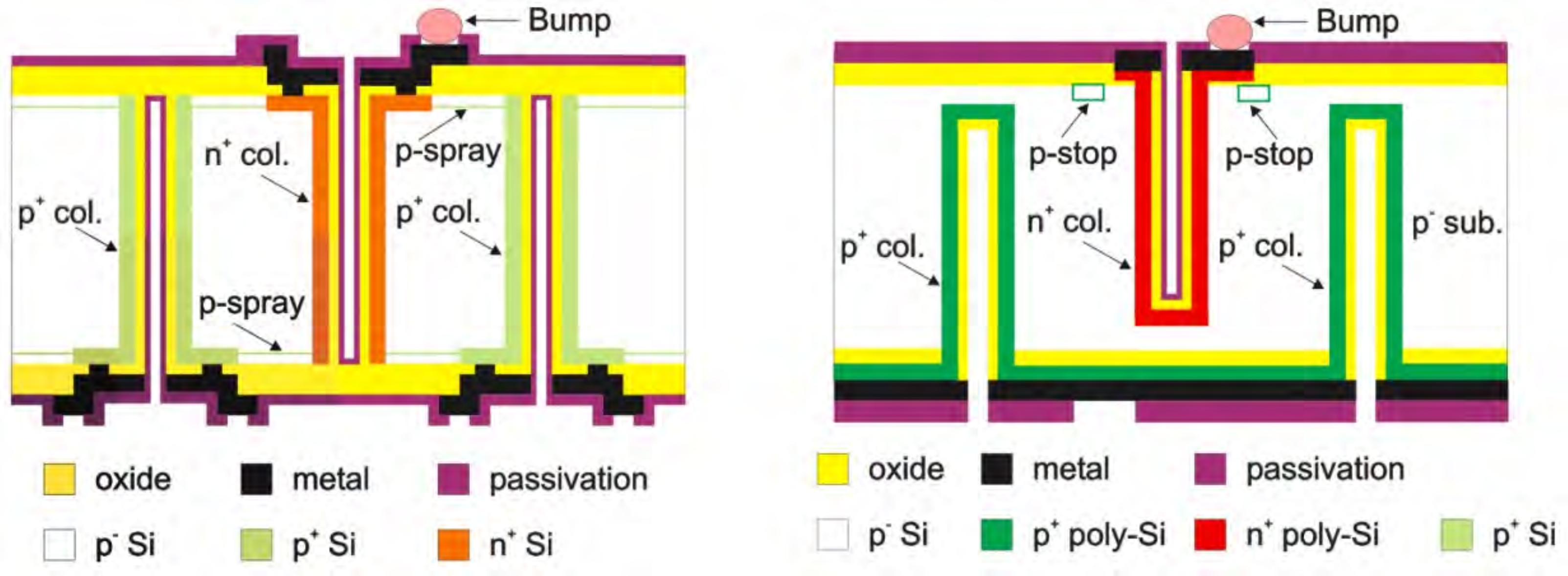}
	\begin{subfigure}[t]{0.45\textwidth}
		\caption{}
	\end{subfigure}
	\begin{subfigure}[t]{0.45\textwidth}
		\caption{}
	\end{subfigure}
\caption{\textbf{}Design of the columns of (a) FBK and (b) CNM 3D sensors. This sketch is for illustration only and is not to scale.}
\label{fig:3d_design}
\end{figure}

The IBL 3D sensors were fabricated at FBK\footnote{Fondazione Bruno Kessler, Povo di Trento (Italy).} and CNM\footnote{Centro Nacional de Microelectronica, Barcelona (Spain).} with a double-sided technology~\cite{ds-cnm,ds-fbk}. 
Starting from p-type Float Zone wafers  of \SI{100}{\milli\meter} diameter and \SI{230}{\micro\meter} thickness, with 
high-resistivity (10 to \SI{30}{\kilo\ohm\centi\meter}) crystalline silicon and $<\!\!100\!\!>$ orientation, columnar electrodes of \SI{12}{\micro\meter} diameter were obtained by Deep Reactive Ion Etching (DRIE) and dopant diffusion from both wafer sides, without the presence of supporting wafers. By doing so, the substrate bias can be applied from the back side (p$^+$), as in planar devices. Figure~\ref{fig:3d_design} shows details of the 3D column layout.

Each pixel contains two read-out (n$^+$) columns (two-electrode configuration), with an inter-electrode spacing between n$^+$ and p$^+$ columns of $\approx$\SI{67}{\micro\meter}. In order to maintain a reasonable yield, each wafer contains eight sensor tiles of 
dimension \SI{20.4}{\milli\meter} $\times$ \SI{18.7}{\milli\meter}, 
rather than the four larger sensor tiles of the planar design. 
A \SI{200}{\micro\meter} wide region separates the active pixel area from the physical edge of the tile.

The main differences between FBK and CNM 3D sensors are the following:
\begin{itemize}
\item[-] {FBK sensors have pass-through columnar electrodes~\cite{GIACOMINI_TNS_13}; in CNM sensors, on the other hand, electrode etching is stopped $\sim$\SI{20}{\micro\meter} before reaching the opposite side~\cite{PELLEGRINI_NIMA_13};}
\item[-] {in FBK sensors, the surface isolation between n$^+$ electrodes is obtained by a p-spray layer on both wafer sides, whereas in CNM sensors, p-stops are used on the front side (n$^+$) only;}
\item[-] {the edge isolation in FBK sensors is based on multiple rows of ohmic columns stopping the lateral spread of the depletion region~\cite{Povoli:2012zz}, whereas in CNM sensors a 3D guard-ring, surrounded by a double row of ohmic columns, is used to sink the edge leakage current.}
\end{itemize}
Table~\ref{tab:sensors} summarises the main parameters of the IBL sensors.

\begin{table}[htb]
\centering
\begin{tabular}{lccc}
\hline \hline
Parameter & Planar & 3D FBK & 3D CNM \\ \hline 
Tile dimension [\SI{}{\micro\meter\squared}]    & 41315 $\times$ 18585 % 41315 x 18585 (nominal: 41300 x 18600)
                                                & 20450 $\times$ 18745 % 20450 x 18745 (nominal: 20400 x 18700)
                                                & 20450 $\times$ 18745 \\
Sensor thickness [\SI{}{\micro\meter}]  & 200 & 230 & 230  \\
Sensor resistivity [\SI{}{\kilo\ohm\centi\meter}] & \SI{}{2-5}	&  \SI{}{10-30} & \SI{}{10-30} \\ 
Pixel size (normal) [\SI{}{\micro\meter\squared}]
                                                & 250 $\times$ 50 & 250 $\times$ 50 & 250 $\times$ 50 \\
Pixel size (tile edge) [\SI{}{\micro\meter\squared}]
						& 500 $\times$ 50 & 250 $\times$ 50 & 250 $\times$ 50 \\
Pixel size (tile middle) [\SI{}{\micro\meter\squared}]
						& 450 $\times$ 50 & - & - \\
Edge isolation       & Guard-rings & Fences & 3D guard-ring, fences \\
Pixel isolation      & p-spray     & p-spray &	p-stop on n-side \\ \hline
Nominal operating bias voltage [\SI{}{\volt}]
                                   & \SI{}{-80} / \SI{}{-1000} & \SI{}{-20} / \SI{}{-160}  & \SI{}{-20} / \SI{}{-160} \\
(non-irradiated / \SI{5e15}{\nq})  & & \\
Maximum operational power [\SI{}{\milli\watt\per\centi\meter\squared}] 
                     &    90
                     &    15
                     &    15  \\
(\SI{-15}{\celsius} and \SI{5e15}{\nq})  & &  \\ \hline \hline					
\end{tabular}
\caption{Summary of the main design specifications for the planar and 3D sensors of the IBL detector.}
\label{tab:sensors}
\end{table}

\subsubsection{Sensor production and quality assessment}
\label{sec:sensor_qa}

The electrical quality of the sensors was evaluated from the measurement of the current-voltage (I-V) dependence, as this is sensitive to bulk and surface defects. Tiles satisfying the selection criteria described below were chosen for hybridisation (connection between the sensor and the FE-I4B read-out electronics). 

The planar design includes a grid structure that allows biasing of the entire sensor by means of a punch-through technique~\cite{Lutz:1999wg}. This bias grid was used to evaluate the quality of the tiles before the sensors were connected to the read-out electronics with the bump-bonding process. After bump-bonding the pixels were biased through the FE-I4B chip while the bias grid, connected to ground via a special bump in the periphery of the pixelated region, was not in operation.

The leakage current of the planar tile, evaluated at an operating voltage (V$_{\SI{}{op}}$) \SI{30}{\volt} below the depletion voltage (V$_{\SI{}{dp}}$), was required to be I(V$_{\SI{}{op}}$)~<~\SI{1}{\micro\ampere},
and the slope of the I-V curve was limited to 
I(V$_{\SI{}{dp}}~-~\SI{30}{\volt}$)/I(V$_{\SI{}{dp}}$)~<~1.6.  
Wafers with two or more planar tiles that satisfied this requirement were sent for under-bump metallisation (UBM) and dicing at IZM\footnote{Fraunhofer IZM, Gustav-Meyer-Allee 25, 13355 Berlin, Germany.}. The yield of the planar production (the percentage of planar tiles satisfying the above criteria) before under-bump metallisation and dicing was \SI{90.6}{\percent}. 

% 3D
Due to the difficulty of implementing a bias grid structure compatible with the 3D design, alternative evaluation methods were developed for 3D sensors:
\begin{itemize}
\item[-] {FBK sensors include a metal grid connecting all pixels in each column to a pad located in the periphery of the active region. By measuring the I-V curves of the 80 columns with a specially designed probe card, the quality of each sensor on the wafer can be evaluated. The metal layer was removed by chemical etching 
after the I-V measurement 
and the wafers with three or more selected tiles were sent to IZM for UBM and dicing. The sensors that passed the selection criteria were bump-bonded to read-out chips. The sensors were required to have a breakdown voltage V$_{\SI{}{bd}}$~<~\SI{-25}{\volt}, V$_{\SI{}{dp}}$~>~\SI{-15}{\volt} and I(V$_{\SI{}{op}}$)~<~\SI{2}{\micro\ampere} where V$_{\SI{}{op}}$~=~V$_{\SI{}{dp}}-\SI{10}{\volt}$. The slope of the I-V curve was also constrained to satisfy I(V$_{\SI{}{op}}$)/I(V$_{\SI{}{dp}}+\SI{5}{\volt}$)~<~2. The sensor yield of the FBK production on the selected 
wafers %AGC-100817 wafers (tiles?) end 
was \SI{57}{\percent}.} 
\item[-] {The CNM sensor selection criteria were initially based on the leakage current measured through the 3D guard ring structure surrounding the pixelated area. While the p-side of the wafer was biased, the 3D guard ring was connected to ground via a dedicated pad, and the I-V curve was measured for each sensor before the wafer dicing. After hybridization, the 3D guard ring was connected to ground through two special bumps of the FE-I4B chip. The CNM sensors were required to satisfy  V$_{\SI{}{bd}}$~<~\SI{-25}{\volt}, V$_{\SI{}{dp}}$~>~\SI{-15}{\volt} and I$_{\SI{}{GR}}$(V$_{\SI{}{op}}$)~<~\SI{200}{\nano\ampere} with V$_{\SI{}{op}}$~=~V$_{\SI{}{dp}}-\SI{10}{\volt}$. I$_{\SI{}{GR}}$ is the leakage current measured on the 3D guard ring.  The slope of the I-V curve was required to satisfy I(V$_{\SI{}{op}}$)/I(V$_{\SI{}{dp}}+\SI{5}{\volt}$)~<~2. Wafers with at least three sensors passing the selection criteria were sent to IZM for UBM and dicing.
Initial studies indicated a good correlation between V$_{\SI{}{bd}}$ measured through the 3D guard ring structure and that  after detector assembly~\cite{IBL_mod_proto}.  However, during module assembly, the correlation proved to be poor, with several CNM 3D modules showing a low V$_{\SI{}{bd}}$. 
This was because of defects located in the central volume of the sensor that do not affect the region probed by the 3D guard ring.Once this lack of correlation was established, all CNM sensors that were not assembled were re-tested on a probe station. 
The n-side of the sensor was placed in contact with a grounded chuck via the under-bump metallisation (Section~\ref{sec:mod_bumping}), while the p-side was connected to the bias potential. Those sensors satisfying V$_{\SI{}{bd}}$~<~\SI{-25}{\volt} were selected for hybridisation. The sensor yield of the CNM production on the selected wafers, as measured with the 3D guard ring method, was \SI{72}{\percent}.  However, after re-testing, the final CNM production yield was similar to that for FBK wafers.}
\end{itemize}

The typical sensor I-V behaviour of prototype sensors was previously detailed before and after radiation~\cite{IBL_mod_proto, WittigPhd}.  Typical I-V curves for each sensor type are shown after module assembly in Section ~\ref{sec:module_IV_qa}. 

\subsection{On-detector electronics}
\label{sec:overview_electronic}

\subsubsection{The FE-I4 front-end chip}
\label{sec:FEI4_electronic}

The FE-I4B front-end chip was developed for the IBL read-out.
A first version, the FE-I4A~\cite{FE_I4:2010, Hemperek:2009}, was fabricated in 2010 and used to develop and validate the IBL module design~\cite{IBL_mod_proto}. The FE-I4A was not intended for the final detector and the  pixel matrix was non-uniform to allow  performance comparisons between various analog circuit design choices. 
The FE-I4B chip was first fabricated in 2011~\cite{FE_I4B:2012,FE_I4B:2013} and  tailored to fully meet the IBL requirements. In addition to selecting the analog design and making the pixel matrix uniform, specific powering choices were made and data acquisition features added. 

The FE-I4A and FE-I4B both contain read-out circuitry for \SI{26880} hybrid pixels arranged in 80 columns of \SI{250}{\micro\meter} pitch by 336 rows of \SI{50}{\micro\meter} pitch. Each FE-I4 pixel  contains a free running clock-based amplification stage with adjustable shaping, followed by a discriminator with an independently adjustable threshold. 
The chip keeps track of the time stamp for each discriminator as
well as the 4-bit Time over Threshold (ToT)\footnote{
The Time over Threshold is defined as the time the amplifier output signal stays above threshold, measured in units of the LHC clock (25\,ns). This quantity is related to the collected charge.}. 
Information from all firing discriminators is kept in the chip for a latency interval programmable up to 255 LHC clock cycles of 25\,ns, and is retrieved if a trigger is supplied within this latency. 
The IBL data output is a serial Low Voltage Differential Signal (LVDS), 8b/10b encoded at a rate of \SI{160}{\mega\bit\per\second}.
The chip has many  configurable settings that are stored in triple-redundant  registers providing the required radiation hardness to  single event upsets (SEU)~\cite{Menouni:2013}. 

Because of space and material limitations, the IBL FE-I4B chips are powered from a single DC supply over long cables providing a resistive load.
The single voltage feeds two 
Shunt-LDO\footnote{A shunt combined with a \textbf{L}ow-\textbf{D}rop-\textbf{O}ut regulator.} 
voltage regulators~\cite{Karagounis:2009,Gonella:2012} drawing a minimum standing current of \SI{270}{\milli\ampere} even when the chip is neither clocked nor configured. This limits the amplitude of voltage transients resulting from current changes on the resistive supply lines, 
particularly important since the difference between the nominal and maximum input voltage ratings is small. The regulators and attendant voltage references have an input voltage limit of \SI{2.5}{\volt}, compared with nominal operation at \SI{1.8}{\volt}. Once the chips are configured and clocked, and their internal current draw exceeds \SI{270}{\milli\ampere}, the regulator shunt elements shut off and draw no additional current. 
The chip operates internally with two voltage rails generated by the regulators, nominally \SI{1.4}{\volt} for the analog circuitry and \SI{1.2}{\volt} for the digital circuitry. Both voltages are adjustable with a hard-wired maximum around \SI{1.5}{\volt} (which varies slightly from chip to chip). The voltage references use a combination of a programmable current reference (feeding a poly-silicon resistor) and a fixed voltage reference.
This combination was chosen to allow reliable start-up at low temperature (as low as \SI{-40}{\celsius}), as well as excellent stability (<~\SI{\pm2}{\percent}) up to high radiation dose (\SI{250}{\mega Rad}).

Several features important for IBL operation were introduced in the FE-I4B design following experience with prototype FE-I4A modules.
Some details of the analog bias distribution and charge injection were
changed to correct for degradations observed in the FE-I4A after the expected IBL lifetime dose, particularly at low temperatures. A programmable event-size limit 
was introduced to avoid data acquisition time-outs from occasional pathologically large events.
Bunch crossing and trigger counters 
%AGC-090817 have been 
were increased to respectively 13 and 12 bits, to avoid ambiguities in tracking the state of each chip. Improved diagnostics 
%AGC-090817
were implemented to count and report any skipped triggers (the chip will skip any triggers received when the 16-bit trigger buffer is full).
%end

\subsubsection{FE-I4B production and quality assessment}
\label{sec:mod_electronic}
\begin{figure}[h]
	\centering
	\begin{subfigure}[t]{0.49\textwidth}
		 \includegraphics[width=\textwidth]{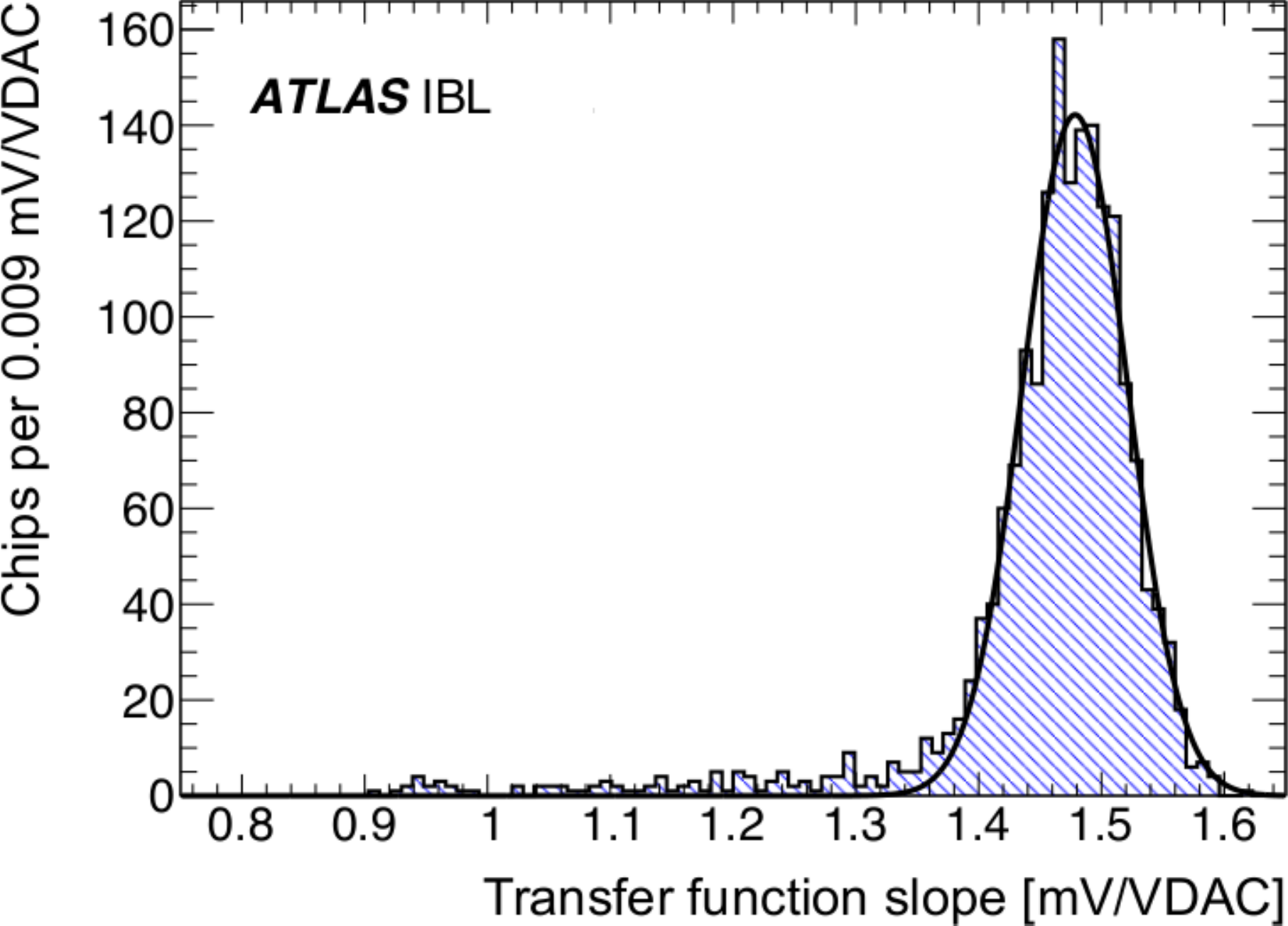}
		\caption{}
		\label{fig:reference_current}
	\end{subfigure}
	\begin{subfigure}[t]{0.49\textwidth}
		 \includegraphics[width=\textwidth]{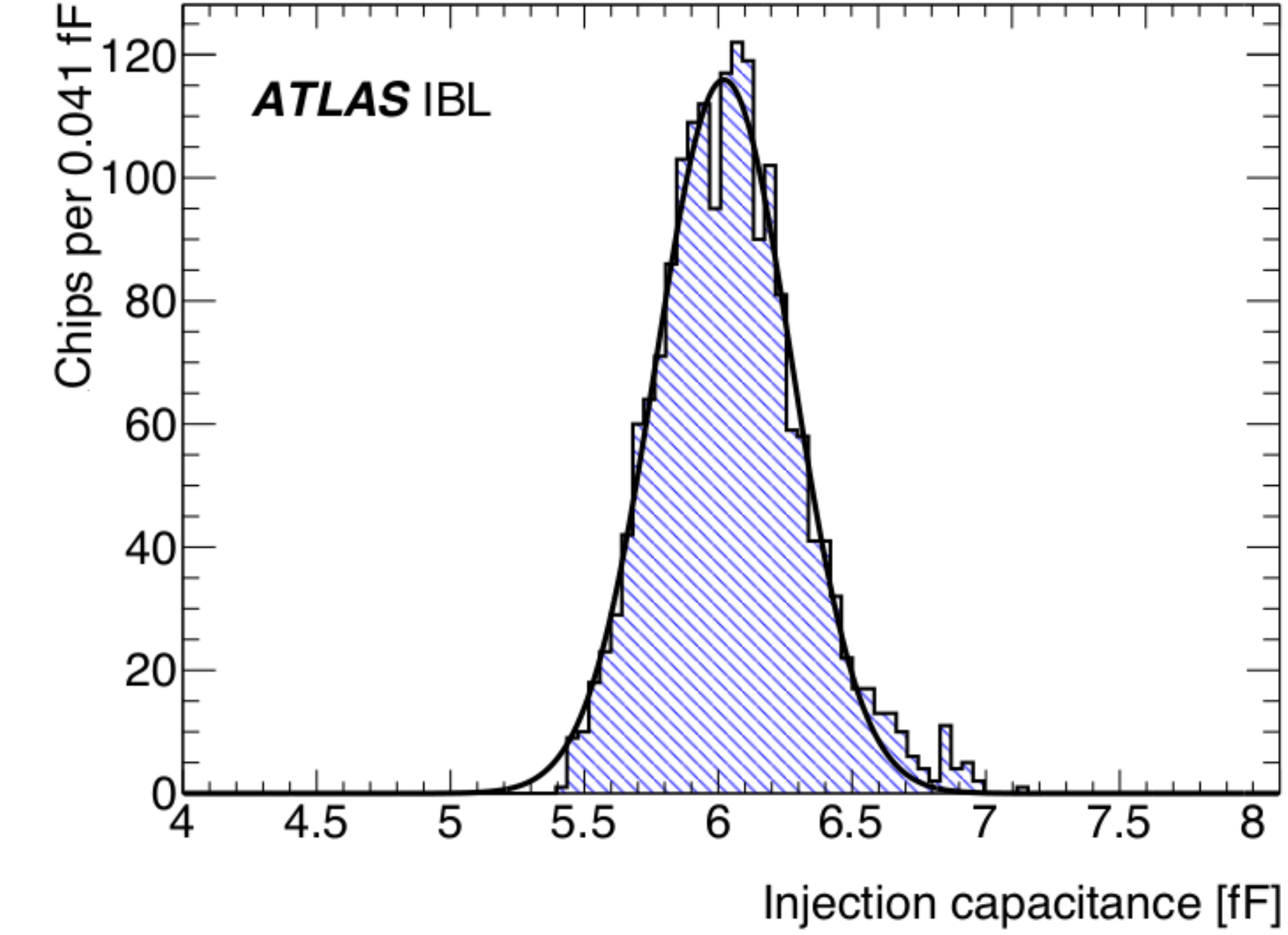}
		\caption{}
		\label{fig:injection_capacitance}
	\end{subfigure}
  \caption{\textbf{}Calibration constants of the internal charge injection circuit for accepted FE-I4B chips. The circuit distributes a voltage step to injection capacitors on board each pixel. The superimposed curves are Gaussian fits. (a) The slope of the transfer function between the signal voltage and the value VDAC of the DAC setting. (b) The measured injection capacitance.}
	\label{fig:calibration_results}
\end{figure}

For the IBL production, \SI{3060}~FE-I4B chips 
 on fifty-one \SI{200}{\milli\meter} diameter wafers were tested.
The data acquisition and handling were performed with a custom read-out system~\cite{Aad:2008zz,USBpix}. 
A custom PCB was used to interface the read-out system hardware with a probe card, establishing electrical contact with 108 FE-I4B pads. 
All 60 chips of a wafer were probed with an average measurement time of ~2.5 days. The goal was to identify FE-I4B chips that were suitable for the IBL and to measure their calibration constants. For chip calibrations it was necessary to make contact with dedicated FE-I4B pads not wire-bonded on the IBL read-out circuit. The chip calibrations were therefore only possible at wafer level. Two calibration constants of the internal charge injection circuit are shown in Figure~\ref{fig:calibration_results}. The circuit distributes a voltage step to injection capacitors present in each pixel and is needed to tune the FE-I4B chips during IBL operation. On average for the accepted chips, the injected charge changes with the value (VDAC) of the injection circuit digital-to-analog converter (DAC) setting approximately as
%\begin{equation}
\[
\frac{\Delta Q}{\mathrm{VDAC}} = 6.05\,\mathrm{fF} \cdot 1.45\,\frac{\mathrm{mV}}{\mathrm{VDAC}} = 55\,\frac{\mathrm{e}}{ \mathrm{VDAC}}
\]
%\end{equation}
%AGC 200118
to provide a transfer function between the value of the DAC setting and the signal.

% using format as in chapter 6

More than 50 %AGC- different 
tests were used to evaluate the %AGC-add chip
chip response to charge injection, the functionality of digital hit processing, the chip configurability, and the power consumption. %AGC-About replaced by approximately 
Approximately 18000 values were recorded per wafer. 
%{\textit WaferAnalysis}, a 
A custom made software designed for wafer and module tests of the IBL production was used to automatically determine the chip status. The selection criteria were defined after the distributions of the first ten wafers were studied. %AGC-A complete list of the tests and a detailed description of the cut scheme can be found in~\cite{WaferAnalysis}
A detailed description of the tests and selection criteria is available elsewhere~\cite{WaferAnalysis}. 

\begin{figure}[t]
	\centering
		 \includegraphics[width=0.95\textwidth]{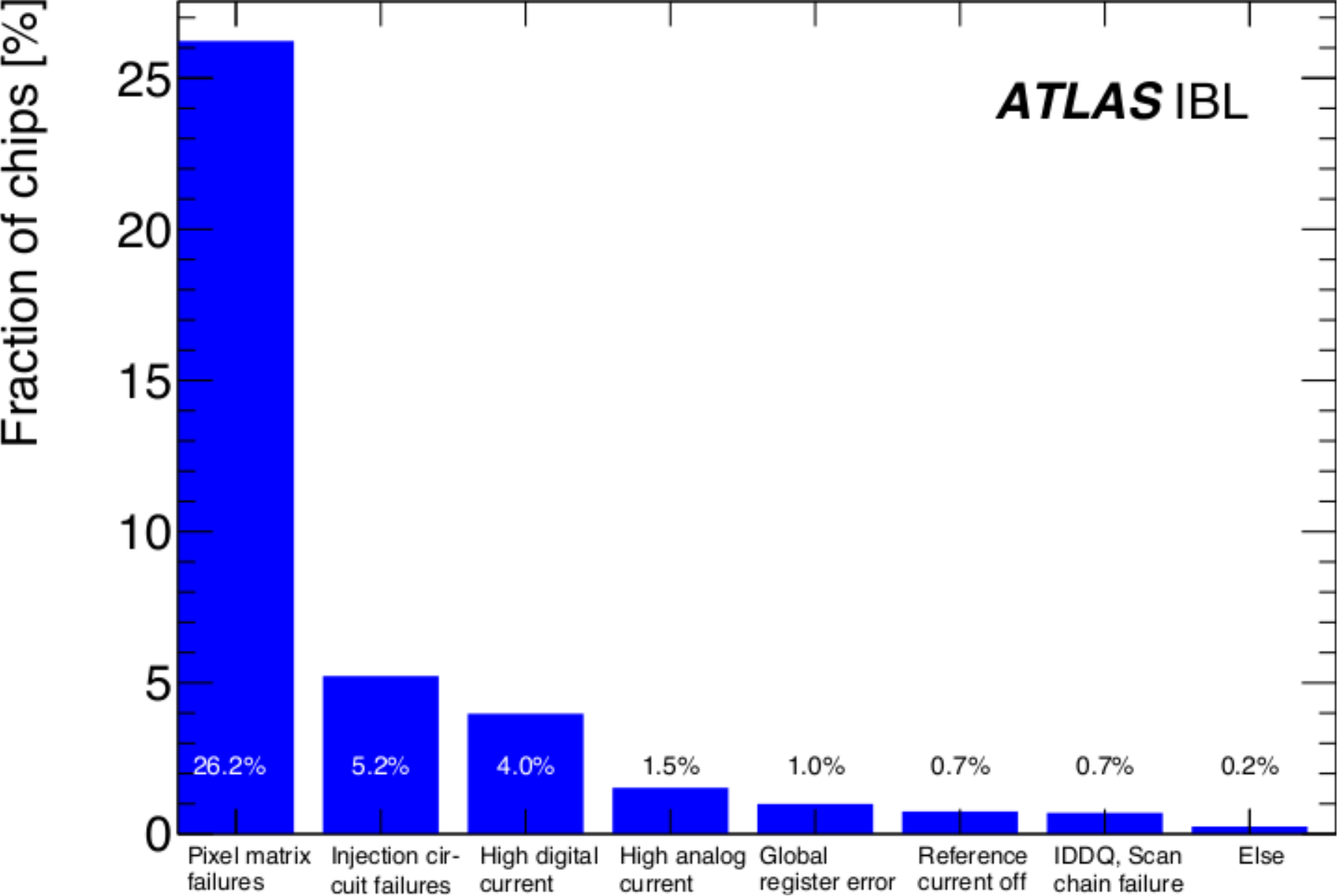}
	\caption{\textbf{}Failure modes leading to a rejection of FE-I4B chips before module assembly for 2814 fully probed chips. The bin \textit{Pixel matrix failures} groups chips where the number of bad pixels were too high ($>$\SI{0.2}{\percent} failing pixels or $>$\SI{20} pixels per column). The bin \textit{Injection circuit failures} groups failures (e.g. low maximum voltage, a non-configurable injection delay) that prevent using the charge injection for calibration during IBL operation. The bins \textit{High analog/digital current} combine current measurements in different chip states (un-configured, configured, high digital activity). The remaining bins list the rate for chips failing the global register tests, the reference current generation tests and the Scan chain tests, respectively. All failure modes that are not explicitly mentioned contribute only \SI{0.2}{\percent} and are included in the bin \textit{Else} The failures are non-exclusive and are evaluated as a percentage of the probed chips.}
\label{fig:failures}
\end{figure}

Test results of 2814 fully probed chips are listed in Figure~\ref{fig:failures}. In addition, 246 chips (\SI{8}{\percent} of all chips) were not fully probed because of an anomalous high current at start-up (dead-shorts). 
Additional IDDQ\footnote{Measurement of the supply current (Idd) in the quiescent state.}, Scan chain and Shmoo plot\footnote{A plot showing the range of conditions (voltages, temperatures and inputs) in which the chip operates.} tests were made by an external company for the first $\sim$\SI{20} wafers, but the failure rate was low~(less than \SI{0.5}{\percent}). In total, 1821 chips (\SI{59.5}{\percent}) were qualified for IBL module assembly.

Since the powering scheme was not finalised at the time of wafer testing, the on-chip power regulators of the FE-I4B chips were only tested after the module assembly (Section~\ref{sec:module_qa}).

\subsection{Module assembly}
\label{sec:module_assembly}
\subsubsection{Hybridisation of the FE-I4B chip and the sensor}
\label{sec:mod_bumping}

The connection between sensor and electronics was achieved using fine-pitch bump-bonding and flip-chip technology. This was already used with a \SI{50}{\micro\meter} pitch for the construction of the Pixel detector modules~\cite{Aad:2008zz}. The IBL modules use a similar electroplated (SnAg) bumping process provided by IZM. The bumping process is divided into three steps: 
under-bump metallisation (UBM) on the sensor and FE-I4B wafers; solder bump deposition on the FE-I4B wafers; and a flip-chip of the diced FE-I4B chips and sensors. 
The UBM is necessary due to the non-solderable aluminium pads on the sensors and FE-I4B chips; 
%the UBM metal stack, consisting of Ti/W and Cu, is sputtered on both the sensor and FE-I4B wafers. 
the UBM metal stack consists of electro-deposited Cu on top of a sputtered Ti/W adhesion layer.
Solder bumps are then deposited on the FE-I4B wafers using electroplating only. The flip-chip operation follows the dicing of the sensor wafers. The FE-I4B chip is placed on the sensor substrate with high accuracy and the assembly is soldered to form the electrical and mechanical interconnection in a reflow soldering process. The sensor bonded to the FE-I4B chip(s) is commonly referred to as a bare module. %AGC \textit{bare module}.

The procedure was modified with respect to that used for Pixel modules to suit the dimensions of the IBL module components.
The FE-I4B chip covers an area of 20.27 $\times$ \SI{19.20}{\milli\meter\squared} and was thinned to \SI{150}{\micro\meter} before bump-bonding. 
Unconstrained, the thinned FE-I4B would undergo a distortion exceeding \SI{40}{\micro\meter} during the high temperature reflow soldering phase, which would result in unconnected bumps especially in the outer areas of the assemblies. To avoid this, a temporary \SI{500}{\micro\meter}-thick sapphire glass 
handle wafer was bonded to % AGC 261117support was mounted on 
the FE-I4B chip before UBM. A polyimide bonding technique allowed a laser-induced debonding of the glass carrier at room temperature after dicing and flip-chipping. This debonding process used an UV excimer laser with a wavelength of \SI{248}{\nano\meter} traversing the glass carrier to the bonding interface. The glass carrier was optimised to ensure that the laser light was fully absorbed in the polyimide bonding layer, thus releasing the FE-I4B chips.

Only \SI{2}{\milli\meter} of the chip length is dedicated to End-of-Column (EoC) logic outside the active pixel matrix. The size is determined by the need to wire bond the I/O and power pads to the read-out chip with the bump bonded sensor in place. The chip-level logic and global configuration occupy less than \SI{20}{\percent} of the periphery.
Once bonded, most of the EoC part extends beyond the sensor area so that the wire bonding pads at the output of the EoC logic are still accessible to connect the read-out chip via aluminium-wire wedge bonding.

\subsubsection{Module flex hybrid}
\label{sec:mod_flex}

\begin{figure}[htbp]
\centering
\includegraphics[width=0.9\textwidth]{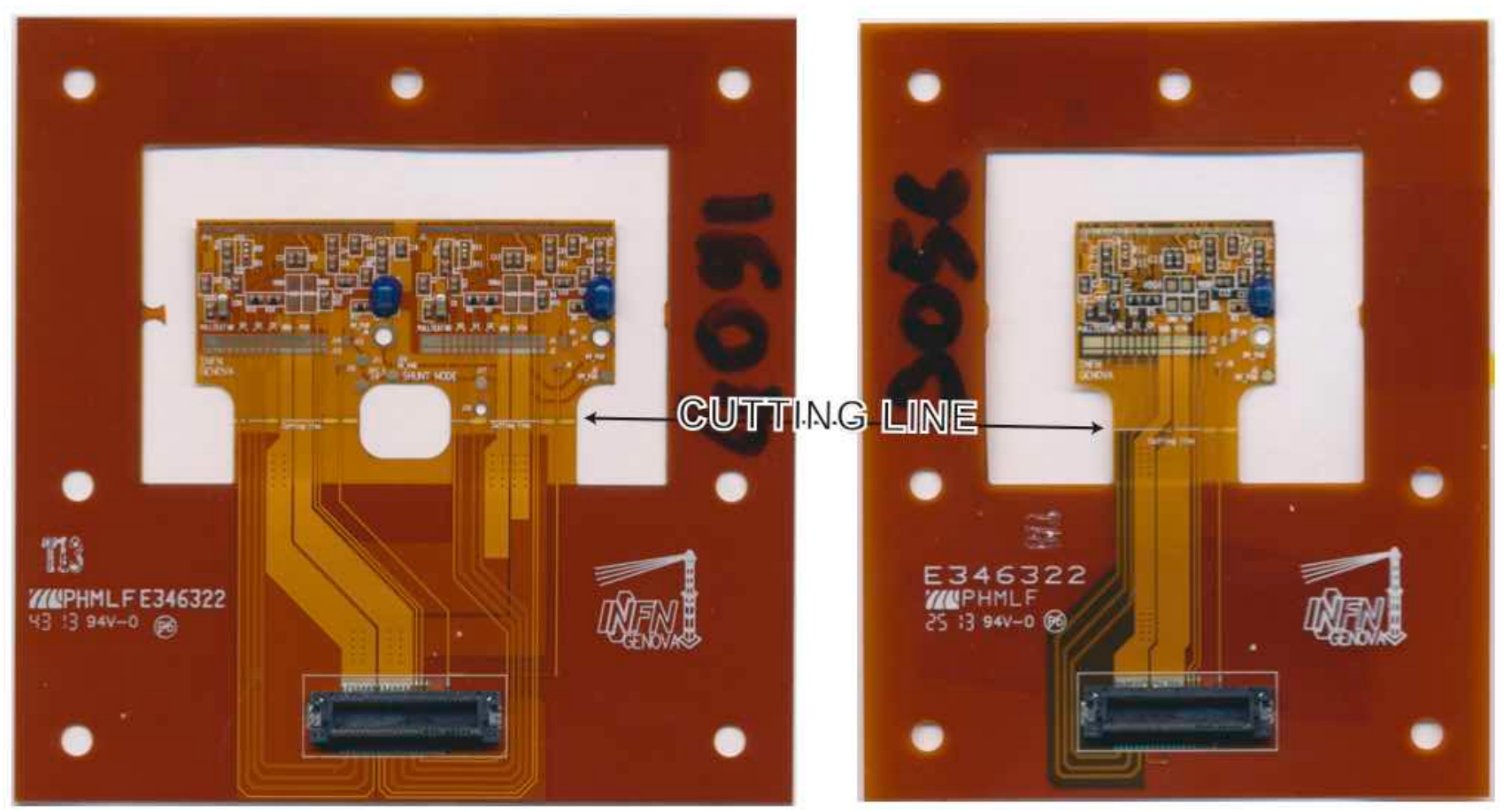}
	\begin{subfigure}[t]{0.5\textwidth}
		\caption{}
	\end{subfigure}
	\begin{subfigure}[t]{0.4\textwidth}
		\caption{}
	\end{subfigure}
\caption{Photographs of (a) a double-chip and (b) a single-chip module flex hybrid. The frame and flex extensions allow testing of the module before stave loading. The hybrid cutting line (see text) is visible as a white trace slightly outside the module envelope. 
}
\label{fig:module_flexes}
\end{figure}

The module flex hybrid is a double-sided, flexible printed circuit board which routes the signal and power lines between the 
%AGC-100817 Type 0 cables (stave flex hybrid) 
stave flex hybrid
%AGC end
and the FE-I4B chips, % AGC 261117distributing the bias voltage to sensors via copper traces. % AGC-180817 Two types of 
holds the required passive components, and routes the bias voltage to the sensor via Cu traces.
Figure~\ref{fig:module_flexes} shows a photograph of the module flex hybrids for single-chip and double-chip modules. The envelope of the module flex hybrid is defined by the sensor dimensions and it is slightly narrower than the sensor width. 

The module flex hybrids
are glued to the back side of the sensor and connected to the longitudinal stave flex, which is located at the back side of the stave, via thin transversal wings, one per read-out chip (Section~\ref{sec:overview_staveflex}). The \SI{130}{\micro\meter}-thick flex stack consists of two \SI{18}{\micro\meter}-thick copper layers embedded in dielectric polyimide 
sheets, glued with acrylic adhesive. Passive components are soldered on the module flex hybrid for the FE-I4B chip decoupling, power supply and HV filtering, and for terminations of the signal traces. The module temperature monitoring and interlock is made via 
a Negative Temperature Coefficient thermistor (NTC)
mounted on the module flex hybrid. All passive components are soldered on the top layer of the module flex hybrid. Special emphasis is given to HV routing and filtering since the flex hybrid must be functional up to \SI{1000}{\volt}. To avoid HV discharges, wider spacing between the HV traces and the data %signal 
and LV traces is introduced. The HV capacitor is encapsulated with a polyurethane resin and \SI{27}{\micro\meter} thick Kapton\textsuperscript{\textregistered}\footnote{Kapton\textsuperscript{\textregistered} is a Dupont Corp. trademark for polyimide films, see http://www.dupont.com.} cover layers are used on the top and bottom of the flex hybrid.

All signal and power traces of the module flex hybrid are routed to a connector on a frame outside the module area that is used during the module production QA. A temporary wire bond connection is necessary to connect all signal and power lines from the flex to the connector on the frame\label{pag:wirebondbridge}. 
Prior to the loading of the module to a stave the connector area is cut away. The cutting line is approximately \SI{1.5}{\milli\meter} from the sensor. 
%end-AGC

The module flex hybrids were produced by Phoenix S.r.l.\footnote{Phoenix S.r.l., Via Burolo 22, 10015 Ivrea (Torino), Italy.} and the surface mount component loading %AGC-180817 (SMD\footnote{Surface Mounted Device.}), 
and encapsulation was made by Mipot S.p.A.\footnote{Mipot S.p.A., Via Corona 5, 34071 Cormons (Udine), Italy.}. Basic QA operations such as testing of line integrity for open and shorted connections were made by the vendors and were followed by more detailed tests at the two module assembly sites. These procedures included HV standoff tests at \SI{1.5}{\kilo\volt}, visual inspection and dedicated cleaning to allow for high-quality wire bonding.

\subsubsection{Final module assembly}
\label{sec:mod_assembly}

%The final (dressed) module assembly  was performed at the module production sites by following a procedure of four major steps:
%\begin{itemize}
%\item[-] {The preparation and cleaning of the module flex hybrid, including a detailed visual inspection and an electrical test of all hybrid components;}
%\item[-] {The preparation and visual inspection of the bare module for scratches or other damage, including a re-measurement of the I-V for planar double-chip modules to check the sensor quality, and a measurement of the bare module weight. 
%Thirteen planar modules (\SI{3.2}{\percent}) and eight 3D modules (\SI{2.9}{\percent}) were rejected;}
%\item[-] {Alignment of the module flex hybrid to the bare module using an assembly tool, followed by glue deposition and curing;}
%\item[-] {Wire bonding of the FE-I4B chip to module flex hybrid and of the wire bond bridge to the test connector (Section~\ref{pag:wirebondbridge}), using \SI{25}{\micro\meter}-thick aluminium wire (at least three wire bonds were applied to the low- and high-voltage pads for redundancy and safety).}
%\end{itemize}

The final (dressed) module assembly was made at two module production sites in the period 2012 to 2014, following four assembly steps described 
below. 

A detailed visual inspection of the module flex hybrid was initially made, together with electrical tests of the line and pad integrity, and the hybrid components. 
To ensure a good wire bonding performance, the flex hybrid was then cleaned in an ultrasonic bath, rinsed with distilled water, and dried. The visual inspection was then repeated. 

A visual inspection of the bare module was made to identify scratches or other damage. For planar double-chip modules a re-measurement of the I-V was 
made to check the sensor quality. Thirteen planar modules (\SI{3.2}{\percent})  and eight 3D modules (\SI{2.9}{\percent}) were rejected. 

The key assembly step is the alignment and attachment of the bare module and the module flex hybrid. The module flex is glued on the sensor back-side. 
For this reason, it is necessary to visually access the sensor alignment marks, and to be able to wire-bond to both the FE-I4 chip and the flex wings. An alignment precision of order \SI{100}{\micro\meter} is required. 
The alignment and gluing procedure differed slightly between the production sites, and the detailed jig designs were developed autonomously.
Separate alignment jigs were developed for the planar double-chip and 3D single-chip modules. 
Several jig sets were made to ensure production capacity, but the module assembly rate was in fact determined by the component supply. 
Both the module flex and the bare module were initially aligned on separate jigs using alignment marks, and fixed in place via vacuum. The module flex was then removed with a special jig, maintaining the alignment position but allowing access for the deposition of glue. The jig was designed to protect the hybrid components. 
Glue patterns were then deposited on the flex hybrid: a double tape strip 
(PPI RD-577F\footnote{PPI RD-577F\textsuperscript{\textregistered}, PPI Adhesive Products GmbH, see www.ppi-germany.de.} 
or ARclad\footnote{ARclad\textsuperscript{\textregistered}, Adhesives Research Corp., see www.adhesivesresearch.com.}) 
was placed underneath the FE-I4B wire bond pads, and epoxy glue 
patterns 
(UHU EF 300\footnote{UHU EF 300\textsuperscript{\textregistered}, UHU GmbH, see www.uhu.com.}
or Araldite 2011\footnote{Araldite\textsuperscript{\textregistered} is a trademark of Huntsman Advanced Materials, see \SI{}{http://www.huntsman.com/advanced\underline{{ }}materials/a/Home.}}) 
were placed under the remainder of the module flex hybrid, especially under the wire bond bridge area and the HV connection pads. The jigs were then aligned and brought into contact. Pressure was applied on the assembly, and in particular around the critical wire bond regions, during curing. 

The final step was the wire bonding of the FE-I4B chip and sensor to the module flex hybrid and of the wire bond bridge to the test connector (Section~\ref{pag:wirebondbridge}), using \SI{25}{\micro\meter}-thick aluminium wire (at least three wire bonds were applied to the low- and high-voltage pads for redundancy and safety). Wire bond pull tests were consistently recorded to ensure the bond integrity.

At each stage of the assembly, details of the module components, as well as metrology and bonding information, were recorded. No site dependence of the module quality was identified. 

\begin{figure}[htbp]
	\centering
	\begin{subfigure}[t]{0.64\textwidth}
		 \includegraphics[width=\textwidth]{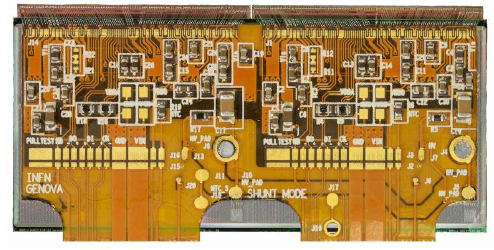}
		\caption{}
		\label{fig:DC_picture}
	\end{subfigure}
	\begin{subfigure}[t]{0.325\textwidth}
		 \includegraphics[width=\textwidth]{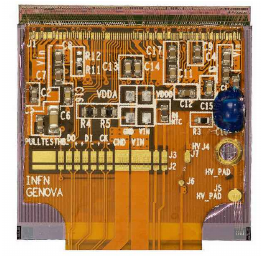}
		\caption{}
		\label{fig:SC_picture}
	\end{subfigure}
  \caption{Photographs of (a) an IBL planar module and (b) an IBL 3D module after the removal of the module handling frames. The HV encapsulation step is not yet made on the planar module.}
	\label{fig:module_pictures}
\end{figure}

Fully dressed planar (double-chip) and 3D (single-chip) modules are shown in Figure~\ref{fig:module_pictures}. Table~\ref{tab:module_x0} summarizes the material budget (units of radiation length for normal incidence) of the IBL modules; the contributions of the different components are averaged over the %whole AGC-180817
active module area. 

A total of 688 fully dressed modules (410 planar, 162 3D CNM and 116 3D FBK) were delivered for module testing, tuning and characterisation.

\begin{table}[htb]
\centering
\begin{tabular}{llll}
\hline \hline
	 && X/X$_0$(\SI{}{\percent})\\
\hline
Total FE-I4B + bumps && 0.21 \\
FE-I4B chip	   &0.20& \\
Bump-bonds & $\lesssim$0.01 & \\
Planar (3D) sensor && 0.24 (0.27) \\
Total module flex hybrid  && 0.13 \\
Cu traces	& 0.076 & \\
Polyimide/epoxy & 0.028 & \\
SMD components & 0.025 & \\
\hline
Total planar (3D) module && 0.58 (0.61) \\
\hline \hline
\end{tabular}	
\caption{Material budget of the IBL module components in units of radiation length (X/X$_0$)\, for normal incidence. The contributions of the different components are averaged over the active module area. Sub-components are shown in a separate column.}
%end AGC
\label{tab:module_x0}
\end{table}

\subsection{Module performance and quality assurance}
\label{sec:module_qa}

\begin{table}[htb]
\centering
  \small
\begin{tabular}{llc}
\hline \hline
  \textbf{Test Name} & \textbf{Test Purpose} & \textbf{Test stage} \\
  \hline 
  Sensor current vs. applied bias voltage & Sensor operability and & Q, F\\ & reference characteristics measurement &\\ 
  Power-up test & Chip supply voltage and current test & Q, F\\ 
  Shunt-LDO I/V scan & Shunt-LDO warm/cold calibration & Q, F\\ & and functionality test & \\ 

  Generic ADC test & Generic ADC warm/cold calibration and & Q, F\\ & Shunt-LDO calibration verification &\\ 
 
  Sensor bias ON & Ramp-up of sensor depletion bias & Q, F\\ 
  Digital test (high threshold) & Pixel read-out chain functionality & Q, F\\ 
  Module tuning & Multi-step threshold and feedback & Q, F\\ & current adjustment (global and pixel level) &\\
  %Tuning (-15\degree) & Multi-step threshold and feedback & F\\ & current adjustment (global and pixel level) &\\ \hline
  Digital test (operation threshold) & Pixel digital read-out chain functionality & Q, F\\ 
  Analog test (operation threshold) & Pixel analog read-out chain functionality & Q, F\\ 
  Threshold and noise measurement & Threshold and noise (ENC) measurement & Q, F\\ 
  ToT verification at the charge & Charge measurement verification & Q, F\\ of \SI{16000}{\e} (MIP) & & \\ 
  Crosstalk test & High analog charge injection & Q, F\\ & for inter-pixel crosstalk test &\\ 
  t$_0$-tuning & Injection timing fine adjustment & F\\ 
  In-time threshold measurement & Threshold for hit detection within & F\\ & single bunch crossing measurement &\\ 
  ToT calibration & Full range ToT to charge calibration & F\\ 
  Noise occupancy measurement & Noise hit probability at operation threshold & F\\ & and noisy pixel masking & \\ 
  Source scan & $^{241}$Am high statistics source scan & F\\ & for bump connectivity test &\\ 
  Low threshold operability test & Noise occupancy as function & F\\ & of threshold measurement & \\ 
  Sensor bias OFF & Ramp-down of sensor depletion bias & Q, F\\ 
  Threshold and noise measurement & Threshold and noise (ENC) measurement & Q, F\\ & with undepleted sensor to detect defective bumps &\\
  \hline \hline
\end{tabular}
\caption{Test flow of the initial electrical test after module assembly (Q) and the intense module functionality and performance validation test (F). The tests labelled Q are made at room temperature.}
\label{tab:testsQA}
\end{table}

Prior to the loading of modules onto a stave, each module was tested to ensure its mechanical and electrical functionality, its tolerance to environmental stress, and its electrical performance. The performance validation identified both module-level failures and pixel-level failures. Modules were selected for stave loading on the basis of this validation and only modules passing all tests were selected. At the pixel level, any pixel that failed at least one electrical test was recorded and modules with a bad pixel count of more than \SI{1}{\percent} were rejected. The bad pixel category also included bad pixels resulting from nonconformities such as re-work, sensor scratches or the chipping of electronic components.

An initial electrical verification, including the sequential room-temperature tests labelled Q in Table~\ref{tab:testsQA},  
was performed after the module assembly. Modules accepted by this initial test were then subjected to an
environmental stress test of \SI{10}{} thermal cycles between \SI{-40}{\celsius} and \SI{40}{\celsius}. The modules were not powered during the thermal cycles.
Seventy-three modules (\SI{10.6}{\percent} of those delivered) were rejected at this stage because of major mechanical or electrical failure, a large fraction being due to a defective on-chip power regulator on the FE-I4B chips. As noted in Section~\ref{sec:mod_electronic}, the power regulators were not tested at the wafer level. 
%All modules failing this initial electrical test were rejected. 

An extensive validation stage was then made for each module at both the module and individual pixel level. This included the sequential tests labelled F in Table~\ref{tab:testsQA}. The different measurements are described in Sections~\ref{sec:module_IV_qa} through~\ref{sec:module_bump_bond_connect_qa}. 

A measurement of the sensor I-V was initially made at room temperature with a requirement on the breakdown voltage (V$_{\SI{}{bd}}$) depending on the module type. Modules failing the V$_{\SI{}{bd}}$ requirement were rejected.
A detailed electrical calibration and characterisation was then made at the foreseen detector operation temperature of \SI{-15}{\celsius}. 
This included the module timing and threshold calibration and validations of the operational range (for example the low threshold operability). 
Finally, pixel-level failures, for example threshold tuning failures, bump-bond failures and noisy pixels, were identified and recorded. Forty-one modules (\SI{6}{\percent} of the delivered modules) were rejected following this detailed electronic validation. Accepted modules were ranked according to their quality, and as discussed in Section~\ref{sec:stavrank}, an additional penalty was applied to a module in case of mechanical rework or any other problems in the production and testing procedure.

\subsubsection{Module I-V characteristics}
\label{sec:module_IV_qa}

\begin{figure}[h]
	\centering
	\begin{subfigure}[t]{0.49\textwidth}
		 \includegraphics[width=\textwidth]{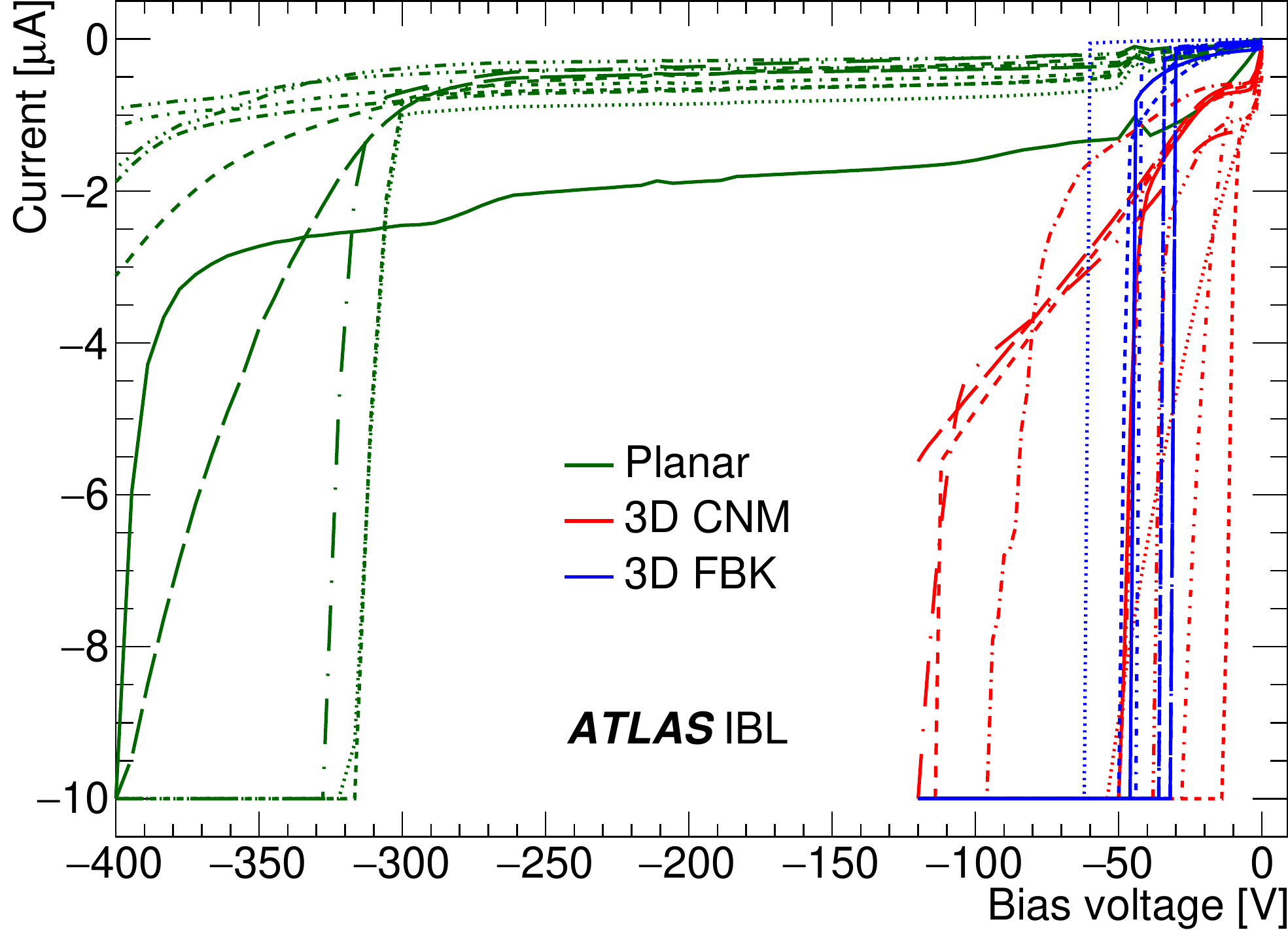}
		\caption{}
		\label{fig:IVten}
	\end{subfigure}
	\begin{subfigure}[t]{0.49\textwidth}
		 \includegraphics[width=\textwidth]{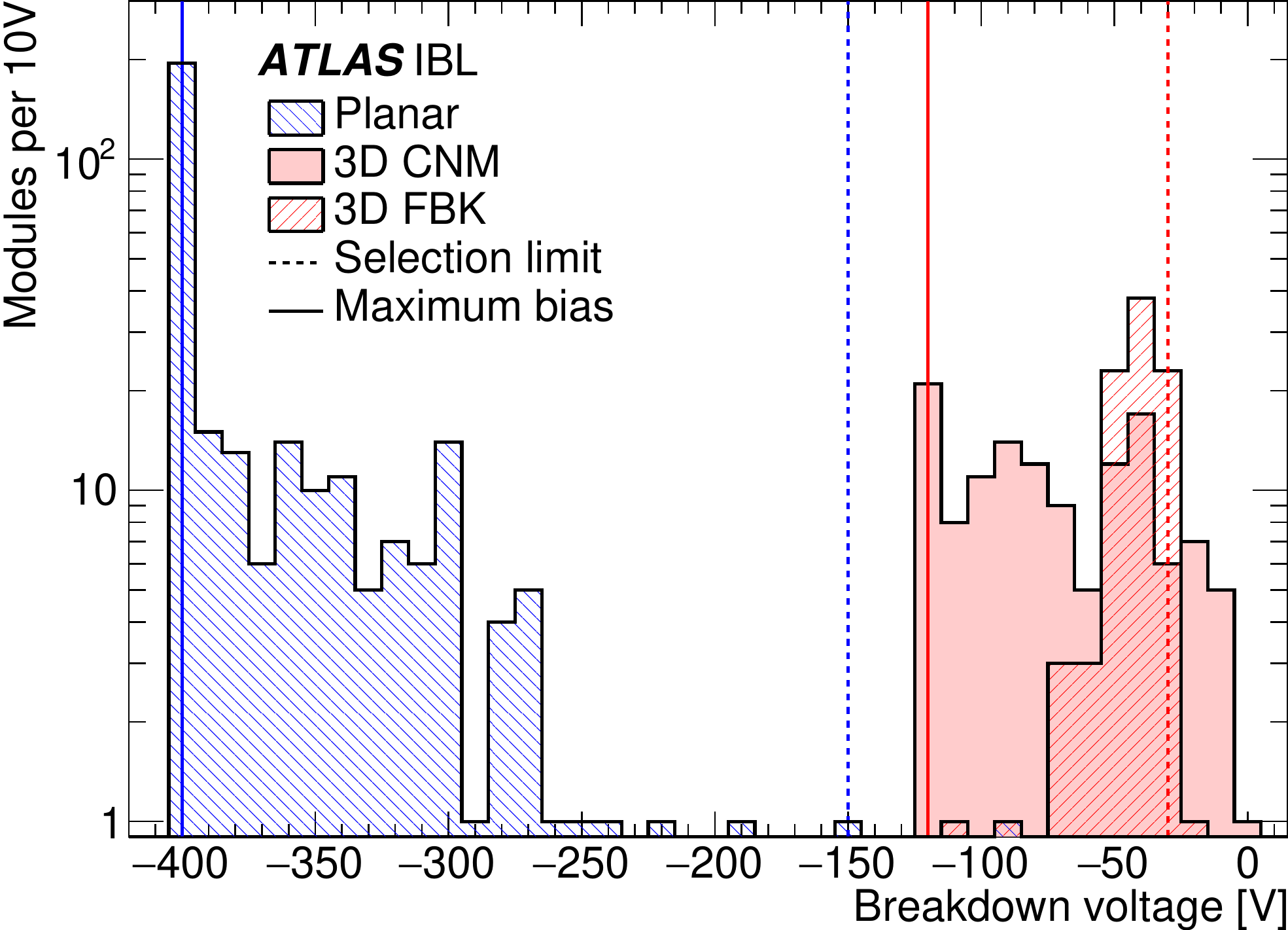}
		\caption{}
		\label{fig:VbdLogY}
	\end{subfigure}
  \caption{\subref{fig:IVten}  The I-V characteristics of ten randomly chosen modules of each sensor type. The I-V curves were measured at approximately \SI{20}{\celsius} and a current limit of \SI{-10}{\micro\ampere} was used to protect the modules. The FE-I4B chips were not powered. \subref{fig:VbdLogY} The breakdown voltage (V$_{\SI{}{bd}}$) distributions of the tested IBL modules (313 planar, 128 3D CNM and 93 3D FBK modules). The solid vertical lines indicate the maximum measurement point of \SI{-400}{\volt} for the planar modules and \SI{-120}{\volt} for the 3D CNM and FBK modules. The V$_{\SI{}{bd}}$ acceptance criteria of below \SI{-150}{\volt} (blue) for the planar modules and below \SI{-30}{\volt} (red) for 3D modules are also indicated as dashed lines. 
 }
	\label{fig:modVbd}
\end{figure}

The leakage current was measured as a function of the sensor bias voltage (I-V characteristic) during both the initial room-temperature electrical test and the full performance %validation 
test. The breakdown voltage (V$_{\SI{}{bd}}$) was used as an acceptance criterion for the modules.
Example I-V curves for ten randomly chosen modules of each module type are shown in Figure \ref{fig:IVten}. From these curves, V$_{\SI{}{bd}}$  can be determined. The I-V behaviour was measured with the FE-I4B chip unpowered and at approximately \SI{20}{\celsius}. A current limit of \SI{-10}{\micro\ampere} was used to protect the modules.

%AGC-The planar modules fulfill the cut criteria if there is a difference of more than \SI{70}{\volt} between the breakdown voltage and the operational voltage of \SI{-80}{\volt}. 
As shown in Figure \ref{fig:VbdLogY}, V$_{\SI{}{bd}}$ depends on the module type.
The V$_{\SI{}{bd}}$ of planar modules was required to be less than \SI{-150}{\volt}, \SI{70}{\volt} below the nominal bias voltage, V$_{\SI{}{op}}$~=~\SI{-80}{\volt}.
 %(V$_{\SI{}{op}}$) of \SI{-80}{\volt}). 
Only four of the dressed planar modules failed the sensor breakdown voltage criterion.
The V$_{\SI{}{op}}$ of the 
%93 FBK and 128 CNM modules 
FBK and CNM modules is \SI{-20}{\volt}, and for this reason all 3D modules having a V$_{\SI{}{bd}}$ above \SI{-30}{\volt} at the final performance test were rejected. As already noted  (Section~\ref{sec:sensor_qa}), the sensor test procedure at wafer level was significantly different for the CNM and the FBK modules.
For the 3D FBK modules, only one module was rejected. However, for CNM modules, the correlation between V$_{\SI{}{bd}}$ at wafer level and for dressed modules was poor. For this reason, the wafer-level criteria were relaxed and 27 of the dressed CNM modules failed the V$_{\SI{}{bd}}$ requirement. Furthermore, in some cases, the early soft breakdown is thought to originate in the p-stop region around the n-columns. After radiation, the leakage current due to radiation damage in the silicon bulk dominates.
%\clearpage

\subsubsection{Module time-walk and threshold tuning}
\label{sec:module_elec_noise_tw_qa}

During detector operation, the IBL modules digitise the measured hits with respect to the master clock, which is synchronised to the LHC clock. Only hits that are recorded within one clock cycle, i.e. within a sensitive time of \SI{25}{\nano\second}, can be assigned to the correct bunch crossing of the LHC. The in-time hit detection probability is significantly influenced by the time-walk effect of the charge sensitive amplifier. Small signal charges at the input of the amplifier cross the discriminator threshold with some time delay with respect to a large reference charge and therefore the knowledge of this time-walk is important for the IBL operation.

\begin{figure}[htbp]
\centering
    \begin{subfigure}[t]{0.49\textwidth}
        \includegraphics[width=\textwidth]{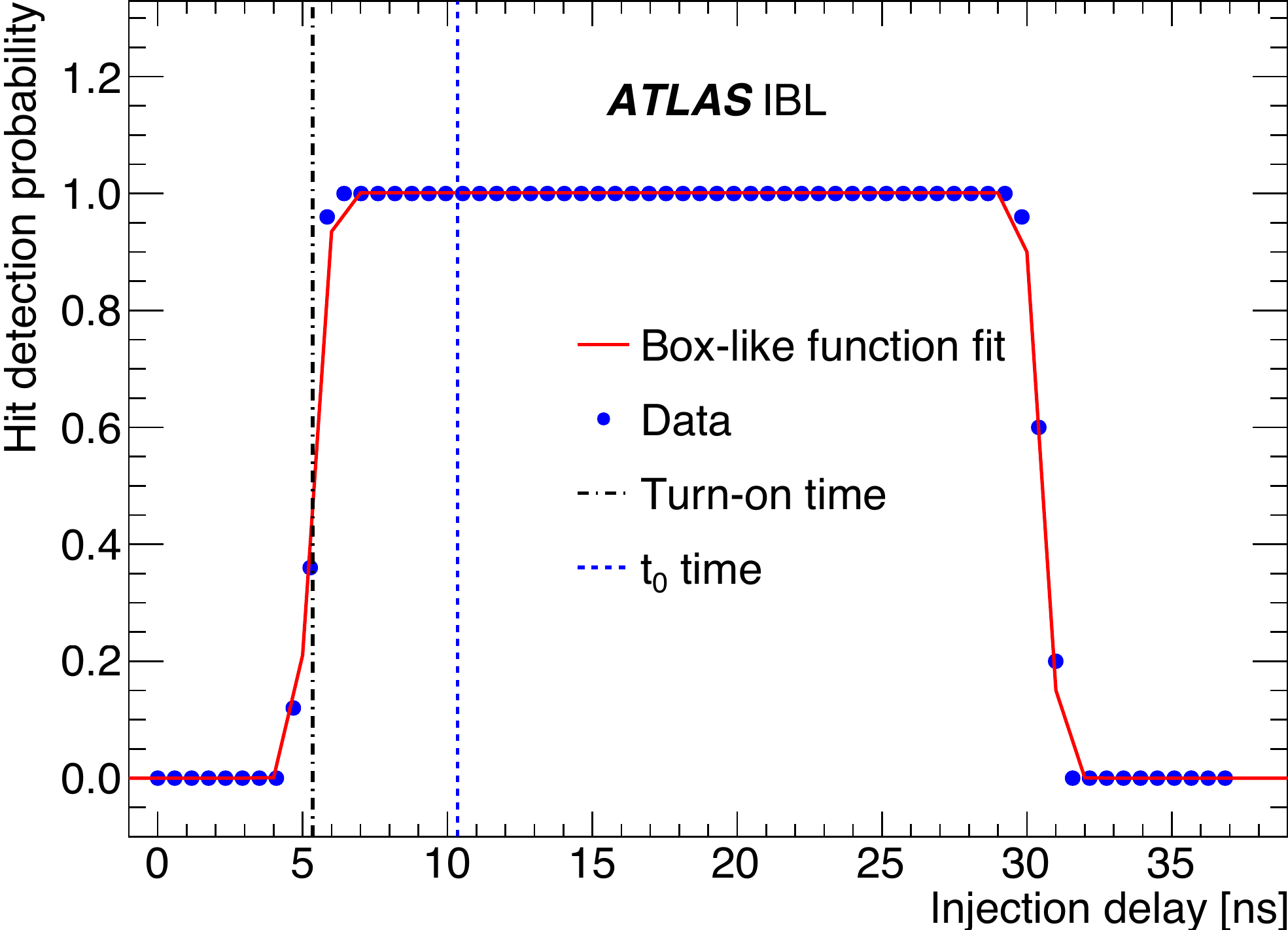}
        \caption{}
		\label{fig:T0_Scan}
	\end{subfigure}
	\begin{subfigure}[t]{0.49\textwidth}
    \includegraphics[width=\textwidth]{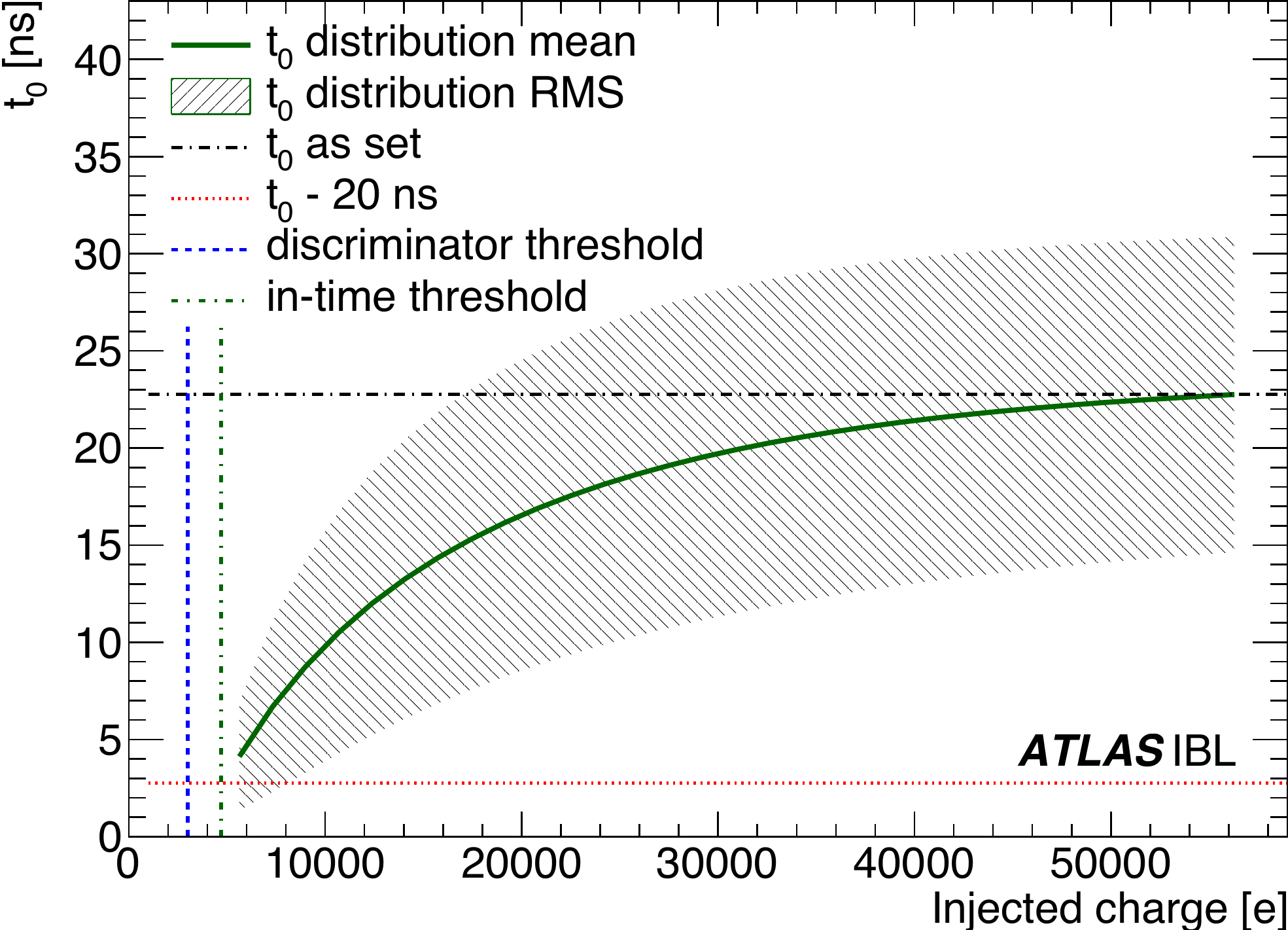}
        \caption{}
		\label{fig:TimewalkMeasure}
	\end{subfigure}
\caption{\textbf{}(a) The single pixel hit detection probability of one pixel of a typical chip during a t$_{\SI{}{0}}$ scan with a high injected charge (approximately \SI{10}{\kilo\e}), and (b) the measured mean t$_{\SI{}{0}}$ as a function of the charge of the entire chip. Analog injections with high charge are performed as a function of the on-chip injection delay. Hits are digitised during a single cycle of the master clock of the chips. The hit detection time is determined using a box fit convoluted with a Gaussian shape (box-like). The t$_{\SI{}{0}}$ is set to the hit detection time plus a safety margin of \SI{5}{\nano\second}. For the full pixel matrix the measured mean t$_{\SI{}{0}}$ is shown as a function of the injected charge. The effect of the time-walk is visible for small charges.}
    \label{fig:modules:Timing_Scans}
\end{figure}

To measure the time-walk, the time difference between the arrival time of the signal charge at the input of the amplifier and the time at which the amplifier output voltage crosses the discriminator threshold %AGC 190118 (the hit detection time) 
needs to be determined. During this measurement, the charge is generated by the on-chip charge injection circuitry and thus the signal charge arrival time is 
determined by %AGC 220118 given by 
the charge injection time, which needs to be precisely measured and adjusted. The FE-I4B chip has adjustable on-chip chip-level injection delay circuitry that is used to tune the charge injection timing. The circuitry delays the injection timing globally with respect to the chip master clock, thus decreasing the time 
%AGC 220118 add "difference"
difference 
between the charge injection and the digitisation time window. The injection delay is scanned and the hit detection probability is measured as a function of the delay setting for a large injected charge, using a single clock cycle digitisation window. This results in a box-shaped function as shown in Figure~\ref{fig:T0_Scan} for a single pixel, in this case for an injected charge of approximately \SI{10}{\kilo\e}. 
%AGC 190118 The hit detection time is defined by the \SI{50}{\percent} hit detection probability. 
The time difference between the master clock and the charge injection (i.e. injection delay) for a \SI{50}{\percent} hit detection probability is 
defined to be the hit detection time.
%AGC 220118The hit detection time is defined to be the injection delay for a \SI{50}{\percent} hit detection probability. 
The difference between the hit detection time for two consecutive digitisation windows is known to be \SI{25}{\nano\second} (one bunch crossing) and this is used to calibrate the step-width of the delay circuitry. The step-width of this particular FE-I4B chip is \SI{0.58}{\nano\second}. 
The mean hit detection time of the full pixel array is measured and the time t$_{\SI{}{0}}$ is defined to be the mean hit detection time of the chip plus a safety margin of \SI{5}{\nano\second} to ensure that early pixel timings are not excluded.
 
The calibration of the internal injection timing was verified with reasonable agreement using charge pulses induced in a planar sensor using a picosecond \SI{671}{\nano\meter} laser. The scanning procedure measuring t$_{\SI{}{0}}$ is similar to that based on internal injection but with the injection time being controlled outside the FE-I4B chip, thus providing an important cross check.

As shown in Figure~\ref{fig:TimewalkMeasure} the value of t$_{\SI{}{0}}$ measured as a function of the injected charge reveals the effect of time-walk. For high charges the mean t$_{\SI{}{0}}$ saturates at a fixed delay. For small charges, closer to the discriminator threshold, t$_{\SI{}{0}}$ is smaller, i.e. the time between charge injection and the digitisation time window is larger. Given the \SI{5}{\nano\second} 
%AGC 200118
safety margin 
added for early signals, the time-walk must not exceed \SI{20}{\nano\second} for an injected charge to be detected in the correct digitisation window, i.e. the correct bunch crossing. The time-walk can be related to the input charge in electrons using Figure~\ref{fig:modules:Timing_Scans}. The charge corresponding to \SI{20}{\nano\second} time-walk is called the overdrive, and the in-time threshold is the sum of threshold plus overdrive. Injected charges greater than the in-time threshold will be detected in the correct bunch crossing. Smaller hits will be 'out-of-time'. Out-of-time hits can be recovered with on-chip processing. 
In the FE-I3 chips of the Pixel detector, 
there is a function to duplicate all hits below a programmable ToT value to the prior bunch crossing, at the cost of a significantly increased data volume.
The FE-I4B chip contains a more sophisticated recovery method that limits the impact on the data volume. Hits with a ToT value of 1 or 2 (optional and programmable) can be replicated in the previous bunch crossing assignment, if they are adjacent to a larger hit. This exploits the fact that low-charge hits are mostly due to charge sharing, since charged particles are unlikely to produce very low-charge single-pixel clusters.

\begin{figure}[htbp]
	\centering
	\begin{subfigure}[t]{0.49\textwidth}
		 \includegraphics[width=\textwidth]{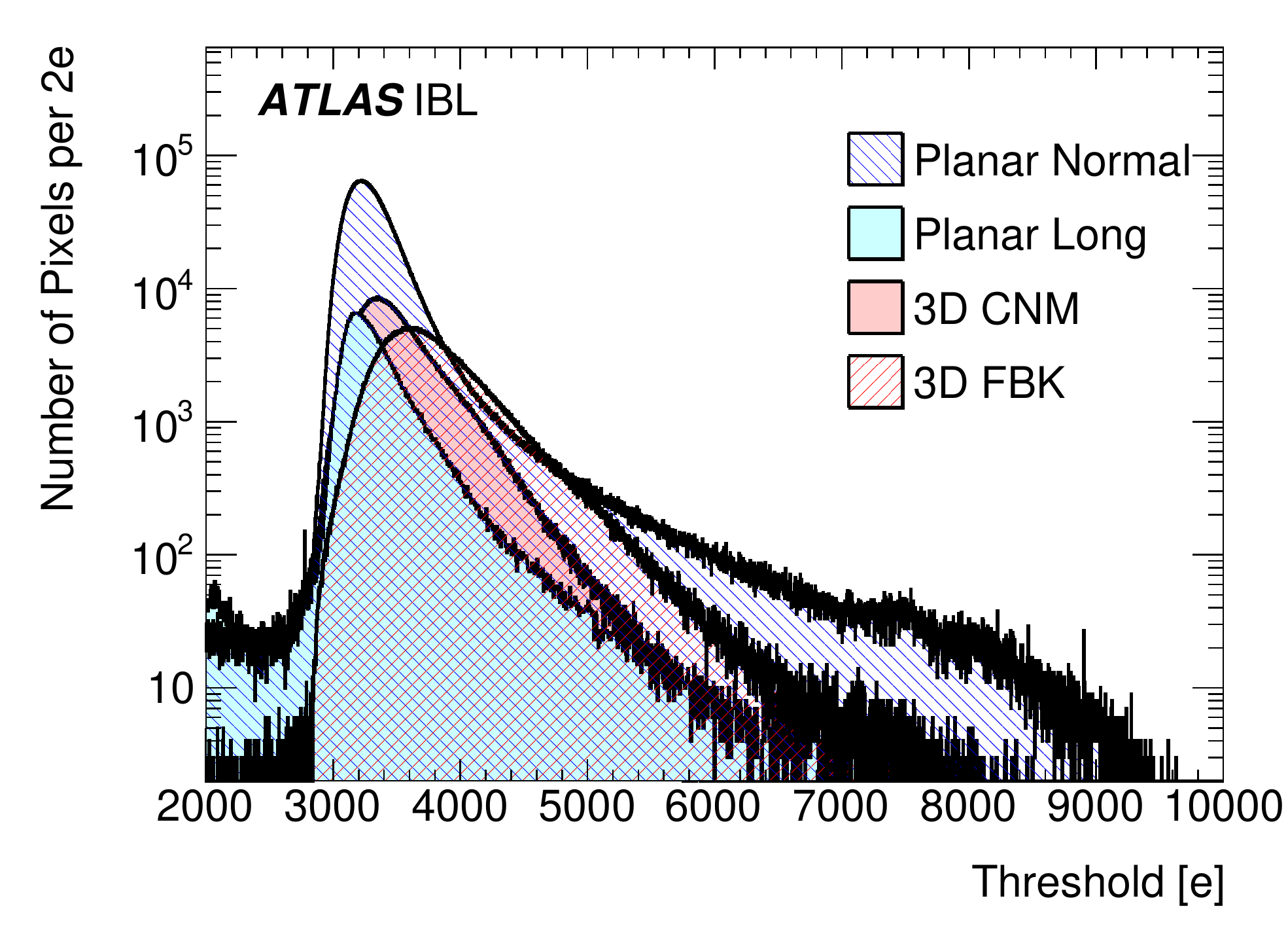}
		\caption{}
		\label{fig:InTimeLogY}
	\end{subfigure}
	\begin{subfigure}[t]{0.49\textwidth}
		 \includegraphics[width=\textwidth]{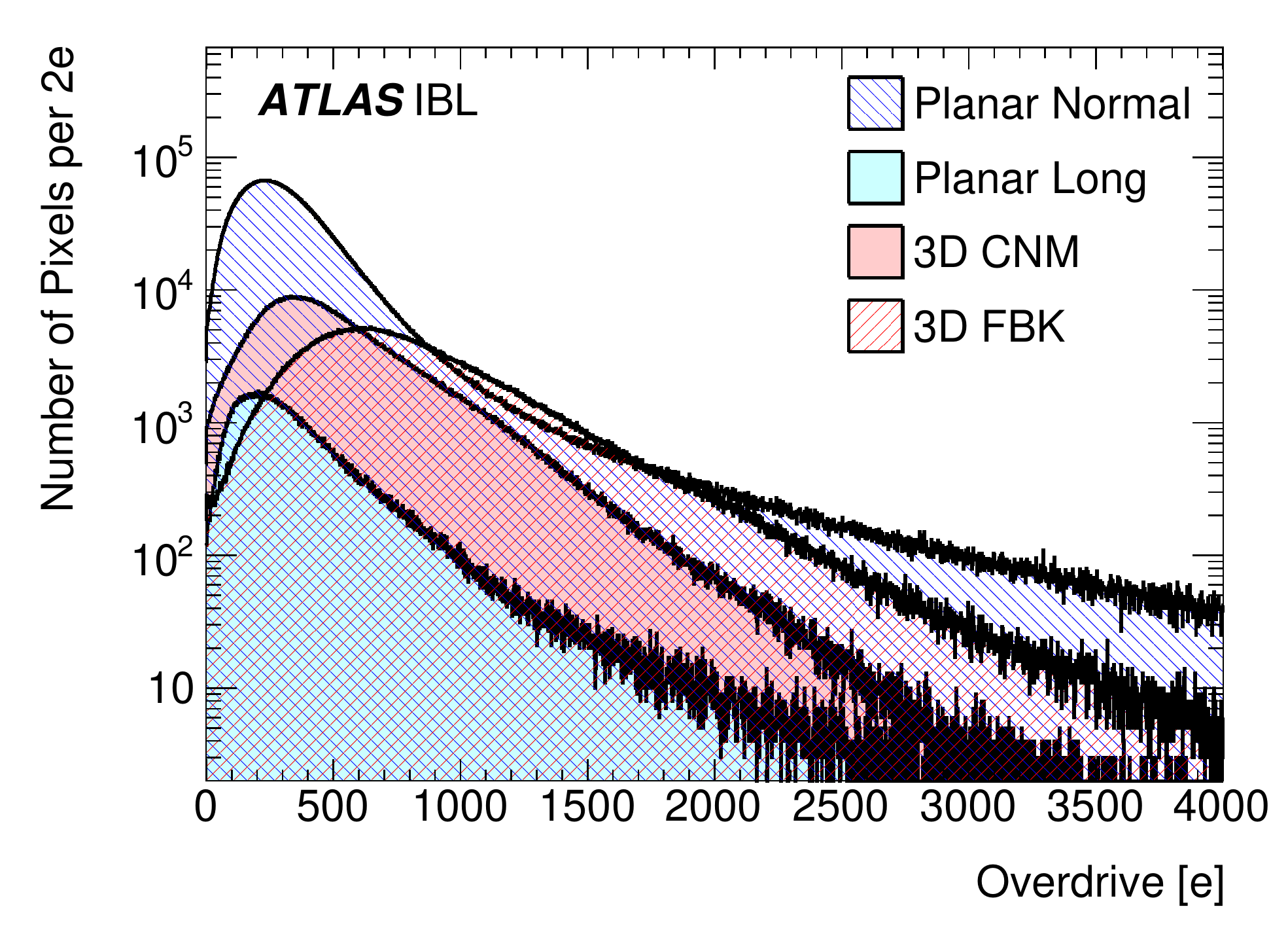}
		\caption{}
		\label{fig:OverdriveLogY}
	\end{subfigure}
  \caption{\textbf{}The distribution for individual pixels of (a) the in-time threshold and (b) the charge overdrive above the discriminator threshold needed for the hit to be detected within one bunch crossing. All modules matching IBL quality criteria are shown. The overdrive is measured after tuning the modules to \SI{3000}{\e} threshold and 10 ToT counts for a charge injection of \SI{16}{\kilo\e} at approximately \SI{-15}{\celsius}. The mean in-time threshold varies between \SI{3354}{\e} 
for the normal planar pixel and  \SI{3820}{\e} for the 3D FBK pixel. 
Pixels failing the measurement are not included in the distributions, and pixel overflows are not shown.}
	\label{fig:modTimewalk}
\end{figure}

The tuning algorithm used to generate the initial module configurations included a global threshold adjustment, a module feedback current  and threshold tuning, and a final iterative pixel-level feedback current and threshold tuning. The threshold was measured using a known injected charge and measuring the \SI{50}{\percent} hit efficiency, initially at the global level and then at the pixel level. The mean value for each FE-I4B chip was tuned to be \SI{3000}{\e} at a nominal temperature of \SI{22}{\celsius}.

The in-time threshold can also be measured using a threshold scan algorithm with a single bunch crossing read-out, following a t$_{\SI{}{0}}$ adjustment. Therefore, the time-walk could be measured during the IBL module production using the so-called 
overdrive measurement (calculated for each pixel as the difference of in-time threshold and the discriminator threshold). Both the in-time threshold and the overdrive distributions are shown in Figure \ref{fig:modTimewalk}. The in-time threshold distributions show the expected dependence on the sensor type; the detector capacitance influences the rise-time of the amplifier and thus the mean time-walk. Similar to the noise distributions for the three module types, the overdrive distribution for planar modules (mean \SI{355}{\e}, RMS \SI{250}{\e}) has a lower mean than for CNM modules (mean \SI{530}{\e}, RMS \SI{351}{\e}) and FBK modules (mean \SI{828}{\e}, RMS \SI{478}{\e}). A small number of pixel channels in the tails of the in-time threshold distribution result from poorly determined threshold fits. These mean time-walk values are well within the out-of-time hit recovery capability of approximately \SI{1500}{\e} for the FE-I4B chip~\cite{FE_I4B:2013}.

\subsubsection{Module ToT-to-charge calibration}
\label{sec:module_tot_calib_qa}

\begin{figure}[htbp]
	\centering
	\begin{subfigure}[t]{0.49\textwidth}
		 \includegraphics[width=\textwidth]{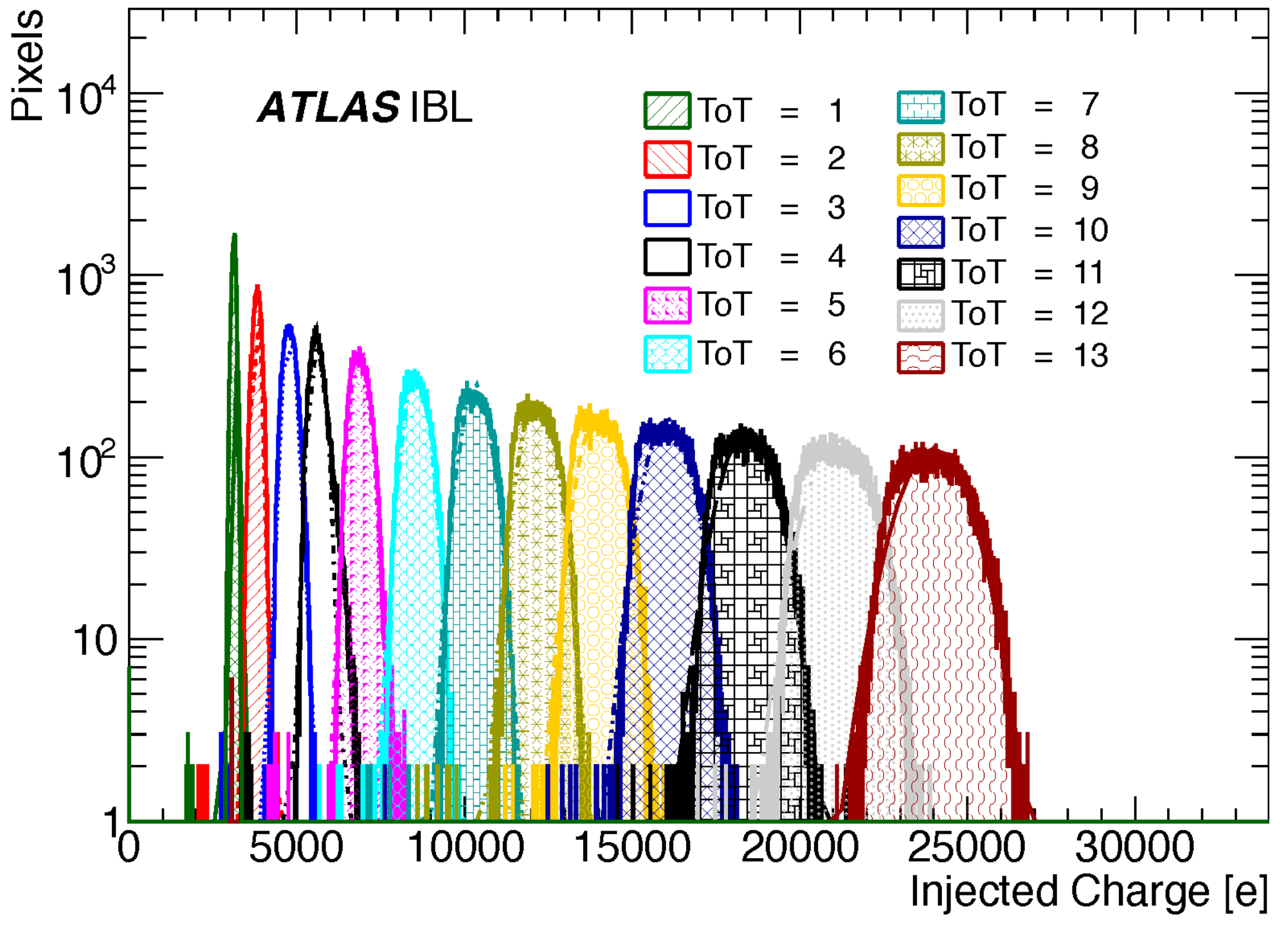}
		\caption{}
		\label{fig:TOT_LUT}
	\end{subfigure}
	\begin{subfigure}[t]{0.49\textwidth}
		 \includegraphics[width=\textwidth]{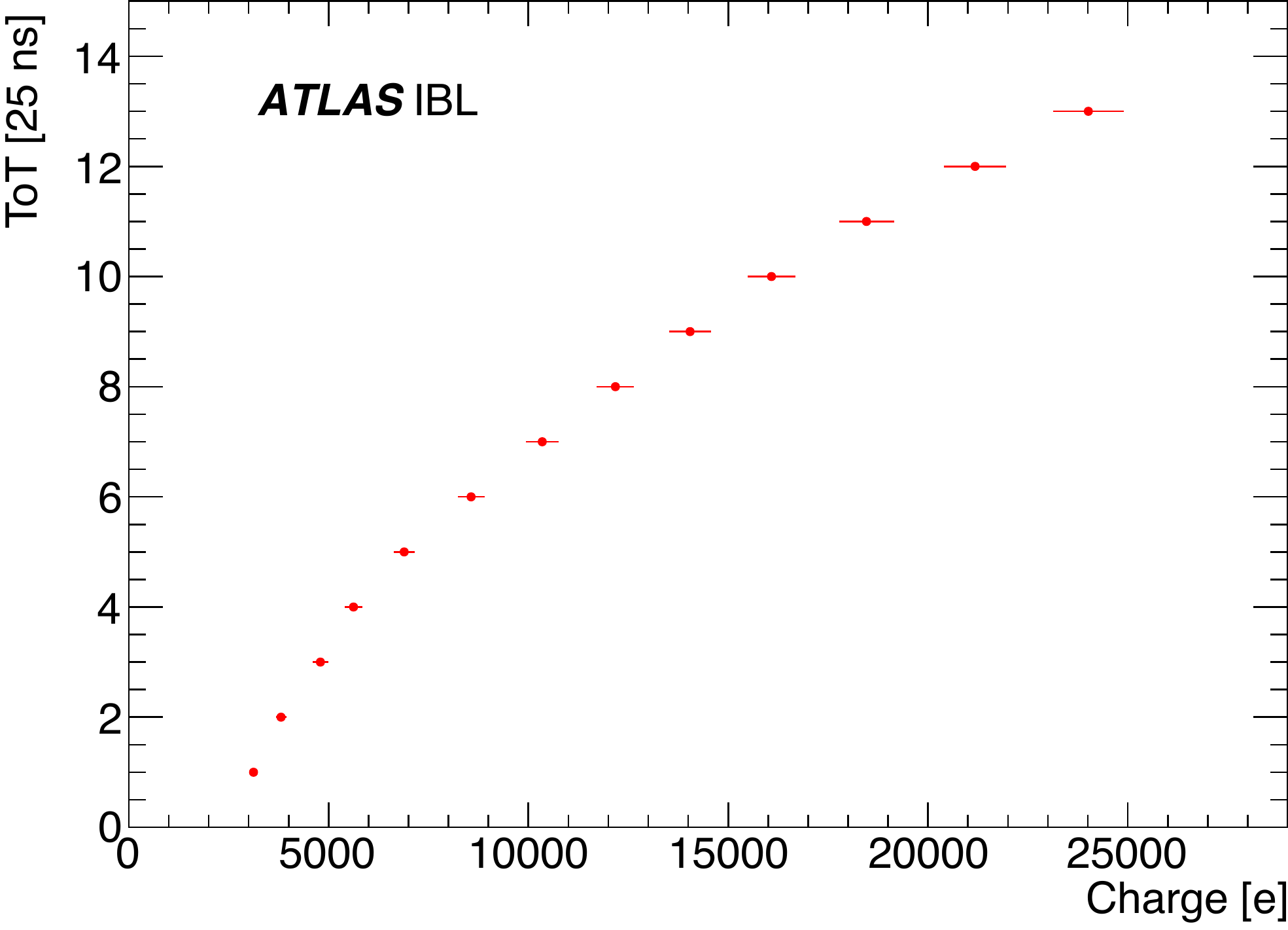}
		\caption{}
		\label{fig:ChargeCalibration}
	\end{subfigure}
	  \caption{\textbf{}Calibration function of the Time-over-Threshold (ToT) to charge: \subref{fig:TOT_LUT} for a given ToT value, the mean injected charge of each pixel, and \subref{fig:ChargeCalibration} the charge-to-ToT calibration function. Measurements of the charge distributions for ToT~$\ge$~14 were biased by the maximum injected charge and were excluded from the calibration.}
	\label{fig:modToTCal}
\end{figure}

The ToT is calibrated 
%AGC tuned 
%AGC tuning is made 
after the discriminator threshold calibration, as that affects the ToT along with the feedback amplifier current, and vice versa. 
The limited available 
%analog hit 
charge information due to 
%AGC the low number of 
only four ToT bits complicates the ToT-to-charge calibration. 
A calibration method was implemented to measure the injected charge histograms for a specific value of the ToT. For each pixel the injected charge value is stored for the pixel that responds with the chosen ToT. This results in charge histograms as shown in Figure \ref{fig:TOT_LUT}. The mean values and the widths of the distributions of injected charge are used as a look-up table for the charge-to-ToT calibration function (Figure \ref{fig:ChargeCalibration}). 
The measurement was made on each IBL module, initially for a given chip and then at the pixel level, and the result was stored for each pixel. Because of the maximum injected charge, the ToT~$\ge$~14 distributions were biased, and excluded from the calibration. 

\subsubsection{Module electronic noise}
\label{sec:module_elec_noise_qa}

\begin{figure}[htbp]
	\centering
	\begin{subfigure}[t]{0.49\textwidth}
		 \includegraphics[width=\textwidth]{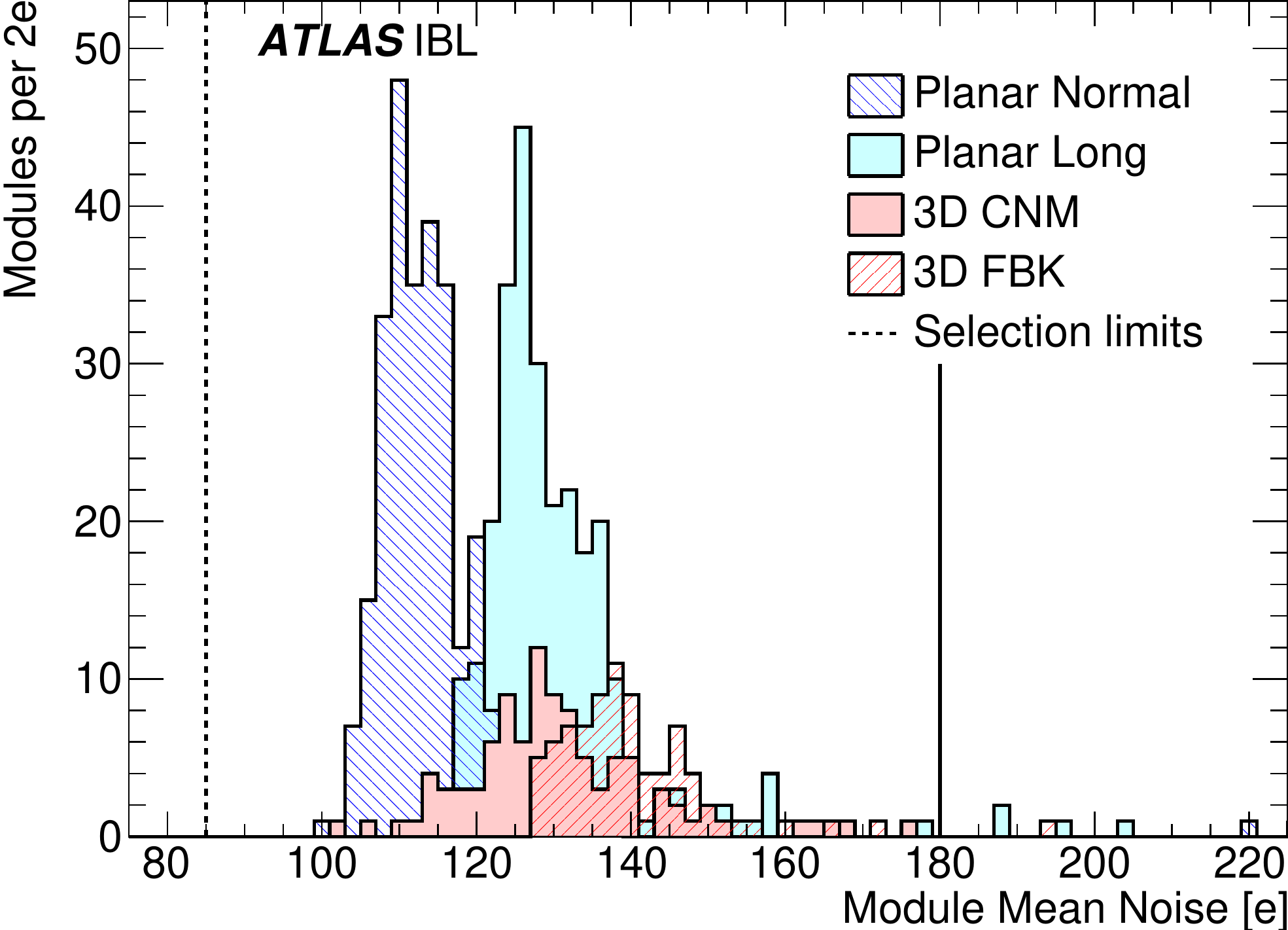}
		\caption{}
		\label{fig:NoiseMean}
	\end{subfigure}
	\begin{subfigure}[t]{0.49\textwidth}
		 \includegraphics[width=\textwidth]{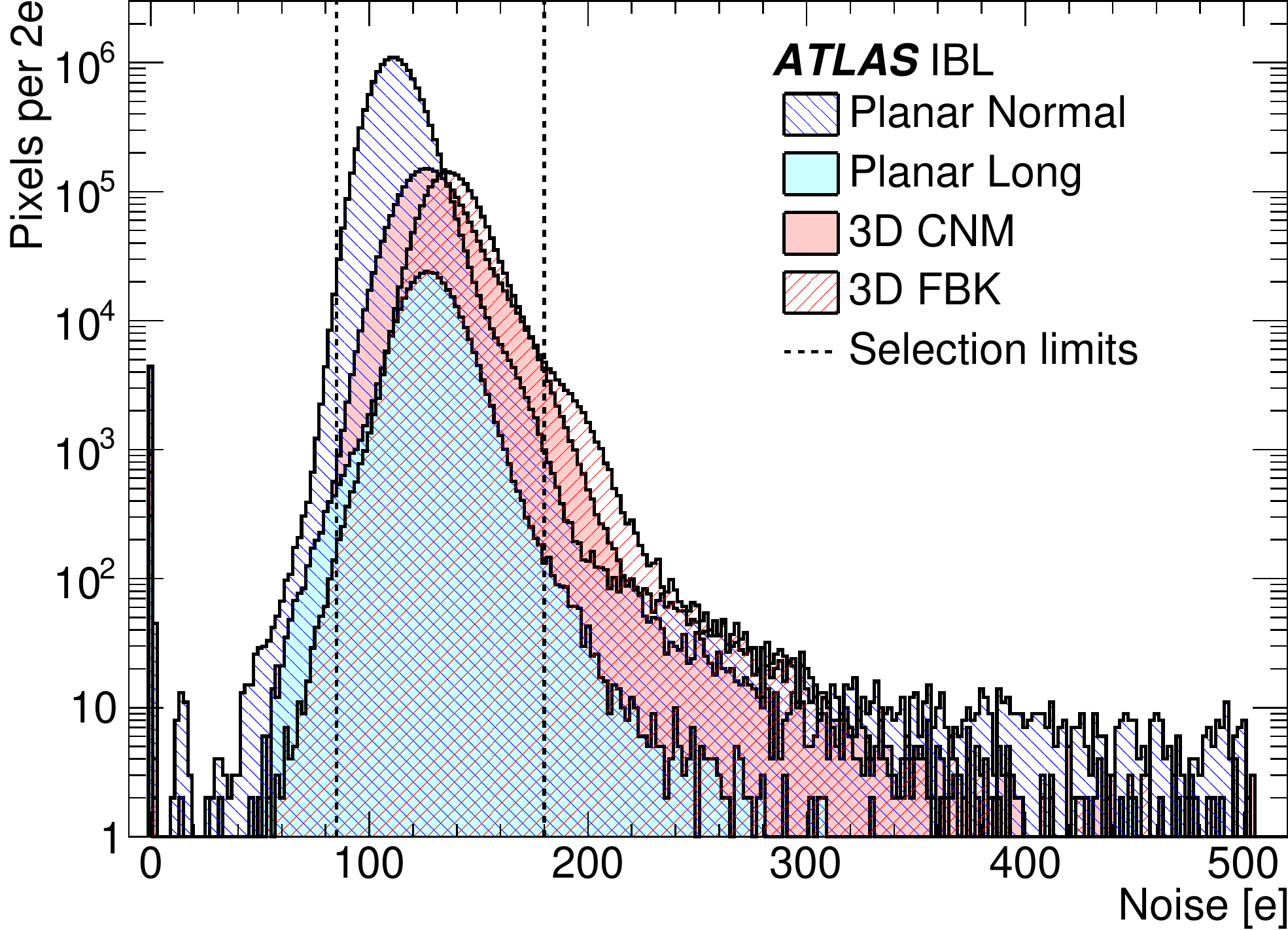}
		\caption{}
		\label{fig:NoiseLogY}
	\end{subfigure}
  \caption{\textbf{}For all modules matching the IBL quality criteria, \subref{fig:NoiseMean} the module mean noise (ENC) distribution, and \subref{fig:NoiseLogY} the individual pixel noise (ENC) distribution. The modules are tuned to a \SI{3000}{\e} threshold, and 10 ToT counts for a charge injection of \SI{16}{\kilo\e}, at a nominal operating temperature of \SI{-15}{\celsius}. 
The minimum and maximum selection limits for the mean module noise (a module selection criterion), and for the noise of individual pixels (a bad pixel criterion), are shown as vertical lines in each figure. Plot overflows are not shown. 
}
	\label{fig:modNoise}
\end{figure}

The average noise of a single pixel is measured in units of %the same output signal as the noise and is called 
Equivalent Noise Charge (ENC)\footnote{Noise is usually measured in charge units and is defined as the charge necessary at the input of an amplifier to generate the same output signal as the observed noise. This quantity is called Equivalent Noise Charge (ENC).}.
%is one of the most important figure of merit for the performance of a pixel detector module AGC-230817
The pixel noise is an important figure of merit for the pixel module performance. It is %The ENC of a pixel is %expected to be 
mainly determined by the capacitance at the input node of the preamplifier and by the leakage current. Both depend on the sensor type. In addition, the noise (ENC) can be affected by external parameters such as the module flex hybrid circuit quality and the power supply stability. No influence of the IBL powering mode using the on-chip regulators is expected~\cite{Gonella:2012}. The noise (ENC) is evaluated from measurements of the pixel occupancy as a function of injected charge in the vicinity of the threshold value (the so-called S-curve method), using the same injected charge circuitry as that used for the pixel ToT calibration and threshold tuning. Figure \ref{fig:NoiseMean} shows 
%the module-to-module mean ENC distributions for all three module types as measured during the IBL QA test. - AGC-230817
the mean noise (ENC) measured for each module during the full electrical test. The minimum and maximum selection limits for the mean module noise are shown by vertical lines. Figure \ref{fig:NoiseLogY} shows the noise (ENC) measured for individual pixels. 
The minimum and maximum selection limits for the noise of individual pixels are shown by vertical lines. 
Only \SI{0.6}{\percent} of the pixels fail this selection and are flagged.
The 3D modules have a higher noise (respectively \SI{140}{\e} and \SI{131}{\e} for FBK and CNM modules) in the same tuning condition as the planar modules (\SI{114}{\e}). This is expected because of the higher pixel capacitance of 3D sensors. The RMS width of all three distributions is comparable and ranges from \SI{30}{\e} to \SI{70}{\e}. 

%\clearpage

\subsubsection{Module bump-bond connectivity and individual pixel failures}
\label{sec:module_bump_bond_connect_qa}

Scans to reveal possible bump failures were made at the level of individual pixels. These scans included threshold and noise measurements without the sensor bias applied, crosstalk measurements and $^{241}$Am source measurements. 
 
 As noted in Section~\ref{sec:prod_yield}, the initial production batches suffered from a very high bump-bonding failure rate, as determined during threshold and noise measurements at the initial test phase. Production was halted until the problem was solved.

\begin{figure}[htbp]
	\centering
		 \includegraphics[width=0.70\textwidth]{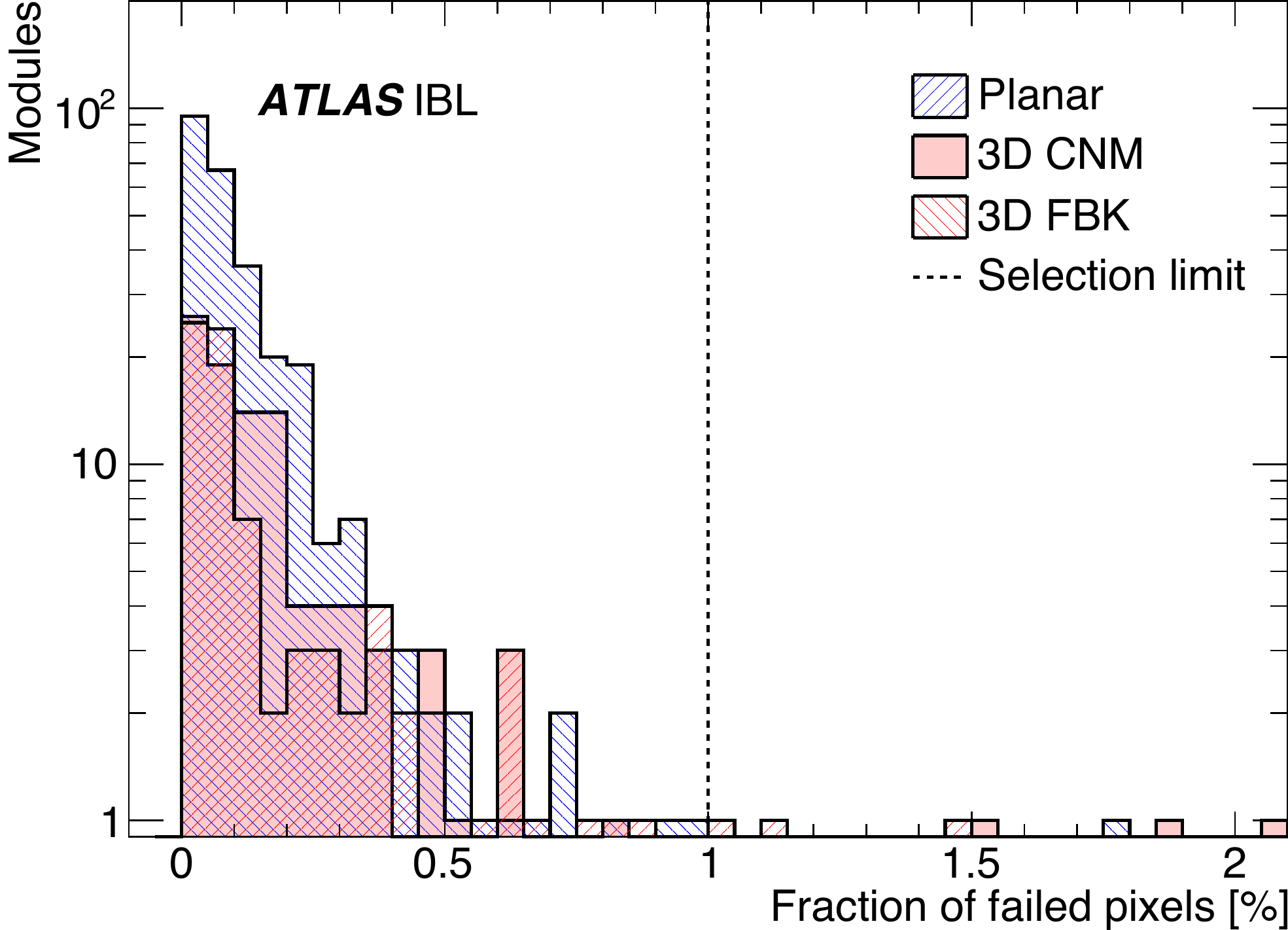}
		\caption{\textbf{}The distribution of the fraction of pixels failing the occupancy criteria in a $^{241}$Am source scan. Only modules otherwise accepted for stave loading are included. The modules are tuned to a \SI{3000}{\e} threshold, and 10 ToT counts for a charge injection of \SI{16}{\kilo\e}, at a nominal operating temperature of \SI{-15}{\celsius}. A pixel is counted as failing if it has less than \SI{5}{\percent} or more than \SI{450}{\percent} of the module's mean pixel occupancy. The selection limit is shown as a vertical line. Overflow entries are not shown.}
%Twelve overflow entries are not shown.}
		\label{fig:modSourceFail}
\end{figure}

An $^{241}$Am source scan was made on each IBL module in the final test sequence and proved to be the most reliable test for %remaining 
open bump-bond detection. 
%To limit the amount of recorded data, the ToT-to-charge conversion, averaged over each FE-I4B chip, was stored. 
Based on the individual pixel hit rate with respect to the average hit rate per pixel for each module type, noisy pixels could be identified. Remaining bump-bonding failures could also be identified as resulting from shorts (or high electronic coupling) or open bumps. 
%The $^{241}$Am source scan occupancy . 
Figure \ref{fig:modSourceFail} shows the fraction of failed pixels for each module in the final source test. Pixels with less than \SI{5}{\percent} or more than \SI{450}{\percent} of the mean pixel occupancy were considered as failing. 
%AGC- 220817 In total only a number of 19 modules have more than \SI{1}{\percent} bad pixel.
A total of 19 modules had more than \SI{1}{\percent} of bad pixels and were rejected.

\begin{figure}[htbp]
	\centering
		 \includegraphics[width=0.7\textwidth]{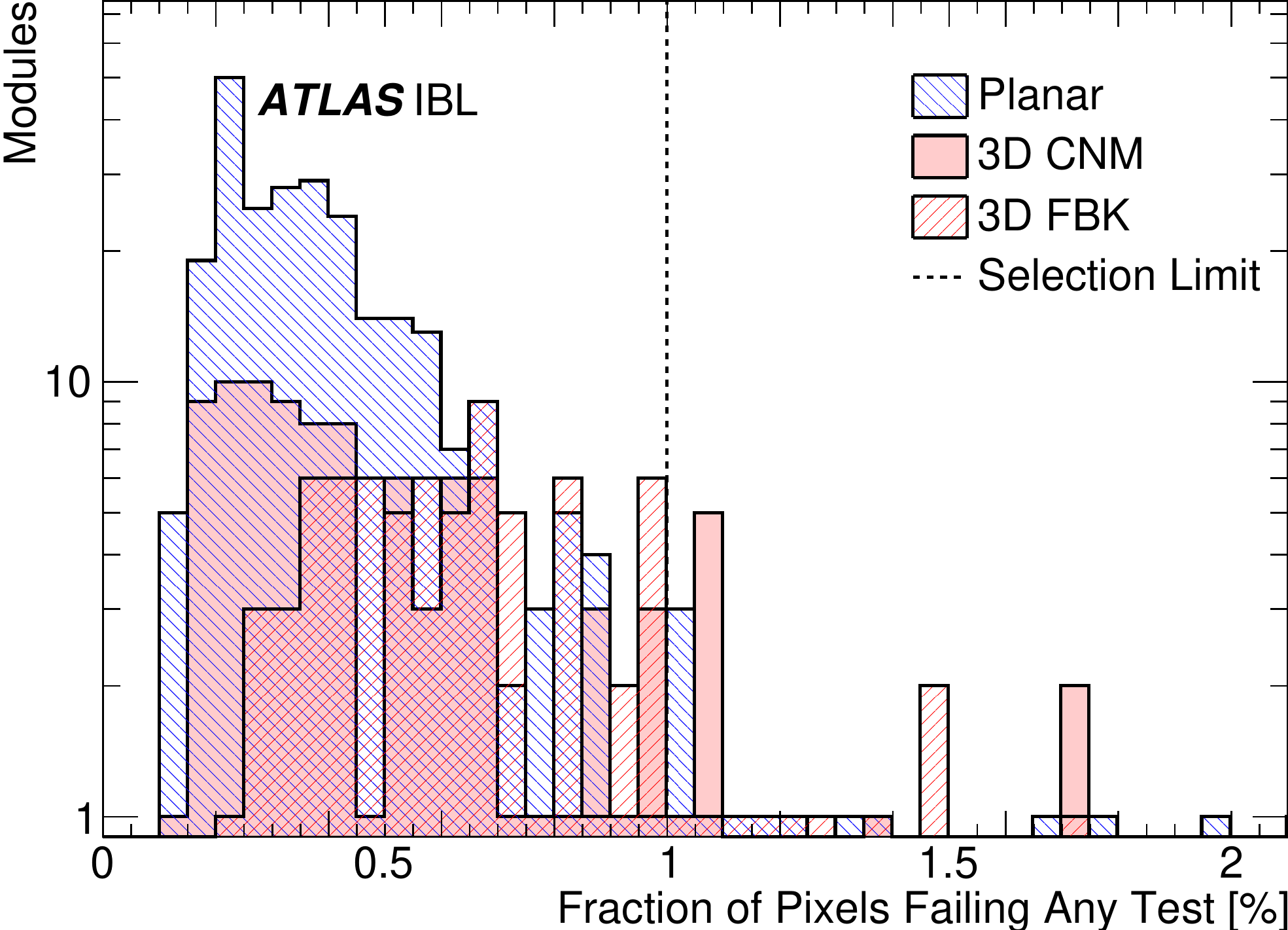}
	\caption{\textbf{}The distribution of the fraction of pixels failing in any test. Only modules otherwise accepted for stave loading are included. The modules are tuned to a \SI{3000}{\e} threshold, and 10 ToT counts for a charge injection of \SI{16}{\kilo\e}, at a nominal operating temperature of \SI{-15}{\celsius}. A pixel is counted if it fails any one of the tests as described in the text. The selection limit is shown as a vertical line. Overflow entries are not shown.}
%Twenty one overflow entries are not shown.}
\label{fig:modTotalPixFail}
\end{figure}

After each test of an individual failure mode, failing pixels were counted (with no double counting for multiple failures). 
The fraction of pixels that failed in any test is shown in Figure \ref{fig:modTotalPixFail} for all modules. This fraction was required to be less than \SI{1}{\percent}  during the module production QA and was used as a basis for the final module selection. The mean fraction of failing pixels for accepted planar modules was \SI{0.56}{\percent}. 
The mean fraction of failing pixels for CNM (\SI{0.44}{\percent}) and FBK (\SI{0.68}{\percent}) modules is comparable to the planar module distribution. 

Occupancy distributions, measured using a $^{90}$Sr source after the loading of modules onto staves, are shown for each sensor type in Section~\ref{sec:source-scans}.

\subsection{Module production and yield}
\label{sec:prod_yield}

The IBL module production was started in July 2012 and completed in April 2014. Bare modules were delivered in batches of about twenty modules for each module type. Fully dressed modules were then assembled, tested in detail and selected according to the quality criteria described in Section~\ref{sec:module_qa}.  Accepted modules were then sent to be loaded on staves as described in Section~\ref{section:staveload}. 

After the production of the first module batches, a high bump-bonding failure rate was observed and the module production was halted for approximately four months until the problem was understood and improved procedures were implemented. Two types of bump defects were identified: large areas of disconnected bumps and a small number of isolated bumps electrically shorted to a neighbour. 
During the production stop, both bump-bonding defects were investigated in detail in close collaboration with the bonding vendor. 
Both %bump-bonding 
problems were traced to the usage of a flux during the soldering process: the disassembled sensors and FE-I4B chips of defective modules showed polymerised flux residuals that acted as a spacer during the reflow, preventing a proper bump connection between the sensor and the FE-I4B pixel. 
For shorted pixels the results were less conclusive but flux residuals in areas with larger number of shorts were also found.
The solder flux applied to the FE-I4B chip prior to the flip-chip and reflow was replaced by glycerin as the new tacking media. With this new flip-chip method neither problem recurred.

\begin{figure}[t]
	\centering
		 \includegraphics[width=0.85\textwidth]{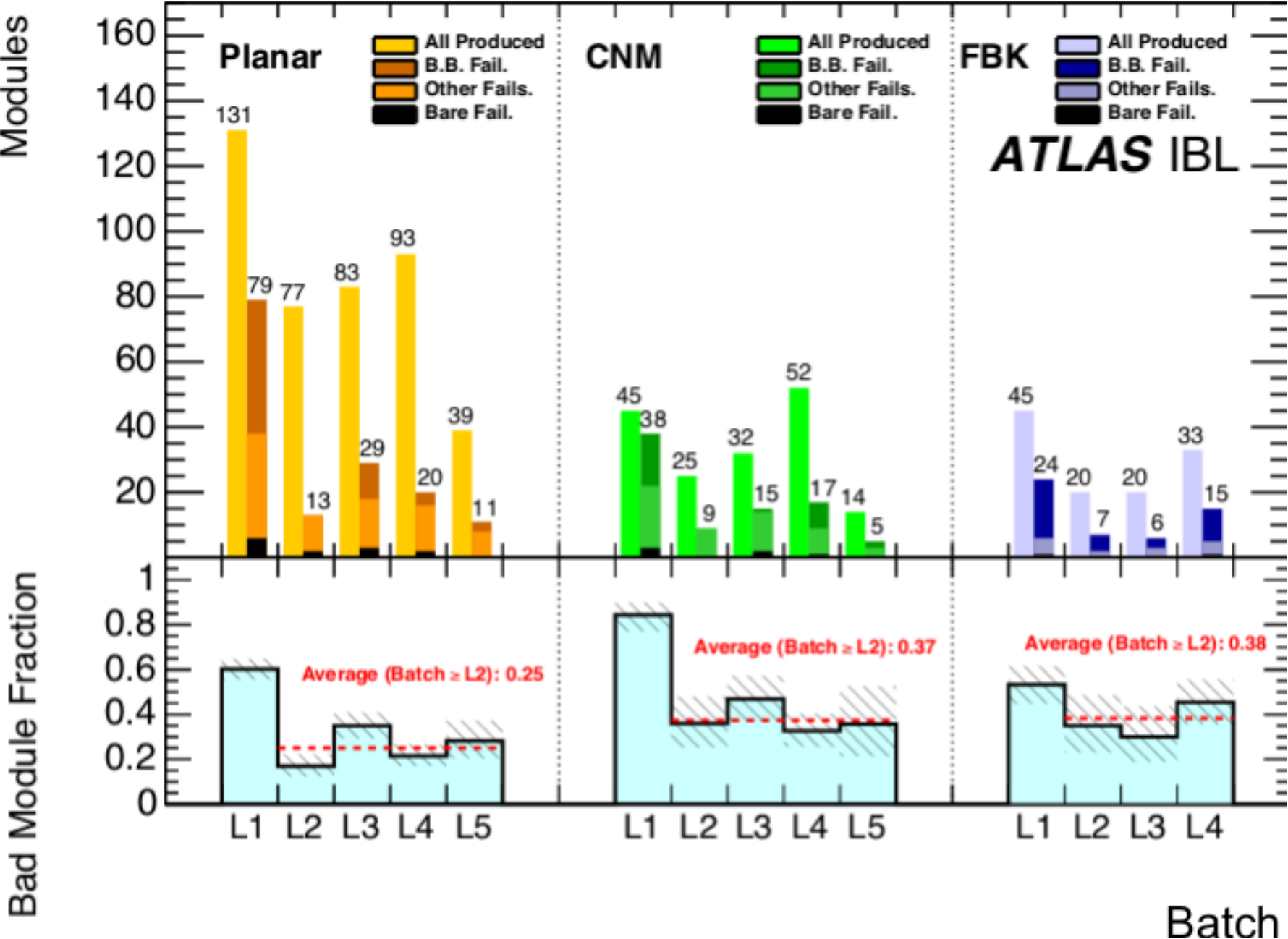}
	\caption{\textbf{}Fraction of rejected planar and 3D (CNM, FBK) IBL modules for the five production batches \SI{}{L1} - \SI{}{L5}. Within a given batch, a similar configuration of the laser vendor and the bump-bonding was applied. In the upper panel,  \textit{B.B. Fail.} denotes bump-bonding failures,  \textit{Bare Fail.} denotes modules that were not assembled due to (mainly) mechanical damage, and \textit{Other Fails.} includes both electrical and sensor failures identified after assembly.  In the lower panel, batch L1 was excluded from the mean rejected module fraction because of bump-bonding problems that were corrected in batches \SI{}{L2} - \SI{}{L5}.  Including L1, the mean rejected module fraction for respectively planar, CNM and FBK  modules was 0.36, 0.50 and 0.44.}
\label{fig:module_yield}
\end{figure}

Figure~\ref{fig:module_yield} summarises the IBL module production yield including the first batches where the bump-bonding problems were identified. In the figure the yield is expressed in terms of rejected modules, separately for planar, CNM and FBK modules. 
The yield is divided into different production batches (L1 to L5) with similar laser de-bonding and flip-chip methods applied. All modules assembled with the initial flip-chip method using solder flux are grouped in L1. 

Another change during the module assembly concerned the laser de-bonding of the glass carrier. The initial vendor was replaced by a second vendor, which was qualified during the production; the first modules from this vendor are in batch L3 for the planar modules, and L4 for the FBK and CNM modules.

The production yield of the first batch L1 was poor for all module types because of the bump-bonding problems. Respectively \SI{40}{\percent}, \SI{20}{\percent} and \SI{35}{\percent} passed the acceptance criteria for planar, CNM and FBK modules. For batches L2-L5,  the yield improved to an average of  \SI{75}{\percent}, \SI{63}{\percent} and \SI{62}{\percent}, respectively. The initial modules coming from the second laser de-bonding vendor showed a higher rate of bump-bonding defects, mainly due to problems with the handling of the thin modules during and after the glass carrier removal. 
The bump-bonding failures result mainly from mis-handling during the laser de-bonding, and not on the sensor type. The mean failure rate was less than \SI{10}{\percent} for planar and CNM modules, and \SI{25}{\percent} for FBK modules.
A summary of the failure rates, separately for the batches L\SI{1} and L\SI{2} - L\SI{5}, is reported in Table~\ref{tab:productionYield}.

\begin{table}[htb]
\centering
\begin{tabular}{lccc}
\hline \hline
            & Planar & 3D CNM & 3D FBK \\ \hline \hline
Batch L\SI{1}   &  &  &  \\  \hline
Delivered bare modules &	131 & 45 & 45 \\ 	
%\textit{Bare module Failure}        & 6      & 3    & 1  \\ 					
Bare module failures        & 6      & 3    & 1  \\ 					
Delivered dressed modules &	125 & 42 & 44 \\ 			
Accepted modules &	52 & 7 & 21 \\  \hline			
%\textit{Bare Fail.}    [\%]       & 5      & 7    & 2  \\ 					
%\textit{B. B.  Fail.}  [\%]        & 31      & 36    & 40  \\ 						
%\textit{Other Fails.}  [\%]       &24     & 42     & 11  \\ 
Bare module failures  [\%]        & 5      & 7    & 2  \\ 						
Bump-bonding (B.B.) failures  [\%]        & 31      & 36    & 40  \\ 						
Other failures  [\%]       &24     & 42     & 11  \\ 
Total  failure rate  [\%]       & 60     & 84 & 53 \\ 
\hline \hline						
Batches L\SI{2} - L\SI{5}  &  &  &  \\ \hline 
Delivered bare modules &	292 & 123 & 73 \\ 	
%\textit{Bare module Failure}        & 7      & 3    & 1  \\ 					
Bare module failures        & 7      & 3    & 1  \\ 					
Delivered dressed modules &	285 & 120 & 72 \\ 					
Accepted modules &	219 & 77 & 45 \\ 	
\hline		
%\textit{Bare Fail.}    [\%]       & 2      & 2    &	1  \\ 					
%\textit{B. B.  Fail.}  [\%]        & 6      & 9    &	25  \\ 						
%\textit{Other Fails.}  [\%]       & 16     & 26 &	12  \\ 
Bare module failures  [\%]        & 2      & 2    & 1  \\ 						
Bump-bonding (B.B.) failures  [\%]        & 6      & 9    &	25  \\ 						
Other failures  [\%]       & 16     & 26 &	12  \\ 
Total  failure rate      [\%]      & 25     & 37 &	38  \\
\hline \hline				
\end{tabular}
\caption{IBL module failures and the failure rate as a percentage of the bare modules delivered to the assembly sites. Failure modes are the same as Figure~\ref{fig:module_yield}. The failure rate is significantly larger for the first batch, as explained in the text.}
\label{tab:productionYield}
\end{table}

Other defects observed during the module production were mainly of mechanical or electrical nature. %origin. 
During the production a constant electrical failure rate of approximately \SI{15}{\percent} was observed, mainly from the on-chip regulators, which were not tested during wafer probing of the FE-I4B wafers. Other electrical problems included failing double columns of the FE-I4B chip, and communications issues, but the rates were low. Several CNM modules in L2 and L3 showed problems with a low sensor breakdown voltage, due to insufficient testing procedures during the sensor wafer QA (Section~\ref{sec:sensor_qa}). CNM sensors used for the batches L4 and L5 were re-tested after UBM deposition and dicing and the yield improved slightly. However, this re-testing introduced new defects on the sensors' pixel side, increasing the bump-bonding failure rate.

The quality of each accepted module was characterised on the basis of:
\begin{itemize}
\item[-] {The number of bad channels, based on source scan results, digital tests and other analog measurements at the operating temperature;}
\item[-] {Electrical measurements such as the mean noise, threshold, noise dispersion, and the regulator voltages;}
\item[-] {Mechanical anomalies or damage including non-conformal glue distribution, invisible alignment marks and reworked wire bonds.}
\end{itemize}
 The highest quality modules were placed closest to the interaction region.

\section{Stave components}
\label{section:stavecomponent}
 
The IBL modules are supported and cooled by 14 staves, cylindrically arranged around the beam pipe axis, as described in Section~{\ref{sec:layout_introduction}}. Two symmetric multi-layer flexible circuits, the stave flexes, are glued on the back of the mechanical stave structure and route electrical services between individual modules and the EoS  region. The design, production and QA of the bare staves and stave flex components, and their assembly until the loading of modules on the stave, are described. 

\subsection{The bare stave}
\label{sec:barestave}

The bare stave should satisfy several important design criteria: a low material budget to improve physics performance; an excellent module thermal performance to optimise the pixel signal-to-noise and to prevent thermal runaway; good mechanical stability for efficient tracking performance; and a low Coefficient of Thermal Expansion (CTE) of component materials to reduce the mechanical stress.

\subsubsection{Bare stave material}
\label{sec:barestavematerial}

\begin{figure}[!htb]
\begin{center}
 \includegraphics[width=0.8\textwidth]{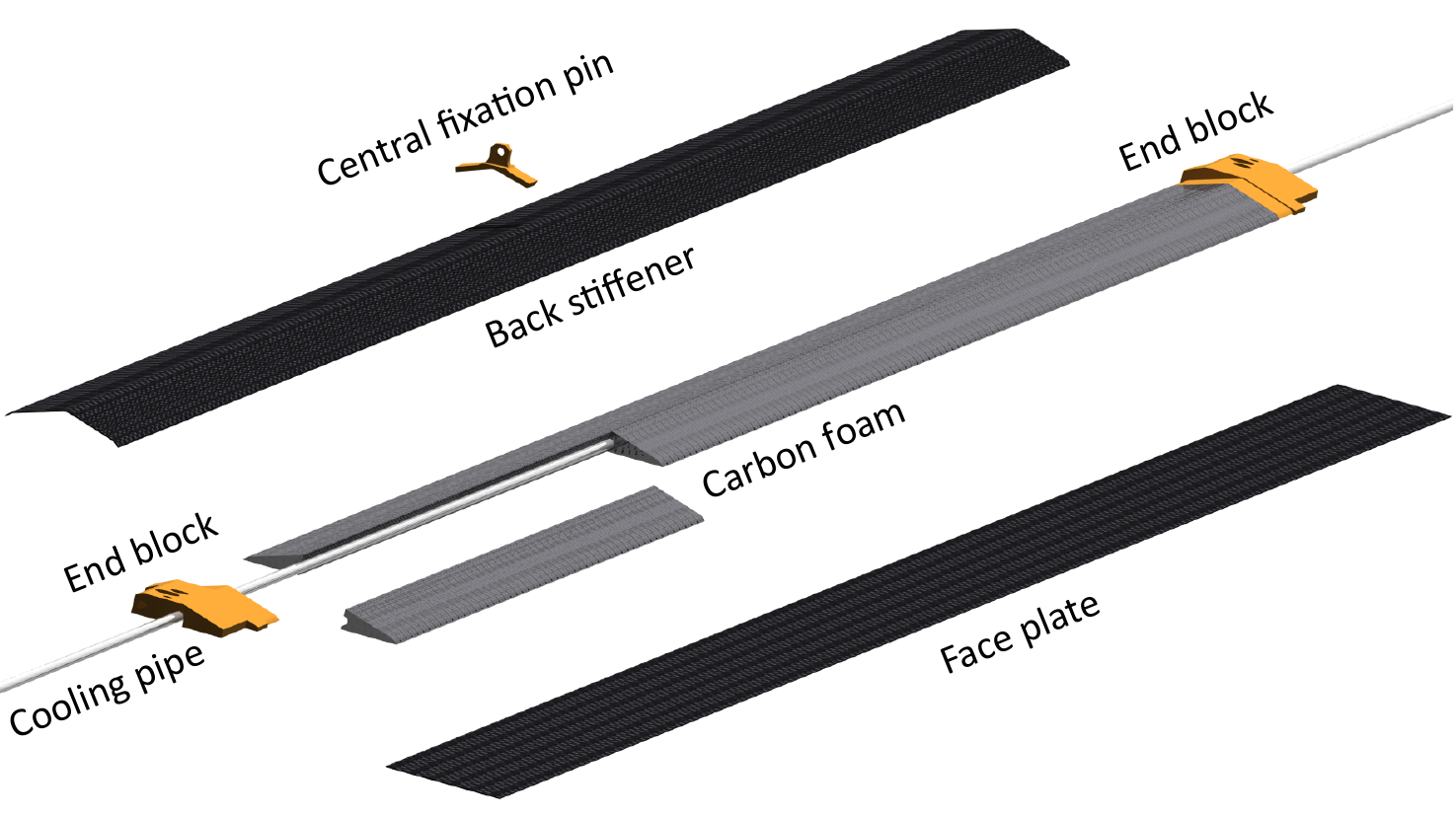}
\caption{Exploded view of the bare stave with its main components: the cooling pipe, the carbon foam, the two carbon laminates (the face plate where modules are loaded and the back stiffener on the opposite side) and  the PEEK elements (the central fixation pin and the end blocks).}
\label{fig:StaveConcept}
\end{center}
\end{figure}

\begin{figure}[!htb]
\begin{center}
 \includegraphics[width=0.8\textwidth]{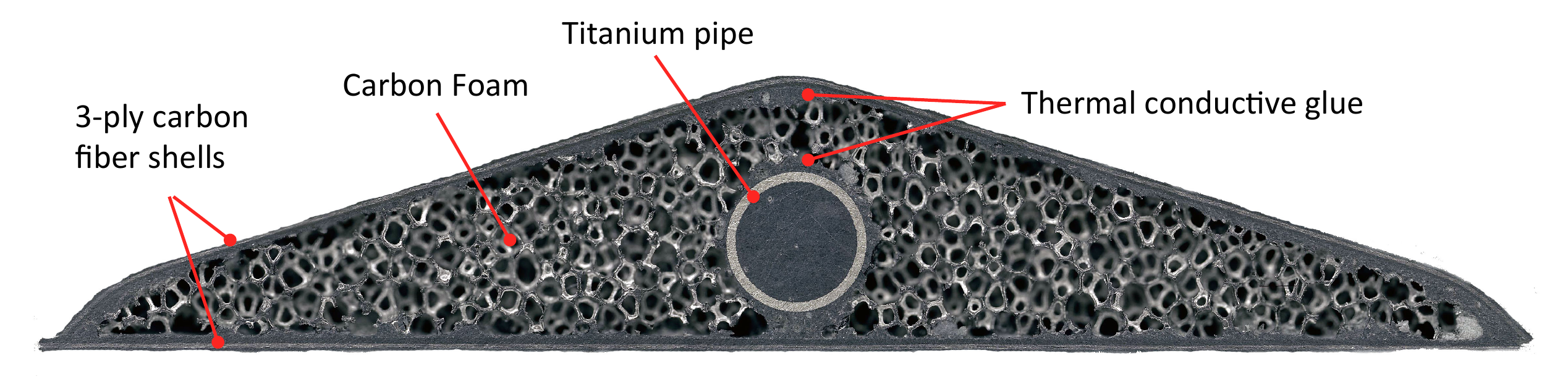}
%\caption{Picture of a bare stave transversal section.}
\caption{The transverse section of a bare stave.}
\label{fig:StaveCrossSection}
\end{center}
\end{figure}

The constraints of low material budget and high structural reliability for the bare stave has resulted in specific design and material choices. The stave is assembled from four main components (Figures~\ref{fig:StaveConcept} and \ref{fig:StaveCrossSection}), together with 
thermally conducting epoxy\footnote{Stycast 2850FT thermally conducting epoxy encapsulant, see http://www.henkel-adhesives.com/electronics.htm}: 
a \SI{1.7}{\milli\meter} external diameter titanium T40 cooling tube; 
a carbon foam section\footnote{K9 Carbon Foam, Allcomp Inc., see http://www.allcomp.net} to drain the heat flux; 
carbon fibre laminates\footnote{RS-3/K13C2U uni-directional prepress, Tencate Advanced Composites, see https://www.tencate.com} to reinforce the stave stiffness; 
and PEEK\footnote{Polyether-ether-ketone, VIRTEX\textsuperscript{\textregistered}  PEEK 450CA40, Victrex plc., see https://www.victrex.com/en/}  elements to fix the stave on the IBL support. 

Each of the stave components requires special manufacturing  and precise machining techniques. For example, the carbon laminates require fabrication in an autoclave %AGC 160118 high pressure ovens 
to obtain the required transverse 
%thermal conductivity (K). 
thermal conductivity ($\kappa$). 
Because of the high critical pressure (\SI{73.8}{\bar} at \SI{31}{\celsius}) of the bi-phase CO$_2$ cooling system, the titanium cooling pipes must fulfill stringent pressure requirements, with a minimal \SI{0.11}{\milli\meter} wall thickness. 
The total material budget of the bare stave is \SI{0.62}{\xzero}. The contribution of each component is  detailed in Table~\ref{tab:staveX0} and the main properties of the materials used are summarised in Table~\ref{tab:stavecomponents}. Each stave has a length of \SI{724}{\milli\meter}, measured between the end block fixation points, and a width of \SI{18.8}{\milli\meter}.    
 
\begin{table}[!htb]
\centering
\begin{tabular}{l c c c c c }
\hline \hline
        Component & Material  & Volume/stave      & Equiv. Thickness & X$_0$ & X/X$_0$  \\
                      &   &  (\SI{}{\centi\meter^3}) &             (\SI{}{\milli\meter})      & (\SI{}{\centi\meter})             & (\SI{}{\percent})      \\
\hline
Back stiffener    &  RS-3/K13C2U prepreg & 2.020 & 0.144 & 21.1 & 0.068 \\
Epoxy layers & Stycast 2850FT & 3.290 & 0.234 & 8.97  &  0.261\\
Carbon foam                             & K9 & 21.600 & 1.536 & 213 &   0.072\\
Cooling pipe & Titanium T40           & 0.411 & 0.029 & 3.56 &  0.082\\
Face plate       &  RS-3/K13C2U prepreg & 1.903 & 0.135 & 21.1 & 0.064 \\
End-of-Stave fixation & PEEK 450CA40 & 2.510   & 0.178 & 25 &  0.071\\
Central fixation          & PEEK 450CA40 & 0.075 & 0.005 & 25 & 0.002 \\
\hline
Total:              &  &  &  &  & 0.621 \\
\hline \hline
\end{tabular}   
\caption{Properties and radiation length (X/X$_0$) of the bare stave components.}
\label{tab:staveX0}
\end{table}

\begin{table}[!htb]
\centering
\begin{tabular}{l c c c c c c }
\hline \hline
        Component & Density  & Thermal      & CTE@\SI{20}{\celsius} & Young's & Tensile  & Compressive  \\
                     &  (\SI{}{\gram /\centi\meter^3})    & Conductivity ($\kappa$) &       (\SI{}{\ppm\per\celsius})            & modulus             & Strength  & Strength  \\
                      &     &(\SI{}{\watt/\milli\kelvin})   &                  & (\SI{}{\giga\pascal})           & (\SI{}{\mega\pascal})                &  (\SI{}{\mega\pascal}) \\
\hline
Titanium T40                        & 4.51 & 16.4 & 8.6 &102 & 430 & 340 \\
K9 carbon foam  & 0.20 & 28.3 & $<1$ & 0.293 & 3.6 & 1.8 \\
RS-3/K13C2U  prepreg  & 1.73   & (\emph{x}) 96   & (\emph{x}) -0.7 & (\emph{x}) 410  & & \\
    				    &         & (\emph{y/z}) 0.5 & (\emph{y/z}) 15 & (\emph{y/z}) 5.6 & & \\
\hline \hline
\end{tabular}   
\caption{The mechanical and thermal properties of the materials as used in finite-element modelling of the bare stave. The x-coordinate for the uni-directional RS-3/K13C2U fibre is along the fibre. The properties of the \SI{0}{^o}/\SI{90}{^o}/\SI{0}{^o} fibre layup (not shown) are calculated from the input of uni-directional fibres, for a given fibre fill factor.
\label{tab:stavecomponents}}
\end{table}

\subsubsection{Bare stave quality control and production}
 \label{sec:bare-stave-qa}

The bare stave production is a 13-step process that requires careful QA to ensure a uniform quality.  The QA included a visual inspection to check for cracks %, strips  
or mechanical non-conformities, a %full 
weight and metrology control, and a \SI{150}{\bar} pressure test and helium leak test for the cooling tubes.  
A total of 33 bare staves were produced, and a similar number of prototypes were built to verify the design performance and to tune the production process. Table~\ref{tab:staveFailureRate} summarises the yield of the bare stave production.

\begin{table}[!htb]
\centering
\begin{tabular}{l c  }
\hline \hline
        Bare stave characterisation & Quantity      \\
\hline
Bare staves produced & 33 \\
Staves used as prototypes & 2 \\
Staves rejected after visual inspection & 7 \\
Staves rejected after metrology & 0 \\
Staves rejected after high pressure and He leak test & 0 \\
Staves accepted for stave flex assembly  & 24 \\
\hline
Bare stave failure rate during QA             & \SI{27}{\percent} \\
\hline \hline
\end{tabular}   
\caption{The number of bare staves produced, and bare stave failures during QA.}
\label{tab:staveFailureRate}
\end{table}

The stave weight was systematically checked to control the manufacturing uniformity. %In fact the weight 
 The average stave weight of the 33 produced staves is \SI{26\pm1}{\gram}.
 
\begin{figure}[!htb]
\begin{center}
\includegraphics[width=1.0\textwidth]{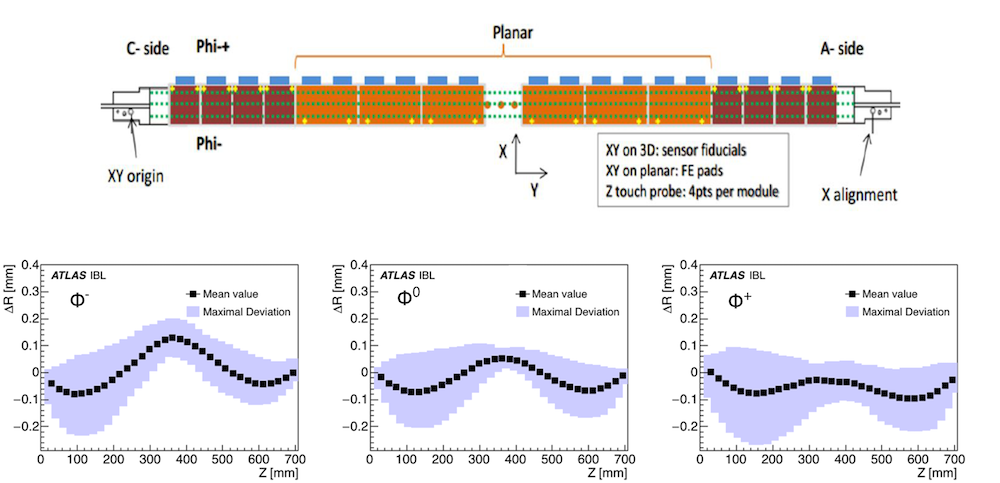}
\caption{Summary of the planarity, $\Delta$R, for the 33 production bare staves in three azimuthal positions as defined in the text: (a) $\phi^{-}$; (b) $\phi^{0}$; and (c) $\phi^{+}$.}
\label{fig:BareStavesPlanarity}
\end{center}
\end{figure}

Given the very tight envelope requirements of the IBL, the bare stave planarity, $\Delta$R, is an important factor that was %AGC 160118 carefully 
controlled during the metrological survey. Deviations from planarity impact the module loading activity and the minimum gap between two adjacent staves around the beam pipe. The design specification required $\Delta$R to be less than \SI{0.25}{\milli\meter} along the stave length. Figure~\ref{fig:BareStavesPlanarity} shows the mean planarity  on all  produced staves, measured at 34 points along the stave profile and in three azimuthal positions as indicated in the stave profile shown in the figure ($\phi^{-}$ and $\phi^{+}$ at the stave edges, and $\phi^{0}$ at the centre of the stave). %Due to the construction procedure, 
The staves have a slight undulated profile. 

\subsection{Stave flex}
\label{sec:overview_staveflex}

The on-stave electrical service lines (the stave flex) connect each module of a stave to the Type 1 internal services at the EoS. 
% AGC 160118 The mechanical and electrical stave flex designs were challenging. These challenges include:
The stave flex design presented unique mechanical and electrical challenges including:
\begin{itemize} 
\item[-] {Space limitations force the services to be tightly integrated with the stave itself, with a
\SI{0.2}{\xzero} design specification for the maximum amount of material averaged over the stave width;}
% AGC-  maximum material specification of \SI{0.2}{\xzero};}
 \item[-] {High-quality transmission of \SI{160}{\mega\bit\per\second} data is required along the stave to the EoS, and subsequently \SI{6}{\meter} from the EoS to the optical module (opto-board) at PP1;}
\item[-] {The operability of the FE-I4B over the full range of drawn currents requires that the round-trip low-voltage drop is less than~\SI{400}{\milli\volt};}  
\item[-] {Via interconnections are required between the Al and Cu layers.}  
\end{itemize}

%\paragraph{Layout}\
\subsubsection{Stave flex layout}
\label{sec:staveflexlayout}

A single stave is served by two stave flexes, one from each side (A and C, see Figure~\ref{fig:IBLLayout}) and  symmetric about the stave centre. The stave flexes run along the back-side of the stave, facing away from the interaction point. Each stave flex serves 16 FE-I4B chips and includes:
\begin{itemize}
\item[-]  {4 low-voltage (LV) supply lines, each serving 4 FE-I4B chips in parallel;}
\item[-]  {4 high-voltage (HV) supply lines, each serving either 2 double-chip  planar  or 4 single-chip 3D modules;}
\item[-]  {16 pairs of LVDS output lines;}
\item[-]  {8 pairs of LVDS data input and clock lines, each connected to 2 FE-I4B  chips in parallel;}
\item[-]  {8 lines reading out 4 equidistant NTC (Negative Temperature Coefficient) sensors located on the module flexes\footnote{All  3D (planar) modules carry one (two) NTC sensors mounted on the module flex hybrid. However, only four of the 16 NTC sensors are read out, to reduce the number of lines on the stave flex.}}.
\end{itemize}

\begin{figure}[!htb]
\begin{center}
 \includegraphics[width=0.9\textwidth]{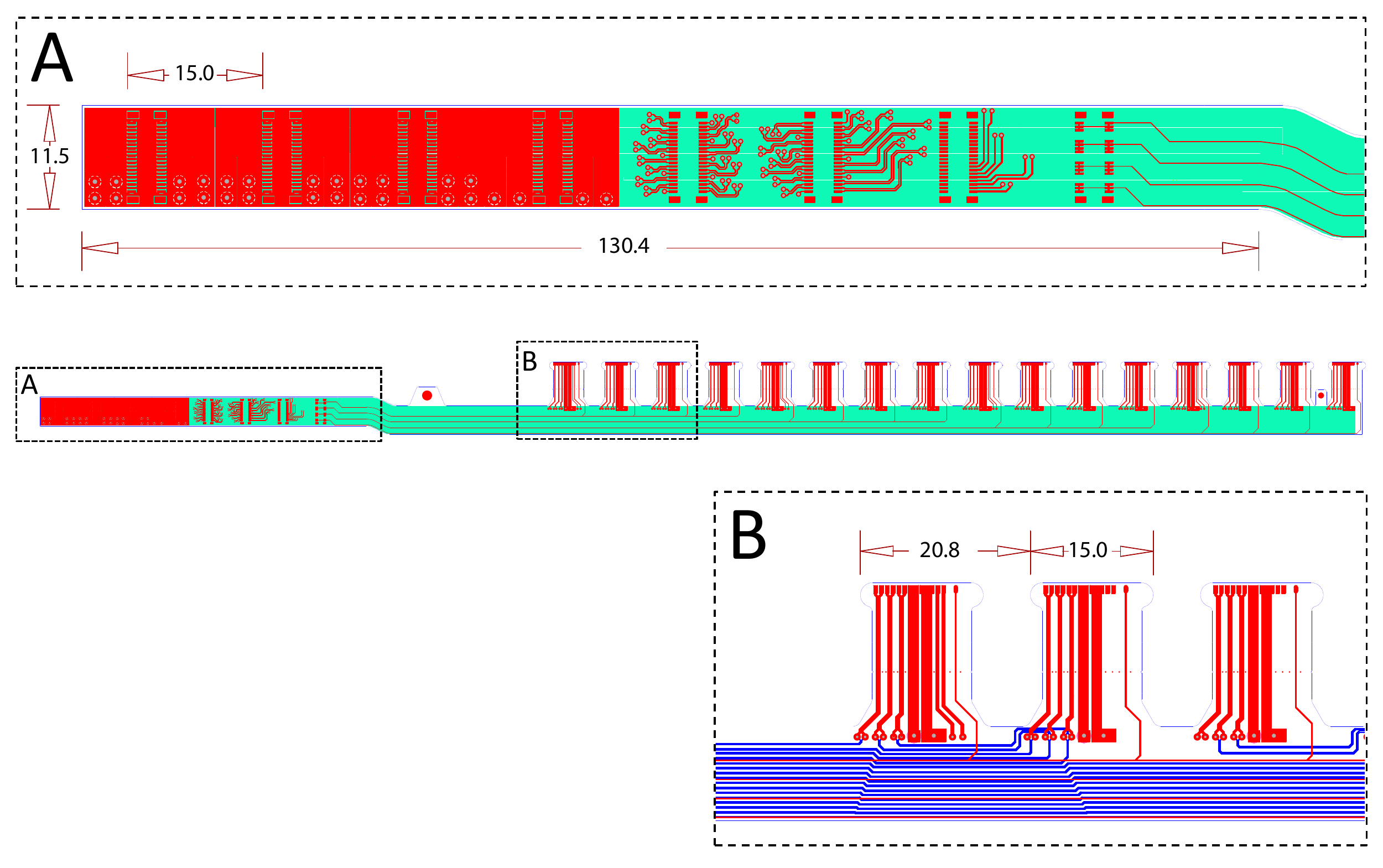}
\caption{The layout of the stave flex: a zoom of the connector region (detail A) and part of the longitudinal section (detail B). The dog-leg section can be seen in between the regions A and B. All units are in mm.}
\label{fig:FlexLayout}
\end{center}
\end{figure}

The stave flex,  shown in Figure~\ref{fig:FlexLayout}, consists of a longitudinal section, a dog-leg part  and a connector region, %of 
with a total length of \SI{528.4}{\milli\meter}.   The longitudinal section,  approximately \SI{350}{\milli\meter} long and \SI{11.5}{\milli\meter} wide, is equipped with 16 identical wings that provide electrical and mechanical connections to the  FE-I4B chips. The wing pitch of \SI{20.75}{\milli\meter} corresponds to the inter-distance between FE-I4B chips in the planar modules, but is approximately \SI{100}{\micro\meter} too short for 3D modules; this mis-alignment is corrected using angled wire-bonding between the wing flex and the module flex (Section~\ref{sec:mod_flex}). Each wing is \SI{17.5}{\milli\meter} long and \SI{12}{\milli\meter} wide. Two ears at the edge of the wing ease the  gluiing of the wing flex to the module during the  integration phase. As the wing needs to be bent by \SI{180}{\degree} to be glued on the module, a row of small holes are drilled between the metallic lines to make the polyimide (Kapton\textsuperscript{\textregistered}) more flexible. The dog-leg region creates a shift in both the radial and azimuthal directions, allowing services to be positioned in the correct location between cooling pipes in the EoS region. The connector region at the end of the stave is equipped with eight Panasonic AXT540124\footnote{Narrow \SI{0.4}{\milli\meter} pitch 40-pin connectors AXT540124 (socket) / AXT640124 (header) from Panasonic Electrical Works, https://www.panasonic-electric-works.com
} SMD sockets.

\begin{figure}[!htb]
\begin{center}
 \includegraphics[width=0.6\textwidth]{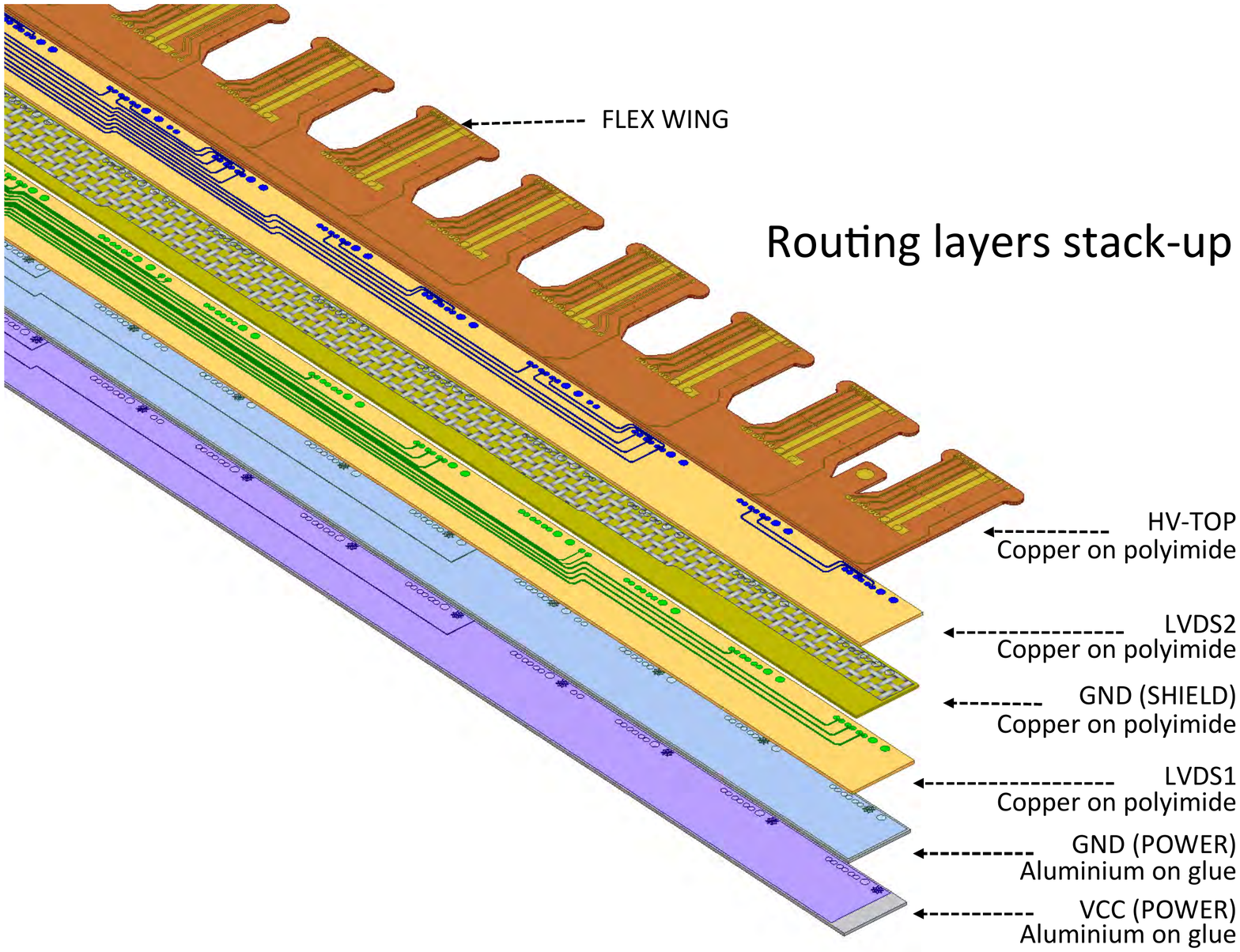}
\caption{An expanded view of the overlapping layers in the stave flex.}
\label{fig:Flex3DView}
\end{center}
\end{figure}

The stave flex has six metal layers of which two are \SI{50}{\micro\meter} thick Al traces used for the power and ground, and four 
%AGC are Cu;
have Cu traces; 
the stack-up is shown in Figures~\ref{fig:Flex3DView} and~\ref{fig:FlexStaveVie_b}. The total thickness is approximately \SI{500}{\micro\meter}. Routing between layers is achieved using vias, which are described in Section~{\ref{sec:flexprocessflow}}. Thin stiffeners are glued on the bottom under the connectors to ensure good connector insertion and removal reliability.  The top metal layer is only used to route the HV signals, in order to maintain sufficient distance between the HV traces to guarantee \SI{1.5}{\kilo\volt} isolation. The signal and NTC lines are distributed on the LVDS1 and LVDS2 layers, but the layout differs for the A- and C-sides to ensure the same layout for the signal connectors at the EoS.  
The routing of LVDS traces has been optimized to avoid cross-talk and to have a controlled differential impedance of \SI{80}{\ohm}. To better control the impedance of the signal traces on the LVDS2 layer, a Cu ground shield is added below it. This \SI{5}{\micro\meter} thick ground shield has a meshed structure optimised to maintain good electrical performance while minimising the additional material budget.
The stave flex thickness averaged over its length and width is approximately \SI{0.26}{\xzero}
 with contributions from Al, Cu, polyimide and epoxy glue. The Cu contribution increases along the stave towards the EoS while the Al is almost constant along the stave length. 
 The quoted material %\SI{}{\xzero}
does not  include  the average $\sim$\SI{0.034}{\xzero} wing contribution. The contribution of the stave flex and the wings to the IBL material budget is approximately \SI{0.21}{\xzero}, once  smeared over the azimuthal angle. Details of the material contribution of the stave flex and wings are reported in Table~\ref{tab:flexstaveX0}.

\begin{table}[!htb]
\small
\centering
\begin{tabular}{c l c c c c c }
\hline \hline

%\multirow{4}{*}{\SI{3000}{\e} at \SI{22}{\celsius}} 	& Planar Normal	& 37 & \num{123\pm 10} & \num{25\pm 2} \\

Component    &  Material  &  Thickness   & Equiv. Thickness &  Average X/X$_0$  &  Central X/X$_0$ & Smeared X/X$_0$\\
                  &   &     (\SI{}{\micro\meter}) &    (\SI{}{\micro\meter})      &   (\SI{}{\percent})  &   (\SI{}{\percent}) &   (\SI{}{\percent})     \\
\hline

\multirow{4}{*}{Stave Flex} &    Cu 		& 61     & 7      & 0.041 & 0.011 & 0.029 \\
					&	Al		& 100   & 92   & 0.104  & 0.101 & 0.072 \\
					&	Polyimide	& 163   & 163 & 0.057  & 0.057 & 0.040 \\ 
					&	Glue		& 173   & 173 & 0.060  & 0.060 & 0.042 \\
\hline

Total Stave Flex     &   & 497  &        &  0.262 & 0.230 & 0.182 \\
\hline \hline

\multirow{3}{*}{Wings} & Cu 		& 19   & 3.8     & 0.024  &  & 0.026 \\
					&	Polyimide	& 38   & 21.4   & 0.007  &  & 0.008 \\ 
					&	Glue		& 10   & 5.6     & 0.002  &  & 0.002 \\
\hline
Total Wings  &      &  67  &        &   0.034 &  & 0.036 \\
\hline \hline

\end{tabular}   
\caption{Properties and radiation length (X/X$_0$) of the stave flex components, separately for the longitudinal section and the stave flex wings. \textit{ Thickness} is the effective thickness of the material layers for normal incidence;  \textit{ Equiv. Thickness}  and  \textit{ Average X/X$_0$} are respectively the thickness and X/X$_0$  normalised to the stave flex length; \textit{ Central X/X$_0$} is the value of the material budget close to the central region, averaged over the stave width but without averaging along the stave length; \textit{ Smeared X/X$_0$} is the  \textit{ Average X/X$_0$} smeared over a cylinder at the mean radius of the stave flex.}
\label{tab:flexstaveX0}
\end{table}

The CTE of the stave flex is approximately \SI{27}{\ppm\per\celsius}, which is significantly different from the almost zero value of the carbon stave structure. 
To ensure the mechanical integrity of assembled and loaded staves during thermal cycling, the stave flex is glued to the carbon structure (Section~\ref{sec:staveflexglueing}).

%\paragraph{Process flow}\
\subsubsection{The process flow}
\label{sec:flexprocessflow}

The  stave flexes were produced at the CERN PCB workshop. The Cu and Al stacks (Figure~\ref{fig:Flex3DView}) were processed separately. The material used for the four Cu layers were two double-sided Cu-clad laminates on polyimide substrates, one used for the TOP/LVDS2 and the other for the GND/LVDS1 routing layers. The Cu and Al stacks are then laminated together.

\begin{figure}[!htb]
\begin{center}
\includegraphics[width=0.6\textwidth]{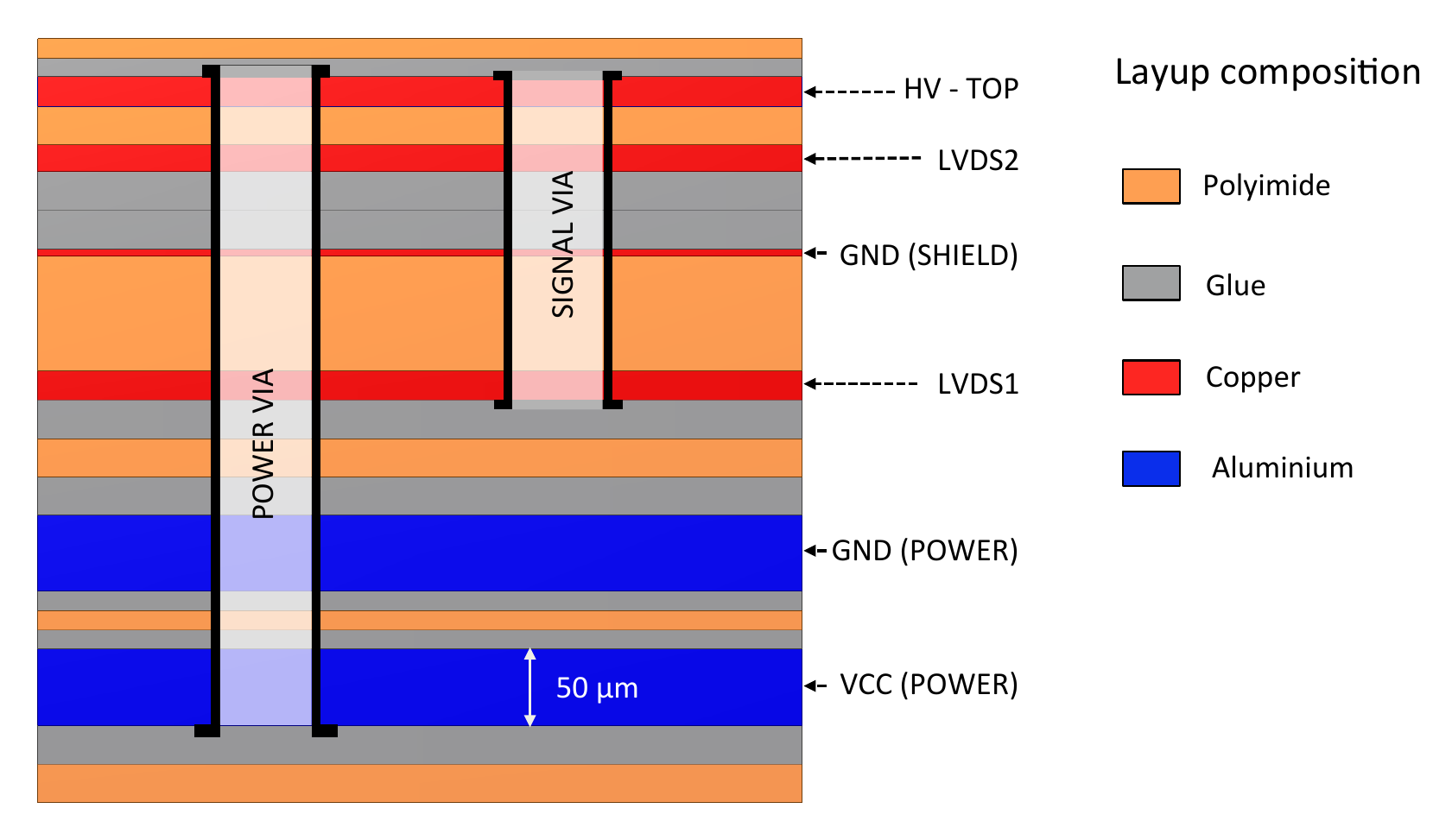}
\caption{Cross-section of the stave flex showing the signal and power vias. }
\label{fig:FlexStaveVie_b}
\end{center}
\end{figure}

%Replace above for paper version as below:
A feature of the stave flex is the use of low-resistivity vias joining the GND/VCC Al layers to the  TOP Cu layer. Following successive etching and deposition steps to create the Al/Cu vias, there is a deposition of \SI{0.2}{\micro\meter} of Cr and \SI{2}{\micro\meter} of Cu in the location of the power vias. 
The stave flex process is completed by: 
\begin{itemize}
\item[-] {pressing together the Cu and aluminum stacks and drilling the power vias;}
\item[-] {\SI{25}{\micro\meter} Cu electroplating of the power vias to connect the TOP layer with GND/VCC, followed by polyimide cover layers on the two stave flex sides;} 
\item[-] {gold plating the bonding and soldering pads and cutting the stave flex to its final shape. }
\end{itemize}  
The power and signal vias are shown in the cross-sectional view of the stave flex in Figure~\ref{fig:FlexStaveVie_b}. The final power vias have a typical resistance of 3-\SI{4}{\milli\ohm}. 

\subsubsection{Quality control and production}
\label{sec:flexprodetqa}

The stave flex production was organised in batches 
%of 2-4
of up to 4 
sheets with three stave flexes each, all of the same type, either for the A-side or the C-side.  After being diced and mounted with connectors, the flexes were cleaned and shipped to the qualification laboratory. The qualification consisted of:

\begin{itemize}
\item[-] {a visual inspection to identify mechanical damage and surface anomalies, and a first electrical test to check resistance measurements, continuity or shorts;}
\item[-]  {the bending of the 16 stave flex wings on themselves (to allow later connection to the module flexes), with Araldite 2011 
gluing on the back-side, followed by mechanical measurements and an electrical continuity test for possible broken lines, at room temperature;}
\item[-]   {ten thermal cycles between \SI{-40}{\celsius} and \SI{40}{\celsius};}
\item[-]   {an extended (one day) HV test at \SI{1500}{\volt} with a current limit at \SI{0.1}{\micro\ampere} and a final electrical verification of resistance, continuity and shorts;}
\item[-]    {a final visual inspection and sign-off for delivery.}
\end{itemize}

Table~\ref{tab:flexstaveFailureRate} reports the stave flex yield.  
Six stave flexes were rejected during production by the vendor because of electrical or mechanical non-conformity, mainly at the beginning of the production. 
Two stave flexes were rejected because of a high resistivity of the LV via connection to the stave flex wings. The resistivity of the LV lines determines the round-trip voltage drop and a high inter-wing resistivity indicates a possible weak via connection. The voltage drop was uniform and typically \SI{\sim320}{\milli\volt}  for I = 2~A in the last batches, well within the \SI{400}{\milli\volt} specification. Two staves were rejected because of non-conformities identified during the visual inspection, and one stave was rejected because of HV non-conformities.
Finally, electrical tests were repeated at CERN before and after gluing the stave flexes on the stave. A small HV resistivity change was observed on one stave, but the stave flex was not rejected. 

\begin{table}[!htb]
\centering
\begin{tabular}{l c  }
\hline \hline
        Stave flexes & Number      \\
\hline
Produced & 72 \\
Rejected during production  & 6 \\
Rejected after visual inspection & 2 \\
Rejected after the electrical test & 2 \\
Rejected after the HV test & 1 \\
Rejected after the thermal cycles & 0 \\
Rejected due to mishandling  & 0 \\
Accepted for loading  & 61 \\
\hline
Total failure rate            & 15\% \\
\hline \hline
\end{tabular}   
\caption{Failure rate of stave flexes during QA.
\label{tab:flexstaveFailureRate}}
\end{table}

\subsection{Bare stave and stave flex assembly}
\label{sec:staveflexassembly}

\subsubsection{Stave flex gluing}
\label{sec:staveflexglueing}

A pair of  stave flexes is directly glued on the stave back stiffener with a procedure that was developed to meet stringent mechanical and thermo-mechanical constraints, for example the constraint of the stave envelope with an assembly accuracy of \SI{\pm300}{\micro\meter} in the longitudinal direction and \SI{\pm100}{\micro\meter} in the transverse direction, the thermo-mechanical characteristics of the two different materials and  the minimisation of the material budget. Furthermore, the flexes  must meet a radiation hardness requirement of \SI{250}{\mega\radian}. 

Following validation of the glue deposition process, radiation hardness and mechanical stiffness, Araldite 2011 glue, together with  a polyimide wet etching for the stave flex surface, were chosen. Based on these tests, a Pyralux LF111
\footnote{Pyralux\textsuperscript{\textregistered} is a DuPont Corp. trademark of a polyimide (Kapton\textsuperscript{\textregistered}) substrate with a laminated or cladded layer of metal/adhesive, used for flexible PCBs, see http://www.dupont.com. In particular, Pyralux\textsuperscript{\textregistered} LF111 coverlay is a polyimide film coated on one side with a proprietary acrylic adhesive.}
bond ply was added as a flex bottom layer in order to apply the glue  to a \SI{12.5}{\micro\meter} acrylic layer after surface treatment, rather than to the polyimide directly.

The above choice was validated with an ageing test using a production IBL stave glued with production stave flexes according to the final gluing procedure. Both thermal and radiation loads were applied to the assembly during the test while using a 10~MeV electron source.
After  approximatively \SI{380}{\mega\radian} and 110 thermal cycles  between \SI{+40}{\celsius} and \SI{-40}{\celsius}, no critical damage was observed on the stave-flex glue joint~\cite{IrradiationStaveNote}. 

\subsubsection{Quality control of stave assembly components}
\label{sec:staveassemblyqa}

During prototyping, one stave flex started to delaminate after several thermal cycles. A carbon clip was added to the design in order to prevent this problem. Out of 22 production assemblies built for the IBL, only one stave encountered a critical problem due to the glue mixture mistake, which led to a polymerisation failure. The remaining assembled staves  underwent  a QA procedure to fully qualify the assembly before module loading. A summary of  the assembled staves and their usage is detailed in Table~\ref{tab:BareStavesAllocation}.

\begin{table}[!htb]
\centering
\begin{tabular}{l c  }
\hline \hline
        Staves & Number      \\
\hline
Staves accepted for stave flex assembly & 24 \\
Staves used for system test prototypes & 2 \\
  & \\
Staves assembled with stave flex  & 22 \\
Staves rejected after stave flex assembly  & 1 \\
Staves qualified for module loading& 21 \\
Staves loaded with modules  & 20 \\
\hline \hline
\end{tabular}   
\caption{Accounting of the produced and qualified bare staves.
\label{tab:BareStavesAllocation}}
\end{table}

\begin{figure}[!htb]
\begin{center}
\includegraphics[width=0.7\textwidth]{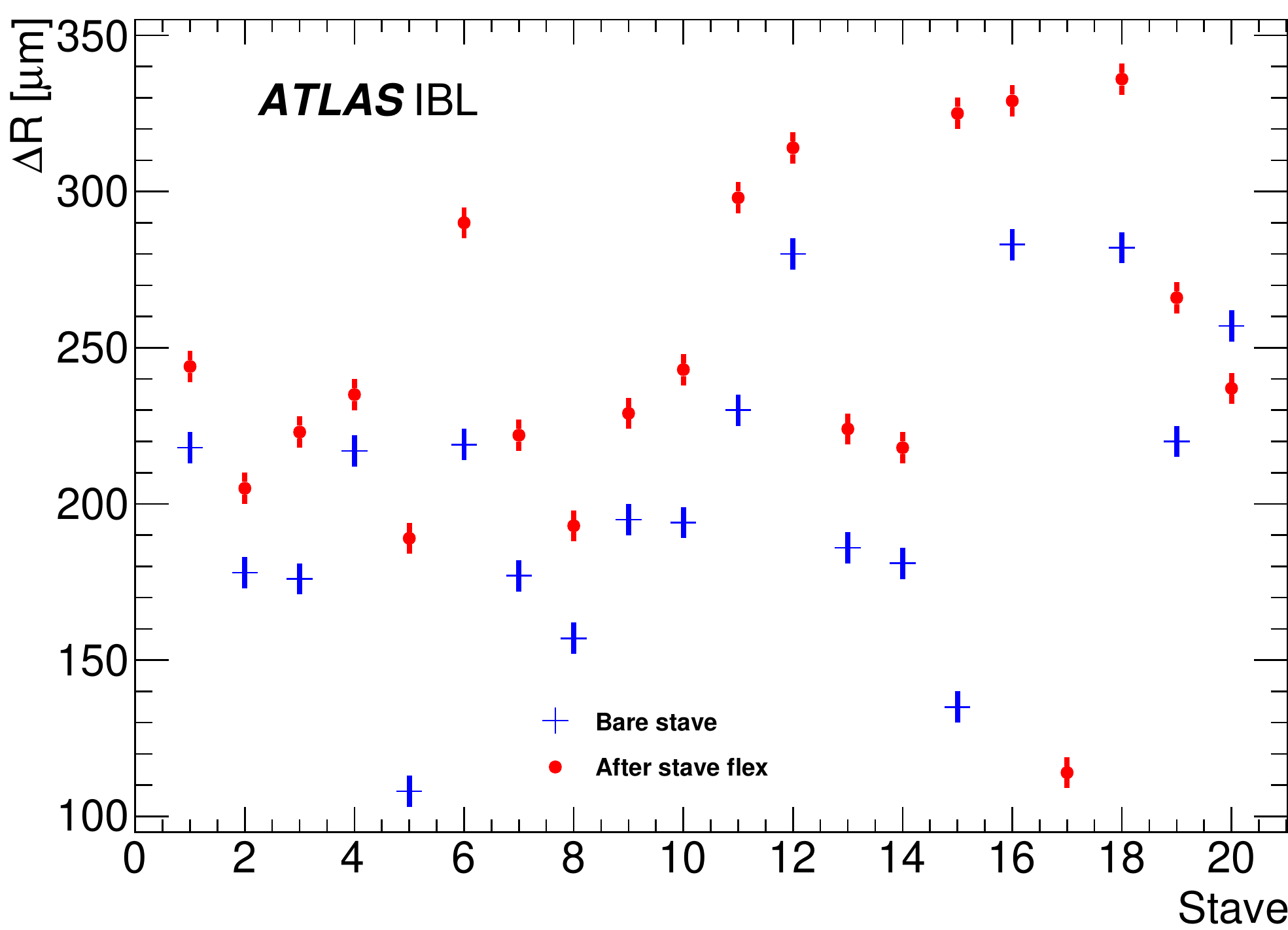}
\caption{The planarity, $\Delta$R, of each production stave, before and after thermal cycling. There is a \SI{5}{\micro m} uncertainty on all measurements. No measurement exists for the planarity of stave 17 before thermal cycling.}
\label{fig:planarity}
\end{center}
\end{figure}

\begin{figure}[h]
        \centering
        \begin{subfigure}[hc]{0.35\textwidth}
                \includegraphics[width=\textwidth]{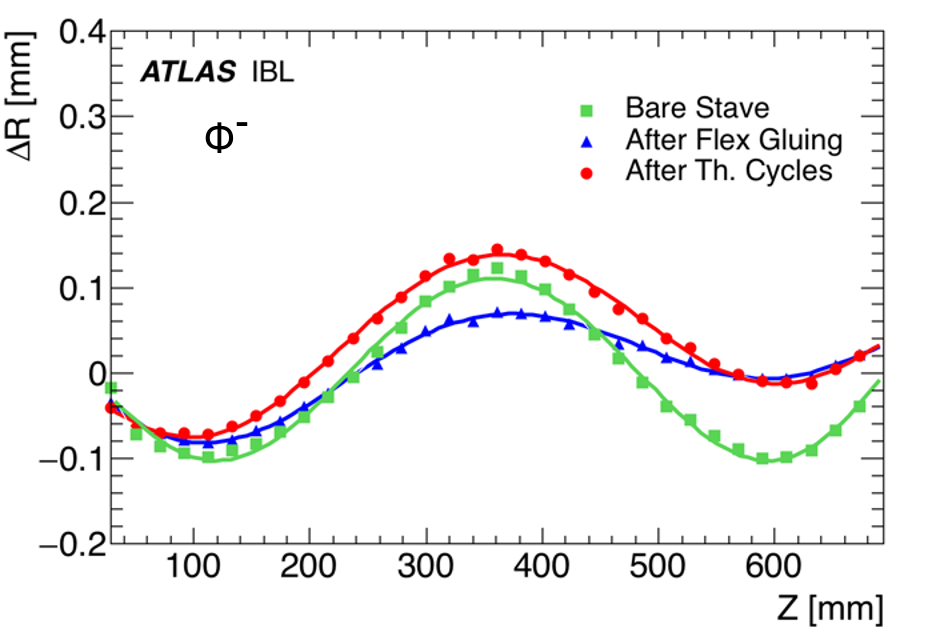}
                \caption{}
                \label{}
        \end{subfigure}%
        \begin{subfigure}[hc]{0.35\textwidth}
                 \includegraphics[width=\textwidth]{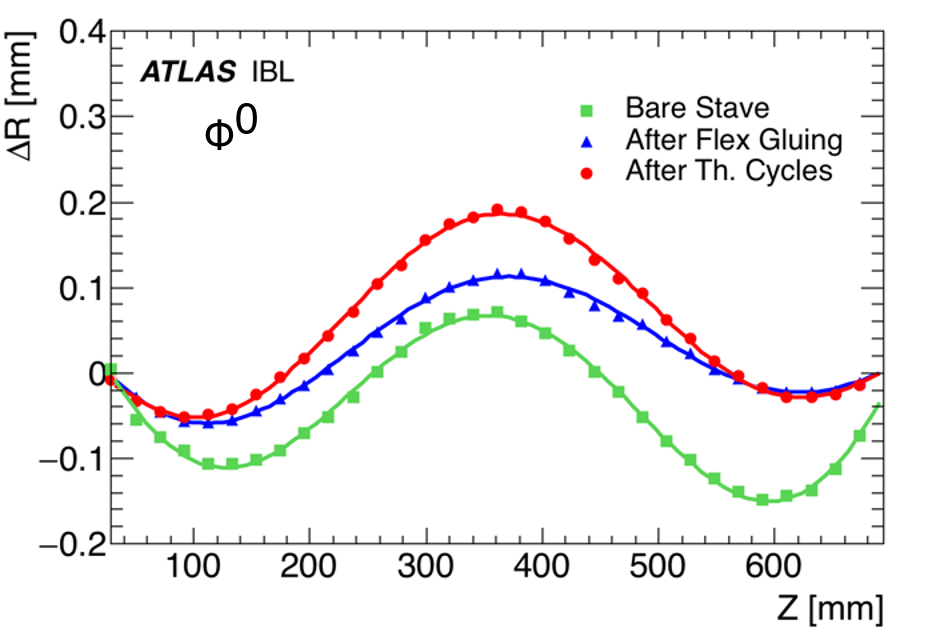}
                \caption{}
                \label{}
        \end{subfigure}%
        \begin{subfigure}[hc]{0.35\textwidth}
                \includegraphics[width=\textwidth]{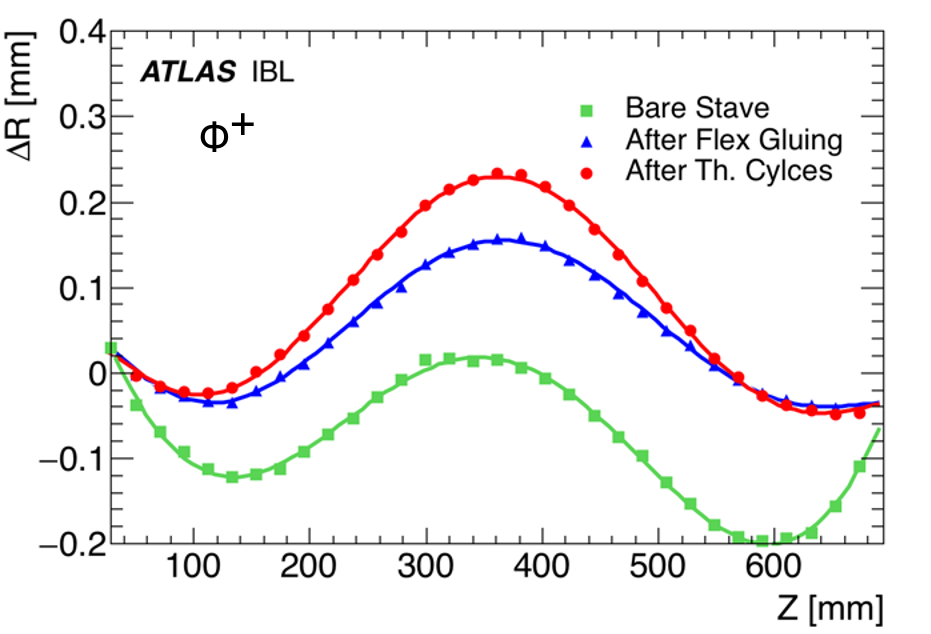}
                \caption{}
                \label{}
        \end{subfigure}
\caption{Stave planarity, $\Delta$R, in three azimuthal positions for a single stave, showing the planarity before stave flex gluing (red), after flex gluing (blue) and after thermal cycling (green): (a) $\phi^{-}$; (b) $\phi^{0}$; and (c) $\phi^{+}$, as defined in the text.} 
\label{fig:StavePlan_VS_FlexGluing}
\end{figure}

As a first qualification step, the staves were visually inspected and the  stave flexes were electrically tested to verify that the services were not damaged during the assembly process.
%AGC 060917 during the loading. 
The staves were then thermally stressed:  
%a program was defined with 
an initial phase at \SI{35}{\celsius} for \SI{1}{\hour} was followed by 10 cycles of \SI{1}{\hour} from \SI{40}{\celsius} to \SI{-40}{\celsius} 
%The program ended with 
and finally a stabilisation phase of \SI{3}{\hour} at \SI{20}{\celsius}~\cite{LoadingNote}.
A metrology survey was made before and after thermal cycling to measure deformations of the mechanical supports by the flex gluing and to verify that the assembly still respected the required envelope. The %production 
stave planarity measurements, before and after thermal cycling, are summarised in Figure~\ref{fig:planarity}~\cite{LoadingNote}. 
The planarity of  one typical stave before and after flex gluing, and after thermal cycling, is shown in
 Figure~\ref{fig:StavePlan_VS_FlexGluing}. As expected, due to the different CTE of the materials and the stave flexes being glued on just one side of the back of stave, the shape of the mechanical supports is affected by the assembly of the flexes and by the thermal cycling. Before thermal cycling, only four staves exceeded the specified bare stave envelope of  \SI{250}{\micro\meter}. After thermal cycling, the planarity is slightly deteriorated, but remains less than \SI{340}{\micro\meter} and within the clearance after integration in the IST of $\sim$\SI{2}{\milli\meter}.

\section{Stave loading and quality assurance}
\label{section:staveload}

Twenty staves were loaded with qualified good modules. Details are provided in reference~\cite{LoadingNote}.  The QA procedure~\cite{StaveQANote} provided a detailed characterisation of the stave including calibration in cold conditions and data taking with radioactive sources. Pixel defects were classified according to the type of failures observed in the FE-I4B chip, in the sensor and in  the bump bonding. Two staves were accidentally damaged by an excess of humidity during the QA procedure (Section~\ref{sec:corrosion}). Of the 18 remaining staves that satisfied the QA procedure, the best 14 were selected for assembly in the IBL detector. 

% Javier
\subsection{Stave loading and rework}
\label{sec:stave_loading}

\begin{figure}[h!]
                \centering
        \begin{subfigure}[t]{0.32\textwidth}
                \includegraphics[width=\textwidth]{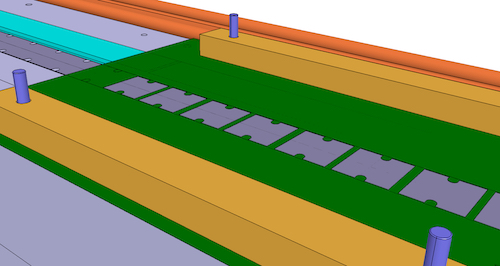}
                        \caption{Positioning of the Al mask.}
                \label{fig:main_2}
        \end{subfigure}
                \centering
        \begin{subfigure}[t]{0.32\textwidth}
                \includegraphics[width=\textwidth]{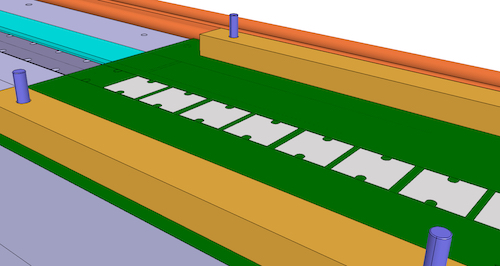}
                       \caption{Thermal grease application.}
                \label{fig:main_3}
        \end{subfigure}
                \centering
        \begin{subfigure}[t]{0.32\textwidth}
                \includegraphics[width=\textwidth]{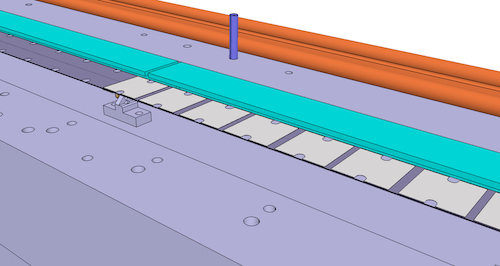}
                \caption{Grease pattern after mask removal.}
                \label{fig:main_4}
        \end{subfigure}
                \begin{subfigure}[t]{0.32\textwidth}
                \includegraphics[width=\textwidth]{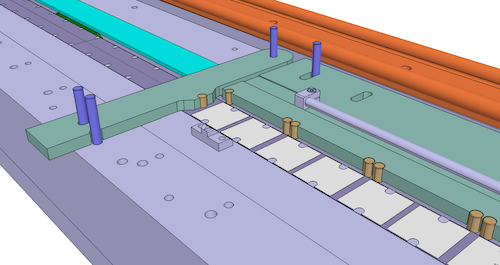}
                        \caption{Installation of the positioning stopper.}
                \label{fig:main_5}
        \end{subfigure}
                \centering
        \begin{subfigure}[t]{0.32\textwidth}
                \includegraphics[width=\textwidth]{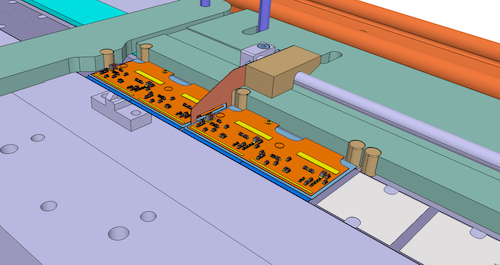}
                       \caption{Positioning of the first module, its spacer, and the second module.}
                \label{fig:main_6}
        \end{subfigure}
                \centering
        \begin{subfigure}[t]{0.32\textwidth}
                \includegraphics[width=\textwidth]{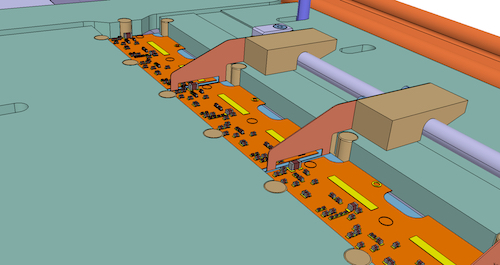}
                \caption{Installation of the last positioning stopper.}
                \label{fig:main_7}
        \end{subfigure}
                \centering
\begin{subfigure}[t]{0.32\textwidth}
                \includegraphics[width=\textwidth]{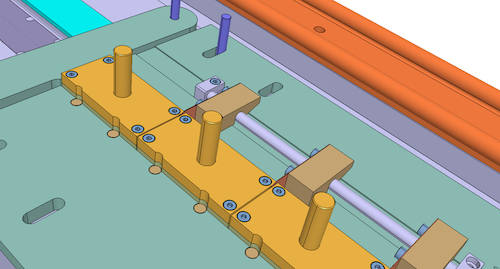}
                        \caption{Installation of the module weights.}
                \label{fig:main_2}
        \end{subfigure}
                \centering
        \begin{subfigure}[t]{0.32\textwidth}
                \includegraphics[width=\textwidth]{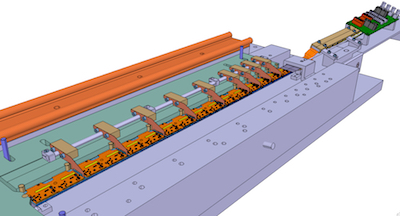}
                       \caption{Tooling dismounting after glue curing.}
                \label{fig:main_3}
        \end{subfigure}
                \centering
        \begin{subfigure}[t]{0.32\textwidth}
                \includegraphics[width=\textwidth]{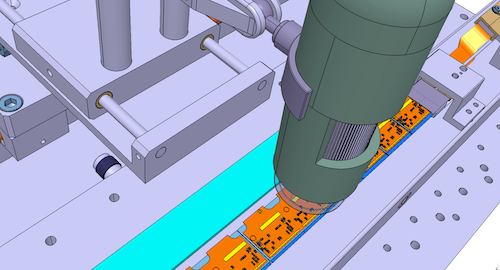}
                \caption{Optical inspection.}
                \label{fig:main_4}
        \end{subfigure}
        \caption{Three dimensional renditions of the main stave module-loading steps. %Main module stave loading steps.
        }
        \label{fig:StaveLoadingSteps}
\end{figure}

Qualified modules were loaded on the staves following the  procedure described below and sketched in Figure~\ref{fig:StaveLoadingSteps}. 
%AGC - moved to previous paragraph. More details can be found in Reference~\cite{LoadingNote}.
Each qualified stave was installed in a loading tool where a \SI{70}{\micro\meter} thick thermal grease\footnote
{HTCP, Electrolube, see https://www.electrolube.com/core/components/products/tds/044/HTCP.pdf.} 
layer was applied to half of its surface using a stainless steel template shim. After removing the template, the positioning tools were installed, and two Araldite 2011 glue drops per FE-I4B chip were %then 
applied with a needle on the thermal grease template openings to later fix the modules on their position. 
% (see the edges of the grease template openings). 
The modules were then installed, one at a time, using spacers to achieve a precise \SI{200}{\micro\meter} gap between neighbouring modules. A load of \SI{\sim20}{\gram} per FE-I4B chip was placed on the modules during the glue-dot curing period. After the removal of  weights and positioning tools,  an optical inspection was performed  %thanks to a built-in sliding camera, 
before repeating the %full 
process  on the other half of the stave.

\begin{figure}[h!]
        \centering
        \begin{subfigure}[t]{0.32\textwidth}
                \includegraphics[width=\textwidth]{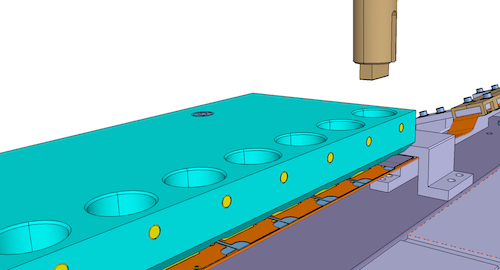}
                \caption{Stamp mask positioning.}
                \label{fig:glu_2}
        \end{subfigure}
                \centering
        \begin{subfigure}[t]{0.32\textwidth}
                \includegraphics[width=\textwidth]{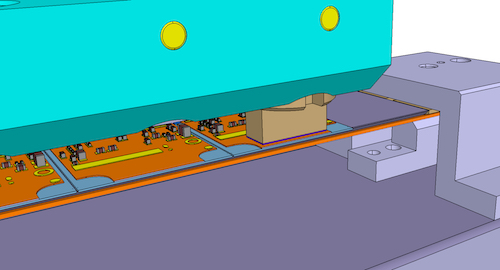}
                \caption{Glue is deposited on each wing position by a stamp.}
                \label{fig:glu_3}
        \end{subfigure}
                \centering
        \begin{subfigure}[t]{0.32\textwidth}
                \includegraphics[width=\textwidth]{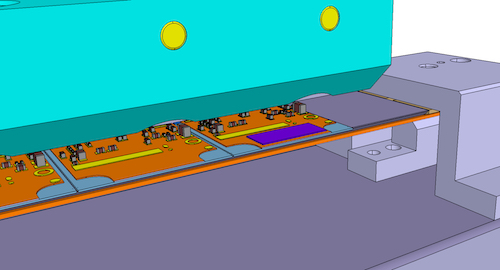}
                \caption{Glue deposits (purple) at the stave wing location.}
                \label{fig:glu_4}
        \end{subfigure}
        \centering
        
                \begin{subfigure}[t]{0.32\textwidth}
                \includegraphics[width=\textwidth]{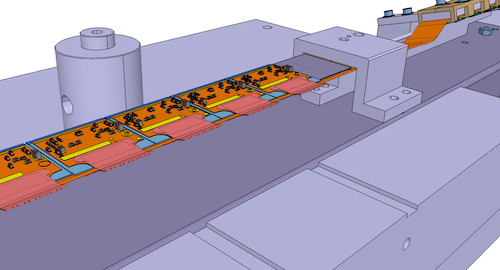}
                \caption{Wing positioning for each FE-I4B chip.}
                \label{fig:glu_5}
        \end{subfigure}
                \centering
        \begin{subfigure}[t]{0.32\textwidth}
                \includegraphics[width=\textwidth]{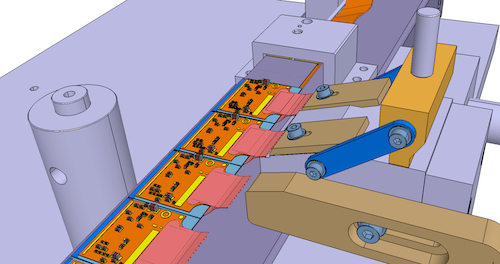}
                \caption{Wing positioning tool installation.}
                \label{fig:glu_6}
        \end{subfigure}
                \centering
        \begin{subfigure}[t]{0.32\textwidth}
                \includegraphics[width=\textwidth]{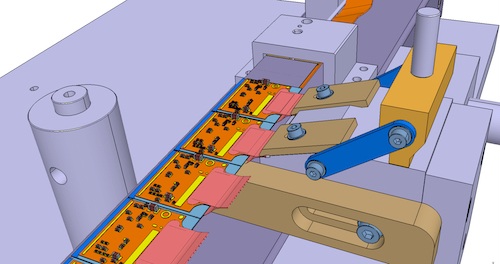}
                \caption{Final stave wing position fixed.}
                \label{fig:glu_7}
        \end{subfigure}
        \centering
        
                \begin{subfigure}[t]{0.32\textwidth}
                \includegraphics[width=\textwidth]{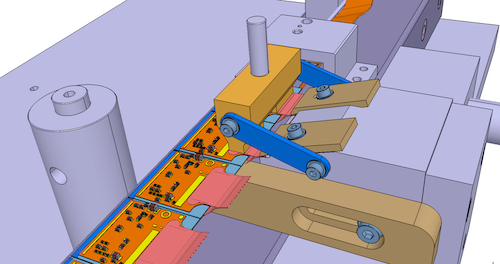}
                \caption{Weight on stave wings during glue curing.}
                \label{fig:glu_8}
        \end{subfigure}
                \centering
        \begin{subfigure}[t]{0.32\textwidth}
                \includegraphics[width=\textwidth]{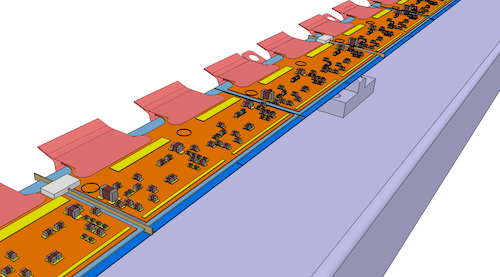}
                \caption{HV insulator placement and bridge gluing.}
                \label{fig:glu_10}
        \end{subfigure}
                \centering
        \begin{subfigure}[t]{0.32\textwidth}
                \includegraphics[width=\textwidth]{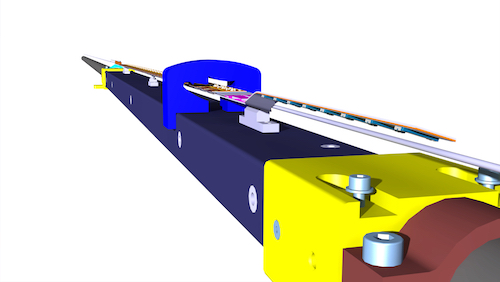}
                \caption{Stave envelope check.}
                \label{fig:glu_11}
        \end{subfigure}
        \centering
        
        \caption{Three dimensional renditions of the main stave-flex wing gluing steps.}
        \label{fig:StaveFlexWingGluing}
       
\end{figure}

After loading,  the modules were connected to the stave flex following the procedure sketched in Figure~\ref{fig:StaveFlexWingGluing}. % (a)-(i). 
First a layer of Araldite 2011 was deposited on each %in correspondence 
of the stave flex wings using a mask.  A tool constrained the wings to stay in their nominal position. A load of \SI{\sim16}{\gram} per FE-I4B chip was applied during the glue curing period. A \SI{\sim100}{\micro\meter}-thick Kapton\textsuperscript{\textregistered} insulator  was then inserted between each powering sector.  Finally, wire-bonds were added to connect the modules to the wings of the service flex, enabling the stave powering and data transmission. 
%As a last step, the stave envelope was checked to fulfil the mechanical constraints of the IBL.
As a last step, the stave envelope was checked to ensure that the mechanical constraints of the IBL were fulfilled.

\begin{figure}[h!]
        \centering
        \begin{subfigure}[t]{0.32\textwidth}
                \includegraphics[width=\textwidth]{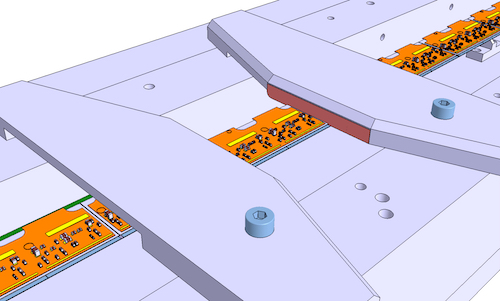}
                \caption{ Installation of spacer and neighbouring module protection.}
                \label{fig:rew_1}
        \end{subfigure}
        \centering
        \begin{subfigure}[t]{0.32\textwidth}
                \includegraphics[width=\textwidth]{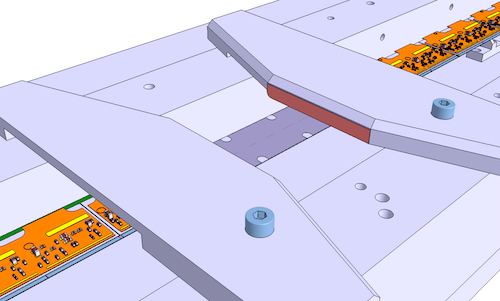}
                \caption{Module ungluing and face-plate cleaning.}
                \label{fig:rew_2}
        \end{subfigure}
                \centering
        \begin{subfigure}[t]{0.32\textwidth}
                \includegraphics[width=\textwidth]{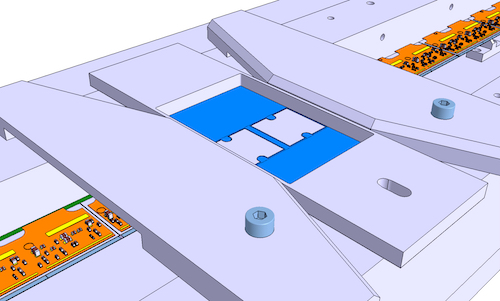}
                \caption{Thermal grease mask installation.}
                \label{fig:rew_3}
        \end{subfigure}
                \centering
        \begin{subfigure}[t]{0.32\textwidth}
                \includegraphics[width=\textwidth]{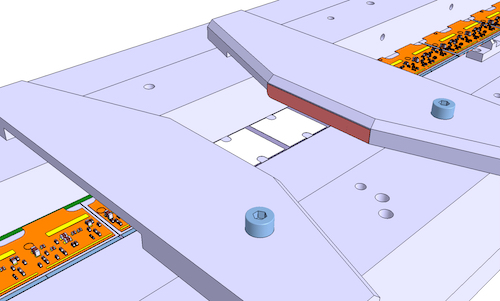}
                \caption{Thermal grease pattern and glue dots.}
                \label{fig:rew_4}
        \end{subfigure}
            \centering
                        \begin{subfigure}[t]{0.32\textwidth}
                \includegraphics[width=\textwidth]{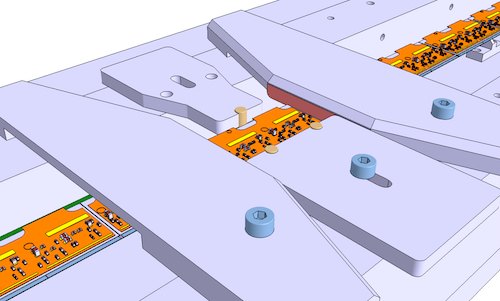}
                \caption{Module placement using positioning tools.}
                \label{fig:rew_5}
        \end{subfigure}
                \centering
        \begin{subfigure}[t]{0.32\textwidth}
                \includegraphics[width=\textwidth]{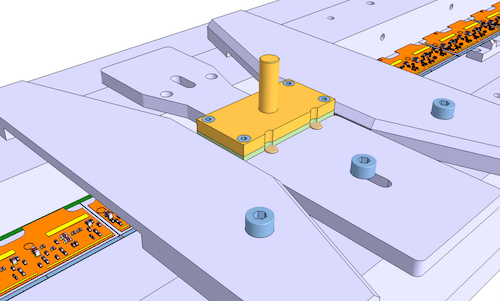}
                \caption{Weight positioning and glue curing.}
                \label{fig:rew_6}
        \end{subfigure}
                \centering

        \caption{Three dimensional renditions of the main re-work steps for replacement of modules.}
        \label{fig:ModuleReplacementStep}
\end{figure}

A module re-work procedure, sketched in Figure~\ref{fig:ModuleReplacementStep}, was established to replace modules 
if required %in case it was needed 
as a result of mechanical damage or functional issues. %that were damaged during the loading procedure or showed electrical or functional issues during the stave tests. 
A \SI{\sim100}{\micro\meter}-thick Kapton\textsuperscript{\textregistered} spacer and  holders were installed at each side of the module to be replaced, protecting the neighbouring modules. The module was %then 
removed using a spatula and the stave face plate was cleaned from grease or glue. A new module was 
%subsequentially 
then loaded following the procedure described above, using a set of dedicated tools. In total, twenty-two modules were re-worked: fifteen modules were replaced 
%re-works were made
due to module damage %being accidentally damaged 
during the module loading procedure that compromised the %, thus compromising the
integrity of the sensor, the FE-I4B chip or the module flex; five modules were replaced because of failures of
the FE-I4B chip after loading; and two modules were replaced because they failed the QA
tests made  before final integration. In addition, due to the cleaning
and re-bonding intervention performed (Section~\ref{sec:corrosion}), six %more 
modules were replaced on the twelve re-bonded staves.

\subsection{Quality assurance of the stave assemblies}

%AGC-060917
\subsubsection{Stave cooling performance}
 \label{sec:bare-stave-cooling}
 
An essential aspect of the stave design is the removal of heat from the IBL modules, under any foreseeable environmental and running condition. The sensor temperatures must be controlled, as an increasing sensor temperature will result in an increased sensor leakage current that may lead to thermal runaway~\cite{kohriki:1996}, damaging the module.
To minimise the effects of reverse annealing and thermal runaway in the sensors, the target temperature of IBL modules during operation is below \SI{-15}{\celsius}.
%An additional constraint is to safely cool down  
Furthermore, the modules must remain cool during beam-pipe bake-out procedures when the beam pipe at the IBL contact points reaches \SI{110}{\celsius}.
%brought at very high temperature to de-gas it  before  beams routinely circulate.% (\emph{{\color{red} Add info and requirements!}}).

The bi-phase CO$_2$ cooling system (Section~{\ref{sec:cooling}) was chosen to achieve the required low temperature module operation while minimising the pipe diameter and the associated radiation length~\cite{1748-0221-12-02-C02064}. 
The \SI{1.7}{\milli\meter} external diameter titanium T40 cooling tubes have a \SI{0.11}{\milli\meter} wall thickness to meet the CO$_2$ critical pressure (\SI{73.8}{\bar} at \SI{31}{\celsius}) requirements.

Several simulations were performed to estimate the %ultimate stave
stave thermal performance, quantified by the Thermal Figure of Merit (\SI{}{TFoM}), i.e. the thermal resistance between the cooling fluid and the sensor surface. Considering the various heat sources and an active sensor area $A_{\SI{}{sensor}}$, the sensor temperature can be expressed as:
\[
%\begin{equation}
T_{\SI{}{sensor}} = T_{\SI{}{coolant}} + \frac{TFoM}{A_{\SI{}{sensor}}} \times  (P_{\SI{}{chip}} + P_{\SI{}{flex}} + P_{\SI{}{sensor}} )
\]
%\end{equation}
where  $P_{\SI{}{chip}}$  is the power dissipated in the chip;
$P_{\SI{}{flex}}$ is the power dissipated in the module flex; and
$P_{\SI{}{sensor}}$ is the power dissipated in the sensor.
In worst-case conditions for $P_{\SI{}{chip}}$, $P_{\SI{}{flex}}$ and $P_{\SI{}{sensor}}$, the thermal runaway was estimated to occur for the 
%$TFoM > \SI{30}{\celsius.\centi\meter^2\per\watt}$.
\SI{}{TFoM} > \SI{30}{\celsius.\centi\meter^2\per\watt}.

Figure~\ref{fig:TemperatureProfile} shows the measurement %measured result 
for a prototype stave during the final design qualification. The measured 
\SI{}{TFoM}  is  \SI{14}{\celsius. \centi\meter^2\per\watt} for $T_{\SI{}{coolant}} = \SI{-22}{\celsius}$, conservatively a factor two  below 
the \SI{}{TFoM} value at thermal runaway. %the allowed value. 
To qualify the thermal performance of the stave during  production, the temperature uniformity was measured for each stave after module loading. All of the installed staves show a temperature dispersion within \SI{0.5}{\celsius}, as shown in Figure~\ref{fig:ModuleTemperature}, which is a good indication of the production uniformity. \\

\begin{figure}[!htb]
        \centering
        \begin{subfigure}[t]{0.47\textwidth}
                 \includegraphics[width=\textwidth]{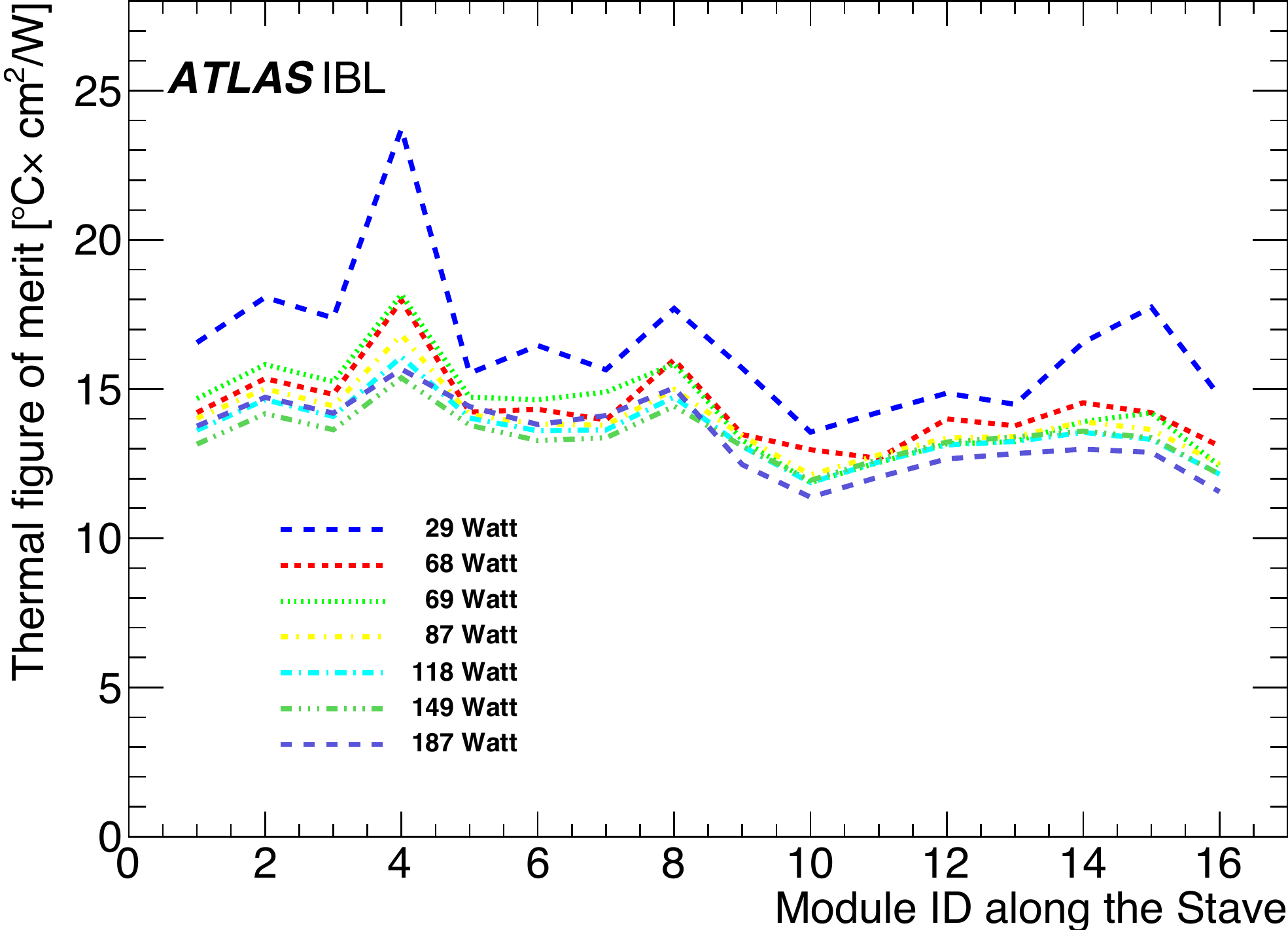}
        \caption{}
    \label{fig:TemperatureProfile}
        \end{subfigure}
        \begin{subfigure}[t]{0.47\textwidth}
                 \includegraphics[width=\textwidth]{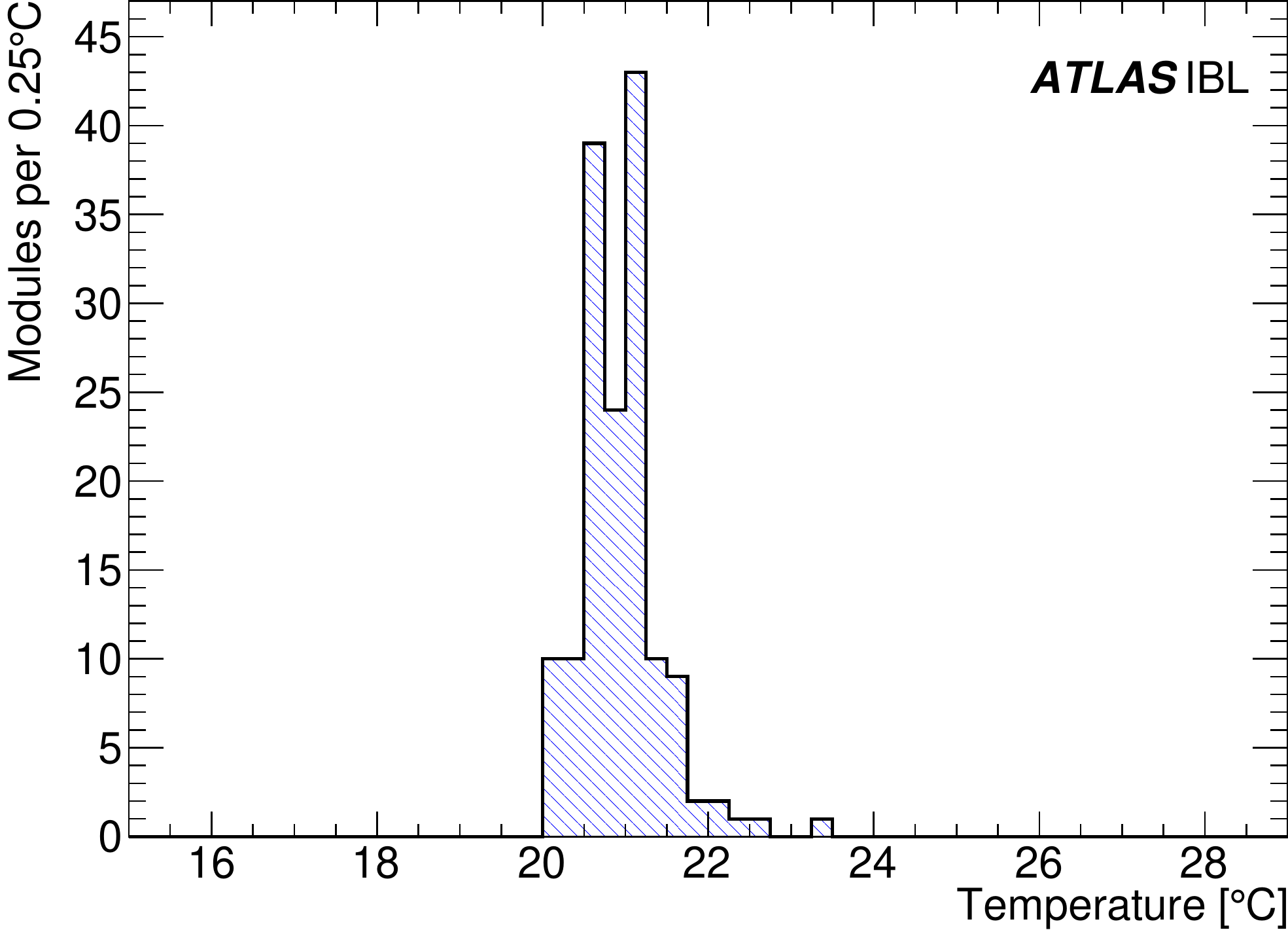}
    \caption{}
    \label{fig:ModuleTemperature}
        \end{subfigure}
  \caption{(a) Measured Thermal Figure of Merit (\SI{}{TFoM}) for modules along a stave, with a \SI{1.7}{\milli\meter} outer diameter titanium cooling tube. (b) The module temperature distribution on produced staves, with configured read-out electronics and a cooling tube temperature of \SI{19}{\celsius}. The histogram has 152 entries, a mean value of %\SI{20.9}{\degrees\celsius} and a RMS of \SI{0.5}{\degrees\celsius}.}
  \SI{20.9}{\celsius} and an RMS of \SI{0.5}{\celsius}.}
  \label{fig:Temperature}
\end{figure}
% Javier
\subsubsection{Metrology survey}
 \label{sec:metrelogy_survey}

An optical survey of the stave assemblies  was made %achieved 
%AGC by exploiting 
using the four fiducial marks  placed on each module.
The error distributions of the module loading position are shown in Figure\,\ref{fig:loadingpre} for the $\phi$ and $z$ axes\footnote{For the definition of the $\phi$ and $z$ directions refer to Figure~\ref{fig:BareStavesPlanarity}.}: an RMS precision of \SI{50}{\micro\meter} is measured for planar modules in both directions, and for 3D modules an RMS precision of \SI{56}{\micro\meter} in $z$ and \SI{33}{\micro\meter} in $\phi$ is achieved.
%Mean value for the $x$ axis is 50 with a RMSD of $TBC$ and the mean value for the $y$ axis is $TBC$ with a RMSD of $TBC$.

\begin{figure}[h!]

\centering
        \begin{subfigure}[t]{0.45\textwidth}
                \includegraphics[width=\textwidth]{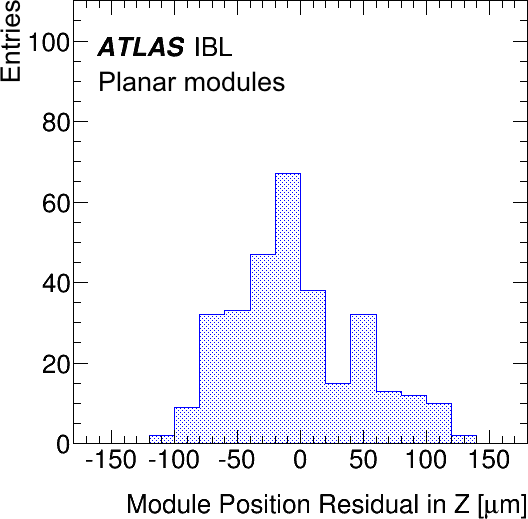}
                \caption{}
                \label{}
        \end{subfigure}
        \begin{subfigure}[t]{0.45\textwidth}
                \includegraphics[width=\textwidth]{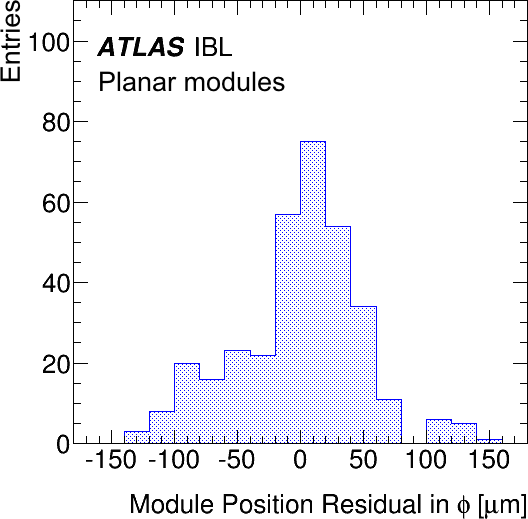}
                \caption{}
                \label{}
        \end{subfigure}
        \begin{subfigure}[t]{0.45\textwidth}
                \includegraphics[width=\textwidth]{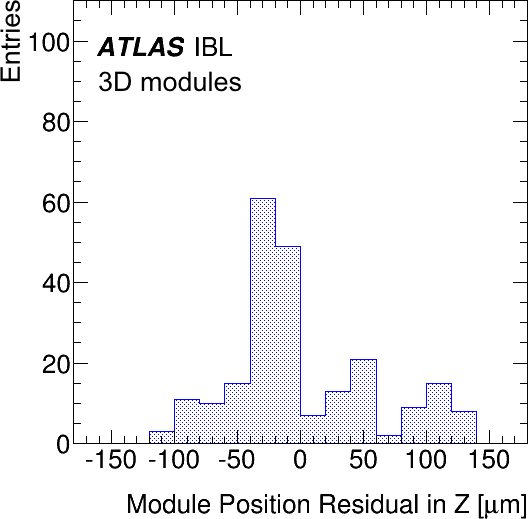}
                \caption{}
                \label{}
        \end{subfigure}
        \begin{subfigure}[t]{0.45\textwidth}
                \includegraphics[width=\textwidth]{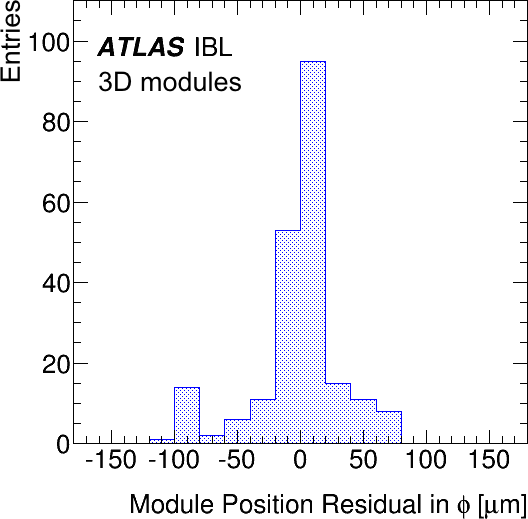}
                \caption{}
                \label{}
        \end{subfigure}

\caption{Fiducial mark residuals in the directions $z$  and $\phi$ for (a,b) double-chip planar modules and (c,d) single-chip 3D modules.}
\label{fig:loadingpre}
\end{figure}

% Antonello, Francesco, Jenny, Michael
\subsubsection{Functional qualification}
 \label{sec:func-qualif}

Following the metrology survey, the staves were transported to CERN where the final stave qualification was carried out. Results from the electrical qualification~\cite{StaveQANote} were an important input for the selection of the 14 staves that were integrated in the final detector.

The electrical QA test stand could operate two staves simultaneously. 
%An overview of the setup is shown in Figure~\ref{fig:setup}.
The staves were installed in an environmental box, which was flushed with dry air to maintain the humidity below \SI{3}{\percent},
corresponding to a dew point of \SI{-53}{\celsius} for the minimum box temperature of \SI{-20}{\celsius}. 
The temperature on the staves was controlled  by a TRACI\footnote{
Transportable Refrigerator Apparatus for CO$_2$ Investigation, https://ep-dep-dt.web.cern.ch/co2-coolingplants/traci-geneve.}
CO$_2$ cooling plant. 

The Detector Control System (DCS) and the Data Acquisition (DAQ) for the test stand emulated the DCS and DAQ of the installed IBL detector, described in Sections~{\ref{sec:system_dcs}} and {\ref{sec:system_daq}}. They were connected to the stave using custom EoS PCBs. The  read-out system 
%was based on 
used the Reconfigurable Cluster Elements (RCE) architecture based on ATCA technology~\cite{1748-0221-11-01-C01059}.
As for the installed IBL detector, the DCS granularity was a group of four read-out chips; it was not possible to record and control individual modules. The nomenclature of the FE-I4B chips as used for the stave QA, and of the DCS read-out for a stave, are shown in Figure~\ref{fig:stave_chipname}.    
\begin{figure}[!htb]
    \centering
                \includegraphics[width=\textwidth]{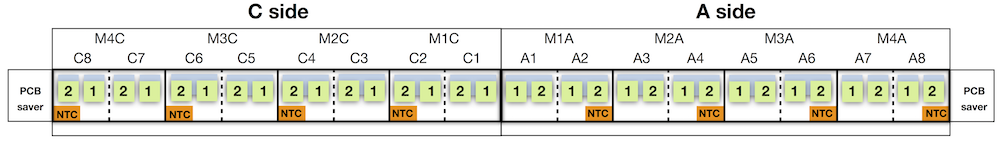}
    \caption{Schematic for a full stave of the module and FE-I4B layout, as well as the DCS and powering modularity. The FE-I4B chips are labelled for the stave and integration QA as A(C)X-Y where X runs from 1 to 8 towards the PCB cards, and Y is 1 or 2. A(C)7 and A(C)8 are single-chip 3D modules.  The DCS modularity is shown as M1A(C) through M4A(C). The NTCs measure the temperature of each module group.}
    \label{fig:stave_chipname}
\end{figure}

%%OPTICAL INSPECTION
A detailed optical inspection of each stave was made before and after the QA procedure. Photographs of each chip were taken with a high resolution camera to identify any major damage to the chip, as well as debris that might have been left on the stave during assembly or wire bonding. A detailed microscope inspection of the stave was then made. 
%The full two-step optical inspection was repeated after all remaining QA tests were completed to ensure that nothing had been damaged during the QA process.

%Before running calibration measurements, the basic electrical functionality of a stave must be verified. 
The electrical functionality of each stave was next verified, before making calibration measurements. This included power-up studies, the verification of set voltages and currents in un-configured and configured FE-I4B chip states, and I-V characteristics of the sensors.
The powering behaviour of the modules was verified during ten LV power cycles. The time dependence of current and voltage from this test  are shown for stave ST12 in Figure~\ref{fig:LVcycle}. Most modules %tend to 
show a stable current consumption for every cycle; current fluctuations can be explained by   improperly reset chip registers. 
After a first successful power-up of the stave, the temperature, the set voltage reading, and the currents before and after read-out configuration
of each DCS group, were recorded. The expected values during the stave QA  are listed in Table~\ref{tab:rec_test}.
\begin{table}[!htb]
\centering
\begin{tabular}{lcc}
\hline\hline
Quantity & Expected Value \\
\hline
Temperature & \SI{22}{\celsius} \\ %[when powered?]
Voltage & \SI{2.1}{\volt}\\
Unconfigured current & \SI{1.1}{\ampere}\\
Configured current & \SI{1.5}{\ampere}\\
%Leakage current & \SI{<20}{\micro\ampere}\\
Leakage current & Compatible with module QA\\
\hline\hline
\end{tabular}
%\caption{Data recorded for each DCS group, and their expected values, in the QA qualification of loaded staves after their reception at CERN.}
\caption{Quantities recorded for each DCS group (four read-out chips) and their nominal expected values in the QA qualification of loaded staves after their reception at CERN.}
\label{tab:rec_test}
\end{table}
%
%\subsubsection{IV scans}
An I-V scan was performed to  characterise the sensor quality. The sensor HV was ramped in 20 steps from 0 to \SI{100}{\volt} for 3D sensors and from 0 to \SI{200}{\volt} for planar sensors.  
The I-V characteristics of the DCS module group of four read-out chips was required to be compatible with the measurements performed on modules before the stave loading. Results for stave ST12 are shown in Figure~\ref{fig:IVcurves}. 
For this particular stave, one DCS group (M4C) indicated a 3D sensor I-V breakdown, above the operating voltage and still within the QA specification.

\begin{figure}
	\centering
	\begin{subfigure}[t]{0.47\textwidth}
%AGC		 \includegraphics[width=\textwidth]{figures/chapter06_LoadingQA/ST12_LVcylces.pdf}
		 \includegraphics[width=\textwidth]{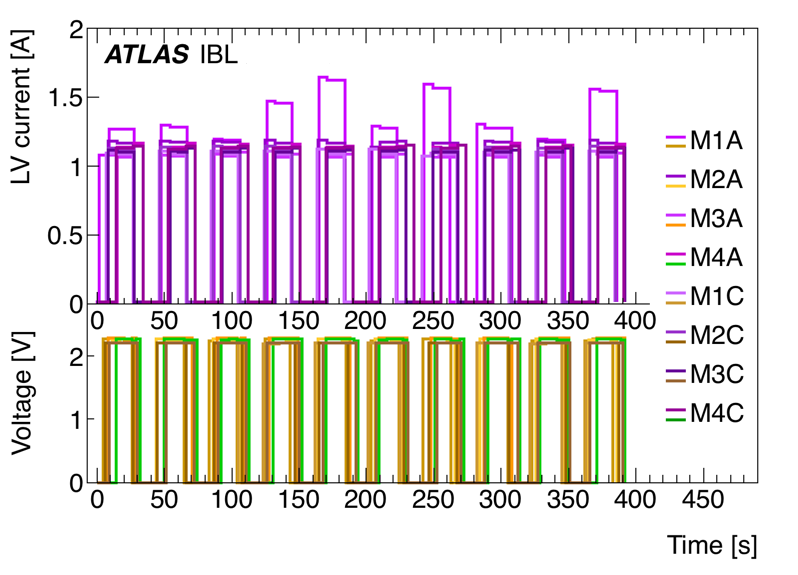}
		\caption{}
		\label{fig:LVcycle}
	\end{subfigure}
	\begin{subfigure}[t]{0.47\textwidth}
%AGC		 \includegraphics[width=\textwidth]{figures/chapter06_LoadingQA/ST12_IVcurves.pdf}
		 \includegraphics[width=\textwidth]{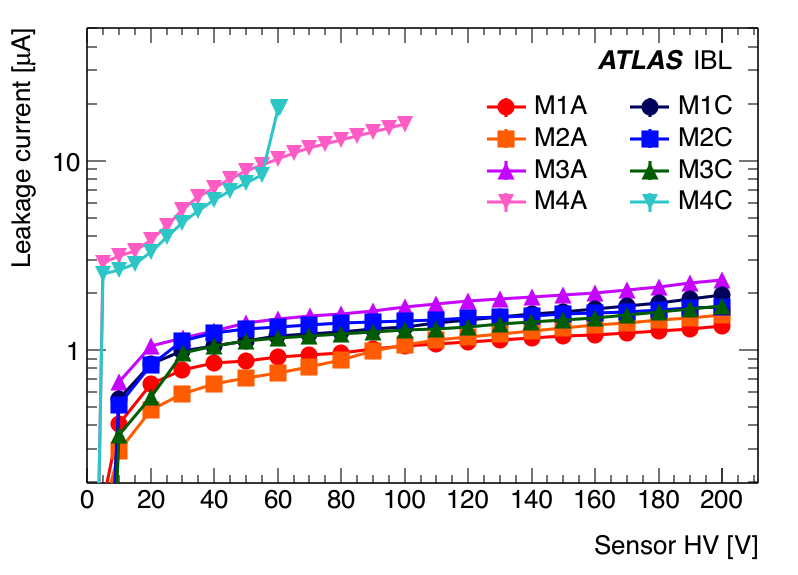}
		\caption{}
		\label{fig:IVcurves}
	\end{subfigure}
	\caption[]{ \subref{fig:LVcycle} LV cycles and \subref{fig:IVcurves} sensor I-V characteristics for the 8 DCS groups of stave ST12. The DCS granularity is a group of four read-out chips (two planar modules or four 3D modules). }
	\label{fig:PoweringPlots}	
\end{figure}

A series of basic functionality  tests, including  the digital functionality, the t$_{\SI{}{0}}$ calibration and threshold, ToT  and crosstalk scans, were also made, using the module configurations recorded at previous module QA testing sites (Section \ref{sec:module_qa}). 
This  allowed a direct comparison of the stave functionality, and discrepancies indicating possible damage induced during module loading or transportation to CERN.

A comparison of threshold scans with and without HV allows the identification of areas of disconnected bumps, since a pixel that is no longer connected to the sensor is not affected by the higher noise present when the sensor is not fully depleted. Nevertheless, a source scan is needed to fully identify all disconnected bumps as described in Section~\ref{sec:source-scans}. 

\subsubsection{Stave calibration}
 \label{sec:stave-calib}

Following a t$_{\SI{}{0}}$ calibration, the discriminator threshold and the ToT parameters of the FE-I4B chips must be calibrated and tuned to distinguish charged particle signals from electronic noise, and to ensure that the charge determination is uniform over all IBL pixels. The selected threshold should be as low as possible to ensure maximal detector efficiency for charged particles (especially when the charge is shared between pixels), and to ensure the best possible ToT tuning. 
%On the contrary, it should be sufficiently high to discriminate against electronic noise. 
It should also be sufficiently high to discriminate against electronic noise. 

%During the module QA (Section~\ref{sec:module_qa}), a mean module threshold of \SI{3000}{\e} at a module temperature of \SI{22}{\celsius} was chosen. To compare with those results the same calibration working point was retained. However, at lower operational temperatures, the mean module threshold can be lowered because of the reduced electronic noise and a second calibration working point of \SI{1500}{\e} at \SI{-12}{\celsius} was chosen to test the IBL calibration capabilities in realistic operational conditions~\cite{StaveQANote}. The mean threshold can be tuned for each FE-I4B chip using charge injection circuitry. In each case the ToT was tuned to 10 units of \SI{25}{\nano\second} for a charge of \SI{16000}{\e}, corresponding to a minimum ionising particle at normal incidence. 
During the module QA (Section~\ref{sec:module_qa}), a mean module threshold of \SI{3000}{\e}. To compare with those results, the same calibration working point was retained. A more realistic operating condition is to use a lower pixel threshold as noted above. A second calibration working point of \SI{1500}{\e} at an operating temperature of \SI{-12}{\celsius} was chosen~\cite{StaveQANote}. The mean threshold can be tuned for each FE-I4B chip using charge injection circuitry. As for the QA of individual modules, the pixel noise (ENC)\footnote{ENC is defined in Section~\ref{sec:module_elec_noise_qa}.} is evaluated from a measure of the pixel occupancy as a function of injected charge, using the FE-I4B charge injection circuitry (the S-curve method). 

In each case the ToT was tuned to 10 units of \SI{25}{\nano\second} for a charge of \SI{16000}{\e}, corresponding to a minimum ionising particle at normal incidence. 
Figure~\ref{fig:totstaves} shows the one-dimensional ToT distribution for all pixels when tuned to 10 ToT, and the mean ToT as a function of chip number along the stave. The data are shown for a mean module threshold of \SI{1500}{\e} at \SI{-12}{\celsius}. 

The threshold calibration results for the complete set of staves are summarised in Table~\ref{table:calib_summary}. 
%The threshold RMS of approximately \SI{40}{\e} is compatible with the accuracy of the injection circuit. 
The pixel-to-pixel RMS of the threshold distribution (threshold RMS), of approximately \SI{40}{\e}, is compatible with the accuracy of the injection circuit. 
%The behaviour exhibited by FBK modules is driven by  two noisy modules and by the instrumental setup, as explained in ~\cite{StaveQANote}. 
Figure~\ref{fig:ThresholdPerPixel} shows the threshold and noise (ENC) distributions for individual pixels, in the case of a \SI{3000}{\e} threshold tuning at \SI{22}{\celsius}. Figure~\ref{fig:ThrPerPix} indicates that all of the modules could be tuned to a mean threshold of  \SI{3000}{\e}, with only a few individual pixels that could not be tuned to that value. 
%%%%%%%%%%
%
\begin{figure}[h!]
        \centering
                \includegraphics[width=1.1\textwidth]{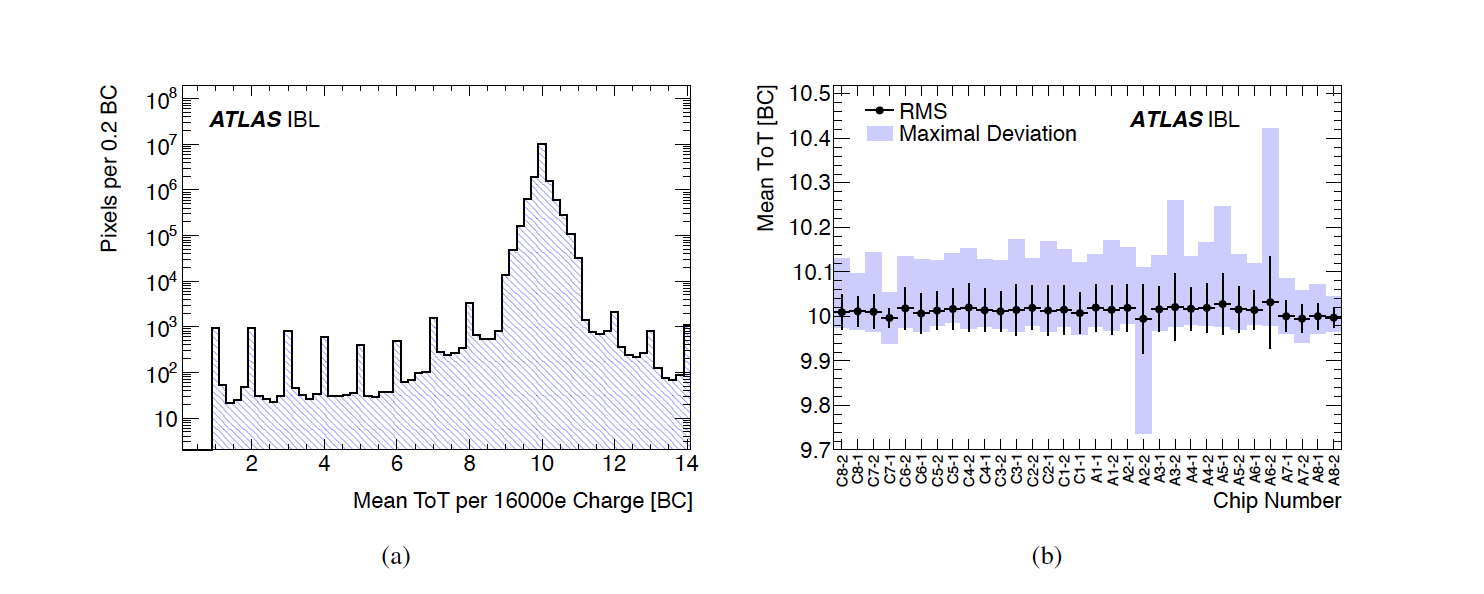}
        \caption{(a) One-dimensional Time-over-Threshold (ToT) distribution for all pixels when tuned to 10 ToT, using an injected charge of \SI{16000}{\e}. The spikes result from injection failures where the selected ToT value is not 10. Non-integer mean ToT values result when successive injections to evaluate the mean select differing ToT values. (b) The average ToT distribution as a function of the chip number (position) on the stave. The data are from all 18 qualified staves using a mean module threshold of \SI{1500}{\e} at \SI{-12}{\celsius}.}
         \label{fig:totstaves}
\end{figure}
%%%%%%%%%%
% THIS EXPLANATION SHOULD NOT BE NECESSARY ANYMORE IF WE ONLY TALK ABOUT 14 STAVES!
%
\begin{table}[!htbp]
\renewcommand{\arraystretch}{1.2}
\centering
\begin{tabular}{llccccc}
\hline\hline
Tuned Threshold & Pixel Type & Threshold RMS [\SI{}{\e}] & ENC [\SI{}{\e}] & Threshold-over-Noise\\
\hline
\multirow{4}{*}{\SI{3000}{\e} at \SI{22}{\celsius}} 	& Planar Normal	& 37 & \num{123\pm 10} & \num{25\pm 2} \\
		                & Planar Long	& 58 & \num{146\pm 15} & \num{21\pm 2}\\
                                & 3D FBK		& 39 & \num{171\pm 25} & \num{18\pm 2}\\
                                & 3D CNM		& 40 & \num{149\pm 15} & \num{20\pm 2}\\
\hline
\multirow{4}{*}{\SI{1500}{\e} at \SI{-12}{\celsius}} 	& Planar Normal	& 42 & \num{129\pm 13} & \num{12\pm 1}\\
                                & Planar Long	& 47 & \num{149\pm 16} & \num{10 \pm 1}\\
                                & 3D FBK		& 46 & \num{171\pm 25} & \num{9\pm 1}\\
                                & 3D CNM		& 41 & \num{146\pm 16} & \num{10\pm 1}\\
\hline\hline
\end{tabular}
\caption{Calibration summary for the 18 qualified staves. Long pixels of planar sensors are listed separately to indicate a higher noise because of their larger pixel size.}
\label{table:calib_summary}
\end{table}
\begin{figure}[h!]
        \centering
        \begin{subfigure}[t]{0.48\textwidth}
                \includegraphics[width=\textwidth]{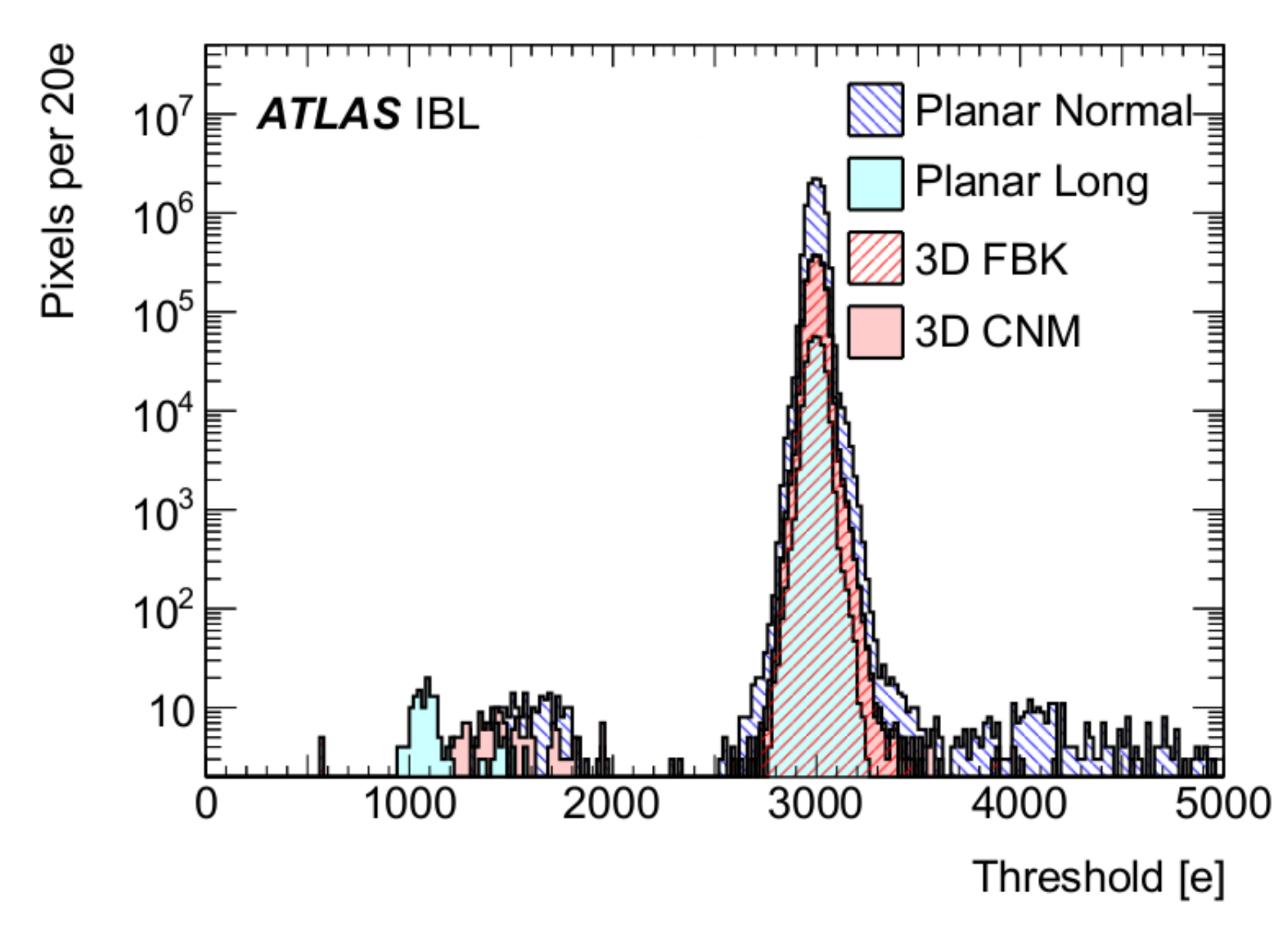}
                \caption{}
                \label{fig:ThrPerPix}
        \end{subfigure}
        \begin{subfigure}[t]{0.48\textwidth}
                \includegraphics[width=\textwidth]{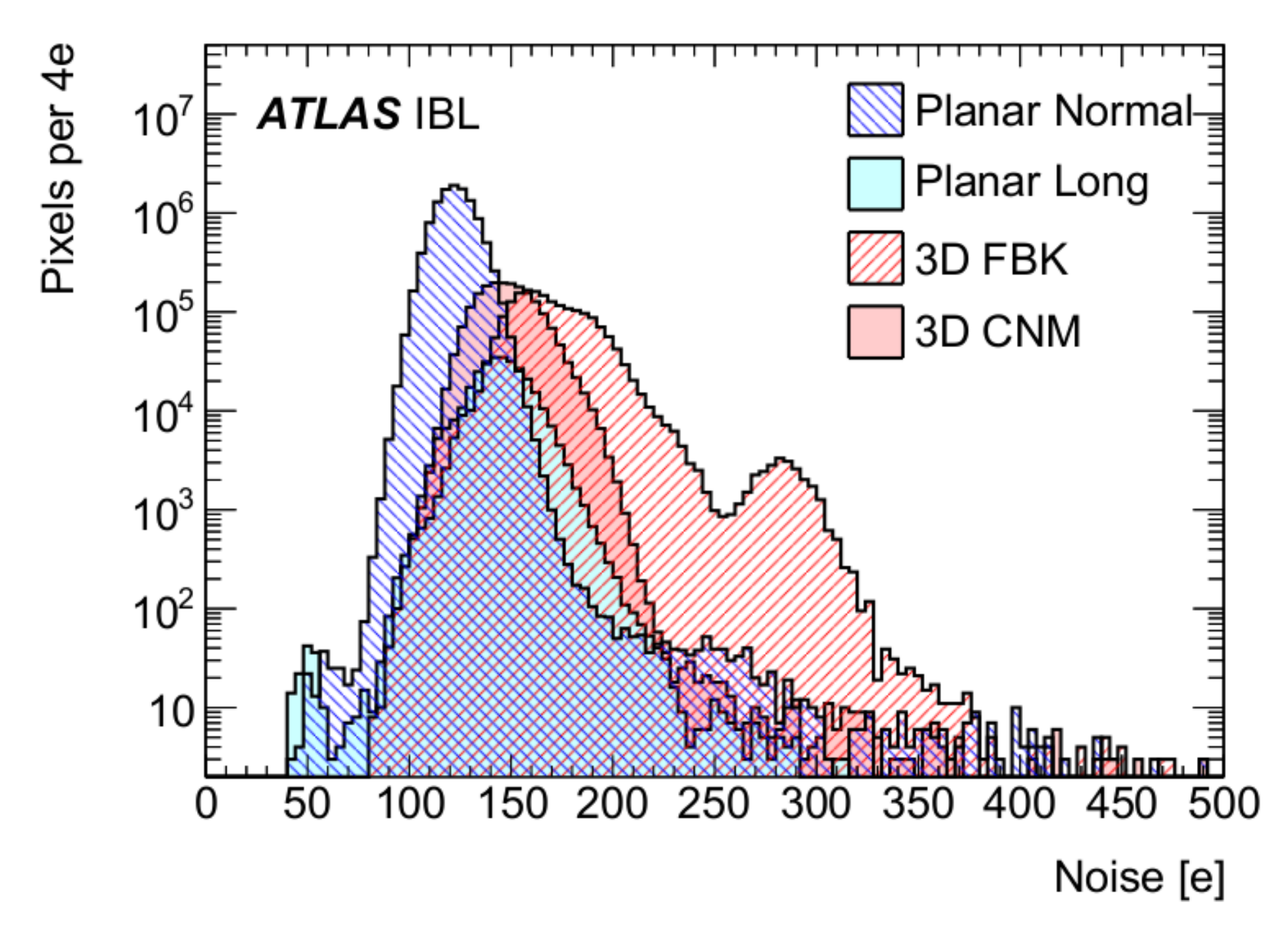}
                \caption{}
                \label{fig:ThrSigPerPix}
        \end{subfigure}
        \caption{\subref{fig:ThrPerPix} Threshold and \subref{fig:ThrSigPerPix} noise (ENC) distributions of individual pixels for a \SI{3000}{\e} threshold tuning at \SI{22}{\celsius}. The threshold width is primarily determined by the injection circuit, and only a few individual pixels were badly tuned.}
         \label{fig:ThresholdPerPixel}
\end{figure}

The threshold-over-noise distributions for \SI{3000}{\e} and \SI{1500}{\e} tunings are shown in Figure~\ref{fig:ThresholdOverNoise}; this quantity is the key parameter in determining the quality of the IBL modules with respect to their operability at a given discriminator setting.  
The physics occupancy in the ATLAS Pixel B-Layer was approximately \num{5e-4} hits per pixel per bunch crossing unit of \SI{25}{\nano\second} at the end of Run~1 and that for the IBL was expected to be approximately \num{e-3} hits per pixel per bunch crossing unit at the beginning of its operation. Pixels with a noise occupancy rate higher than \num{e-6} hits per pixel per bunch crossing unit were considered to be noisy and  were disabled from data taking. 
%This ensured a noise contamination in collisions of less than \SI{0.5}{\percent}. 
A threshold-over-noise value larger than 5  ensured a noise contamination in IBL physics hits of less than \SI{0.1}{\percent}. For the \SI{1500}{\e} reference tuning at \SI{-12}{\celsius} module temperature, 
the observed fraction of noisy IBL pixels was less than \SI{0.03}{\percent}.

\begin{figure}[h!]
        \centering
        \begin{subfigure}[t]{0.48\textwidth}
                \includegraphics[width=\textwidth]{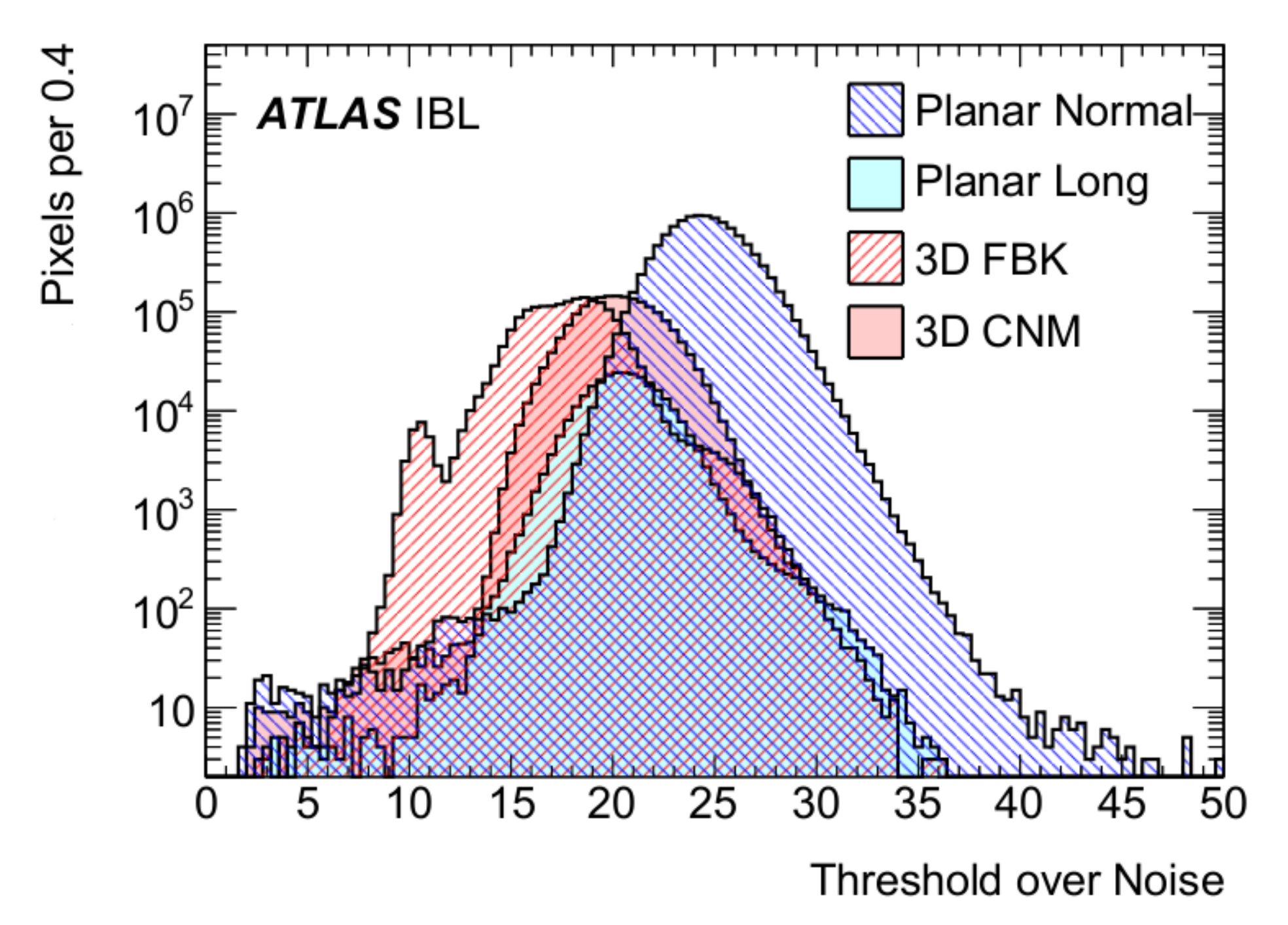}
                \caption{}
                \label{fig:ThrOverSigPerPix}
        \end{subfigure}
        \begin{subfigure}[t]{0.48\textwidth}
                \includegraphics[width=\textwidth]{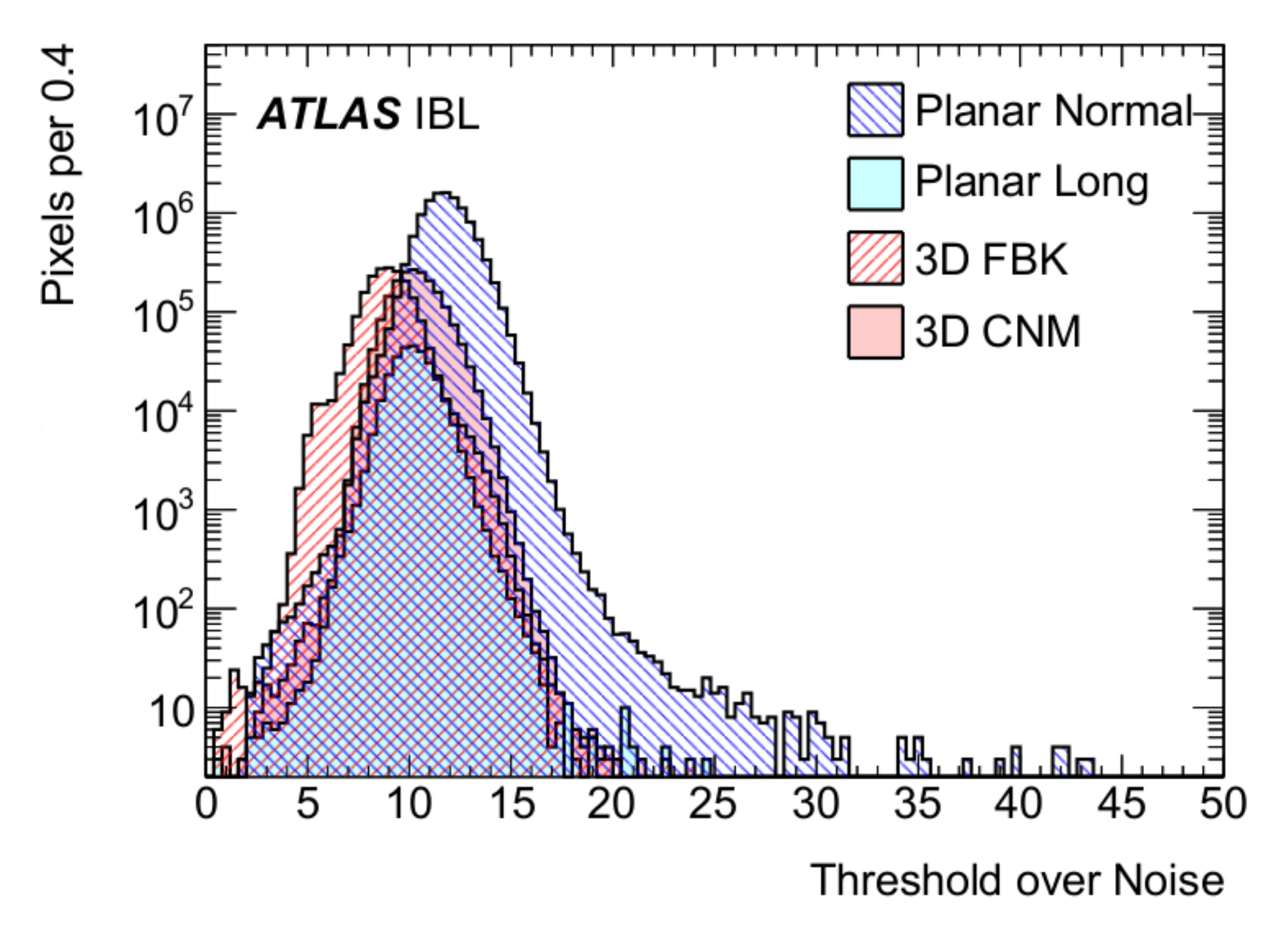}
                \caption{}
                \label{fig:ThrOverSigPerPix_1500e}
        \end{subfigure}
        \caption{Threshold-over-noise distribution of pixels for threshold tunings of \subref{fig:ThrOverSigPerPix} \SI{3000}{\e} at \SI{22}{\celsius} and \subref{fig:ThrOverSigPerPix_1500e} \SI{1500}{\e} at \SI{-12}{\celsius}.}
         \label{fig:ThresholdOverNoise}
\end{figure}

Figure~\ref{fig:thresh} %and~\ref{fig:thresh_sigmamean} 
shows the threshold and noise distributions  averaged over all 18 qualified staves as a function of chip number for the \SI{1500}{\e} reference tuning.  
Figure~\ref{fig:thresh_mean} shows that it was possible to tune all production staves  to \SI{1500}{\e} within  \SI{40}{\e}. The average noise was approximately \SI{130}{\e} for planar sensors and  less than about \SI{170}{\e} for 3D sensors. Slightly higher noise was observed on the A-side of the setup:  this was due to a combination of increased noise on the HV lines of the setup and the fact that FBK modules, which are more sensitive to external noise, represented the majority of the 3D sensors on this side. 
\begin{figure}[h!]
        \centering
      \begin{subfigure}[t]{0.48\textwidth}
                \includegraphics[width=\textwidth]{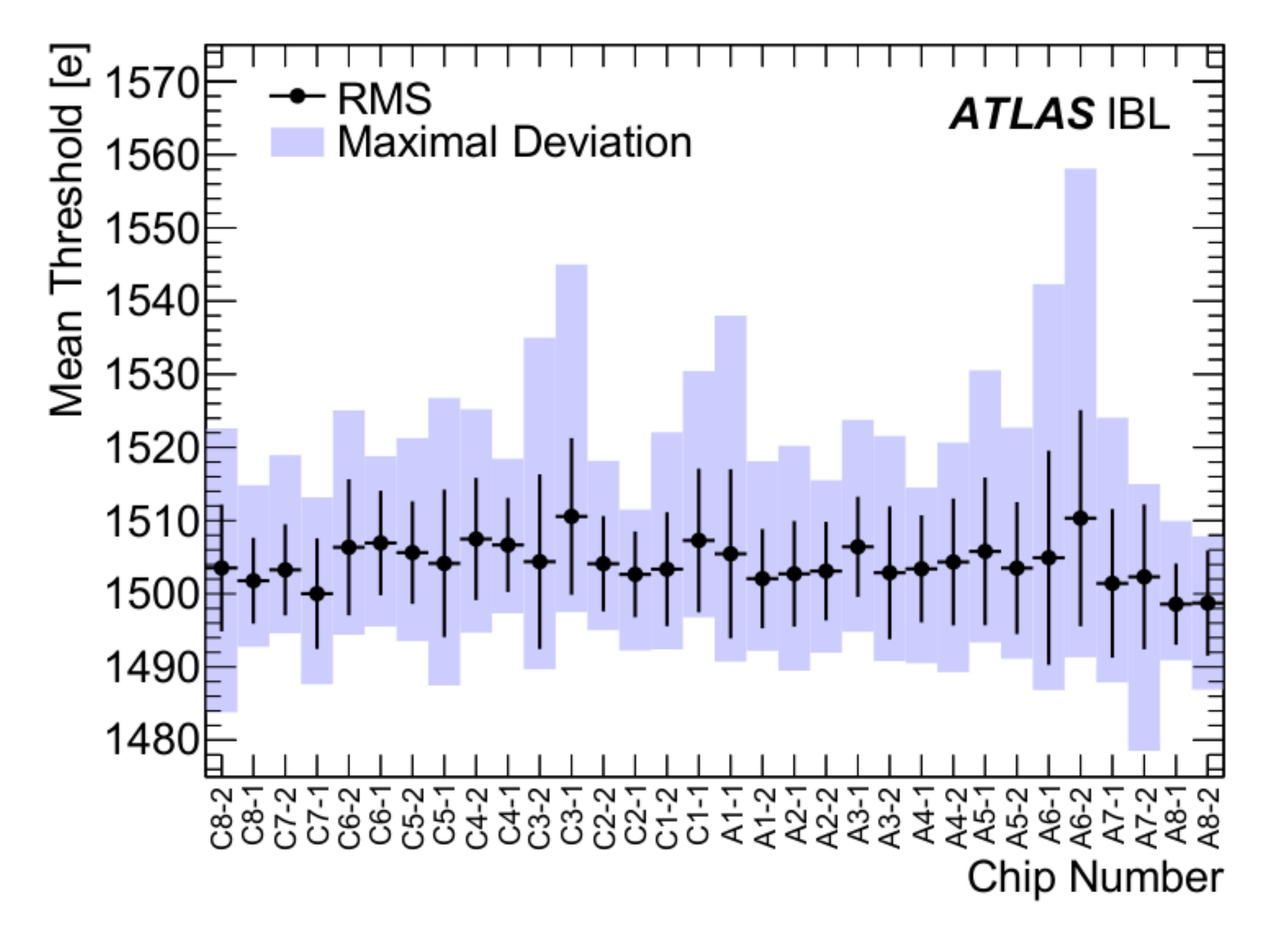}
                \caption{}
                \label{fig:thresh_mean}
        \end{subfigure}
        \begin{subfigure}[t]{0.48\textwidth}
                \includegraphics[width=\textwidth]{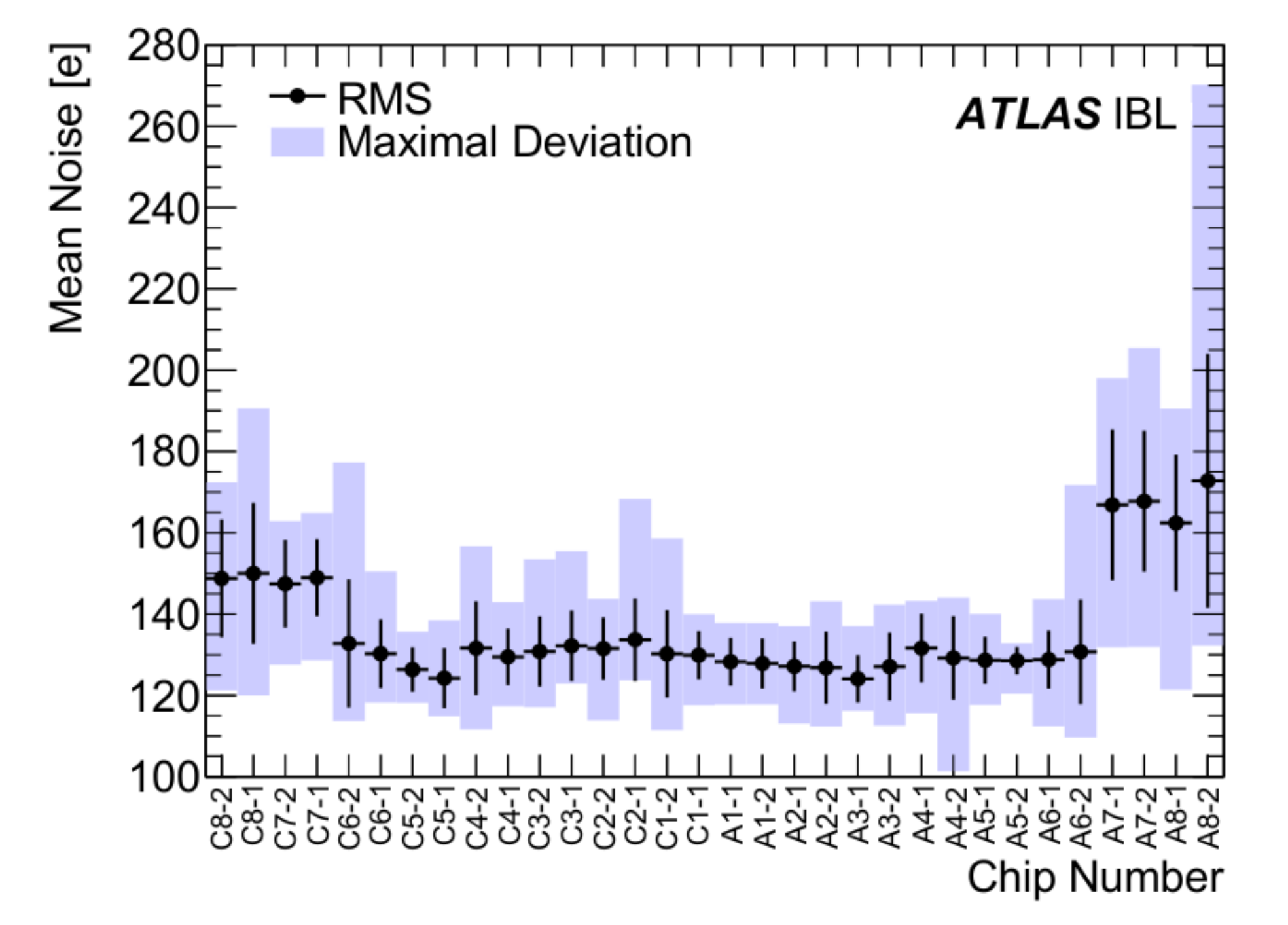}
                \caption{}
                \label{fig:thresh_sigmamean}
        \end{subfigure}
        \caption{\subref{fig:thresh_mean} Mean threshold and  \subref{fig:thresh_sigmamean} mean noise (ENC) distributions for all 18 qualified staves as a function of the chip number (position) on the stave, for a \SI{1500}{\e} threshold tuning at \SI{-12}{\celsius} . The error bars represent the RMS value accounting for  the differences between staves whereas the solid colour describes the full range covered by data.  }
         \label{fig:thresh}
\end{figure}

%%%%%%%%%%%%%%
\subsubsection{Source scans}
 \label{sec:source-scans}

\begin{figure}[!htb]
        \centering
 \includegraphics[width=0.8\textwidth]{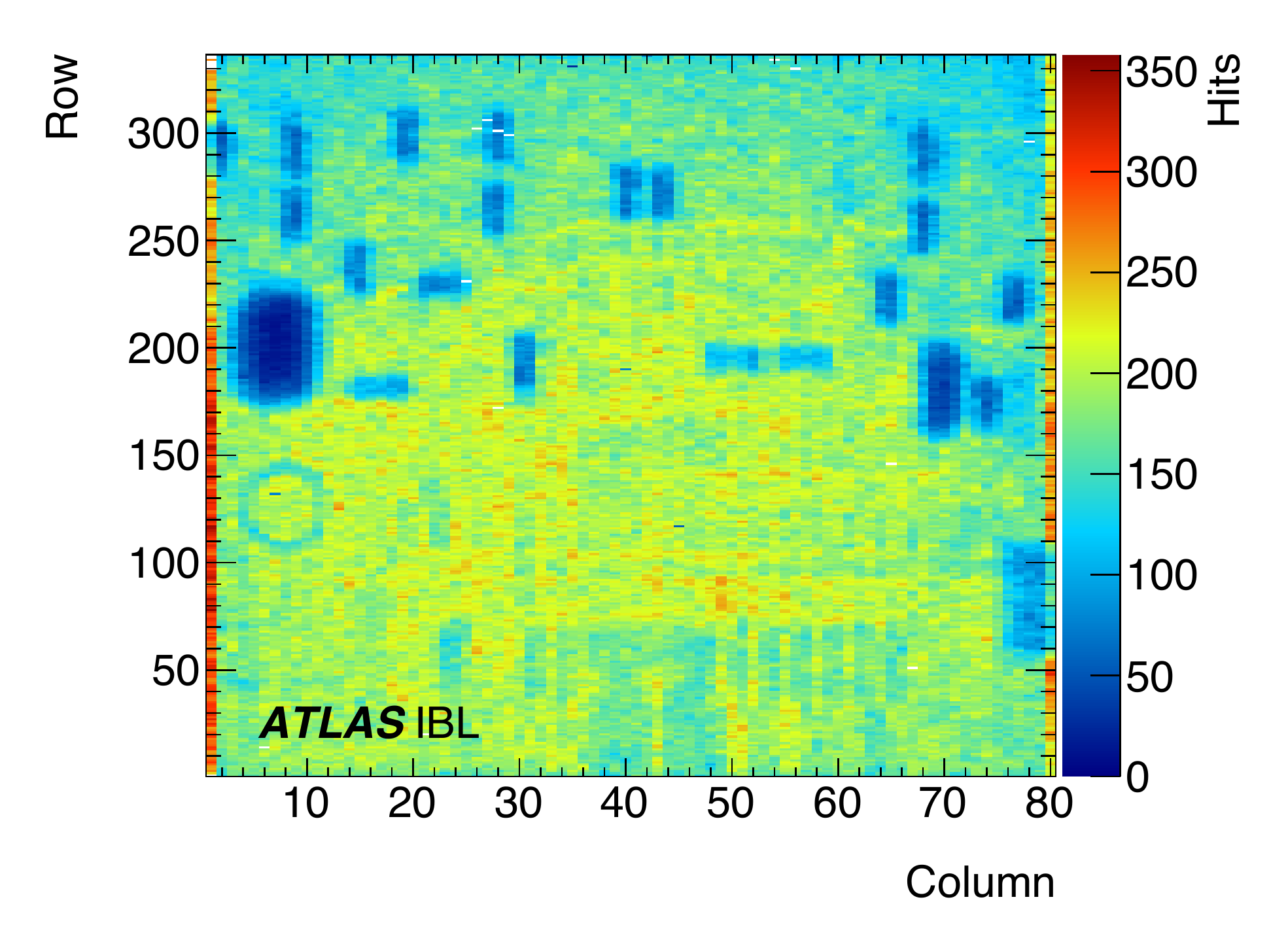}
\caption{Typical source scan hit map of a 3D FBK module during the stave QA. Regions with a lower number of hits match passive components mounted on the module flex. The module threshold is tuned to \SI{3000}{\e} at an operating temperature of \SI{22}{\celsius}.}
\label{fig:source_occ}
\end{figure}

\begin{figure}[h]
        \centering
        \begin{subfigure}[t]{0.48\textwidth}
                \includegraphics[width=\textwidth]{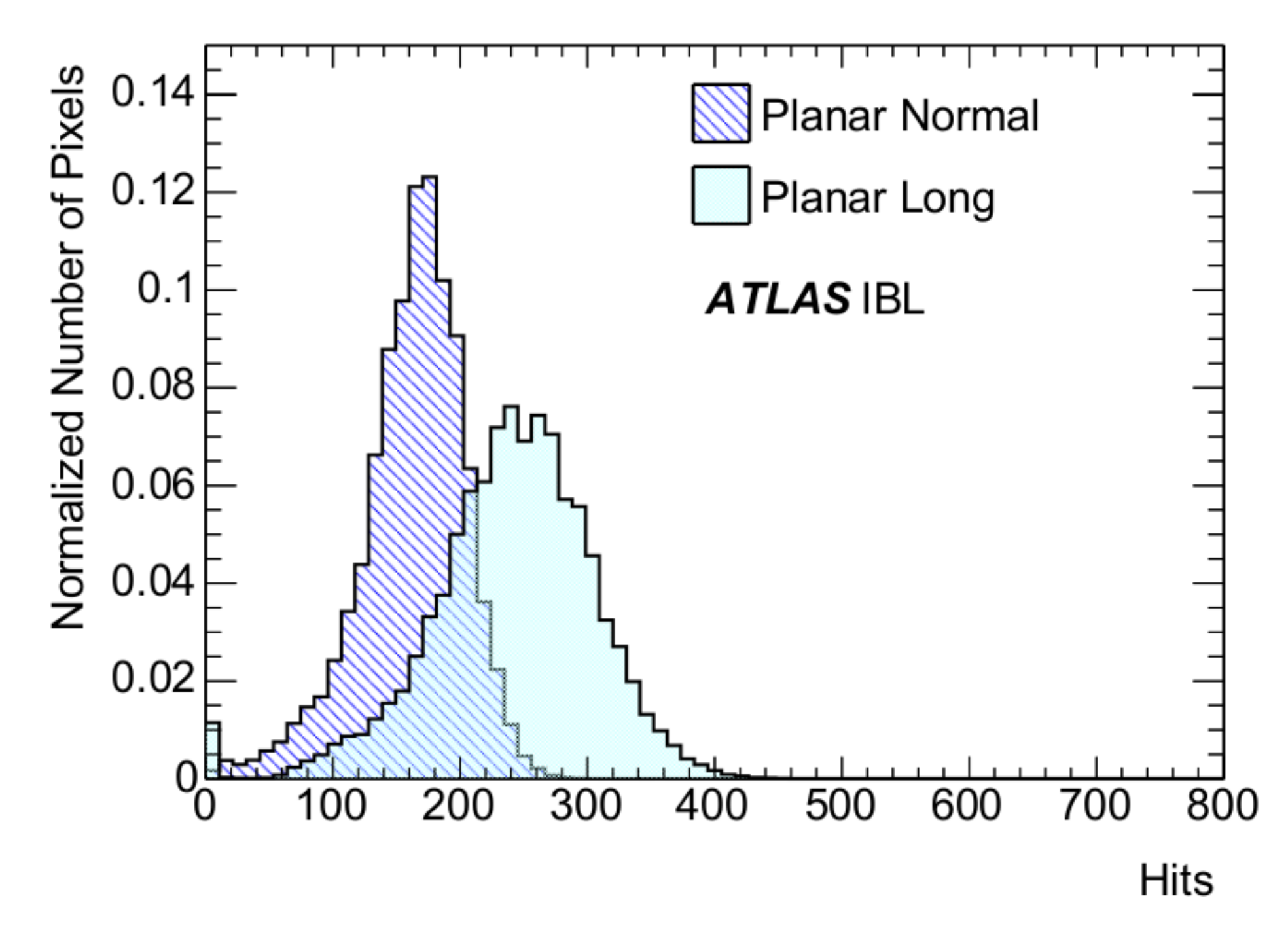}
                \caption{}
                \label{fig:source_occ_plan}
        \end{subfigure}
        \begin{subfigure}[t]{0.48\textwidth}
                \includegraphics[width=\textwidth]{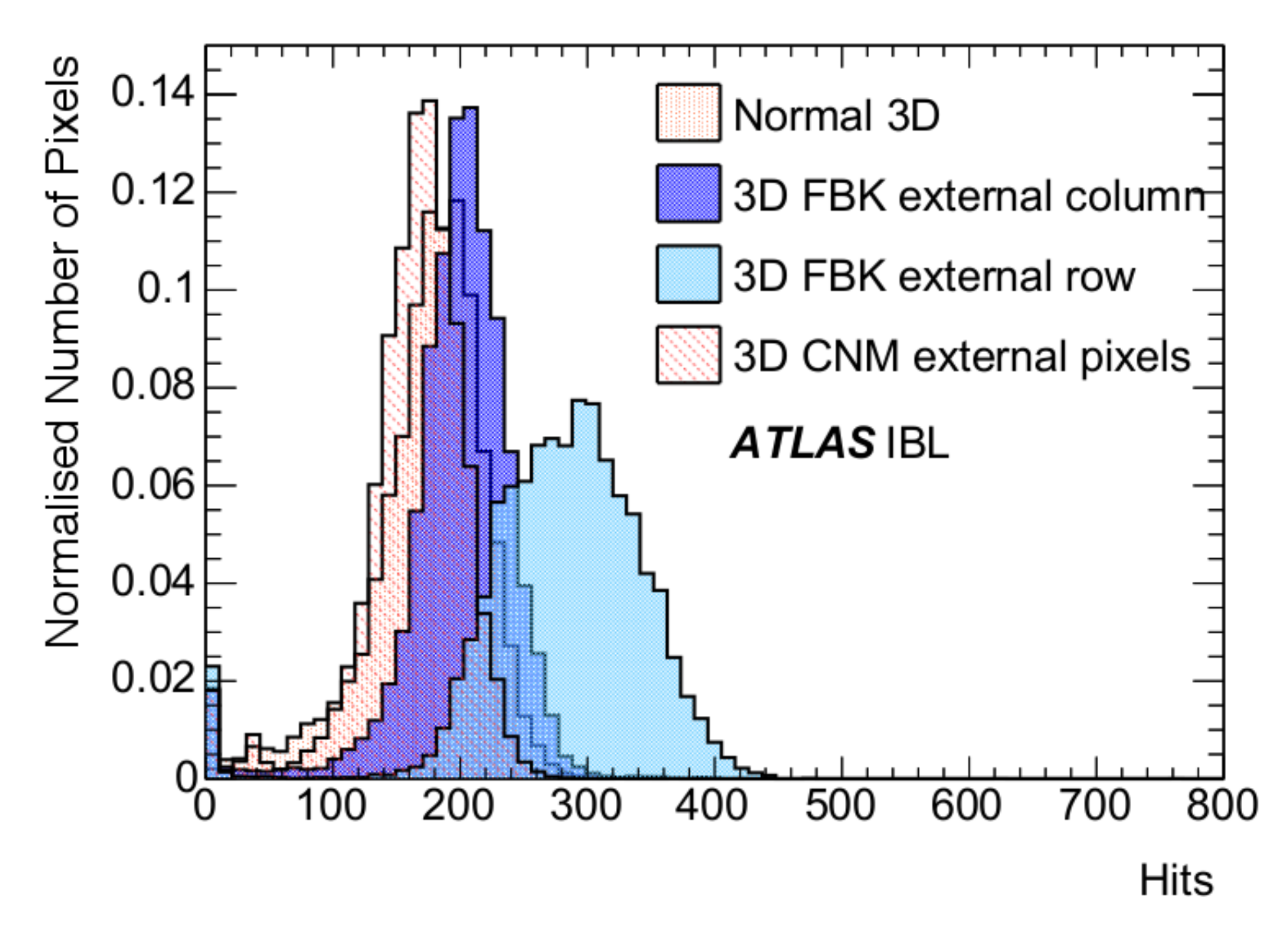}
                \caption{}
                \label{fig:source_occ_3D}
        \end{subfigure}
        \caption{Occupancy distributions for \subref{fig:source_occ_plan} planar modules and \subref{fig:source_occ_3D} 3D CNM and FBK modules collected using a $^{90}$Sr source and an internal self trigger mechanism during IBL stave QA. The content of each category of source measurement is normalised to unity in the figures. The module threshold is tuned to \SI{3000}{\e} at an operating temperature of \SI{22}{\celsius}.}
         \label{fig:sourcescanmod}
\end{figure}

Source scans were performed  with a  \SI{3000}{\e} threshold configuration using a  $^{90}$Sr source and an internal self trigger mechanism. An example of a source scan hit map for a single 3D FBK module is shown in Figure~\ref{fig:source_occ}. Regions with a lower number of hits are clearly visible and match the passive components mounted on the module flex. Each normal pixel collected approximately $150 - 200$ hits. An increased hit occupancy in the edge columns and the edge rows is due to an increased active area for FBK pixels resulting from 
%the guard ring structure. 
the fence structure at the edge.
Figure~\ref{fig:source_occ_plan} shows the hit occupancy for planar pixels, separating the categories of normal pixels and long pixels. The mean hit occupancy is less than that for 3D modules because of the reduced sensor thickness  (\SI{200}{\micro\meter} instead of  \SI{230}{\micro\meter}). Figure~\ref{fig:source_occ_3D} shows the occupancy for normal and edge pixels of CNM and FBK modules. 
Because of their guard-ring design CNM sensors do not show an excess in the edge rows and columns.
Although not shown, CNM sensors also have a slightly lower occupancy than FBK sensors for normal pixels because of the smaller depth of their columnar electrodes.

Source scans were mainly used to identify disconnected bumps, but it was also possible to check the average charge using the $^{90}$Sr source. 
This used the ToT distribution of all clusters with more than one hit\footnote
{Clusters are formed by groups of hits collected by adjacent pixels.}. 
Figure~\ref{fig:SourceClus} shows the  most probable value (MPV) of the Landau-Gauss fit of such distributions as a function of FE-I4B position, averaged over all staves; the %one dimensional 
MPV distribution of each chip is shown in Figure~\ref{fig:SourcePix}.  The different behaviour between 3D and planar modules is due to their different sensor thickness. The difference with respect to the calibrated mean of 10 ToT is small. 
 
\begin{figure}[h!]
        \centering
        \begin{subfigure}[t]{0.48\textwidth}
                \includegraphics[width=\textwidth]{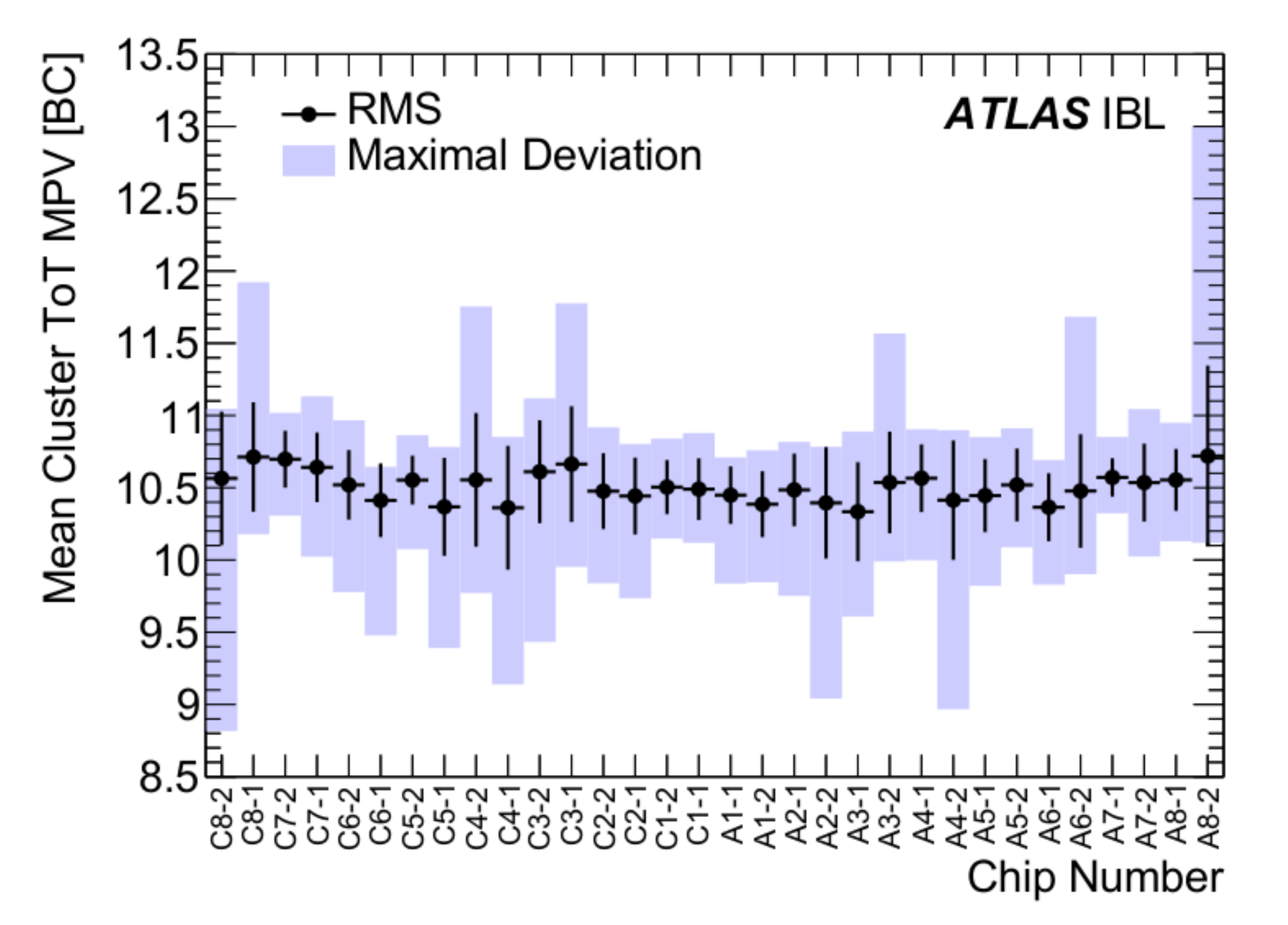}
                \caption{}
                \label{fig:SourceClus}
        \end{subfigure}
        \begin{subfigure}[t]{0.48\textwidth}
                \includegraphics[width=\textwidth]{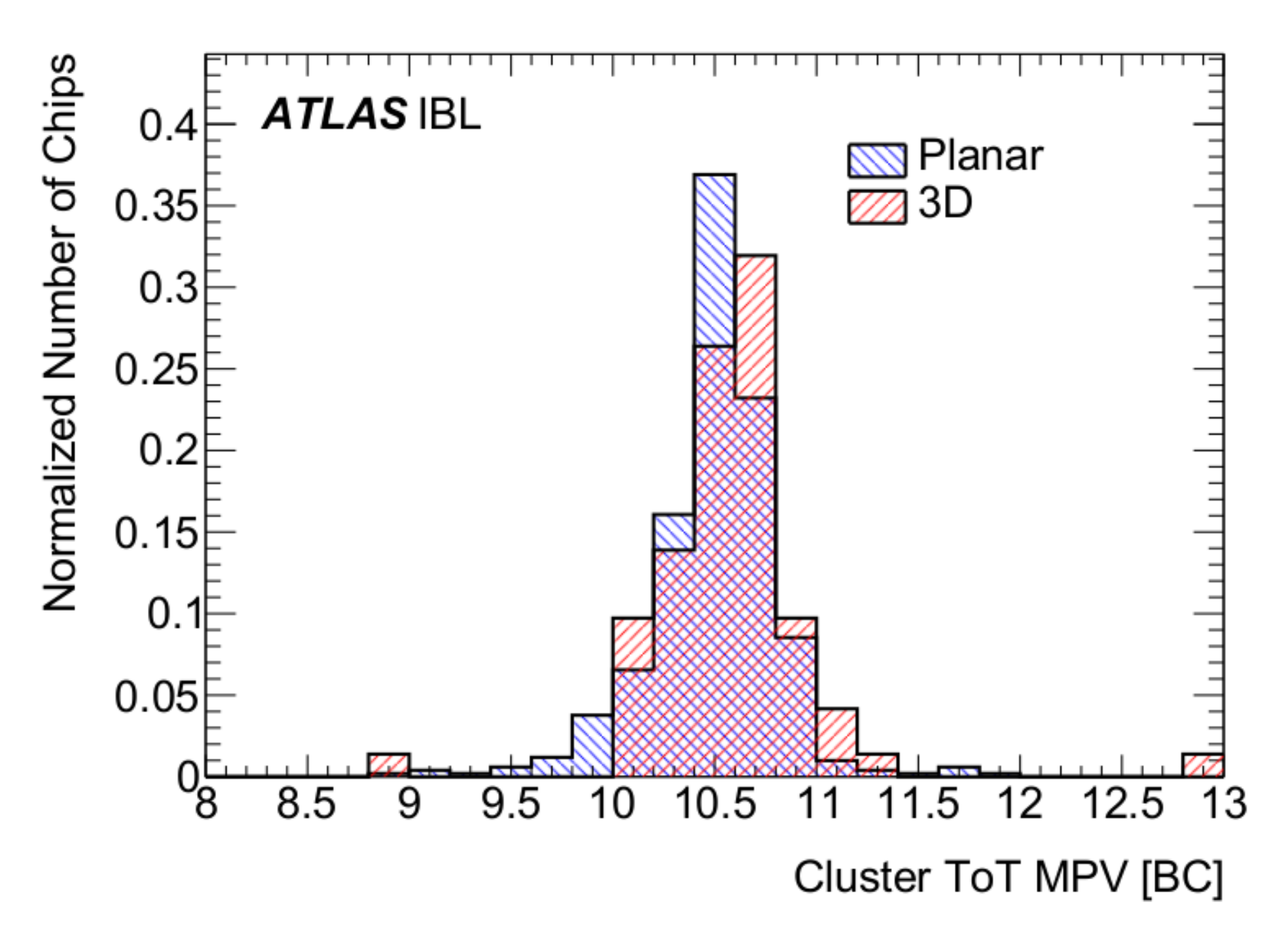}
                \caption{}
                \label{fig:SourcePix}
        \end{subfigure}
        \caption{Most probable value (MPV) of the Time-over-Threshold (ToT) distribution for all clusters with more than one hit, fit with  a  Landau-Gauss function:  \subref{fig:SourceClus} the mean MPV as a function of chip number (position) on the stave and \subref{fig:SourcePix} the distribution of the mean MPV for all chips. In \subref{fig:SourceClus} the error bars represent the RMS value accounting for  the differences between staves whereas the solid colour describes the full range covered by data. The module threshold is tuned to \SI{3000}{\e} at an operating temperature of \SI{22}{\celsius}.
        }
         \label{fig:SourceToT}
\end{figure}

%%%%%%%%%%%%%
\subsubsection{Pixel defects}
 \label{sec:pixel-defects}
 %%%%%%%%%%%%%%%

Faulty pixels were classified on the basis of 
 calibration and source scan results and assigned to a single %only one of the categories 
 category listed in Table~\ref{tab:badpix}. This table indicates the failure type, the method used to identify the failure, and the detailed selection criteria.
\begin{table}[!htb]
    \centering
    \begin{tabular}{lllll}
    \hline\hline
        Failure Name & Scan type & Criteria && \\
        \hline
        Digital dead & Digital scan & Occupancy & \SI{<~1}{\percent} of signal injections \\
        Digital bad & Digital scan & Occupancy & \SI{< 98}{\percent} or \SI{> 102}{\percent} of signal injections \\
        Merged bump & Analog scan & Occupancy & \SI{< 98}{\percent} or \SI{> 102}{\percent} of signal injections\\
        & Crosstalk scan & Occupancy & \SI{>80}{\percent} of \SI{25}{\kilo\e} signal injections  \\
        Analog dead & Analog scan & Occupancy & \SI{< 1}{\percent} of signal injections  \\
        Analog bad & Analog scan & Occupancy & \SI{< 98}{\percent} or \SI{> 102}{\percent} of signal injections \\
        Tuning failed & Threshold scan & s-curve fit & Fit failure   \\ %(threshold = 0~$e$)
         & ToT test & ToT response & 0 or 14 ToT units of \SI{25}{\nano\second}& \\
        Noisy pixel & Noise scan & Occupancy &  \num{>e-6} hits per \SI{25}{\nano\second} bin &\\
        Disconnected bump & Source scan ($^{90}$Sr) & Occupancy & \SI{< 1}{\percent} of mean occupancy & \\
        High crosstalk & Crosstalk scan & Occupancy & \num{> 0} with \SI{25}{\kilo\e} signal injection & \\
        \hline\hline
    \end{tabular}
    \caption{Classification of pixel failures\label{tab:badpix}.}
\end{table}
\begin{figure}[h!]
        \centering
        \begin{subfigure}[t]{0.48\textwidth}
                \includegraphics[width=\textwidth]{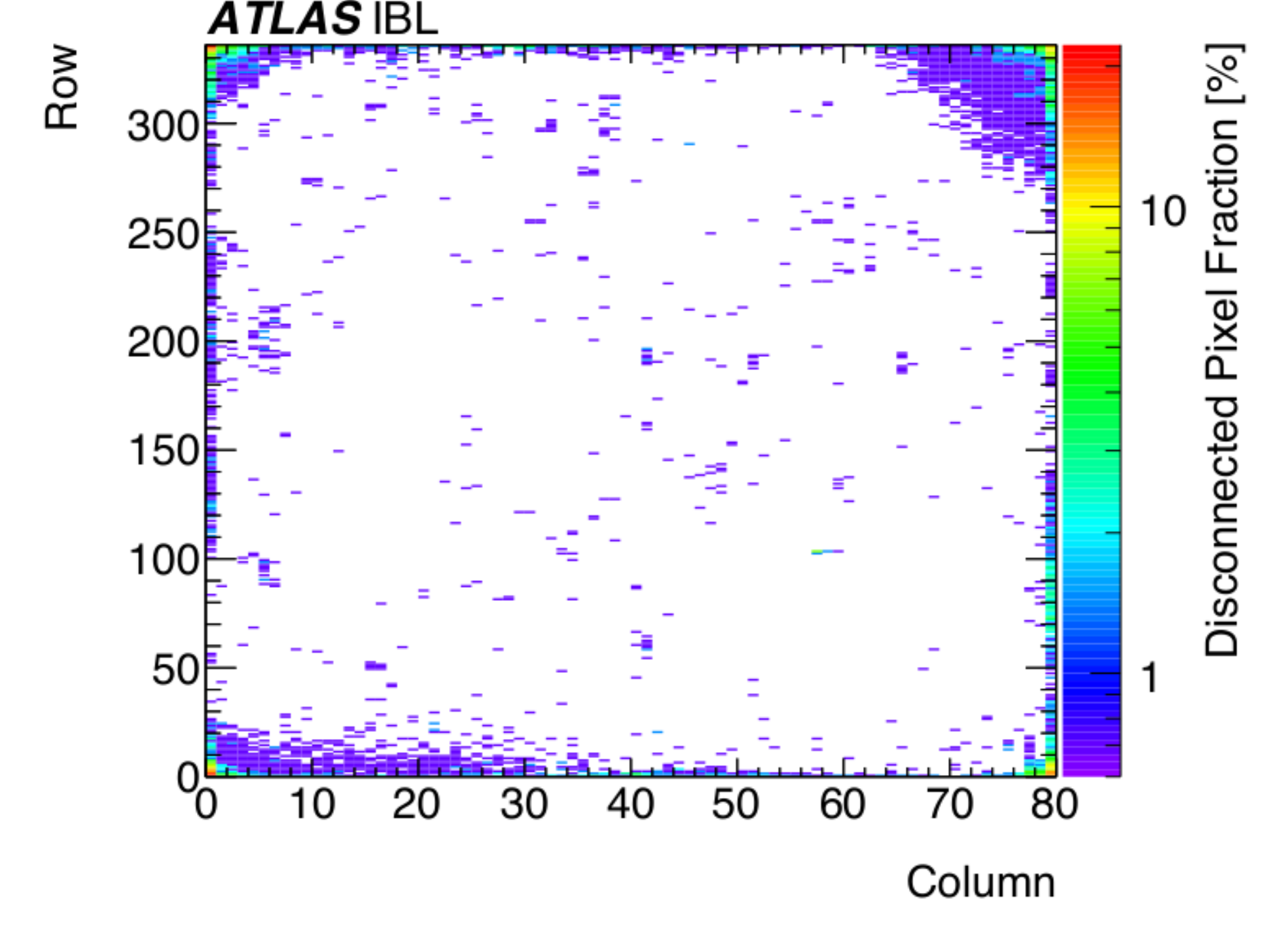}
                \caption{}
                \label{fig:disc3D}
        \end{subfigure}
        \begin{subfigure}[t]{0.48\textwidth}
                \includegraphics[width=\textwidth]{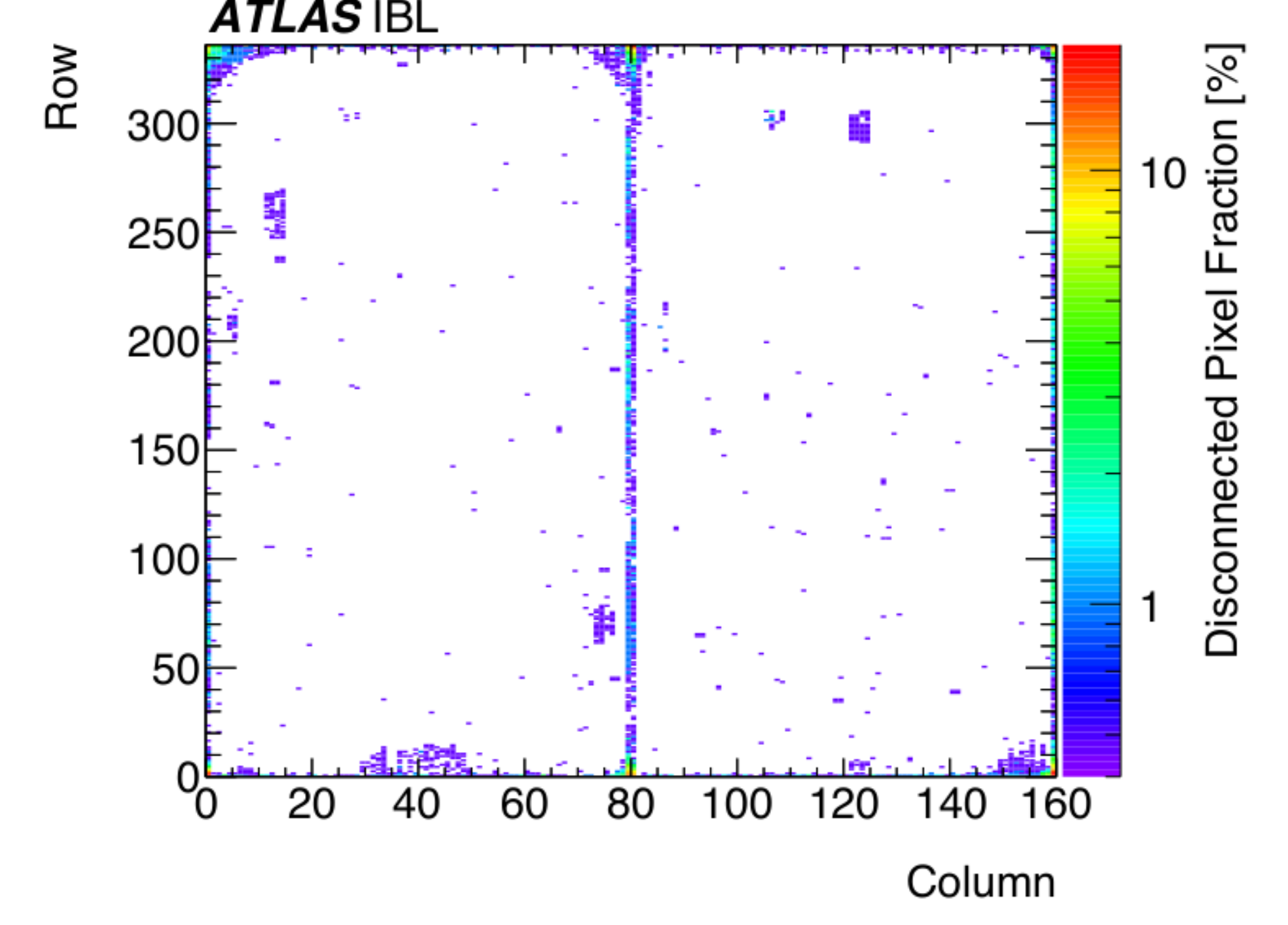}
                \caption{}
                \label{fig:discplanar}
        \end{subfigure}
        \begin{subfigure}[t]{0.48\textwidth}
                \includegraphics[width=\textwidth]{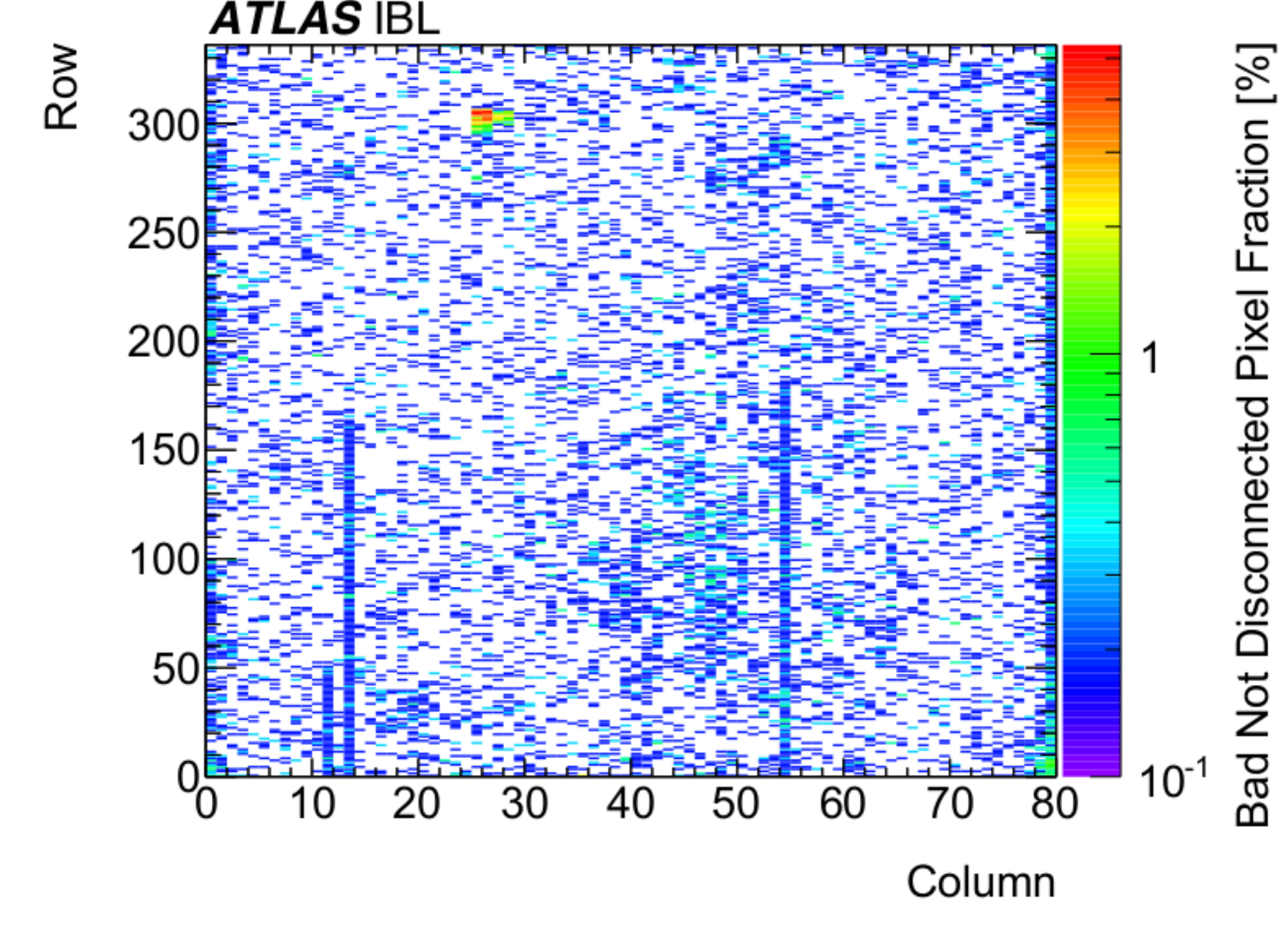}
                \caption{}
                \label{fig:badnodisc}
        \end{subfigure}
        \begin{subfigure}[t]{0.48\textwidth}
                \includegraphics[width=\textwidth]{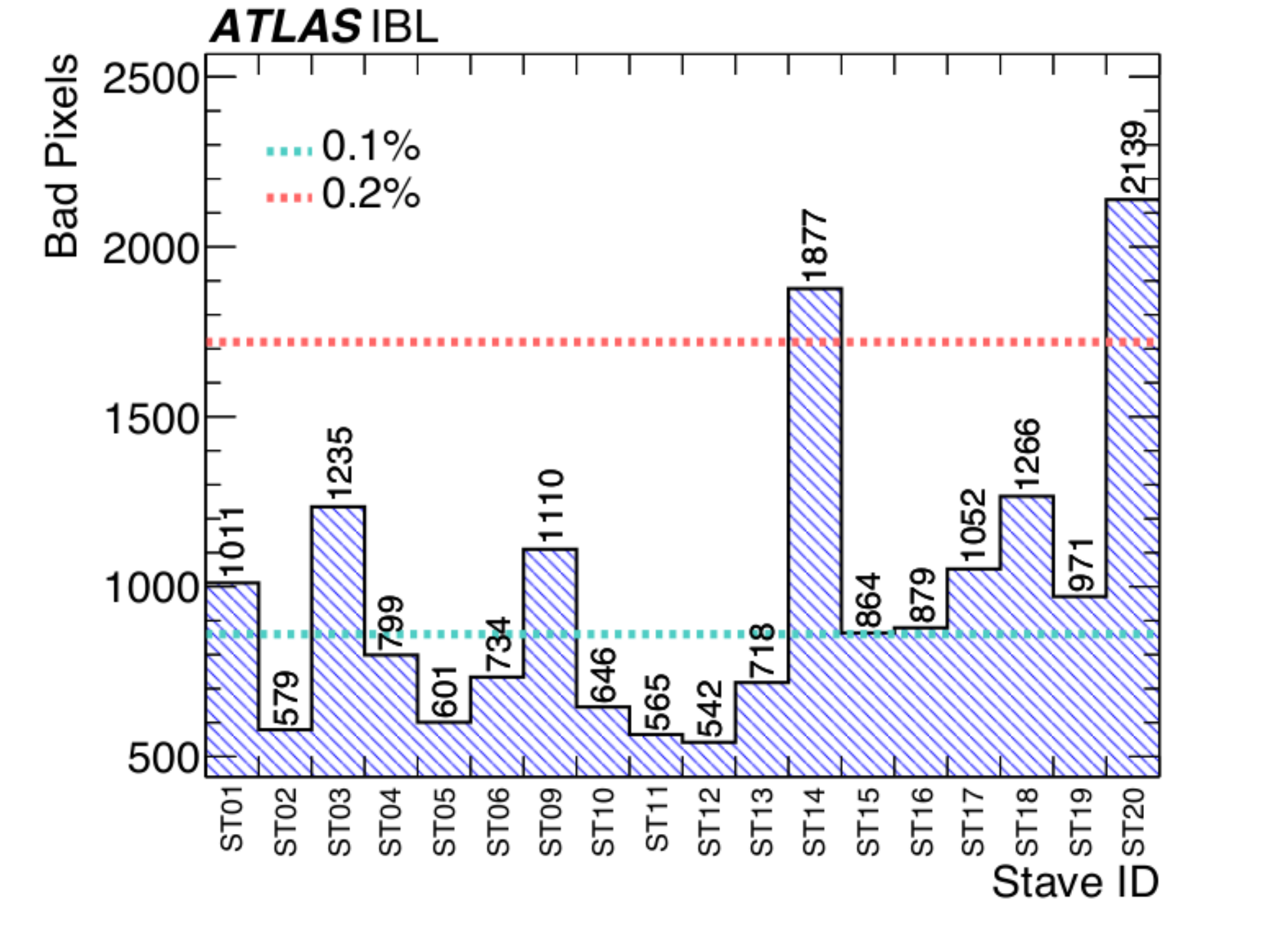}
                \caption{}
                \label{fig:bad_per_stave}
        \end{subfigure}
        \caption{Map of the fraction of disconnected pixels for \subref{fig:disc3D} 3D modules and \subref{fig:discplanar} planar modules . \subref{fig:badnodisc} Map of the total fraction of bad but not dis-connected pixels (both module types). \subref{fig:bad_per_stave} The total number of bad pixels per stave for all 18 qualified staves. The horizontal bars in \subref{fig:bad_per_stave} indicate the \SI{0.1}{\percent} and \SI{0.2}{\percent} bad pixel levels. 
        }
         \label{fig:baddiscfull}
\end{figure}

Most of the failures relied on the module showing an excess or deficit of the hit rate. The digital and analog functionalities can be directly tested with respectively digital and analog scans. Dead digital and analog pixels are common electronic failures. However, the digital and analog bad categories only appear in high numbers for the case of a low ohmic connection between pixels or the identification of a merged bump as occurred in early production batches (Section~\ref{sec:module_qa}). 
The merged bump category exists when two solder bumps connecting the sensor to the read-out chip are merged and manifests itself in an analog failing pixel which still gives a response in a crosstalk scan. 
A pixel is classified as untunable if the threshold or ToT cannot be tuned at all; nevertheless a high discrepancy from the tuning target is accepted as these pixels can still be used for operation. 
% even if just to a limited extent. 
The noise occupancy, which indicates how many hits per BC are produced due to noise, is a very important diagnostic and operational quantity. 
The easiest way to identify a disconnected bump is to analyse the response from a source scan. If a pixel shows zero or only very few hits in the source scan data, the bump is assumed to be disconnected (Section~\ref{sec:module_qa}). The high crosstalk pixel category is not directly related to the performance of the pixel but if a sensor pixel shows significant charge sharing with a neighbour, this can influence the precision of the offline reconstruction.

The total fraction of bad pixels, averaged over all modules, is shown in Figure~\ref{fig:baddiscfull}. 
%AGC No clear correlation between the chip topology, other than edge regions, and measured disconnected bumps, were found. 
No clear correlation was found between the chip topology (other than edge regions) and the measured disconnected bumps. There was a clear increase in the number of bad pixels towards the stave ends because of the choice to load the best modules into the central region of a stave. The number for each bad pixel category is collected for each stave in Table~\ref{table:badpixsummary}. 

\begin{table}[!htbp]
\renewcommand{\arraystretch}{1.0}
\centering
\begin{tabular}{lccccccccc}
\hline \hline
Stave & Digital & Analog & Disconnected & Merged & Untunable & Noisy & High & Total \\
          &  fault    & fault     & pixel               & bump     &                    & pixel   & crosstalk &  \\
\hline
ST01 & 6 & 389 & 272 & 3 & 232 & 11 & 98 & 1011 \\
ST02 & 10 & 255 & 54 & 3 & 117 & 15 & 125 & 579 \\
ST03 & 6 & 375 & 473 & 0 & 182 & 21 & 178 & 1235 \\
ST04 & 2 & 201 & 254 & 0 & 275 & 8 & 59 & 799 \\
ST05 & 2 & 207 & 172 & 0 & 183 & 4 & 33 & 601 \\
ST06 & 6 & 206 & 337 & 0 & 147 & 9 & 29 & 734 \\
ST09 & 8 & 360 & 476 & 3 & 167 & 8 & 88 & 1110 \\
ST10 & 16 & 179 & 304 & 0 & 141 & 3 & 3 & 646 \\
ST11 & 10 & 196 & 159 & 0 & 155 & 8 & 37 & 565 \\
ST12 & 15 & 172 & 169 & 0 & 166 & 7 & 13 & 542 \\
ST13 & 9 & 127 & 205 & 0 & 336 & 6 & 35 & 718 \\
ST14 & 4 & 161 & 1364 & 0 & 330 & 7 & 11 & 1877 \\
ST15 & 5 & 222 & 350 & 0 & 259 & 20 & 8 & 864 \\
ST16 & 1 & 237 & 414 & 1 & 187 & 15 & 24 & 879 \\
ST17 & 2 & 214 & 598 & 0 & 229 & 5 & 4 & 1052 \\
ST18 & 13 & 161 & 902 & 1 & 178 & 2 & 9 & 1266 \\
ST19 & 10 & 163 & 543 & 0 & 228 & 11 & 16 & 971 \\
ST20 & 14 & 224 & 1051 & 0 & 535 & 13 & 302 & 2139 \\
\hline \hline
\end{tabular}
\caption{Overview of the number of different bad pixel categories for the 18 qualified staves. Digital and Analog failure modes include both dead and bad categories as defined in Table~\ref{tab:badpix}. }
\label{table:badpixsummary}
\end{table}

A production cut of 100 bad pixels per chip was initially applied after the module assembly. This cut was relaxed at the end of the production because of a shortage of 3D sensor assemblies; \SI{73}{\percent} of all chips loaded onto staves had less then \SI{0.1}{\percent} bad pixels.

The total number of failing pixels per stave is shown in Figure~\ref{fig:bad_per_stave}. The dashed lines indicate the  \SI{0.1}{\percent} and \SI{0.2}{\percent} marks; the specification required a stave to satisfy $<$~\SI{1}{\percent}. All staves were well within the specification; \SI{80}{\percent} of the staves had $<$~\SI{0.2}{\percent} failing pixels and \SI{50}{\percent} of those staves had $<$~\SI{0.1}{\percent}. Approximately \SI{50}{\percent} of all failures were due to disconnected bumps, the other \SI{50}{\percent} were distributed between a pixel with failing analog functionality, or a bad ToT tuning.

Correlations between  the defects observed at this stage and  in the module production were extensively studied~\cite{StaveQANote}. Although similar selections to classify a pixel %as bad 
were used in the two test stages, a more intense test procedure was applied during module production and more bad pixels were detected per FE-I4B chip. However, there was a good correlation between the number of bad pixels per FE-I4B chip in the two tests. 
Moreover it was verified that there was no significant increase of bad pixels in specific geographical areas, for example the chip edges, thus excluding damage resulting from the module transport, handling and stave loading. %\textit{CG: removed 4 plots on correlation and differences with module production and simplified text}

\subsection{Stave ranking and layout assignment}
\label{sec:stavrank}

Of the 18 qualified loaded staves, 14 staves were selected for integration as part of the IBL. The stave quality was scored by  a geometrical  inefficiency based on a  $\eta$-weighted bad pixel fraction, 
together with an algorithm developed to minimise the clusterisation of bad pixels in $\eta - \phi$. In addition, two artificial constraints were applied: the first and the last integrating staves were required to have the best planarity in order to simplify assembly;  and staves that were re-worked as a result of corroded wire bonds (Section~\ref{sec:corrosion})  were loaded alternately.

Table~\ref{table:rankingsummary} summarises the position of each stave in the IBL loading map, %the rank 
and other characteristics of the staves considered in the selection. Figure~\ref{fig:badpix_dist_eta} compares the average bad pixel fraction for the 14 installed staves with the four non-installed staves, as a function of %pseudo-rapidity 
$\eta$.  
%The former achieving better performance in terms of acceptance. 
The average bad pixel fraction for the integrated IBL staves is \SI{0.07}{\percent} for $|\eta|<2.5$ and \SI{0.09}{\percent} over the full $\eta$ range. %For comparison, 
The corresponding fractions for the four non-installed staves are respectively  \SI{0.16}{\percent} and \SI{0.18}{\percent}. Figure~\ref{fig:badpix_etaphi} shows the distribution of bad pixel fraction as a function of $\eta$ and $\phi$. The stave overlap is taken into account in this figure.  

\begin{table}[htbp]
\renewcommand{\arraystretch}{1.0}
\centering
\begin{tabular}{lccccc}
\hline \hline
Position & Stave & Number of bad pixels  & Planarity [\SI{}{\micro\meter}] & Reworked because of corrosion \\
\hline
01 & ST17  & 1052 & 114 & no \\
02 & ST02  & 579   & 205 & yes \\
03 & ST19  & 971   & 266 & no\\
04 & ST09  & 1110 & 229 & yes \\
05 & ST18  & 1266 & 336 & no\\
06 & ST04  & 799  & 235 & yes \\
07 & ST13  & 718  & 224 & no\\
08 & ST10  & 646  & 243 & yes \\
09 & ST11  & 565  & 298 & no \\
10 & ST12  & 542  & 314 & yes \\
11 & ST16  & 879  & 329 & no \\
12 & ST06  & 734  & 290 & yes \\
13 & ST15  & 864  & 325 & no \\
14 & ST05  & 601  & 189& yes\\ \hline
n/a & ST01 & 1011 & 224 & yes\\
n/a & ST03 & 1235 & 223 & yes\\
n/a & ST14 & 1877 & 218 & no\\
n/a & ST20 & 2139 & 237 & no\\
\hline \hline
\end{tabular}
\caption{Loading order overview of the 14 IBL staves. The position is sequential around the beam pipe. The cooling pipe of the stave in position 01 is at $\phi = \SI{-6.1}{\degree}$, subsequent staves are displaced by \SI{25.7}{\degree} in $\phi$. The planarity shows the difference between the minimum and maximum height of a stave. The last column indicates whether a stave has been reworked at the CERN wire bonding laboratory because of the corrosion issue. For completeness, the last four lines show the same parameters for those staves that were not selected for installation. 
}
\label{table:rankingsummary}
\end{table}

\begin{figure}[!htb]
                \centering
        \begin{subfigure}[t]{0.49\textwidth}
                \includegraphics[width=\textwidth]{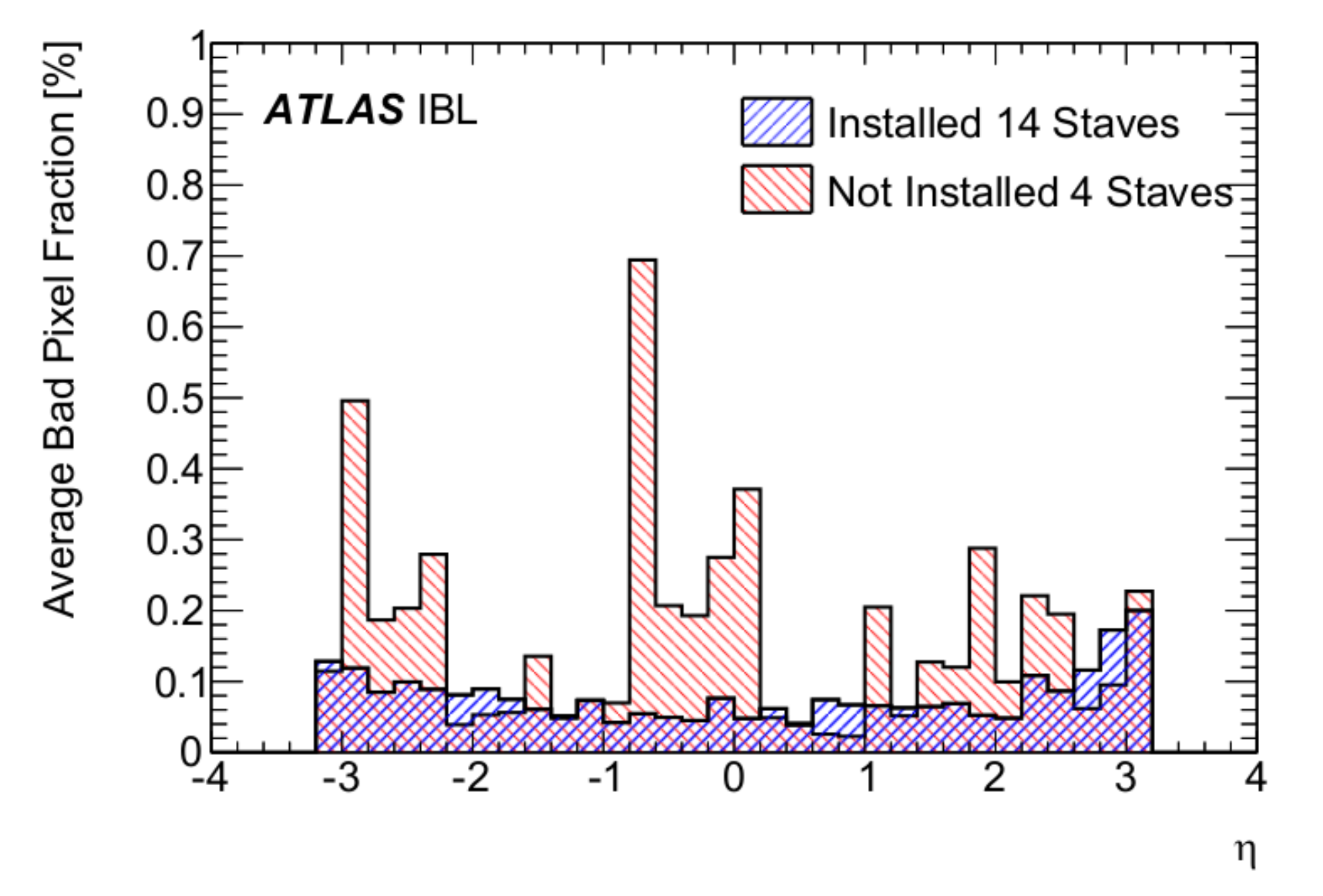}
                \caption{}
                \label{fig:badpix_dist_eta}
        \end{subfigure}
                \centering
        \begin{subfigure}[t]{0.49\textwidth}
                \includegraphics[width=\textwidth]{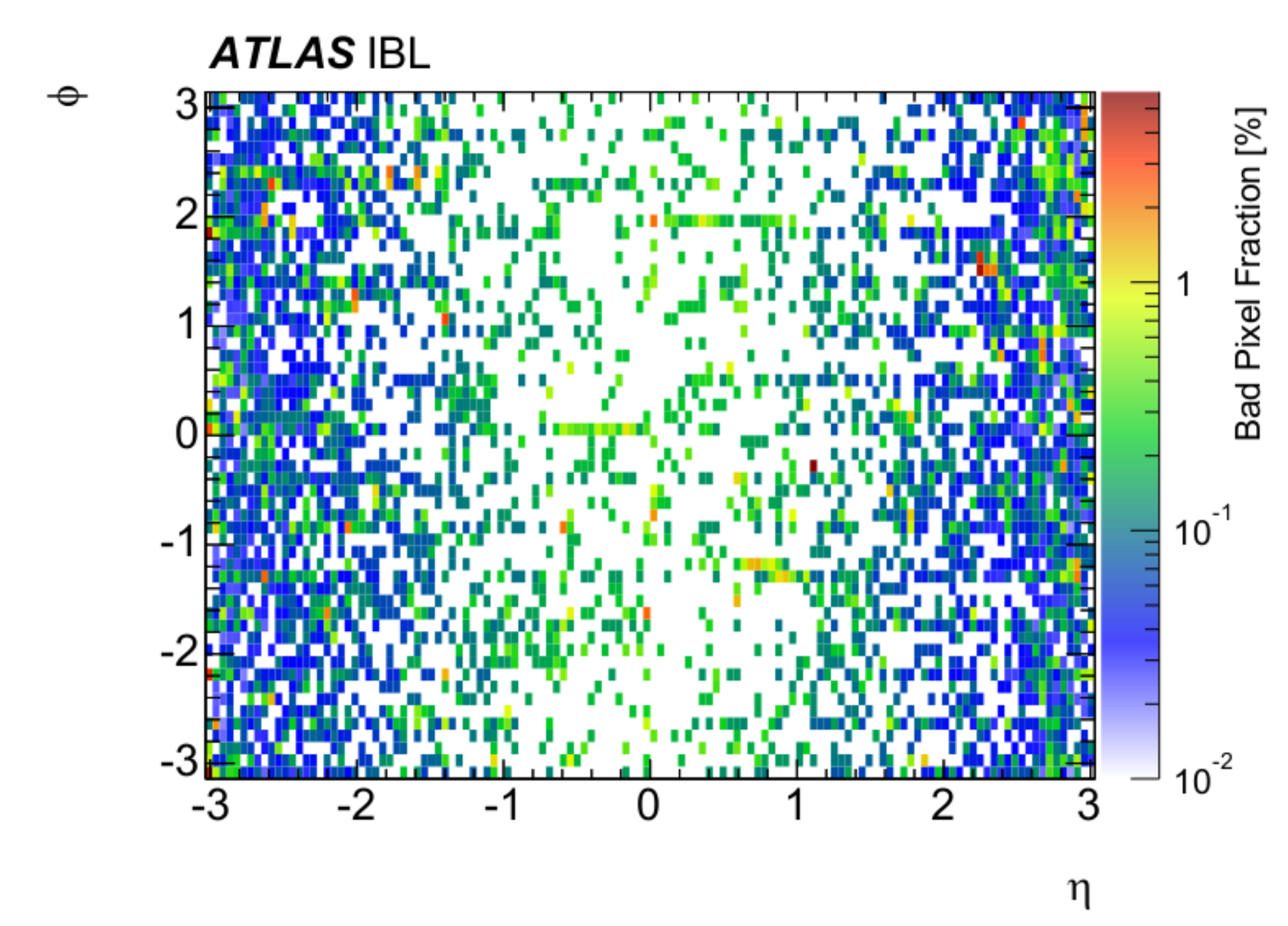}
                \caption{}
                \label{fig:badpix_etaphi}
        \end{subfigure}
    \centering
   \caption{Average bad pixel fraction \subref{fig:badpix_dist_eta} as a function of $\eta$ for installed and non-installed production staves, and \subref{fig:badpix_etaphi} in the $\eta-\phi$ plane for the 14 selected staves. The stave overlap is taken into account and the fraction is computed as the number of bad pixels divided by the total number of pixels in a given bin.}
    \label{fig:badratio_load_unload}
\end{figure}

\subsection{Wire bond corrosion}
\label{sec:corrosion}
During the IBL production, two production staves  were damaged during testing inside the QA environmental box. While the staves were in operation  and cooled at \SI{-20}{\celsius}, ice was identified  around the coldest part of the staves. 
The two staves were carefully inspected under a microscope and it was discovered that most of the aluminium wire bonds were corroded (Figure\,\ref{fig:wire-corrosion}), with a  few broken. A white residue around the bond foot  %could be barely seen with normal incident light 
could be easily identified with ring lighting. 
The other  staves were re-inspected  and, of the 12 staves already produced, 11 staves were identified to suffer from bond corrosion. 

 \begin{figure}[h!]
        \centering
        \begin{subfigure}[t]{0.49\textwidth}
                \includegraphics[width=\textwidth]{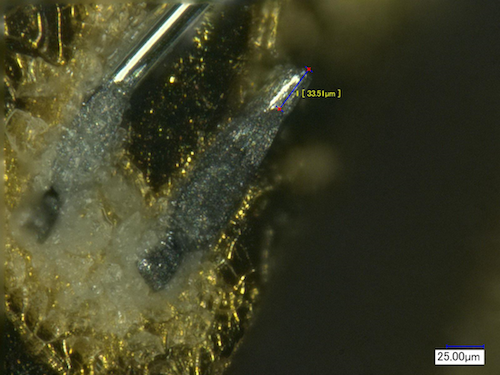}
                \caption{}
                \label{fig:wire-corrosion_25a}
        \end{subfigure}
        \centering
        \begin{subfigure}[t]{0.46\textwidth}
                \includegraphics[width=\textwidth]{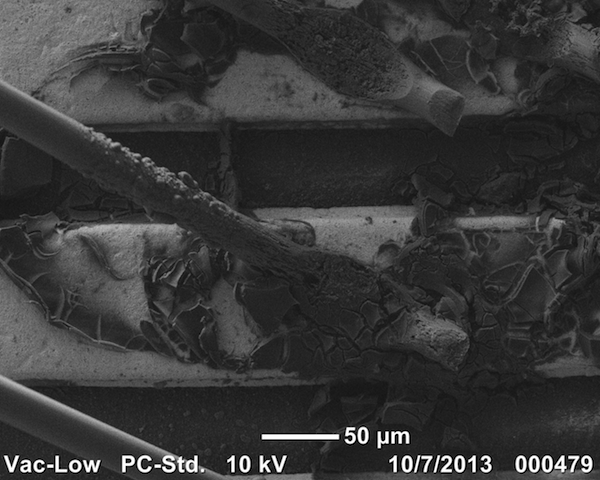}
                \caption{}
                \label{fig:wire-corrosion_25b}
        \end{subfigure}
        \caption{\subref{fig:wire-corrosion_25a} White residue of corrosion on staves as observed under a microscope. \subref{fig:wire-corrosion_25b} A Scanning Electron Microscopy (SEM) image of corroded wires and the corrosion residue at the foot of the bond.}
        \label{fig:wire-corrosion}
\end{figure}

Production was halted until the problem was understood. A number of successive actions were taken:
\begin{itemize}
\item[-] {Identify the origin of the corrosion and identify corrective actions such as to improve the QA procedure, before resuming the production;}
\item[-] {Investigate the evolution of the corrosion and identify  possible preventive actions, for example cleaning and coating;}
\item[-] {Organise a re-work centre to clean all the wires and to re-bond the 11 defective staves;}
\item[-] {Launch an additional module production with remaining components to ensure a sufficient number of IBL staves for the integration.}
\end{itemize}

The investigation initially considered the possibility of humidity in the QA environmental boxes, and it was found that 2 setups were concerned.
The first concerned the \SI{1.6}{\meter\cubed} climate chamber used to qualify loaded staves by temperature cycling in the range \SI{-40}{\celsius} to  \SI{+40}{\celsius}.
As shown in Figure~\ref{fig:TCplot}, the dew point was reached for a few minutes %1-2 minutes 
in the proximity of the stave modules during the fast temperature ramp-up because of local restrictions to the dry air flow, even though the volume was flushed with dry air  and the chamber humidity control was activated~\cite{LoadingNote}.
The stave electrical and mechanical integrity was not affected by the corrosion: this was confirmed by electrical characterisations and metrology surveys.  For the remainder of the production, the newly loaded staves were not thermally cycled.
The second problematic environmental box  was that used for the stave electrical qualification at low temperature. 
Upgrades to improve the stave dryness  and the reliability of the environmental control were made by adding dedicated interlock actions on the cooling and the power supply. 

 \begin{figure}[h!]
         \centering
         \includegraphics[width=0.7\textwidth]{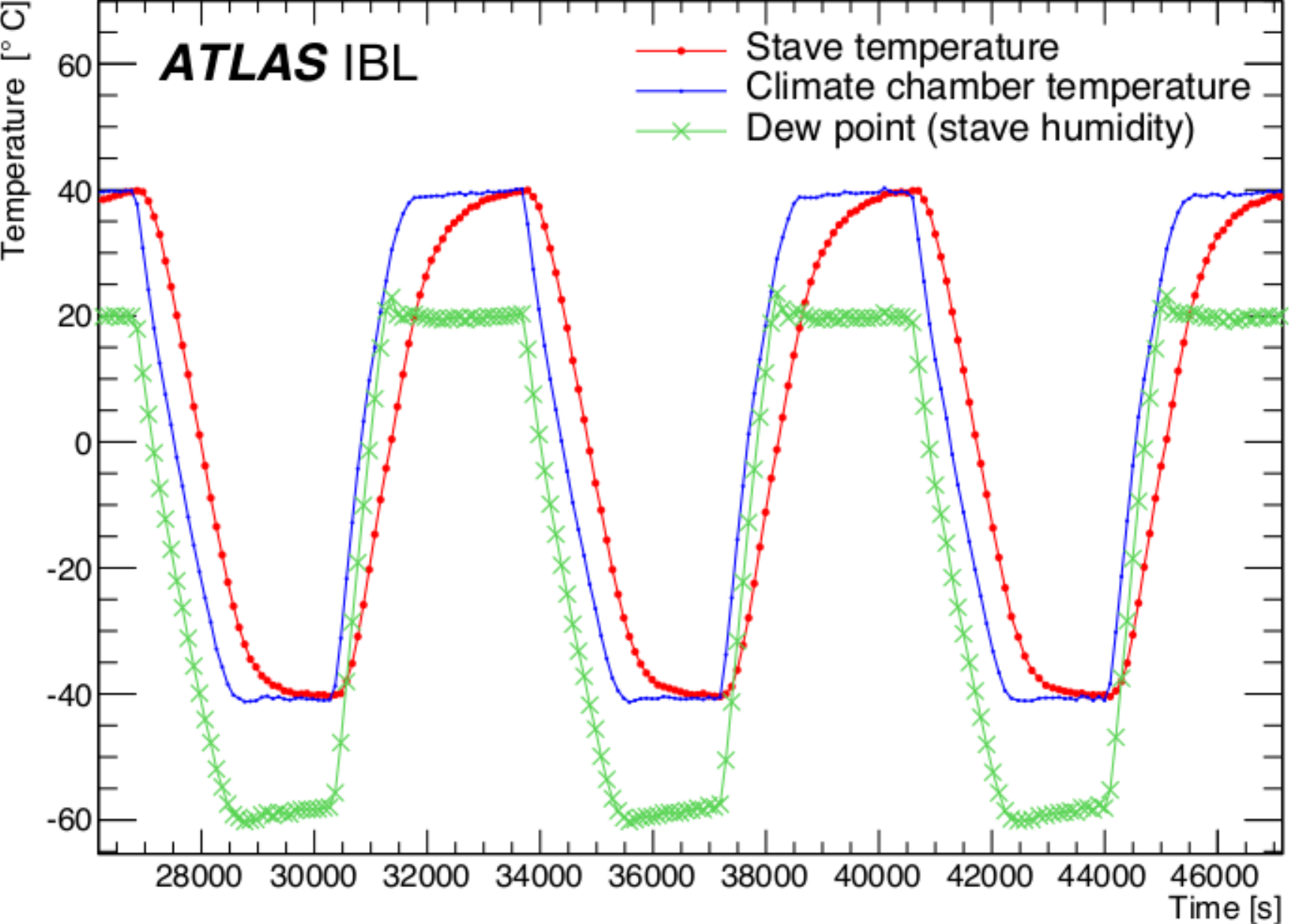}
         \caption{Temperature and dew point monitoring in the vicinity of stave modules during thermal cycling in the \SI{1.6}{\meter\cubed} climate chamber. When increasing the temperature the dew point (green curve) rises faster than the stave temperature and surpasses it for a few minutes.
}
        \label{fig:TCplot}
\end{figure}

\subsubsection {Investigations of the Al wire corrosion process}

The interconnection of Al wires with the  Ni/Au bonding pads remains an issue even  at room temperature, because of the galvanic coupling  between  Au and Al. 
During the ultrasonic wire bonding the  Au layer is locally removed and the final metal contact is between Al and Ni.
In addition the Al wire is normally protected by a thin oxidation layer that is formed in a few hours and that  stabilises at a thickness of about 5\,nm. This protective layer can be damaged in the presence of water or because of mechanical or chemical attack. If this occurs, as during the stave cold test, a corrosion process will be initiated.

\begin{figure}[h!]
        \centering
        \begin{subfigure}[t]{0.49\textwidth}
                \includegraphics[width=\textwidth]{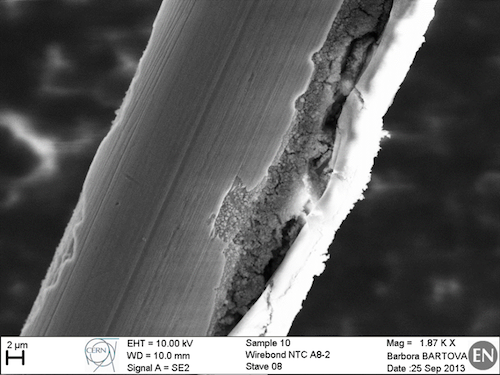}
                \caption{}
                \label{fig:wire-corrosion_25c}
        \end{subfigure}
        \centering
        \begin{subfigure}[t]{0.49\textwidth}
                \includegraphics[width=\textwidth]{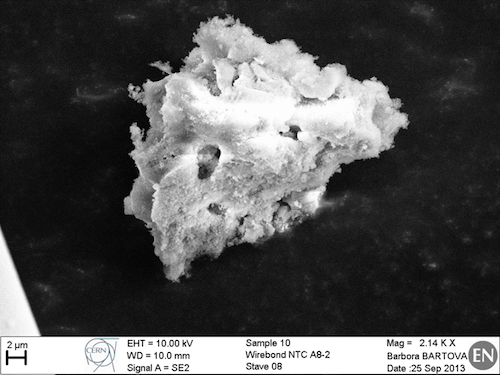}
                \caption{}
                \label{fig:wire-corrosion_25d}
        \end{subfigure}
        \caption{Images of \subref{fig:wire-corrosion_25c} a corroded Al-wire and \subref{fig:wire-corrosion_25d} residue taken from an affected stave. The images were taken with an Energy Dispersive X-Ray Spectroscopy (EDS) analysis setup.}
        \label{fig:wire-corrosion_EDS}
\end{figure}

The corrosion residues and Al wires were analysed (Figure\,\ref{fig:wire-corrosion_EDS}) with an Energy Dispersive X-Ray Spectroscopy (EDS) technique and the presence of  C, O, Ca, Na, Cl, F at a level of up to a few percent was detected.
The presence of ionic compounds indicated that the cleaning of the module flex after SMD assembly should be improved. The corrosion process was easily reproduced in the laboratory, even in presence of de-ionised water on wire bonded samples. However,  further investigations revealed that the process was even observed on ultra-clean bare flex assemblies.  Additional cleaning procedures such as plasma cleaning proved ineffective in stopping the occurrence of corrosion, although its effect could be mitigated. 

A sample analysis was also performed with an X-ray Photoelectron Spectroscopy (XPS), alternating the measurements with the sputtering of the Au layer with Ar ions. This  probed the atomic  spectrum at the Au surface while removing subsequent  layers  until reaching the Ni interface. The procedure  was applied to several flex circuits delivered by two different producers
\footnote{Laboratory studies showed that the susceptibility of the flexes to corrosion was vendor dependent. In particular, the flexes used for the Pixel detector were 
much less sensitive to corrosion.
}. 
On one sample fluorine was detected at a significant level (up to \SI{14}{\percent} at a depth of \SI{7}{\nano\meter}). This presence  could not be understood nor reproduced in other flex circuits from the same producer.

Options considered to protect against corrosion included the potting of the bond foot and the use of spray coating such as polyurethane, but neither were possible because of the tight production and integration schedule. All wire bonds showing signs of corrosion were replaced. It was decided to leave the wire bonds unprotected, but to ensure that a safe humidity level would always be maintained during production, testing, integration and operation. Tests of the susceptibility of bond pads to corrosion are recommended before and during the production process, for all future projects.

\section{Off-detector electronics and services}
\label{section:electronicinterfaces}

Beyond the EoS cards located at %Patch Panel 0 (PP0) at
the end of the detector, off-detector electrical and optical cabling connects each half-stave
to the off-detector electronics in the USA15 electronics cavern (see Figure~\ref{fig:sec7_1_IBL-electrical-services}). Similarly, power to the module sensor and read-out electronics is routed via  electrical  cables from power supplies in USA15. This section summarises the off-detector read-out, control and service components of the IBL.

\subsection{Off-detector electrical cabling} 
\label{sec:ElectricalCables}

The IBL off-detector cabling consists of two parallel paths originating from the EoS region: one path is dedicated to data with a signal bundle that includes the clock and commands; the other path concerns the power distribution, including the LV, HV and DCS lines. Together, they define the Type~1 cables. The data path terminates at an opto-box that contains the 
opto-boards~\cite{Gan:optoboard:2015} at the outer edge of the ID end-plate region, while the power bundle is routed to the patch panel PP1 and then to a second patch panel PP2 located at  the ATLAS periphery,  as illustrated in Figure~\ref{fig:sec7_1_IBL-electrical-services}. 

As also noted in Sections {\ref{sec:electronic_introduction}} and {\ref{sec:surface-integration}}, the data and power lines are initially routed from the EoS to a cable board by a set of six intermediate flexes (2 LV, 1 clock/commands , 1 data, 1 DCS, 1 HV) that are stacked vertically. %The intermediate flexes 
These flexes are pre-shaped into a corrugated form to partially compensate for the thermal mismatch of the cables and the supporting carbon fibre structure, given temperature excursions of up to \SI{80}{\celsius}, from cold operation and the beam-pipe bake-out. The intermediate flexes connect to the Type 1 cables at the cable board.

The signal transmission in the data bundle uses LVDS data transmission over twisted pair cables with radiation hard polyimide insulation. 
The control lines, feeding the command signals and the \SI{40}{\mega\hertz} clock,  rely on a twisted pair of thin AWG36 copper wires each serving two FE-I4B chips. 
The \SI{160}{\mega\bit\per\second}  read-out lines use one twisted pair of AWG28 copper wires per FE-I4B chip. Both the control and read-out twisted pairs adopt double quad insulation to better match the stave flex impedance and a tight twist of 4-5~twists/inch is used for better transmission quality. The data bundle runs for approximately \SI{5}{\meter} directly to the opto-box with no intermediate junction, again to avoid transmission degradation. Due to this relatively long distance, an LVDS common-mode reference voltage control is necessary.

Each Type 1 power bundle is split into two parts:  an inner cable running for approximately \SI{3.5}{\meter} to PP1 and a \SI{9}{\meter} length outer cable  leading to PP2. Custom 67-pin AXON\footnote{Axon' Cables, see http://www.axon-cable.com.} connectors of \SI{21}{\milli\meter} diameter are  used at PP1 to join the two  sections.

The LV is delivered to the stave via a dedicated regulator located at PP2. These LV supply wires dominate the service material, because of the low resistance required to avoid an excessive voltage drop. FE-I4B Shunt-LDO regulators (Section \ref{sec:FEI4_electronic}) are used, with an input  voltage range of \SI{1.8}{\volt} to \SI{2.5}{\volt}. The resistance and voltage drops 
%AGC-090817 of the implemented design 
for a single LV channel serving four FE-I4B chips are shown in Table~\ref{tab:LVdrop}. A minimum  shunt current of \SI{270}{\milli\ampere} per FE-I4B chip is set to prevent excessive transient over-voltage due to a sudden current drop. The inner HV section shares a common ground return with the LV to reduce the number of EoS connectors. 

Each Type 1 power bundle includes fourteen or fifteen lines for seven DCS signals. For each group of four FE-I4B chips one NTC is read out (that is 4x2 lines per half-stave). These signals are routed via the intermediate flex to the cable board. 
Three more signals originate from the cable board region: one NTC near the cable board itself; one NTC on the cooling pipe next to the cable board; and either a third NTC or a humidity sensor. The humidity sensor is mounted on the cable board itself and requires three lines. In the case of a third NTC, this is placed to measure the temperature of the cable bundle upstream of the cable board.

\begin{table}[h]
\begin{center}
\begin{tabular}{lcc}
\hline
\hline
Section            &  Resistance (\SI{}{\ohm}) & Voltage Drop (\SI{}{\volt}) \\
\hline
Module             &  0.017                 &  0.038  \\
Stave flex         &  0.170                 &  0.381  \\
Intermediate flex  &  0.035                 &  0.078  \\
Cable board        &  0.033                 &  0.074  \\
Type 1 inner cable        &  0.139                 &  0.311  \\
Type 1 outer cable        &  0.091                 &  0.204  \\
LV regulator crate &  0.046                 &  0.103  \\
\hline
\hline
\end{tabular}
\caption{Room temperature resistance and voltage drop for a single LV channel serving four FE-I4B chips at the maximum current of \SI{0.56}{\ampere} per chip (\SI{2.24}{\ampere} per LV channel).}
\label{tab:LVdrop}
\end{center}
\end{table}

\subsection{The detector control, interlock and power supply systems}
\label{sec:system_dcs}

A schematic of the IBL Detector Control System (DCS) and interlock hardware and functionality~\cite{DCS:2011} is shown in Figure~\ref{fig:DCS_Overview}. The inputs to the DCS and interlock systems from the power supplies, the cooling plant and the environmental monitoring are also shown.

\begin{figure}[tb]
\begin{center}
 \includegraphics[width=0.8\textwidth]{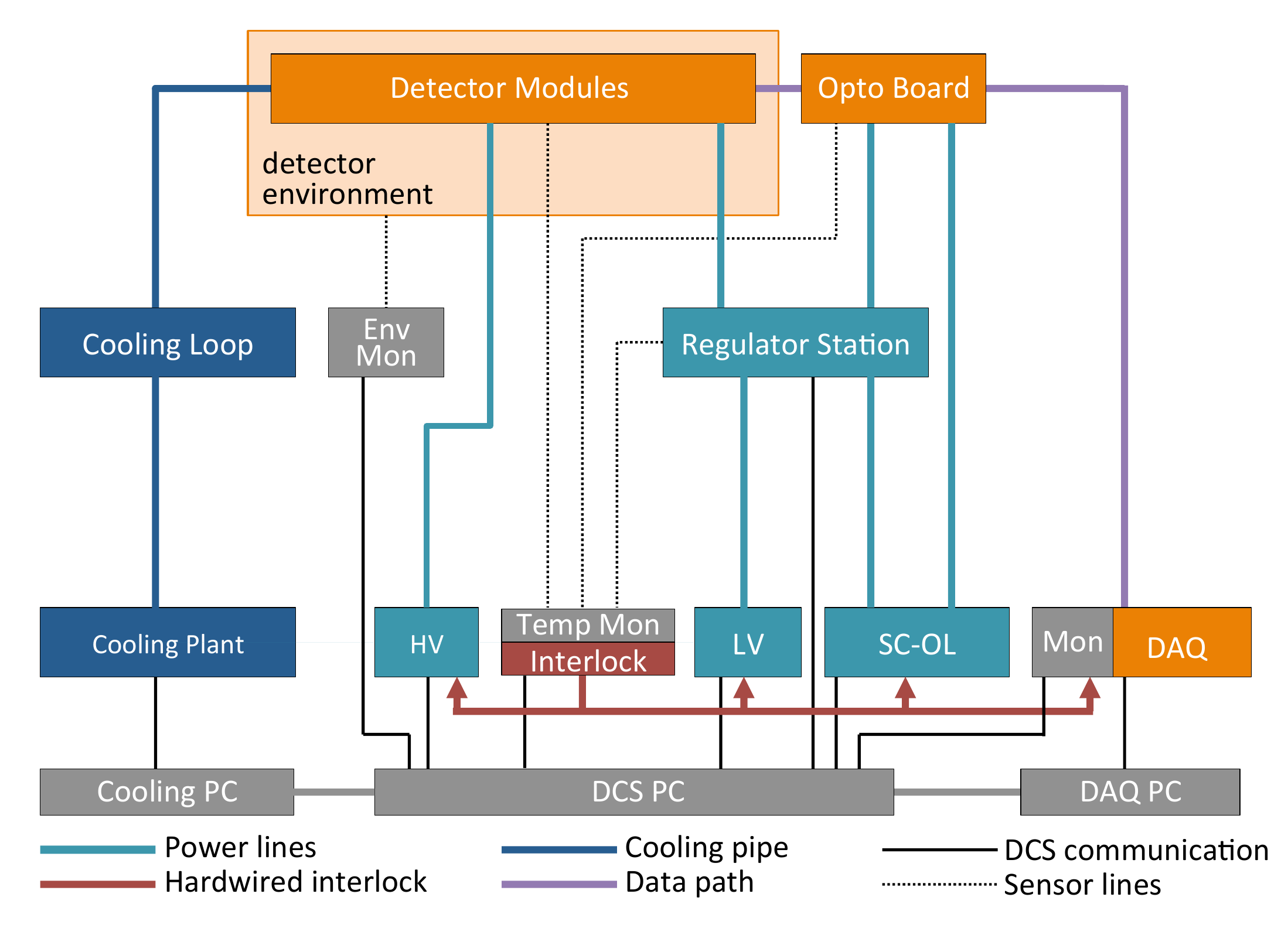}
\caption{Schematic of the DCS control and monitoring functions on the sensor (HV), front-end electronic (LV) and opto-board (SC-OL) power supplies, as well as temperature and humidity monitoring. The schematic also indicates an independent hardware interlock system used for detector and operational security.}
\label{fig:DCS_Overview}
\end{center}
\end{figure}

\subsubsection{The DCS and interlock systems}
\label{sec:DCS-and-Interlock}

The DCS has three main functions: to control the detector, the opto-board and the power supply operation; to monitor all operational and environmental aspects of the detector system; and to provide inputs to the power supply and interlock systems as needed for detector and operational security. Most parts of the DCS are provided by dedicated PCs located in the counting room, about \SI{100}{\meter} from the ID. %Inner Detector. 
The regulator station at the PP2 panel and selected monitoring units are installed inside the detector cavern.

The main component of the IBL DCS software is a Finite State Machine (FSM) that is fully integrated into the DCS of the Pixel detector and ATLAS. The tree structure of the ATLAS FSM nodes reflects the 
structure  
of the detector. Commands are sent from the top node to its children, then the status information is sent back to the top node and informs the operator about the success of a command. 
As all modules of a half-stave are read out through a single opto-board, the half-stave is the key element of the IBL FSM. Because of the service modularity, four FE-I4B chips with their two (planar) or four (3D) sensors are the smallest units that can be separately steered by the DCS.

The interlock system is an independent hardware implementation complementary to the DCS. The core component of the interlock system is a flash FPGA with an internal Electrically Erasable Programmable Read-Only Memory (EEPROM) . This avoids the need of a program loading at power-on. A negative logic is implemented, which means that a missing cable or lost power causes an interlock automatically. Power supply and environmental (temperature and humidity) data are fed to the DCS, and the temperature data are independently fed to the interlock system. Additional information from the laser protection system, the cooling system, the LHC or other external systems is included into the interlock matrix. As this system is completely hardware based, maximum safety is provided.

\subsubsection{The IBL power supplies}
\label{sec:power-supplies}

The detector modules and the opto-boards require dedicated powering, well adapted to the corresponding electrical loads. Common to all components is the use of floating power supplies with variable output voltages. The power supplies are controlled and monitored by the DCS, but essential security data such as over-current or high-temperature alerts are also transferred to the interlock system.  

To deplete the sensors an HV power supply is required. While the planar sensors will require up to \SI{1000}{\volt} after irradiation, the 3D sensors can be operated with significantly lower voltages of a few hundred volts.

The FE-I4B chips are powered by the LV power supplies. Because of  the large currents in the FE-I4B chip, the voltage drops on the services are non-negligible. The FE-I4B chips themselves require a nominal input voltage of \SI{1.8}{\volt}, however the LV supply is able to deliver up to \SI{15}{\volt}.  Since the FE-I4B chips would be destroyed by voltages of greater than \SI{2.5}{\volt}, a voltage regulator close to the detector is installed to protect the chips against transients. 
The output voltage of each regulator can be remotely programmed to deliver precise voltages to the FE-I4B chips, via a Controller Area Network (CAN) bus in 100 steps between \SI{1.2}{\volt} to \SI{2.2}{\volt}. The maximum deliverable current per channel is approximately \SI{3.4}{\ampere} and each channel is protected against sense line interruption compatible with the maximum voltage drop allowed in the system. The IBL regulator station is based on the design used in the other layers of the Pixel detector~\cite{Aad:2008zz}.

The Supply and Control of the Opto Link (SC-OL) is used to power the opto-boards and requires three different low voltages. For the main supply voltage, 
V$_{\SI{}{VDC}}$, 
the SC-OL provides a maximum voltage of \SI{10}{\volt} at a maximum current of \SI{800}{\milli\ampere}. As the chips on the opto-board must also be protected, V$_{\SI{}{VDC}}$ is routed through the regulator station in the same way as the LV. 
The second supply voltage, V$_{\SI{}{pin}}$, biases the receiver PiN diodes. The supply voltage of up to \SI{20}{\volt} provides a normal operation voltage of V$_{\SI{}{pin}}$ of 5-10 V. 
The third supply voltage, V$_{\SI{}{Iset}}$, with a maximum of \SI{5}{\volt}, controls the current in the Vertical-Cavity Surface Emitting Lasers (VCSEL).
Additionally, a reset signal is provided, which can be sent to the opto-board in case the decoder is stuck.

\subsubsection{Temperature and humidity monitoring}
\label{sec:temp-humidity}

The environmental monitoring is handled independently of the other monitoring tasks. Temperature sensors (NTCs) are installed in many different locations to protect detector components that might be damaged by overheating. Each Type~1 power bundle, serving a half-stave, includes fourteen or fifteen lines sending seven DCS signals to both the DCS monitoring units and the interlock system. 
Since the FE-I4B chips and silicon sensors can be permanently damaged by the overheating of detector modules, each sensor is also equipped with an NTC. A comparator sets a logical signal in case of overheating. In the same way, the opto-boards and the regulator station are equipped with NTCs. NTCs are also mounted on the cooling pipe and near the cable board. Finally, either a humidity sensor mounted on the cable board, or an NTC mounted on the cable bundle, is read out.  

Several voltage and current diagnostic measurements as well as temperature sensors (diodes) are also built into the FE-I4B chips. An on-chip \SI{10}{\bit} Analogue to Digital Converter (ADC) associated to an 8-to-1 analog Multiplexer (MUX) can be used to select and read out the temperature, power supply voltages, voltage references, detector leakage current, and other DCS analog voltages. On demand in calibration mode, this information can be sent to the DCS or through the standard data path. This monitoring strategy is not fully implemented for the IBL but remains a promising approach for future DCS developments, to reduce the material inside the detector volume.

\subsection{Data Acquisition System (DAQ)}
\label{sec:system_daq}

\begin{figure}
%\begin{center}
	\centering
\includegraphics[width=0.8\textwidth]{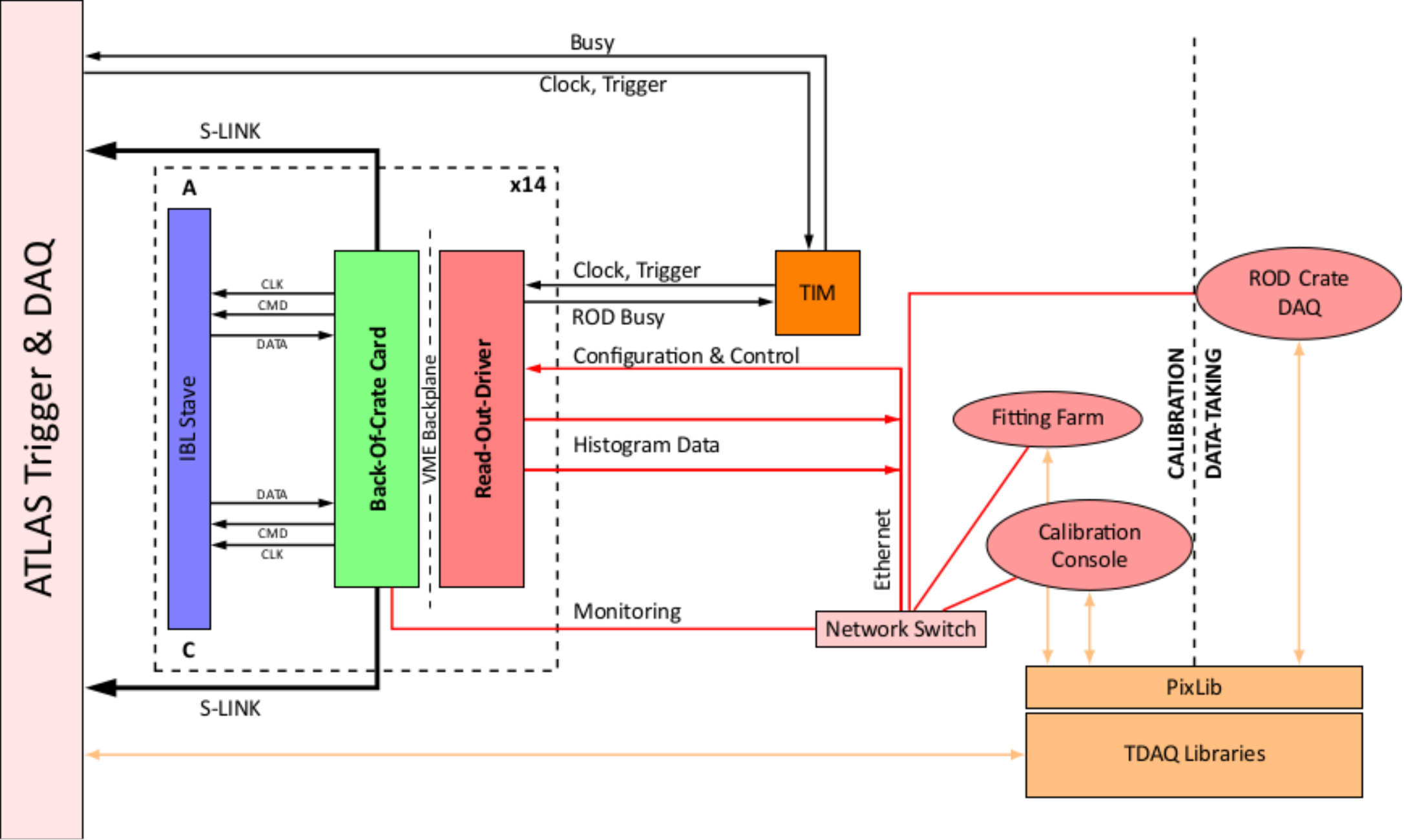}
\caption{\textbf{}Schematic of the IBL read-out system.}
\label{fig:IBL_Read-out_System}
%\end{center}
\end{figure}

The IBL read-out system~\cite{Polini:2011wna}, shown in Figure~\ref{fig:IBL_Read-out_System}, 
is based on the Pixel detector read-out~\cite{Aad:2008zz}. Each IBL half-stave is connected via the opto-board~\cite{Gan:optoboard:2015} 
%as shown in Figure~\ref{fig:optoboard} 
and a fibre bundle to the off-detector electronics boards: the Back-Of-Crate card (BOC)~\cite{BOC:2014, Wensing:2012ff} and the Read-Out-Driver (ROD)~\cite{Gabrielli:2015}.

\subsubsection{Optical link}
\label{sec:system_optical_link}

The opto-board, 
shown in Figure~\ref{fig:optoboard}, 
is connected to the counting room via approximately \SI{80}{\meter} of optical fibre. On the BOC card clock and data signals are encoded into one Bi-Phase-Mark (BPM) signal, running at \SI{40}{\mega\bit\per\second}, which is sent to the opto-board via a single optical link serving as TTC link. At the same time, the detector modules generate data streams at \SI{160}{\mega\bit\per\second} using 8b/10b encoding. The data are then sent by the opto-board via one optical link per FE-I4B chip to the BOC card.

\begin{figure}[tb]
%\begin{center}
	\centering
 \includegraphics[width=0.7\textwidth]{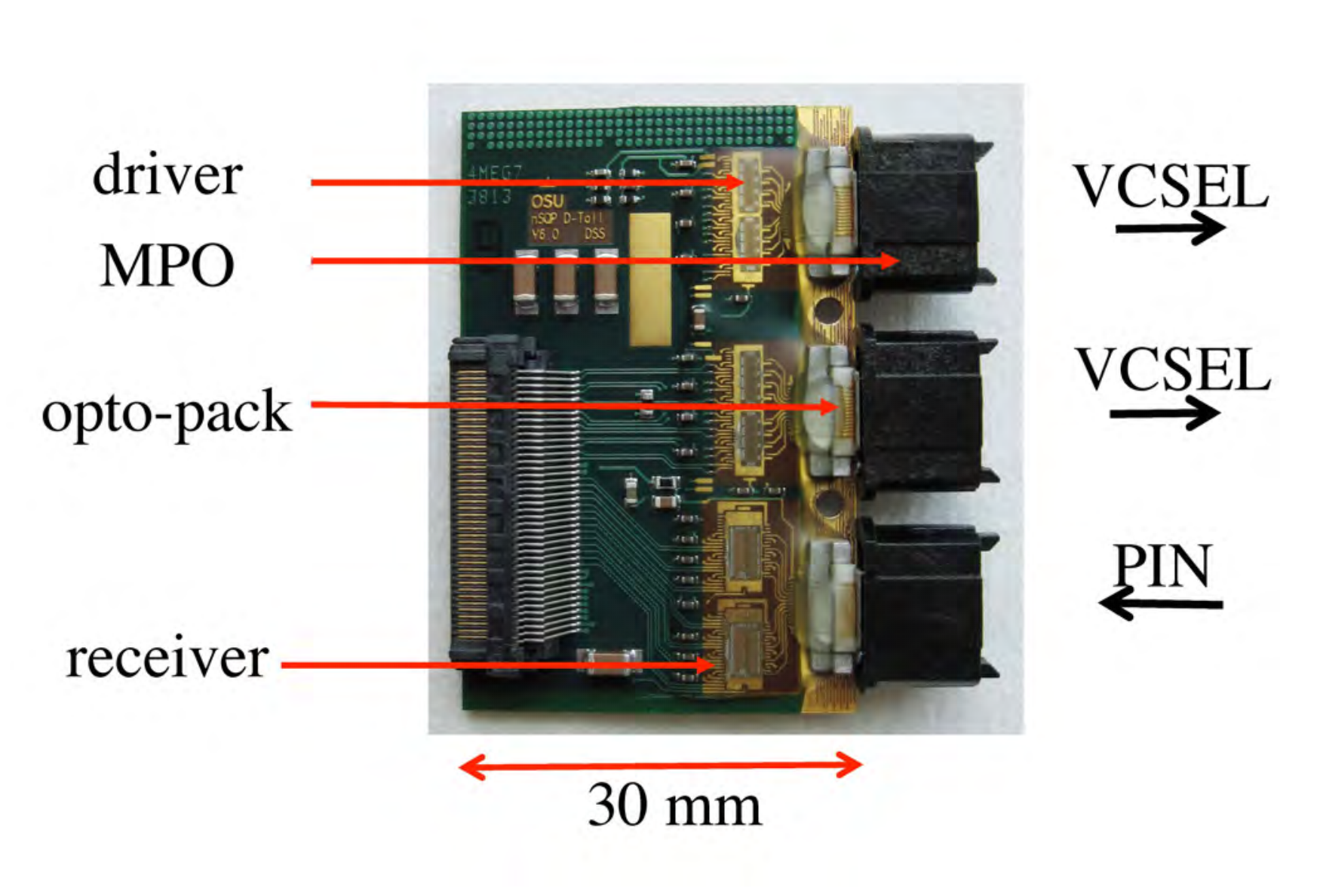}
\caption{\textbf{}Photograph of the IBL opto-board. 
}
\label{fig:optoboard}
%\end{center}
\end{figure}

The opto-board handles data in both directions and provides eight receiver and 16 transmitter channels. Hence, each opto-board serves a half-stave %holding 
(six planar modules and four 3D modules). 
Each opto-board therefore contains one PiN diode array and two VCSEL arrays, with each array containing 12 channels but only the inner 8 channels are used.  
The PiN diode array is paired to two 4-channel Digital Optical Receiver Integrated Circuits (DORIC). The PIN diode converts an optical signal into an electrical signal that the DORIC  decodes the into clock and data signals. Both signals are then transferred in LVDS format to the module.
Two 4-channel VCSEL Driver Chips (VDC), driving one VCSEL array with tunable current levels, are used to route the signals via the VCSELs to the BOC card. 
While the DORIC is a self-adjusting chip, the VDC requires an externally tunable voltage to steer the drive current and hence the optical output power of the VCSEL. 
The IBL uses 28 opto-boards and two additional boards are used for a Diamond Beam Monitor (DBM)~\cite{Diamond-C02026}.

\subsubsection{Off-detector read-out electronics (ROD/BOC)}
\label{sec:ROD/BOC}

A ROD/BOC card pair is shown in Figure~\ref{fig:readoutcards}. As shown in Figure~\ref{fig:IBL_Read-out_System}, data communication between the detector and  the ROD/BOC cards is bi-directional: one control and two parallel data processing paths link the ROD/BOC pair to an IBL half-stave.  

\begin{figure}[htbp]
	\centering
		 \includegraphics[width=0.9\textwidth]{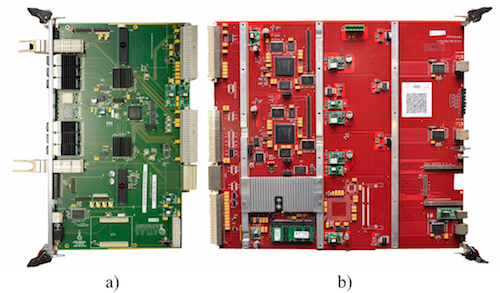}
\caption{\textbf{}The IBL read-out cards located at the off-detector side of the optical link. (a) The BOC card  and (b) the ROD card are paired in a VME crate via its back-plane.}
\label{fig:readoutcards}	
\end{figure}

The ROD controls the detector operation for both the calibration and data-taking modes. All of the front-end commands, including the trigger and clock signals, are generated in the ROD using a Virtex-5 FPGA with an embedded PowerPC (PPC) processor. The trigger and clock signals are distributed to the ROD by the Trigger Timing Control (TTC) link from the TTC Interface Module (TIM). The ROD control signals are transmitted to the detector via the BOC. 
An FPGA running a MicroBlaze processor on each of the ROD and BOC cards handles the Ethernet connection between them. 
The BOC card is interfaced to the detector and to the ATLAS read-out system: the detector interface uses commercial SNAP12 optical transmitters and receivers\footnote
{The SNAP12 Multi-Source Agreement (May 2002) outlines specifications for the mechanical, electrical and optical interfaces of 12-channel pluggable parallel optical transmitter and receiver modules.}; the read-out interface is via 
S-LINK connections\footnote{
S-LINK, for simple link interface, is a high-performance data acquisition standard developed at CERN.}. 

Each data processing path handles 16 FE-I4B chips and contains two FPGAs\footnote
{Unless otherwise specified, all FPGAs are from the XiLinx Spartan-6 family.}: 
one on the BOC card, responsible for synchronising, decoding and processing the signal coming from the FE-I4B, and one on the ROD, which builds event fragments and packages them for transmission to the ATLAS read-out via QSFP transceivers\footnote
{The Quad Small Form-factor Pluggable (QSFP) is a hot-pluggable transceiver allowing data rates of 4x10 Gbit/s.
} 
on the BOC card. 
The FPGA on the ROD card is also responsible for generating the histograms used to calibrate the detector. These histograms are transmitted using the Ethernet protocol to a fitting farm that uses commercial PC processors.  The boot and reset of the ROD is controlled by the Program Reset Manager (PRM) FPGA, directly mounted on the ROD.

A total of 15 ROD/BOC card pairs, 14 for the IBL staves and one for the DBM, are installed in a VME crate together with the TIM module which distributes the LHC clock and the ATLAS trigger signals. The loading of the ROD firmware and software as well as the transmission of control signals and data between the ROD and BOC is performed via the VME back plane; contrary to the Pixel detector read-out, the configuration of the cards and the transmission of calibration data are performed via Ethernet. As noted above, the fitting of calibration histograms is performed using a farm of PCs instead of the on-ROD Digital Signal Processors (DSPs)
used for the Pixel detector read-out.
This solution provides the scalability that is required to deal with the higher bandwidth of Run~2.

The IBL read-out hardware and system architecture is being implemented in stages to read the other Pixel detector layers because of the higher allowed bandwidth.
The readout of the Pixel B-Layer is equipped with twice the number of optical links, and the upgrade of Pixel Layer 1 was made at the end of Run 1.

\section{Interfaces and integration}
\label{section:integration}
% Chapter coordination: Claudia, Eric
The insertion of the IBL pixel layer was made possible by the reduction of the ATLAS beam-pipe diameter~\cite{Capeans:1291633}. The inner radius was reduced from \SI{29}{\milli\meter} to \SI{23.5}{\milli\meter}, allowing sufficient radial space for the IBL and its mechanical support structure. The new beryllium beam pipe is described in Section {\ref{sec:beampipe}}, and the support structure is described in Section {\ref{sec:IBL-mechanics-envelope}}. Prior to insertion in the ATLAS experiment, the full IBL as well as its services were assembled around the beam pipe on the surface (Section {\ref{sec:surface-integration}}) and electrically tested (Section {\ref{sec:eltests-integration}}) at room temperature. Once installed in the ATLAS cavern, electrical and environmental connections were made. In particular, the 2-phase CO2 cooling system as well as its connection to the IBL package and its subsequent performance is described in Sections {\ref{sec:cooling}} and {\ref{sec:coolingline}}.

\subsection{The beryllium beam pipe}
 \label{sec:beampipe}

The new beam pipe was designed and fabricated 
with a length of \SI{7300}{\milli\meter}, including the flanges (which were unchanged). The beryllium section is \SI{7100}{\milli\meter} long with a nominal wall thickness of \SI{0.8}{\milli\meter}. It is installed symmetrically with respect to the interaction point.
At each end of the beam pipe, split aluminium flanges of \SI{100}{\milli\meter} are welded, %in order to be
compatible with an insertion inside the IBL Inner Positioning Tube (IPT).
The inner surface of the beam pipe is treated with a Non-Evaporable Getter (NEG) thin film coating
 to optimise the vacuum quality. The NEG coating is activated by heaters that are wrapped along the beam-pipe length for the bake-out. 
A bake-out is made each time the ultra-high vacuum is established. The heater temperature is monitored and controlled by thermo-couples installed along the beam pipe. To reduce possible damage to the IBL layer, the infra-red emissivity
of the beam pipe is reduced by a surface layer of aluminium.

\begin{figure}[htbp]
	\centering
		 \includegraphics[width=0.70\textwidth]{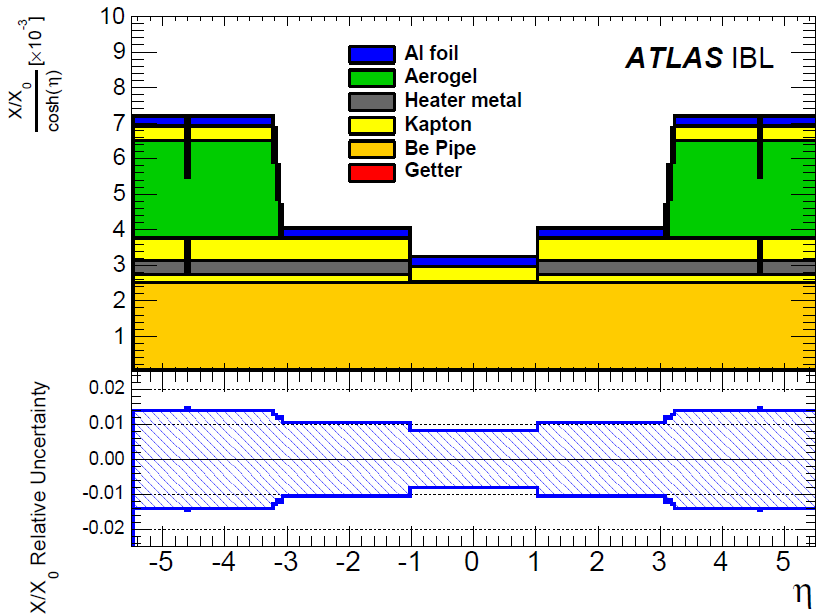}
	\caption{The calculated radiation length X/X$_{0}$ of the new beam pipe normalised by $\cosh^{-1}(\eta)$ as a function of $\eta$. The lower plot shows its relative uncertainty. The Getter (NEG) contribution is small and is not visible.}
	\label{fig:Section7_2_BPMaterialBudget}
\end{figure}

The composition and radiation length of the new beam pipe were both carefully validated. The thickness of the beryllium pipe was measured with a precision of \SI{\pm2.5}{\micro\meter} for eight positions along each section of the pipe, and the average thickness was measured to be \SI{0.8683\pm 0.0032}{\milli\meter}. The other components of the beam pipe (Kapton\textsuperscript{\textregistered}, heater metal, aerogel, aluminium foil) were individually measured by $^{109}$Cd (\SI{22}{\kilo\ev}) X-ray absorption and their nominal composition and thickness were confirmed to within a few percent precision. 
The total uncertainty of the radiation length considers not only the material composition, but also fabrication details such as wrinkles of the Al foil.

 As a result, the total thickness of the beam pipe was calculated to be \SI{0.32}{\xzero} at the centre of the beam pipe and up to \SI{0.72}{\xzero} at \SI{1.5}{\meter} from the centre, as shown in Figure~\ref{fig:Section7_2_BPMaterialBudget}. The main reduction of material thickness was the removal of the aerogel insulating layer over a length of \SI{622}{\milli\meter} at the centre of the beam pipe.

As a cross-check, the fabricated beam pipe was scanned using X-rays at \emph{z} positions of \SI{2.2}{\centi\meter}, \SI{25}{\centi\meter}, and \SI{90}{\centi\meter} along the beam pipe. The scanning equipment, shown in Figure~\ref{fig:Section7_Xray_jig.pdf}, was mounted on a multi-purpose container (MPC; see Section {\ref{sec:surface-integration}}) supporting the beam pipe. 
\begin{figure}[htbp]
	\centering
		 \includegraphics[width=0.50\textwidth]{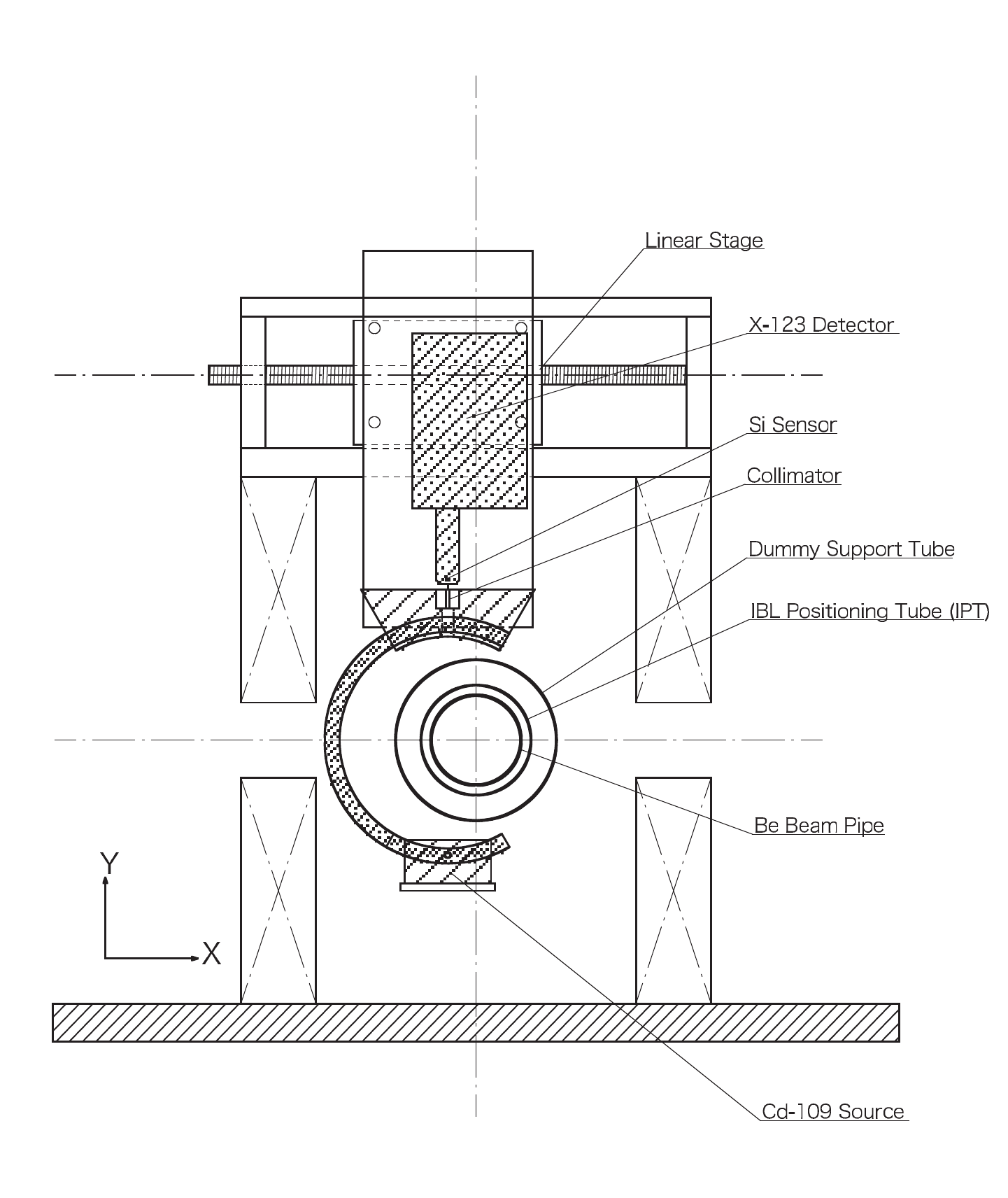}
	\caption{Beam pipe X-ray scanning equipment.}
	\label{fig:Section7_Xray_jig.pdf}
\end{figure}
The X-ray assembly included a fully-depleted Si-PIN X-ray counter (Amptek X-123\footnote{Amptek X-123 Si-Pin X-ray counter, see www.amptek.com}) and a vertically collimated $^{109}$Cd source. The source and detector could slide together horizontally, maintaining a constant X-ray flux. The relative  X-ray absorption at \emph{z} = \SI{2.2}{\centi\meter} is shown in Figure~\ref{fig:Section7_Xray_scan_z22mm.pdf} as a function of the horizontal position.
\begin{figure}[htbp]
	\centering
		 \includegraphics[width=0.70\textwidth]{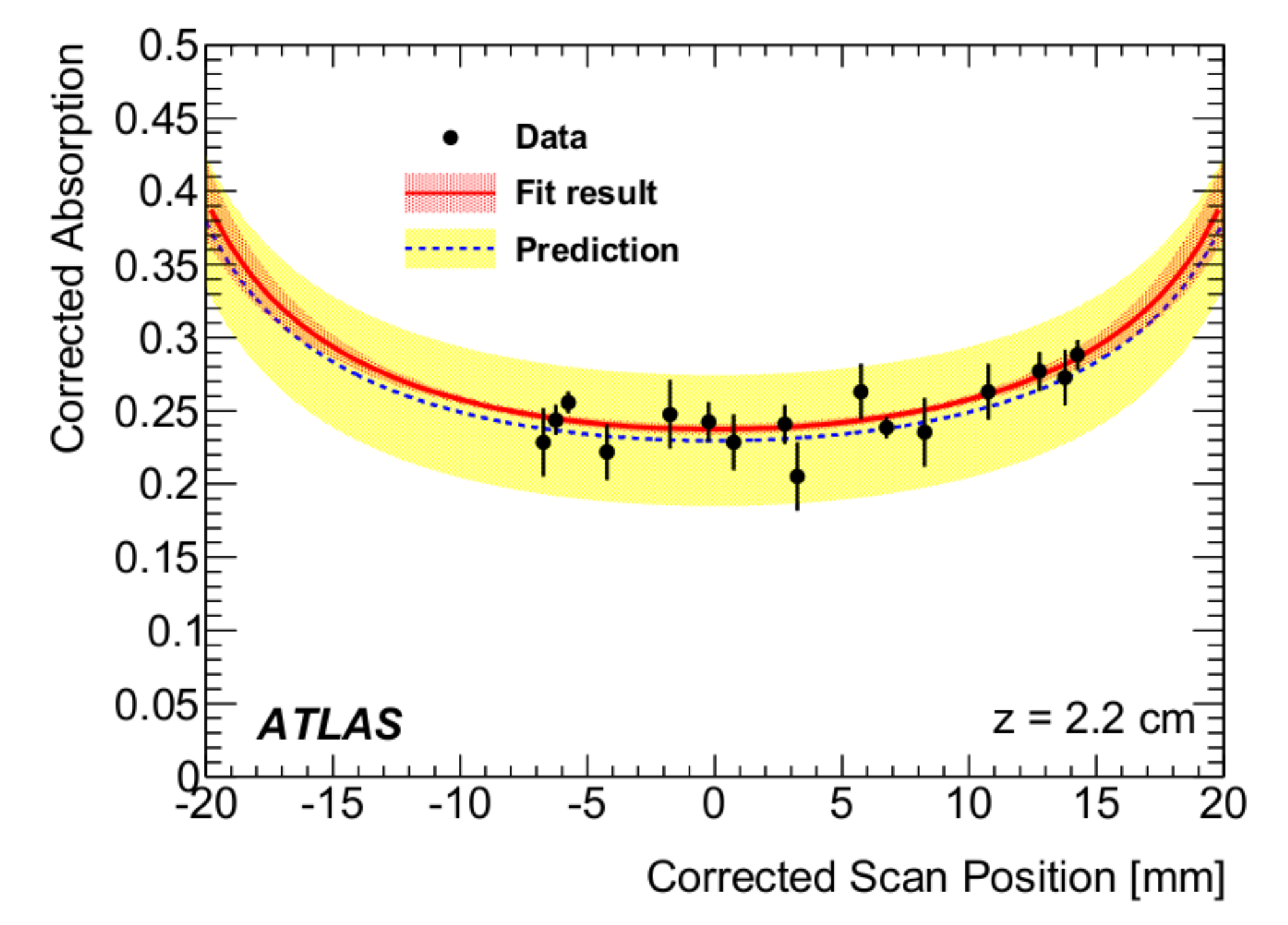}
	\caption{\textbf{}Relative amount of X-ray absorption in the beam pipe as a function of the corrected transverse scan position, at \emph{z} = \SI{2.2}{\centi\meter}.}
	\label{fig:Section7_Xray_scan_z22mm.pdf}
\end{figure}
An  X-ray absorption model, parameterising the absorption coefficient of the layers and the absolute X-ray flux, is in good agreement with the data. Similar validation results were obtained for \emph{z} =  \SI{25}{\centi\meter} and \emph{z} =  \SI{90}{\centi\meter}.

\subsection{The inner mechanical structure of IBL and its external envelope}
 \label{sec:IBL-mechanics-envelope}

As shown in Figures~\ref{fig:IBLView2} and ~\ref{fig:IBLLayout}, the IBL volume for the stave and services is delimited by the Inner Support Tube (IST), which is fixed on the Pixel structure, and by the IPT, which is a precision mechanical support to hold the staves and the services. A schematic and photograph of the IPT are shown in Figure~\ref{fig:sec7_3_IPTAssembly}.

\begin{figure}[!htb]
	\centering
	\begin{subfigure}[t]{0.47\textwidth}
		\includegraphics[width=\textwidth]
{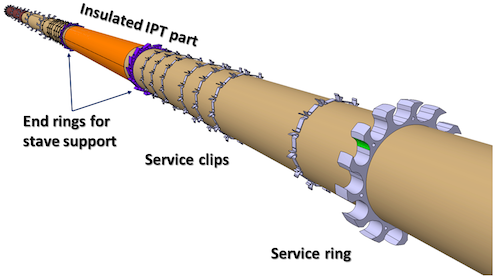}
		\caption{}
		\label{fig:iptsketch}
	\end{subfigure}
	\begin{subfigure}[t]{0.47\textwidth}
		\includegraphics[width=\textwidth]
{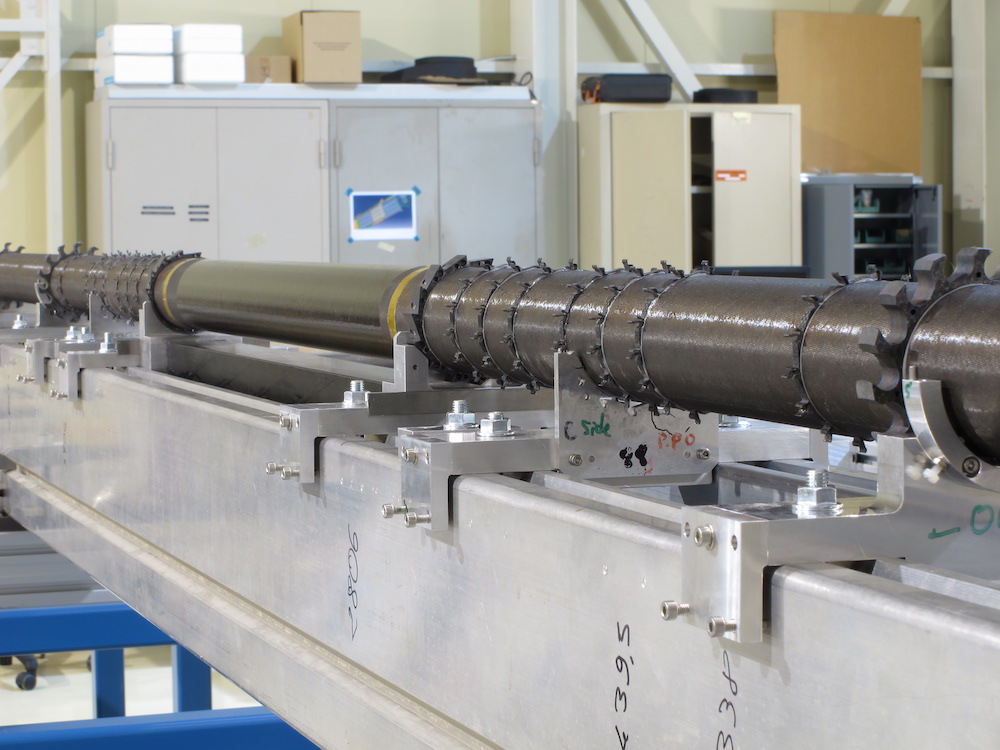}
		\caption{}
		\label{fig:iptphoto}
	\end{subfigure}
%\vspace {-2cm}
\caption{\subref{fig:iptsketch} Sketch and \subref{fig:iptphoto} photograph of  the Inner Positioning Tube (IPT) assembly before the integration of staves and services.}
\label{fig:sec7_3_IPTAssembly}
\end{figure}

The IST  is a \SI{6600}{\milli\meter} long tube with an inner diameter of \SI{85}{\milli\meter} and \SI{0.455}{\milli\meter} wall thickness, containing the full IBL package and the central portion of the beam pipe. 
It is secured  to each end of the Pixel detector frame  in the PP0 area and to the pixel cruciforms in the  PP1 region, located at respectively \SI{728}{\milli\meter} and \SI{3241}{\milli\meter} from the interaction point along the beam axis.
Given the clearance of only \SI{\sim2}{\milli\meter} with respect to the Pixel B-Layer, the IST is designed to have a minimal deflection during its lifetime. It was therefore manufactured using very high modulus carbon fibre (K13), coupled with cyanate ester resin (RS3). The carbon-fibre material consists of seven  \SI{65}{\micro\meter}-thick plies oriented in different directions to minimise the tube deflection. The thermo-mechanical behaviour of the complete package was precisely predicted by finite element analyses and validated with experimental data.

The IPT is a precision assembly of carbon fibre tubes and rings (Figure~\ref{fig:iptsketch}) that aligns the 14 staves with high precision. The five carbon-fibre segments use the same carbon fibre material as the IST, but the central segment uses five plies to minimise the material budget; the resulting thickness is \SI{0.355}{\milli\meter}. In addition, to ensure electrical insulation with respect to the stave module HV, this central segment is co-cured with a   \SI{25}{\micro\meter}-thick polyimide film.
The service rings, each with 14 radial grooves for the stave services, are precisely positioned every \SI{50}{\centi\meter} along the IPT. A  small extra length, estimated to be \SI{\sim3}{\milli\meter}, compensates for the service cable contraction and extension during temperature excursions.

A key feature of the IPT design is the ability to perform a fast insertion or extraction of the IBL and of the beam pipe independently of the rest of the ATLAS ID.
The IPT is locked in z with respect to the Pixel structure at a distance of \SI{3241}{\milli\meter} from the interaction point on the C-side, and is positioned radially within the limited clearance to the IST, which is fixed to the Pixel frame. The beam pipe is positioned in z with respect to the LArg Cryostat at PP1, \SI{3426}{\milli\meter} from the interaction point. It is positioned radially within the limited clearance to the IPT and the Liquid Argon cryostat at PP1. 

Two titanium terminals are positioned at the IPT extremities. Their function is to provide 
shielding from electromagnetic interference, strain relief for the cooling fittings, space for cable bundle integration and mechanical stability during installation.
The gluing of components on the IPT was performed using epoxy glue (Hysol 9394\footnote
{Hysol\textsuperscript{\textregistered} is a trademark of Henkel Corporation, see www.henkel.com.}) that can cure at room temperature and has a maximum service temperature of \SI{177}{\celsius}. This guarantees the integrity of all the parts, even during the beam-pipe bake-out when  the maximum  temperature is expected to reach \SI{110}{\celsius} at the contact points with the beam-pipe rings.

Two units were manufactured for each of the IST and the IPT and were surveyed in a metrology laboratory  to check for geometrical defects and qualification. The best  assemblies  with respect to geometrical specifications were used for the IBL integration.

\subsection{Surface integration and installation}
 \label{sec:surface-integration}

The  elements to be integrated can be grouped in three independent sets: the IPT with the support rings, the staves with their cooling pipes and the services.

The Multi-Purpose Container (MPC), the mechanical support for the IBL integration, was designed to allow  secure and precise stave integration, and to transfer the IBL package to the experimental  cavern. 
Initially, the beryllium part of the beam pipe was inserted inside the IPT. 
A precision tool was then used  to transfer the staves from their holding jigs to the IPT,
preserving the  tight clearances with respect to  the  surrounding structures.
After each stave was connected to the cooling extension (Section~\ref{sec:coolingline}), it was fixed to the integration tool of the MPC (Figure~\ref{fig:sec7_4_StaveIntegration}). Using adjustable screws on the stave integration arm, and IPT rotations referenced with precision pins, each stave could be installed and transferred from the handling frame to the support ring of the IPT with a precision of \SI{\sim50}{\micro\meter} over the \SI{\sim724}{\milli\meter} length.

Once the 14 staves were loaded, a central ring consisting of seven parts was installed and clipped to the central stave support feet,
providing an additional stiffening.  
%Due to tight construction and thermal fixation constraints,  however, it did not eliminate all possible  detector degrees of freedom. 
However, because of stringent construction and thermal fixation constraints, it did not eliminate all possible  detector degrees of freedom. 
It radially stiffens the 14 staves at the centre, while leaving  free movement in the azimuthal direction and along the beam axis.
Operationally, this resulted in an R-$\phi$ distortion of the staves at the level of a few~\SI{}{\micro\meter}/\SI{}{\celsius} (Section~\ref{sec:ibl_in_atlas}).

\begin{figure}[htb]
	\centering
	 \includegraphics[width=1.00\textwidth]{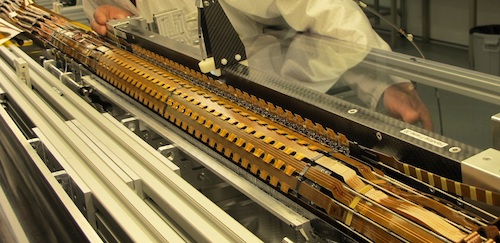}
	\caption{Integration of a stave mounted on a handling frame and the integration arm which is mechanically aligned to the Multi-Purpose Container (MPC).}
	\label{fig:sec7_4_StaveIntegration}
\end{figure}

The services that connect the staves to each ID end, \SI{\sim3.5}{\meter} from the interaction point, were installed after the integration of each stave. The service installation consists of eight intermediate flex circuits 
linking Type~1 cables to the stave flex in the PP0 region (Section~\ref{sec:ElectricalCables}). The Type~1 cables service the HV and LV power, the read-out and the DCS signals and are laced in a single bundle that splits just before exiting the IBL volume into separate data and power bundles.
 After the installation of each stave and their services, electrical qualification tests were performed, before proceeding to the next row. All the staves were successfully integrated without rework or re-installation. 

The next operation consisted of inserting soft sealing rings at the two extremities \SI{\sim3}{\meter} from the interaction region to guarantee a proper environment for the IBL. The IBL is flushed with dry nitrogen gas at a flow rate of up to \SI{450}{\litre\per\hour}. This ensures a dew point below that of the minimal foreseen coolant temperature of \SI{-40}{\celsius}, keeping the detector dry under all operating conditions.  The sealing ring core is moulded as a ring from polyurethane, with holes for electrical and cooling services. During cavern commissioning, the full dry nitrogen circuit, including sealing rings, was leak checked. At the nominal running parameters (\SI{80}{\litre\per\hour} at \SI{20}{\milli\bar} at maximal overpressure), the measured outlet flow was \SI{75}{\litre\per\hour}, equivalent to a nitrogen circuit tightness of better than \SI{90}{\percent}.

Once the integration was completed, the services (including an additional length of  \SI{1}{\meter} for the power bundle and \SI{2.6}{\meter} for the data bundle) were packed to ensure the IBL envelope for installation in the ATLAS cavern. In particular, the Type~1 bundles from PP1 to PP2  were wrapped 
by a spiral wrapping tool around the beam pipe (Figure~\ref{fig:ServiceInstall}) until the IBL package was inserted inside the IST.

\begin{figure}[!htb]
	\centering
	\begin{subfigure}[t]{0.420\textwidth}
		\includegraphics[width=\textwidth]
{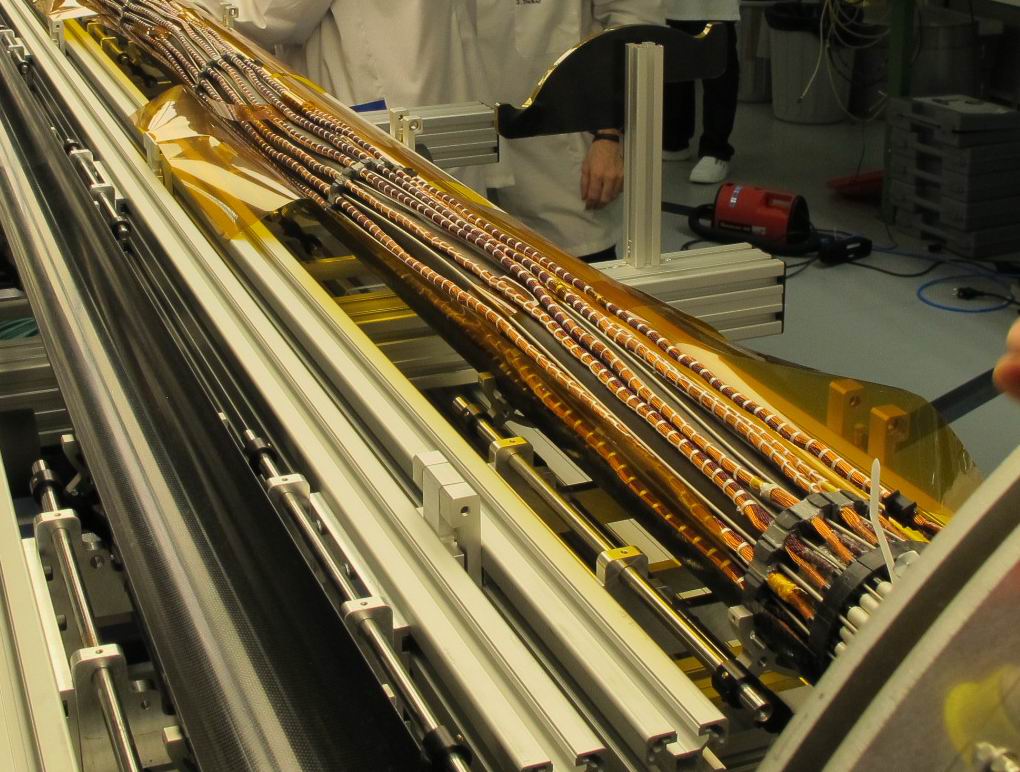}
		\caption{}
		\label{fig:type1ipt}
	\end{subfigure}
	\begin{subfigure}[t]{0.570\textwidth}
		\includegraphics[width=\textwidth]
{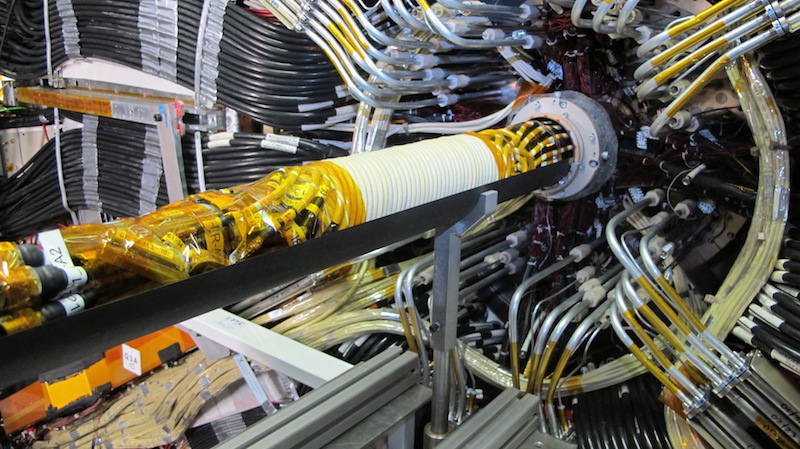}
		\caption{}
		\label{fig:type1wrapped}
	\end{subfigure}
%\vspace {-2cm}
\caption{\subref{fig:type1ipt} Type~1 services integrated on the Inner Positioning Tube (IPT) and Z-stopper locked into the service rings. \subref{fig:type1wrapped} Spiral wrapped Type~1 cables that take up the excess length of the IBL envelope and limit the radial envelope with respect the Inner Support Tube (IST) for the insertion.}
\label{fig:ServiceInstall}
\end{figure}

\subsection{Electrical tests after stave integration. }
 \label{sec:eltests-integration}

The purpose of the stave electrical test after integration of the staves onto the IPT, cabling of the Type~1 cables and insertion inside the IST, was to verify the electrical and functional integrity of the stave components and to test the service chain. 
The stave test included threshold scans with and without sensor bias, a validation of the time-over-threshold (ToT) setting, and  an I-V scan. The tests  were directly compared to results obtained during the stave QA. A sample of these comparisons is shown in Figures~\ref{fig:threshold},~\ref{fig:noise} and~\ref{fig:tot}. Noise and threshold values are larger than in the stave QA because  the detector was not cooled during the stave loading. This is especially true for  3D modules, which are more sensitive to  temperature. With that caveat, no performance degradation was measured.

The  tests also revealed several hardware problems (e.g. broken or shorted lines  in the Type~1 cable or in the intermediate flex)  that were  repaired. The tests were repeated  after  sealing  the IBL volume  with two soft polyurethane disks, to check the correct functionality of all modules before wrapping the services and lowering the detector in the cavern of the ATLAS experiment. No changes with respect to previous tests were measured.

\begin{figure}[htbp]
\begin{center}
\includegraphics[width=0.55\columnwidth]{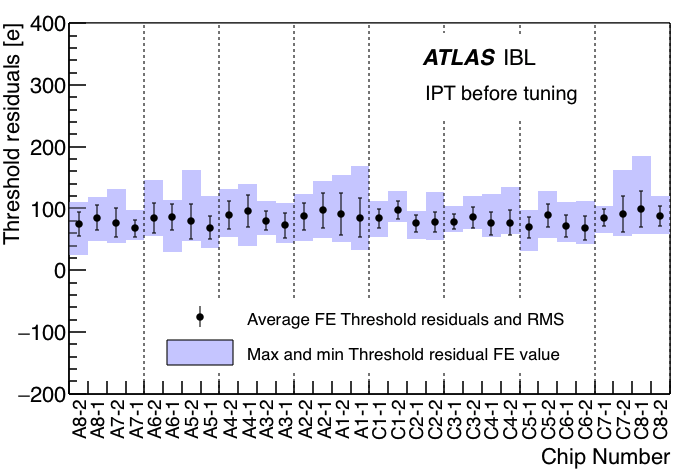}
\end{center}
\caption{Average and RMS of the chip-to-chip FE-I4B threshold difference between the results of the connectivity test after integration around the IPT (IBL positioning tube) and the individual QA values for the 14 staves. The QA configuration setting targeted \SI{3000}{\e} and 10 ToT at \SI{16}{\kilo\e}. Maximum and minimum FE-I4B threshold mean values of the 14 staves are represented for each chip number (position on stave) by the filled area.}
\label{fig:threshold}
\end{figure}

\begin{figure}[htbp]
\begin{center}
\includegraphics[width=0.6\columnwidth]{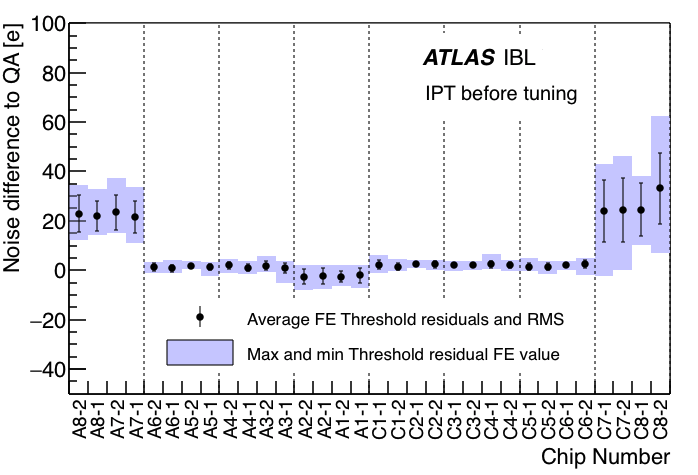}
\end{center}
\caption{ Average and RMS of the chip-to-chip FE-I4B noise difference between the results of the connectivity test after integration around the IPT (IBL positioning tube) and the individual QA values for the 14 staves. The QA configuration setting targeted \SI{3000}{\e} and 10 ToT at \SI{16}{\kilo\e}.  Maximum and minimum FE-I4B noise mean values of the 14 staves are represented for each chip number (position on stave) by the filled area (the four outer chips on each side, using 3D sensor technology, show
 an increased noise behaviour).}
\label{fig:noise}
\end{figure}
 
\begin{figure}[htbp]
\begin{center}
\includegraphics[width=0.6\columnwidth]{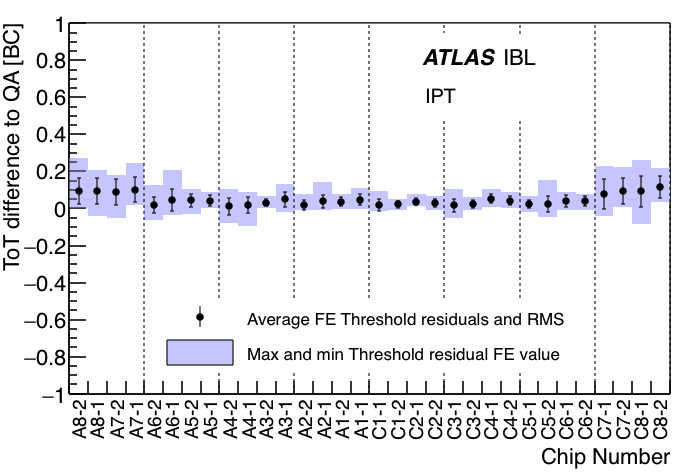}
\end{center}
\caption{Average and RMS of the chip-to-chip FE-I4B ToT mean values between the results of the connectivity test after integration around the IPT (IBL positioning tube) and the individual QA values for the 14 staves. The QA configuration setting targeted \SI{3000}{\e} and 10 ToT at \SI{16}{\kilo\e}.  Maximum and minimum FE ToT mean values of the 14 staves are represented for each chip number (position on stave) by the filled area. }
\label{fig:tot}
\end{figure}

In parallel to the stave integration and test, a system test was prepared using two prototype IBL staves and either production or pre-production  detector services, the full power and DCS chain, but mono-phase cooling. 
The measured noise, threshold and ToT performance were in excellent agreement with the QA.
 Scans were made using $^{241}$Am and $^{90}$Sr sources, as well as cosmic rays, with satisfactory results. The system test also confirmed the functionality of the interlock system. The system test setup continues to be used as a test-bench for operational maintenance and upgrades.

\subsection{CO$_2$ cooling system}
 \label{sec:cooling}

The cooling of the IBL detector is based on CO$_2$, which is circulated in a closed system through the detector with an overflow where part of the liquid is evaporated (approximately \SI{30}{\percent} at \SI{1.5}{\kilo\watt})~\cite{1748-0221-12-02-C02064}. The two-phase liquid-vapour mixture is returned to the cooling plant, which is located in the USA15 service cavern and easily accessible. The cooling plant condenses the returning two-phase CO$_2$ %to liquid 
using a commercial %type 
chiller. The liquid CO$_2$ is pumped back to the manifold system near the IBL detector via a concentric transfer line that bridges the distance between the cooling plant in USA15 and the manifold in the UX15 service cavern. Figure~\ref{fig:sec7_6_Figure1} shows a simplified schematic of the IBL cooling system with the main components of the cooling system highlighted.

\begin{figure}[htb]
	\centering
		 \includegraphics[width=0.80\textwidth]{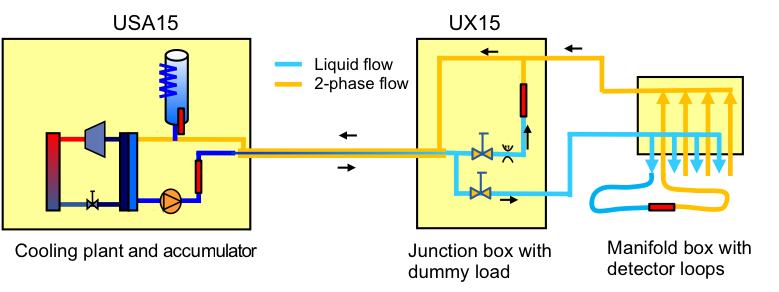}
	\caption{Simplified scheme of the IBL cooling system. USA15 and UX15 refer to the service caverns where the equipment is installed. The junction and manifold boxes are separated from the cooling plant by approximately \SI{80}{\meter}.}
	\label{fig:sec7_6_Figure1}
\end{figure}

\subsubsection{Cooling system operation}

The CO$_2$ arriving in the detector is a saturated liquid, which means that it evaporates directly when heat is applied. The temperature of the arriving saturated liquid is a function of the pressure, which is controlled by the cooling plant in USA-15.
 Changing the temperature of the two-phase mixture in the accumulator will change the pressure in the system and allows
operation with an evaporation temperature between \SI{15}{\celsius} (used for commissioning) and \SI{-40}{\celsius}.

The pressure can also be increased to fully liquify the system.
This is used at start-up to prevent thermal shocks.
 The cool-down temperature ramp is controllable and can be set to  \SI{2}{\celsius} per minute, or to a lower rate.
The preferred inlet condition of the cooling is that the liquid is saturated in the IBL stave region, so that heat needs to be applied to the cold liquid. This heat is taken out of the returning two-phase mixture by a constant thermal contact of the liquid inlet and the two-phase return.
 The inlet and outlet fluids circulate in concentric tubes (the inlet liquid in a \SI{10}{\milli\meter} inner diameter tube, and the outlet fluid circulating in a \SI{21}{\milli\meter} outer diameter tube).
The actual system allows having the same temperature at the detector and at the two-phase temperature controlled by the accumulator.
 The higher pressure on the inlet keeps the CO$_2$
liquified at the inlet to the cooling tubes. It starts boiling in the cooling tubes of the IBL staves once powered.
 This liquid temperature condition works over a large range of operational temperatures and makes the control of the system,
without active elements inside the ATLAS detector,
very reliable in a hard-to-access region. The first tests allowed the system to reach stable temperatures for various heat loads and was tested up to  a thermal load of \SI{3}{\kilo\watt} and down to
\SI{-40}{\celsius} (Figure~\ref{fig:sec7_6_Figure3}).

\begin{figure}[htb]
	\centering
		 \includegraphics[width=0.90\textwidth]{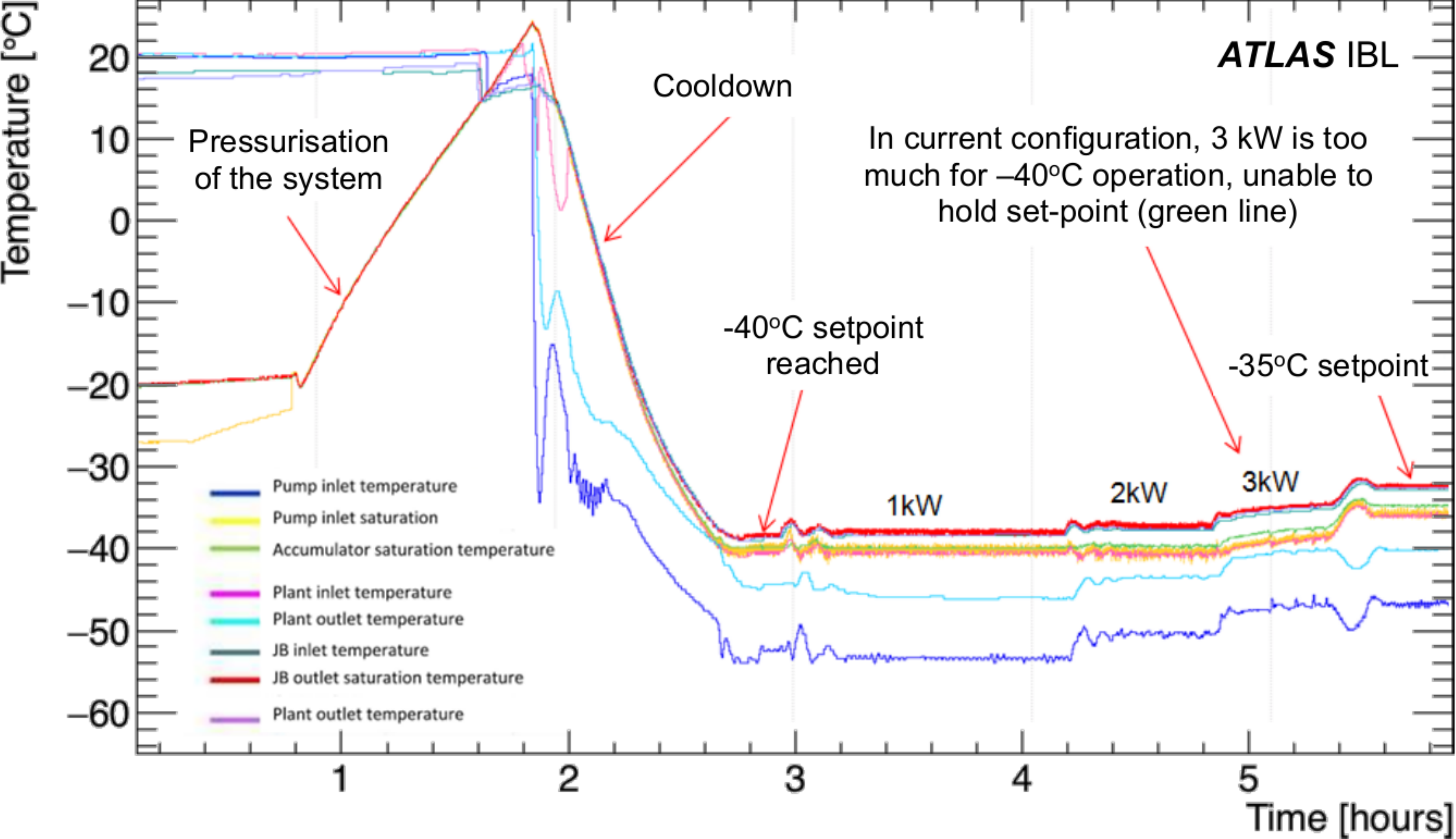}
	\caption{Temperatures measured at start-up and cold operation of the CO$_2$ cooling system with various thermal loads applied at the junction box located several meters %away 
from the IBL.}
	\label{fig:sec7_6_Figure3}
\end{figure}

\subsubsection{Redundant system}

Once irradiated, the IBL must remain cold at all times to 
limit the radiation damage in the silicon sensors. To guarantee a fail-safe solution for the CO$_2$ cooling a 
redundant system has been developed. There are two identical plants where one serves as a full back-up of the running plant. Each plant has its own control 
and sensor
system with a dedicated Programmable Logic Controller (PLC) and power source. The transfer lines and the accumulator are shared with respect to the CO$_2$ volume.
A plant can be disconnected from the main system for interventions. Both chillers are cooled by water provided from the central ATLAS water system. An integrated air cooling condenser is present in each chiller unit to back-up the single source water cooling. During operation, one system can remain on standby for a fast switch-over in case of a system failure. In addition, for greater safety, the two plants can operate in parallel increasing the cooling capacity, but this operational mode is mainly foreseen for the beam-pipe bake-out.

\subsubsection{Detector distribution}

A homogeneous CO$_2$ flow distribution to the 14 IBL staves is achieved by 11\,m long lines and \SI{1}{\milli\meter} inner diameter capillaries (Figure~\ref{fig:sec7_6_Figure4}). These capillaries are routed inside the return tubes of the IBL to be shielded from ambient heating. The manifolds are located in the muon detector area. The total tube length from manifold through the IBL and back to the return manifold is approximately \SI{32}{\meter} ($2 \times \SI{11}{\meter}$ concentric tubing, $2 \times \SI{4}{\meter}$ connection tube and \SI{1}{\meter} stave tube). 
The inlet tubes, the boiling channels and outlet tubes have nominal inner diameters of \SI{1.5}{\milli\meter}, \SI{1.5}{\milli\meter}  and \SI{2}{\milli\meter} respectively.
 The innovative vacuum isolated flexible lines were used for the fluid transfer on long distances ~\cite{1748-0221-12-02-C02064}.
 The concentric return tubes in the flexible transfer line have a diameter of  \SI{3}{\milli\meter}. The \SI{11}{\meter} concentric line is outside of the ID volume and is insulated by multilayer insulation inside a \SI{16}{\milli\meter} diameter vacuum metallic tube. This triple concentric assembly is flexible and is routed similarly to the electrical cabling, through the ID end plate region towards the manifold.
The flex lines inside the ATLAS ID end plate up to the splitter box are shown on Figure~\ref{fig:sec7_6_Figure4}(b).

\begin{figure}[!htb]
	\centering
	\begin{subfigure}[t]{0.8\textwidth}
		\includegraphics[width=\textwidth]
{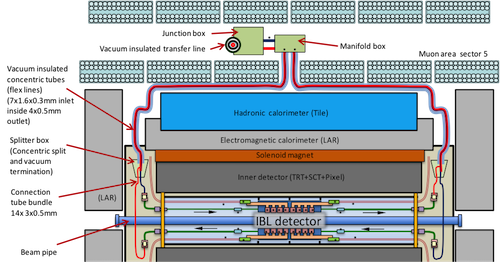}
		\caption{}
		\label{fig:co2distschema}
	\end{subfigure}
	\begin{subfigure}[t]{0.47\textwidth}
		\includegraphics[width=\textwidth]
{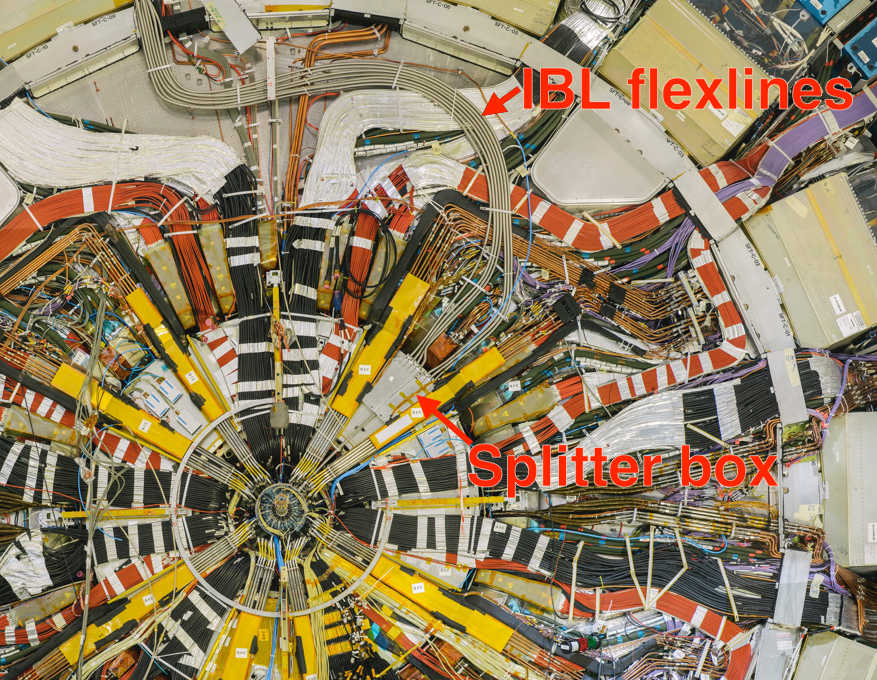}
		\caption{}
		\label{fig:co2distphoto}
	\end{subfigure}
%\vspace {-2cm}
	\caption{\subref{fig:co2distschema} CO$_2$ distribution concept inside the ATLAS detector. \subref{fig:co2distphoto} Flex cooling lines integrated inside the ID %Inner Detector 
	flange and connected to one end of the IBL detector package via Splitter box.}	
	\label{fig:sec7_6_Figure4}
\end{figure}

\subsubsection{Commissioning}

The commissioning of the system started in January 2014 with local circulation of CO$_2$ using the plant internal dummy load. Near the manifold the so-called junction box is present (Figure~\ref{fig:sec7_6_Figure4}) where the flow can be by-passed through a \SI{3}{\kilo\watt} dummy load. The system commissioning was made using this dummy load with a restriction valve having a similar flow resistance to that of the detector loops. One of the  main challenges for the IBL cooling was related to the colder temperatures compared to previous cooling systems. The requirement of cooling to \SI{-40}{\celsius} brings the margin close to the CO$_2$ freezing point (\SI{-56}{\celsius}) %to be small 
and 
hence a very stable primary cooling was needed. 
In the early phase the system was tuned such that under extreme conditions it remained within safe operational boundaries. The measured heat load was approximately \SI{2}{\kilo\watt} at \SI{-40}{\celsius} and \SI{3}{\kilo\watt} at \SI{-35}{\celsius}. The IBL detector was also successfully tested during the commissioning % AC with success 
at various temperatures % AC during the commissioning with nominal power. 
with the nominal power load expected during operation.
The boiling onset inside the IBL cooling loops sometimes had problems to be correctly initiated. %AC show problem to get initiated correctly. 
In this case, super-heated liquid, that is warmer than the boiling temperature and has a worse heat transfer than the desired two-phase flow, could be present in the detector. %AC
As a result, the temperatures of the first modules of a stave were sometimes a few degrees higher. To mitigate this problem, flow restrictions were applied in the inlet manifold to reduce the flow and to keep the inlet pressure high
 ~\cite{1748-0221-12-02-C02064}.

\subsection{On-detector cooling branch and interfaces}
\label{sec:coolingline}

The on-detector cooling branches consist of a \SI{7}{\meter} length straight section of titanium pipe with two brazed junctions at each side of the stave and a dis-mountable fitting at each end (Figure~\ref{fig:sec7_5_SchematicView}). As for the rest of the IBL services, the detector side of the fitting must respect the insertion envelope.
The branch is connected on each side to a short 90{\degree}-bend section of Ti pipe, which is brazed to an electrical break and to a stainless steel pipe at the other end. The stainless steel pipes are then routed on a path specific to each stave to a splitter box where the transition is made to \SI{16}{\milli\meter} flexible vacuum transfer lines of \SI{11}{\meter} length connected to a manifold box further out from the ID.

\begin{figure}[htb]
	\centering
		 \includegraphics[width=1.00\textwidth]{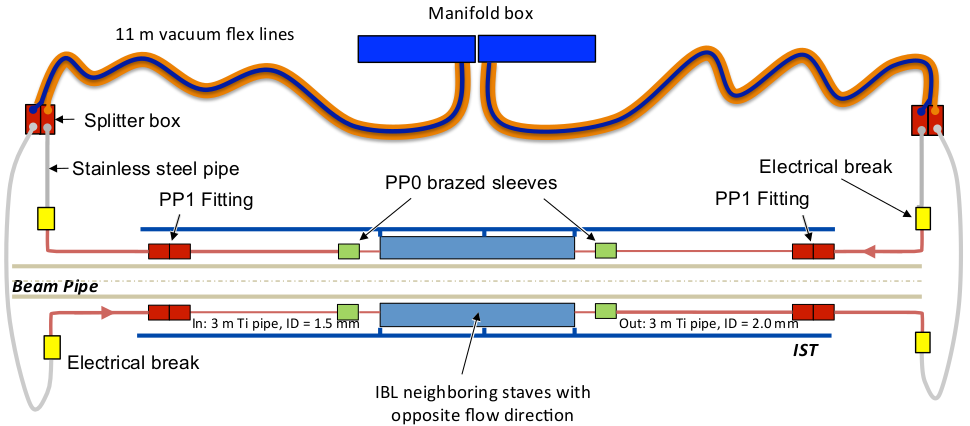}
	\caption{Not-to-scale schematic view of the IBL cooling distribution line inside the ID %Inner Detector 
	volume.}
	\label{fig:sec7_5_SchematicView}
\end{figure}

\subsubsection{Fittings outside the IBL volume}
	
Because of space constraints for the insertion of the IBL inside the IST, the electrical and cooling service envelope is restricted to a maximum external radial space of \SI{4}{\milli\meter}. 
 An industrial fitting of such a small size compatible with the pressure, the radiation hardness, and the reliability requirements of the IBL does not exist; therefore a custom fitting was developed.
 
The use of the CO$_2$ evaporative system (Section~\ref{sec:cooling}) together with titanium as a selected material for both the tube and the fitting required a special engineering and design development to maximise the reliability. The extreme radiation environment excludes the use of organic joints, leading to a metal-to-metal contact solution. The requirement led to the selection of a hard titanium alloy (TA6V, or grade 5) to guarantee the sealing. The tightness is ensured by a cone-sphere junction, with strict requirements on the surface quality.

\begin{figure}[!htb]
	\centering
	\begin{subfigure}[t]{0.57\textwidth}
		\includegraphics[width=\textwidth]
{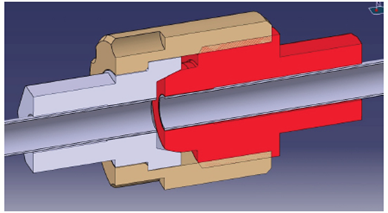}
		\caption{}
		\label{fig:pp1fitting}
	\end{subfigure}
	\begin{subfigure}[t]{0.41\textwidth}
		\includegraphics[width=\textwidth]
{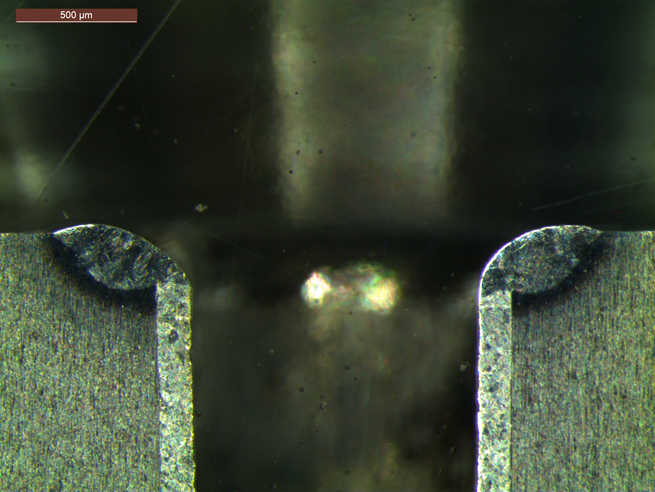}
		\caption{}
		\label{fig:weldsection}
	\end{subfigure}
%\vspace {-2cm}
	\caption{\subref{fig:pp1fitting} Cross-section drawing of the PP1 fittings and tubes. \subref{fig:weldsection} Cross-section photograph of the electron beam welding of the titanium pipe with the fitting gland.}
	\label{fig:sec7_5_FittingCrossSection}
\end{figure}

Prototype fittings were assessed with several batches of $\sim$10 fittings machined in-house and tested to qualify the final design. The final production batches were manufactured in industry and qualified in-house on a large number of test samples. The fitting is connected to the thin titanium pipe 
(\SI{0.11}{\milli\meter} wall thickness)
by electron beam welding at the front face (Figure~\ref{fig:sec7_5_FittingCrossSection}). This welding requires high quality vacuum and was performed in industry. The method imposes strict requirements on the reliability of the fitting which
can not be repaired.
 A large-scale qualification and validation campaign was carried out, with $\sim$100 fittings leak tested, some of them mated and de-mated tens of times. An industrial qualification procedure was applied to fulfil the standard requirements with large statistics pressure cycling. A set of ten fittings, electron-beam welded to short-pipe sections and connected in series, was tightened by the collaboration and sent to a certified ISO qualification lab to perform the pressure cycling (one million water cycles with \SI{1}{\bar} to \SI{100}{\bar} at \SI{\sim1}{\hertz}). The entire set successfully passed the leak tests.
The minimal torque required for reliable tightening was determined during the qualification campaign. A \SI{2}{\newton\meter} torque was sufficient to guarantee the success of tests like fast temperature cycling or thermal shocks using a CO$_2$ blow-off system. A small subset of fittings was tested at higher torque up to \SI{6}{\newton\meter} without visible damage. The final torque used for installation in the cavern was \SI{2.5}{\newton\meter} for which all the 28 IBL connections passed the pressure tests at \SI{100}{\bar}.
%\newline

The pre-series consisted of producing approximately 100 samples for qualification. 
The 28 fittings selected for installation were visually inspected for scratches and dust, and individually leak-tested.

\subsubsection{The cooling line electrical break}

The grounding and shielding scheme of the IBL requires using electrical breaks on the cooling pipes at the PP1 area (Figure~\ref{fig:sec7_5_E-break}) and in the ID end-flange region. Because of space constraints the closest possible location was just after the 90{\degree}-bend
%90 degrees bent 
of the radial section of the pipe. The mechanical stress in this section is significant; a relatively large diameter (\SI{8}{\milli\meter}) ceramic electrical break was chosen for robustness. At this location it is also necessary to make the transition from titanium pipes, which are difficult to bend, to stainless steel pipes, which are easily routed. After testing several options for the titanium to stainless steel transition the most reliable solution found was to braze in the same processing step a stainless steel sleeve on the external side of the ceramic of the electrical break (the detector side of the ceramic being brazed in a titanium sleeve). Due to the mismatch of the thermal expansion coefficient between the ceramic and the stainless steel material the brazing is a delicate process. A full qualification in collaboration with industry was performed to design and produce a reliable junction. The qualification process was the same as for the fitting and a number of destructive tests were made to evaluate the mechanical robustness and the capillary penetration of the brazing material. All the tests were passed successfully and the tensile tests
%AR made
at the electrical break junction revealed that the pipe was weaker and that the junction withstands at least \SI{640}{\bar} internal pressure (limited by the test setup). In addition, a few electrical break samples were irradiated to \SI{250}{\mega\radian} with a \SI{10}{\mega\ev} electron beam, corresponding to the expected maximum end-of-life IBL ionising dose, and found to be leak-tight.

\begin{figure}[htb]
	\centering
	 \includegraphics[width=0.40\textwidth]{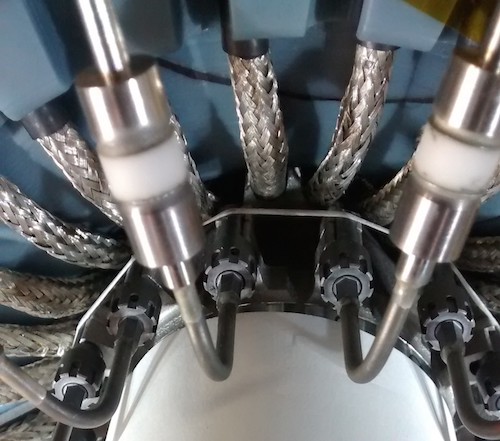}
	\caption{The cooling line electrical break in the ID end flange service region.}
	\label{fig:sec7_5_E-break}
\end{figure}

\subsubsection{The brazing junction of the stave inside the detector volume}

At PP0, the cooling junction connects the staves to the cooling extension running from \SI{705}{\milli\meter} to \SI{3366}{\milli\meter} (PP1 region) from the interaction point. This junction was designed such that the stave production could be made with \SI{1500}{\milli\meter} long objects, easing the module loading task, the testing tool, the handling and the shipment. Due to the limited space around the beam pipe in the PP0 region, and the high level of reliability required inside the detector volume, the use of fittings was not possible.
\newline
One major impact of this design choice was to develop a thin wall
(\SI{0.11}{\milli\meter})
titanium pipe joining technique between the extension and the stave pipe which could be easily connected after module loading and before integration of the staves around the IPT. When such an operation is performed after the module loading, the requirements are not only based on the quality of the welded junction but also the risk of damage to electrical or mechanical components (that should be negligible). Therefore techniques that require excessive heat spread, such as oven brazing, or high and fast current spikes such as orbital welding, were prohibited.
With such a thin wall thickness, titanium 
is not an easy material to weld or braze. Since it is highly sensitive to oxygen, welding or brazing requires an inert environment, e.g. argon, or vacuum. Induction welding was the only technique (Figure~\ref{fig:sec7_5_BrazingStand}) found that uses local heating, does not involve current spikes on mechanical structures, and has a reasonably sized tool that does not risk damage to the front-end electronics.

\begin{figure}[!htb]
	\centering
	\begin{subfigure}[t]{0.770\textwidth}
		\includegraphics[width=\textwidth]
{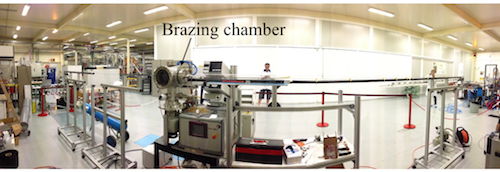}
		\caption{}
		\label{fig:brazingsetup}
	\end{subfigure}
	\begin{subfigure}[t]{0.224\textwidth}
		\includegraphics[width=\textwidth]
{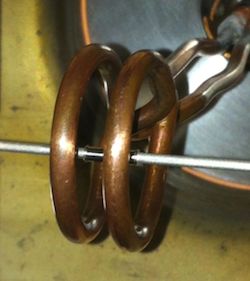}
		\caption{}
		\label{fig:inductorhead}
	\end{subfigure}
	\caption{\subref{fig:brazingsetup} Brazing setup for the extension of the cooling pipe
	     from 1.5\,m to 7\,m. \subref{fig:inductorhead} The inductor head located in the vacuum chamber.}
	\label{fig:sec7_5_BrazingStand}
\end{figure}

The brazing compound or filler used was Palcusil-5 (Ag \SI{68}{\percent}, Cu \SI{27}{\percent}, Pb \SI{5}{\percent}). The brazing process was performed at
\SI{820}{\celsius} to \SI{825}{\celsius} 
in vacuum
 (\SI{< 8.1e-6}{\bar}
) for several seconds, with a well defined ramp-up (Figure~\ref{fig:sec7_5_Induction_Heating}). Prior to the chamber and the pipe vacuum pumping, argon is flushed to minimise the presence of oxygen around the brazing point.

\begin{figure}[htb]
	\centering
	 \includegraphics[width=0.80\textwidth]{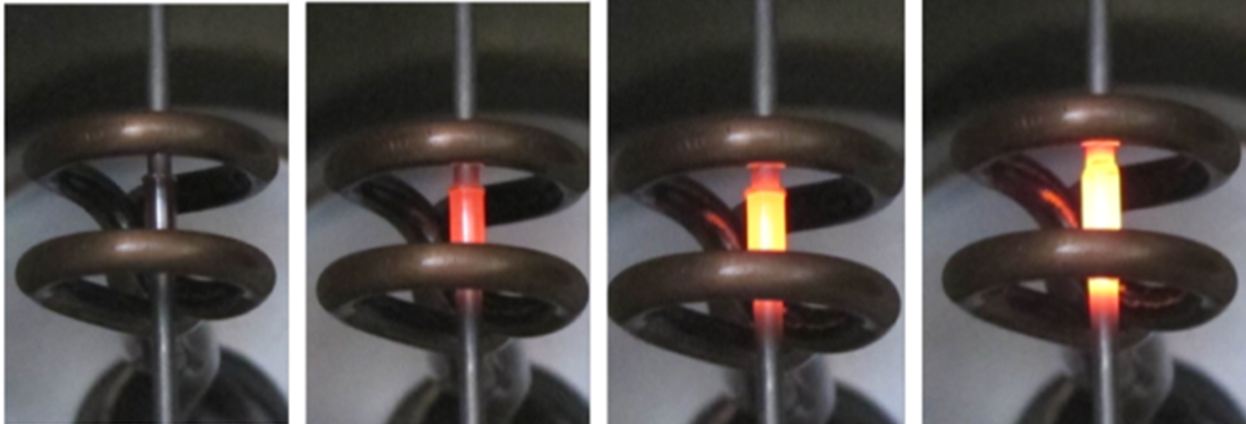}
	\caption{Brazing sequence: local heating of the titanium by an induction head located inside a vacuum chamber, at a temperature of \SI{825}{\celsius} for approximately \SI{3} minutes.}
	\label{fig:sec7_5_Induction_Heating}
\end{figure}

Qualification tests were performed to fine-tune the parameters and to check the quality and reliability of the braze. 
All test samples underwent visual inspection, leak tests, thermal shocks, metallographic inspection, tensile tests and thermal cycling.
The 14 IBL staves were successfully brazed on both sides and no damage was detected on any of the modules loaded on staves. 
The most delicate part of the process was to design and manufacture the feedthrough and sealing parts on the vacuum brazing chamber, given such thin pipes and the high level of vacuum required to complete the junction.

%-------------------------------------------------------------------------------
%\section{Conclusion}
%\label{sec:conclusion}
%-------------------------------------------------------------------------------

%Place your conclusion here.

\section{Final remarks and conclusion}
\label{section:conclusion}

The construction of the IBL detector started in mid-2012 and the completed detector was installed in May 2014. Because of the demanding detector constraints, and the hostile radiation and operational environment, R\&D programs relevant to the IBL started in 2008. IBL commissioning in the ATLAS cavern started in June 2014 and the IBL was fully commissioned from November 2014.

\subsection{The IBL challenges}
\label{sec:ibl_challenges}

Key challenges of the IBL project included:
\begin{itemize}
	\item[-] The R\&D, industrialisation and integration of two sensor technologies (planar and 3D) on one single-stave layout, capable of surviving integrated radiation doses of up to \SI{5e15}{\nq}, and with inactive edges of order \SI{200}{\micro\meter}. The successful development of planar sensors with small inactive edges capable of reliable and efficient operation at full depletion after large radiation doses was an essential requirement for IBL operation. The new 3D sensor technology is being used for the first time in a tracking detector. The radiation tolerance of 3D sensors has been demonstrated, and their reliable and efficient operation in the IBL is a major milestone because of their reduced operating voltage (and power consumption) after high radiation doses;
	\item[-] The bump-bonding of thinned FE-I4B chips to the planar and 3D sensors. This requirement was met by gluing a sapphire glass substrate to the FE-I4B wafer for all the processing steps and  later detaching it using a laser de-bonding technique after the bump-bonding and before further module assembly;
	\item[-] The development of the FE-I4B front-end read-out chip, the largest front-end design in  radiation hard \SI{130}{\nano\meter} CMOS technology for a tracker in particle physics. Its large area maximises the active area and reduces the bump-bonding cost;
	\item[-] The global IBL envelope of less than \SI{10}{\milli\meter} between the Pixel detector and the beam pipe was a challenge for integration, installation and the beam-pipe bake-out. As a result, innovative and custom-made mechanical supports, services and fittings were developed.
	\item[-] The aggressive design minimisation of the stave material budget, and the space constraint above, required that the service bus was tightly integrated with the stave (with implications for thermal expansion mismatches that are noted below). An on-stave flex having two aluminium layers for power and four copper layers for service lines was developed;
	\item[-] The development of Type 1 electrical services for the power distribution, and long (\SI{\sim4}{\meter}) data transmission wires, because of the tight space requirements and the hostile radiation environment. The high-density pin connectors were designed and fabricated specifically to match the limited space for integration;
	\item[-] The optimisation of the thermo-mechanical interface between the module and the local stave support to guarantee the interface reliability and reproducibility, radiation hardness, and the replaceability of a module without damaging neighbouring modules, the stave itself or the service flex integrity;
	\item[-] The design, development and qualification of  cooling pipe connections and electrical insulation for proper grounding and shielding;
	\item[-] The development of reliable CO$_2$ cooling in the detector region while satisfying the space and thermal insulation requirements. This required the R\&D, design and fabrication of flexible vacuum transfer lines; these could not be transferred to industry for logistic reasons and were produced in-house;
	\item[-] Extreme sealing and insulation capabilities, given the low dew-point temperature inside the detector volume, with CO$_2$ cooling close to the CO$_2$ freezing point (\SI{-55}{\celsius});
	\item[-] Installation of the cooling plant and the long vacuum transfer line (\SI{\sim100}{\meter}) to the junction box and a dummy load installed close to the IBL and inside the Muon detector. Two cooling plants, 
each with a \SI{3}{\kilo\watt} maximum load, 
are installed and independently controlled such that they can run either in complementary operational mode, one in standby while the other is operational, or in parallel which was mainly used during the bake-out of the beam pipe.
	\end{itemize}

The successful development of both the planar sensor technology and the new 3D sensor technology, capable of reliable and efficient operation in the IBL after high radiation doses, is a major milestone in the demonstration of their suitability for tracking detectors at the HL-LHC. The lower operating voltage of 3D sensors after high radiation doses is a significant potential advantage for HL-LHC operation. 

During the IBL production and integration, two major issues affected the schedule. The first issue concerned the bump-bonding yield for the initial production batches. An excess of merged and open bumps (Section~\ref{sec:prod_yield}) was identified to result from the solder flux used for bump bonding. 
The problem was resolved after a change of the tacking material and of the flip-chip machine.
The second issue concerned wire-bond corrosion~\cite{StaveQANote}, which was identified on most staves mid-way during the production (Section~\ref{sec:corrosion}). 
This resulted from a combination of extreme susceptibility to corrosion and 
accidental exposure to humidity during the temperature cycling tests after stave loading, and all but two staves could be fully repaired. Two options were considered to protect against potential future
corrosion: the potting of the bond foot or the use of spray coating such as polyurethane, but neither were possible because of the schedule. % (see below). 
It was decided to leave the IBL detector with unprotected wire bonds and to guarantee at all stages of the integration, installation and operation phases a safe humidity level.

\subsection{IBL in ATLAS} 
\label{sec:ibl_in_atlas}

The commissioning of the IBL as part of the ATLAS experiment closely followed the on-surface  
QA procedure. 
Initially, the integrity of the LV sense lines for each module group was ensured. The modules were then powered with their nominal supply voltage. At each step, voltage and current readings were compared to the measurements on-surface QA. After verification of the expected DCS measurements, configuration commands were sent to the FE-I4B chips to establish communication with the modules. These powering and configuration tests were followed by digital, analog, threshold and ToT scans, and finally re-tuning of the threshold and the ToT. The RCE read-out system was used for this initial commissioning in the cavern, to ensure a consistent comparison with the on-surface QA. The results confirmed 100\% damage-free transportation and installation of the IBL before the sealing of the inner detector volume at the end of July 2014.

The commissioning of the ROD/BOC read-out system started in August 2014. Nine of the 14 IBL staves were integrated in the ATLAS experiment for the collection of cosmic ray data in October 2014. Subsequently, the new beryllium beam pipe was heated to \SI{230}{\celsius} to activate the NEG coating necessary to achieve the high vacuum levels required for LHC operation. The CO$_2$ cooling system ensured the safety of the IBL during this bake-out. From November 2014, the IBL was fully integrated as part of the ATLAS experiment.

Details of the ATLAS commissioning, data taking and performance are beyond the scope of this paper~\cite{ATLAS:IBLdata}. 
However, four detector issues related to the design and construction of the IBL are briefly noted below. 
None have affected the quality of data from the IBL, nor the physics performance. 
\begin{itemize}
	\item[-] Since the wire bonds were not encapsulated, and the IBL operates in a 2\,T magnetic field, current changes during the read-out may risk damage from bond oscillations~\cite{AlvarezFeito:2010249}. To avoid oscillation frequencies a Fixed Frequency Trigger Veto was implemented at the DAQ level in the range 3 to 40 KHz;	
	\item[-] As noted in Sections ~\ref{sec:corrosion} and~\ref{sec:ibl_challenges}, wire bond corrosion was identified on most staves during mid-production, because of accidental exposure to humidity at low temperature during temperature cycling. Because of schedule considerations, it was decided to ensure that the staves remain at low humidity. The performance of the staves has not deteriorated following this precaution;
	\item[-] An increase in the current consumption of the FE-I4B chip at low total ionising dose was identified to result from N-MOS transistor leakage currents after the build-up of charge at the SiO$_2$ interface in the 130\,nm CMOS process~\cite{ATL-INDET-PUB-2017-001}. The evolution of this current was evaluated at different temperatures  and annealing procedures were introduced by operating the detector at temperatures around \SI{10}{\celsius};
	\item[-] Distortions resulted from the R-$\phi$ twisting of staves at the level of a few~\SI{}{\micro\meter}/\SI{}{\celsius}, due to the mismatch of the thermal expansion coefficient between the stave and the stave flex, and the asymmetric attachment of the flex that was made necessary by mechanical constraints~\cite{ATL-PHYS-PUB-2015-031}. The impact of this is minimised by ensuring a temperature stability of less than \SI{0.2}{\celsius} 
and by the development of in-run alignment correction procedures.   
\end{itemize}

\subsection{Conclusion} 
The  fabrication and integration of the ATLAS IBL detector is described in this paper. A fully working detector with only 0.09\% of dead channels was successfully installed in ATLAS in May 2014 and fully commissioned as part of the ATLAS detector in November 2014.  The addition of this innermost pixel layer, very close to the interaction point and with a smaller pixel size than other Pixel layers, provides additional redundancy and  significantly improves the ATLAS tracking and vertexing performance.

%-------------------------------------------------------------------------------
\section*{Acknowledgements}
%-------------------------------------------------------------------------------

%\input{acknowledgements/Acknowledgements}
% Acknowledgements for papers with collision data
% Version 06-Mar-2017

% Standard acknowledgements start here
%----------------------------------------------

We acknowledge the support of 
NSERC, NRC and CFI, Canada; 
CERN; 
MSMT CR, MPO CR and VSC CR, Czech Republic; 
IN2P3-CNRS, CEA-DSM/IRFU, France; 
BMBF, HGF, and MPG, Germany; 
INFN, Italy; 
MEXT and JSPS, Japan; 
NWO, Netherlands; 
RCN, Norway; 
MES of Russia and NRC KI, Russian Federation; 
ARRS and MIZ\v{S}, Slovenia; 
MINECO, Spain; 
SRC and Wallenberg Foundation, Sweden; 
SERI, SNSF and Cantons of Bern and Geneva, Switzerland; 
MOST, Taiwan; 
STFC, United Kingdom; 
NSF MRI award PHY-1039175 and DOE, United States of America. 

%------------------------------------------------

In addition,  individual groups and members have received support from the
FP7 Marie Curie Initial Training Network program and the Talent program (http://talent.web.cern.ch/TALENT/index.shtml) 
of the European Union.

A close collaboration with the silicon sensor suppliers, during the R\&D and procurement phases, was essential to the success of the project and is gratefully acknowledged: CiS Forschungsinstitut f$\ddot{\mathrm u}$r Mikrosensorik und Photovoltaik GmbH, Erfurt, Germany; FBK Fondazione Bruno Kessler, Povo di Trento, Italy; CNM Centro Nacional de Microelectronica, Barcelona, Spain. The successful development of new techniques for the bump-bonding of thin wafers by Fraunhofer IZM, Berlin, Germany, was essential to the project and is gratefully acknowledged. The work of a dedicated task force to recommend a production and operation strategy for staves susceptible to wire bond corrosion is gratefully acknowledged. We thank the CERN DSF (Detector Silicon Facility) for their contribution to the rework of IBL staves affected by wire bond corrosion. 

We acknowledge CERN and the ATLAS Collaboration, and in particular the ATLAS TC (Technical Coordination) and CERN EP-DT groups, for their support during the construction and integration of the ATLAS IBL detector. We further acknowledge the support staff from participating IBL institutes, who were essential to the success of the project.

%
%The \texttt{atlaslatex} package contains the acknowledgements that were valid 
%at the time of the release you are using.
%These can be found in the \texttt{acknowledgements} subdirectory.
%When your ATLAS paper or PUB/CONF note is ready to be published,
%download the latest set of acknowledgements from:\\
%\url{https://twiki.cern.ch/twiki/bin/view/AtlasProtected/PubComAcknowledgements}

%-------------------------------------------------------------------------------
\clearpage
%\appendix
%\part*{Appendix}
%\addcontentsline{toc}{part}{Appendix}
%-------------------------------------------------------------------------------

%In a paper, an appendix is used for technical details that would otherwise disturb the flow of the paper.
%Such an appendix should be printed before the Bibliography.

%-------------------------------------------------------------------------------
% If you use biblatex and either biber or bibtex to process the bibliography
% just say \printbibliography here
\printbibliography
% If you want to use the traditional BibTeX you need to use the syntax below.
%\bibliographystyle{bib/bst/atlasBibStyleWoTitle}
%\bibliography{IBLPaper,bib/ATLAS,bib/CMS,bib/ConfNotes,bib/PubNotes}
%-------------------------------------------------------------------------------

%-------------------------------------------------------------------------------
%\clearpage
%\part*{Auxiliary material}
%\addcontentsline{toc}{part}{Auxiliary material}
%-------------------------------------------------------------------------------

%In an ATLAS paper, auxiliary plots and tables that are supposed to be made public 
%should be collected in an appendix that has the title \enquote{Auxiliary material}.
%This appendix should be printed after the Bibliography.
%At the end of the paper approval procedure,
%this information should be split into a separate document -- see \texttt{atlas-auxmat.tex}.

\end{document}